\documentclass[fleqn,10pt]{article}
\usepackage{latexsym, graphicx, epsfig, amsmath,  amssymb,amsfonts}
\usepackage{natbib,amsthm,version}
\usepackage{amsbsy,bm,multirow,enumerate}
\usepackage[titletoc,page]{appendix}
\usepackage[mathscr]{eucal}
\usepackage{mathtools}
\usepackage{color}
\usepackage[utf8]{inputenc}
\usepackage[english]{babel}

\newtheorem{theorem}{Theorem}[section]

\newtheorem{remark}{Remark}

\title{
Wall-Bounded Multiphase Flows of $N$ Immiscible Incompressible Fluids:
Consistency and Contact-Angle Boundary Condition
} 
\author{
  S. Dong\thanks{ Email: sdong@purdue.edu} \\
  Center for Computational and Applied Mathematics \\
  Department of Mathematics \\
  Purdue University 
 } 

\date{} %(\today)}
\begin{document}
\maketitle

%% double space
%\baselineskip 2em %2.2em

%%%%%%%%%%%%%%%%%%%%%%%%%%%%%%%%%%%%%%%%%%%%%%%%%%
%% Abstract

\begin{abstract}

% what are we doing? 
% we are developing a numerical method for wall-bounded multiphase flows
% to simulate contact angle effects etc.
% we are considering the formulations, consistency, contact-angle BCs because of this

We present an effective method for simulating wall-bounded multiphase
flows consisting of $N$ ($N\geqslant 2$) immiscible incompressible fluids
with different densities, viscosities and pairwise surface tensions.
The N-phase physical formulation is based on a modified thermodynamically
consistent phase field model that is more general than in a previous work, 
and it is developed by considering the reduction
consistency if some of the fluid components were
absent from the system. We propose an N-phase contact-angle boundary condition
that is reduction consistent between $N$ phases and $M$ phases ($2\leqslant M\leqslant N-1$).
We also present a numerical algorithm for solving the N-phase governing equations
together with the contact-angle boundary conditions developed herein.
Extensive numerical experiments are presented for
several flow problems involving multiple fluid components and solid-wall boundaries 
to investigate the wettability effects with multiple types of contact angles.
In particular, we compare simulation results with the de Gennes theory
for the contact-angle effects on the  liquid drop spreading on
wall surfaces, and demonstrate that our method produces physically accurate results.

\end{abstract}

%%%%%%%%%%%%%%%%%%%%%%%%%%%%%%%%%%%%%%%%%%%%%%%%%%

\vspace{0.05cm}

\section{Introduction}
\label{sec:intro}

% (1) what is the rationale of paper?
% We have developed an N-phase model that couples hydrodynamics and interfaces
% in  thermodynamically consistent fashion. However, the choice of the free
% energy functional is not good, and that model is not reduction consistent.
% Boyer has discussed reduction consistency extensively. Inspired by Boyer's work,
% we will develop a new model based on that of Dong(2014,2015) 
% so that the new model is  more reduction consistent.
%
% Will further extend Boyer's work, require stronger reduction consistency --
% dynamic equations must reduce from N phases to M phases for any 2<=M<N.
%
% Will further propose a N-phase contact angle condition that is fully reduction consistent.
%
% (2) what is the difference between the consistency conditions used here
%    and in Boyer's paper?
%    the consistency conditions in current paper is stronger!
%
% (3) what is the status of the field?
% (4) what are the outstanding issues?
% (5) what is this paper about? why this study?
% (6) what are the novelties?
% 
% What is the logic of the paper?
% (i) what is the general problem is this paper considering?
% (ii) what is the status of the field?
% (iii) what have we done before?
% (iv) what are the outstanding issues?
% (v) what is this paper about?
% (vi) what are novelties and innovations?
%
% The key issue is reduction consistency, in model, in BCs!
% Give Boyer, X.P. Wang and other researchers proper credit!
%

In the present work we focus on the motion of a mixture of 
$N$ ($N\geqslant 2$) immiscible incompressible fluids 
with different physical properties (such as densities, dynamic viscosities,
and pair-wise surface tensions) within a domain bounded by solid walls.
Moving contact lines form on the solid wall where fluid interfaces
intersect the wall surface, and the wall wettability,
characterized by the contact angles, can significantly influence
the dynamics and the equilibrium state of the system. 
Following the notation of our previous works~\cite{Dong2014,Dong2015},
we refer to such problems as N-phase flows, where N refers to
the number of different fluid components in the system.
Due to the  multitude of different types of fluid interfaces
in the system, wall-bounded N-phase flows can potentially
accommodate a large number of different types of contact lines
and different  contact angles on the wall.
Potential applications of wall-bounded multiphase flows 
are enormous, in e.g.~materials processing, microfluidic devices,
and functional surfaces.

% what is the status of N-phase flows?

We primarily consider N-phase systems involving three
or more fluid components (i.e.~$N\geqslant 3$) in the current work,
and our approach falls into the phase field (or diffuse interface) framework. 
Our attention below and in subsequent sections of this paper will 
therefore be confined to this approach.
For two-phase flows one can refer to
e.g.~\cite{AndersonMW1998,SethianS2003,Tryggvasonetal2001,ScardovelliZ1999}
for a review of this and related approaches.
The developments for three or more fluid phases have been
contributed by a number of 
researchers; 
see e.g.~\cite{KimL2005,BoyerL2006,Kim2009,BoyerLMPQ2010,BoyerM2011,Kim2012,HeidaMR2012,ZhangW2016}.
Among these, two phase field models are developed in \cite{KimL2005,HeidaMR2012}
based on the mass-averaged velocity, which is not divergence free. 
The constitutive relations therein are formulated based on
thermodynamic considerations such as the maximization of entropy
production. The study of a three-phase model in \cite{BoyerL2006} 
signifies the importance in the choice of the free energy form. More interestingly,
the authors thereof have proposed several natural reduction consistency
conditions that three-phase models should satisfy. 

% works of Dong(2014,2015), what have we done so far?

More recently, we have proposed in \cite{Dong2014} a general
phase field model for formulating the motion
of an isothermal mixture of $N$ ($N\geqslant 2$) immiscible incompressible fluids (see also
Section \ref{sec:nphase_model} below).
The model is derived by considering the mass conservations of 
the N individual fluid components, the momentum conservation,
the second law of thermodynamics, and the Galilean invariance principle.
Our model is based on a volume-averaged mixture velocity, which can
be rigorously shown to be divergence free~\cite{Dong2014}.
Therefore, it is fundamentally different from those of \cite{KimL2005,HeidaMR2012}.
This N-phase model involves a free energy density function
and ($N-1$) independent order parameters (or interchangeably phase field variables). 
It is a general model in the sense that the free energy density function
and the set of ($N-1$) independent order parameters remain to be specified. 
Once the form of 
the free energy density function and a set of order parameters are specified,
the model will give rise to a specific physical formulation for the N-phase system.
In \cite{Dong2014,Dong2015}, by employing a specific form for the free energy density function
and choosing a set of order parameters,
we have derived from the general model specific physical formulations
for incompressible N-phase flows. We have further devised associated numerical algorithms
for their simulations.

% the work of Boyer et al (2014), consistency conditions for N-phase models

In an interesting work~\cite{BoyerM2014} Boyer and collaborator have recently generalized
the reduction consistency conditions of~\cite{BoyerL2006} from
three phases to more general N phases, and provided an in-depth discussion of the effects of 
these conditions on the modeling of N-phase systems, in particular,
on the choice of N-phase free energy density function.
More specifically, they have looked into the following three
consistency properties:
\begin{enumerate}[($\mathscr{C}$1):]

\item
The N-phase free energy density function should coincide with the two-phase free energy density
function if $N=2$;

\item
If only a set of $M$ ($2\leqslant M\leqslant N-1$) fluids are present in the system,
then the N-phase free energy density function should reduce to the corresponding
$M$-phase free energy density function;

\item
If $K$ ($1\leqslant K\leqslant N-2$) fluid phases are absent from the initial data, then they
should remain absent in the N-phase solution over time.

\end{enumerate}
%
% what are the implications?
These consistency conditions, and in particular property ($\mathscr{C}$3),
 stringently restrict the form of the free energy density function,
and especially the potential free energy form (i.e.~``multiwell'' potential).
It is shown in \cite{BoyerM2014} that, given an arbitrary set of 
pair-wise surface tension values, it is extremely challenging
to construct an N-phase potential free energy that fully
satisfies ($\mathscr{C}$3) with arbitrary $1\leqslant K\leqslant N-2$.
The existence of such an N-phase potential energy for a general set of
pair-wise surface tensions is still
an open problem. Several possible potential energy forms with interesting
 properties, as well as other related function forms are explored 
 in \cite{BoyerM2014}.

% what is this paper about?

Inspired by the consistency conditions of \cite{BoyerM2014},
we have looked into the N-phase free energy density function
employed in our previous works~\cite{Dong2014,Dong2015}
and the resultant N-phase physical formulations. 
The specific form of the N-phase potential free energy function
employed therein  satisfies the consistency properties
($\mathscr{C}$1) and ($\mathscr{C}$2). However,
when combined with the N-phase governing equations,
it appears to fall short with respect to  ($\mathscr{C}$3).
This inadequacy motivates the work in the current paper.

% what have you done in this paper?
% (i) develop a physical formulation with improved reduction consistency
%     which requires a more general phase field model
% (ii) propose contact-angle BCs that are reduction consistent.
%
% combine consistency idea with the modified model to develop N-phase formulations
% with improved reduction consistency
% (i) modify the phase field model of Dong (2014,2015)
% (ii) combine the consistency idea and the modified model to develop an N-phase
%      formulation to improve reduction consistency
% (iii) present a reduction consistent N-phase contact-angle BC
% (iv) numerical algorithm for solving wall-bounded N-phase flows 

In this paper, we present developments in the following aspects:
\begin{itemize}

\item
We specify, when developing the N-phase model, a different constitutive
relation than that of \cite{Dong2014} 
to ensure the second law of thermodynamics. This leads to a modified phase field model
that is more general than that of \cite{Dong2014}.

\item
We combine the modified N-phase  model and the consistency considerations as discussed in
\cite{BoyerM2014} to develop specific N-phase physical formulations
to improve the reduction consistency.

\item
We propose an N-phase contact angle boundary condition that is reduction consistent
between $N$ phases and $M$ phases ($2\leqslant M\leqslant N-1$).

\item 
We develop a numerical algorithm for solving the 
N-phase governing equations together with the N-phase contact-angle
boundary condition developed herein.

\end{itemize}

% more specific discussions: modified consistency conditions

More specifically, besides the afore-mentioned consistency properties 
($\mathscr{C}$1), ($\mathscr{C}$2) and ($\mathscr{C}$3), we also
consider two additional consistency conditions in the current work:
\begin{enumerate}[($\mathscr{C}$1):]
\addtocounter{enumi}{3}

\item
If only a set of $M$ ($2\leqslant M\leqslant N-1$) fluids are present in the system,
then the $N$-phase governing equations  should reduce to those for
the corresponding $M$-phase system.

\item
If only a set of $M$ ($2\leqslant M\leqslant N-1$) fluids are present in the system,
then the boundary conditions for the N-phase system should reduce to those for
the corresponding $M$-phase system.

\end{enumerate}
Note that the above condition ($\mathscr{C}$4) imposes a stronger consistency requirement
on the N-phase formulation than  ($\mathscr{C}$3) in two aspects:
(i) The condition ($\mathscr{C}$4) includes consistency requirements on the momentum
equations, in addition to the phase field equations.
(ii) The condition ($\mathscr{C}$4) requires not only that the fluid phases initially
absent should remain absent over time, but also that the governing equations for
the fluid phases that are present in the system should reduce to the corresponding ones
for the smaller system excluding the fluid phases that are absent.
One should also note that, in order to satisfy the consistency 
properties ($\mathscr{C}$2)--($\mathscr{C}$5) between the N-phase system
and an $M$-phase system ($2\leqslant M\leqslant N-1$),
it suffices to consider only the reduction from the N-phase system
to the ($N-1$)-phase system, that is, if any one fluid phase is absent from
the N-phase system.

% discuss N-phase contact angles, 
% note that only static contact angles will be considered

Employing the N-phase formulation resulting from the above discussions, we look into
the N-phase contact angle boundary conditions on solid-wall boundaries.
In principle, the interface formed between any pair of these $N$ fluids
can intersect the wall and form a contact angle thereupon.
So potentially $\frac{1}{2}N(N-1)$ different contact angles can
exist on the wall in the N-phase system. However, among them only
($N-1$) contact angles are independent due to the Young's
relations~\cite{Blunt2001}.
Once the ($N-1$) independent contact angles are provided, all the other
contact angles at the wall can be determined based on the Young's
equations. As mentioned earlier, ($N-1$) independent phase field variables 
are involved in our N-phase model, which is consistent with 
the existence of ($N-1$) independent contact angles in the N-phase system. 
We develop a reduction consistent contact-angle boundary
condition by imposing the requirement that the  boundary condition should satisfy 
the  consistency property ($\mathscr{C}$5).
In the current paper our attention will be restricted to equilibrium (or static) 
contact angles only. We note that it is not difficult
to extend these conditions to account for the dynamic effect, by for example incorporating
appropriate inertial terms.
Contact angles in flows with two or three fluid phases are
considered in a number of previous studies; 
see e.g.~\cite{Jacqmin2000,YueZF2010,YueF2011,CarlsonDA2011,Dong2012,ShiW2014,ZhangW2016}.

% what are the novelties? what is new?

The novelties of this paper lie in three aspects: (i) the modified (more
general) phase field model presented herein;
(ii) the N-phase physical formulation resulting from the
modified phase field model and the consistency considerations; 
(iii) the reduction consistent N-phase contact-angle boundary conditions.
While the numerical algorithm for the phase field equations presented herein can also  
be considered new,  the strategies for dealing with
the several numerical issues therein are straightforward adaptations
from those developed in \cite{DongS2012,Dong2014}.

% organization of paper

The rest of this paper is structured as follows. 
In the rest of this section we summarize the key points of
the phase field model developed in \cite{Dong2014}.
Then in Section \ref{sec:method} we present a modified
phase field model, and develop an N-phase physical formulation
by considering the consistency conditions
($\mathscr{C}$1)--($\mathscr{C}$5). We present in Section \ref{sec:cabc}   
 a consistent N-phase contact-angle boundary condition, 
and in Section \ref{sec:algorithm} a numerical algorithm for solving
the N-phase governing equations together with
the contact-angle boundary condition.
After that we demonstrate the performance of the method
developed herein using several flow problems involving
multiple fluid components and solid walls in Section \ref{sec:tests}.
In particular we compare simulations with 
the de Gennes theory~\cite{deGennesBQ2003} to show the
accuracy of the simulation results. We also look into
the effects of the various contact angles on the equilibrium configuration
and the dynamics of the system when multiple fluid components
are involved. Section \ref{sec:summary}
concludes the discussions with a summary of the key points.
In the appendices we provide proofs to several theorems from
the main body of the text, and also
 provide a summary of the numerical algorithm
for solving the momentum equations.

% what else to discuss here?

\subsection{A Phase Field Model for N-Fluid Mixture }
\label{sec:nphase_model}

In \cite{Dong2014} we have derived  a general phase field
model for the isothermal system
consisting of $N$ ($N\geqslant 2$) immiscible incompressible fluids
based on thermodynamic
principles, namely, 
the conservations of mass and momentum,
the Galilean invariance principle,
and the second law of thermodynamics;
see the Appendix of \cite{Dong2014} for the detailed
derivation of this model.
We summarize below several key points in
the development of this model, which are
crucial to  subsequent developments in the current paper.

Consider the mixture of $N$ ($N\geqslant 2$) immiscible incompressible
fluids contained in some flow domain $\Omega$ (domain boundary denoted by $\partial\Omega$). 
Let $\tilde{\rho}_i$ and $\tilde{\mu}_i$
($1\leqslant i\leqslant N$) denote 
the constant densities and constant dynamic 
viscosities of these $N$ pure fluids (before mixing).
Define auxiliary parameters
\begin{equation}
\tilde{\gamma}_i = \frac{1}{\tilde{\rho}_i}, \ 1\leqslant i\leqslant N;
\quad
\Gamma = \sum_{i=1}^N \tilde{\gamma}_i;
\quad
\Gamma_{\mu} = \sum_{i=1}^N \frac{\tilde{\mu}_i}{\tilde{\rho}_i}.
\end{equation}
Let $\phi_i$ ($1\leqslant i\leqslant N-1$) denote
the ($N-1$) independent order parameters,
or interchangeably the phase field variables,
that characterize the system,
and $\vec{\phi}=(\phi_1,\dots,\phi_{N-1})$.
Let $\rho_i(\vec{\phi})$ and $c_i(\vec{\phi})$
($1\leqslant i\leqslant N$)
denote the density and volume fraction of
fluid $i$ {\em within the mixture}, and
let $\rho(\vec{\phi})$ denote the density of
the N-phase mixture. Then we have the relations \cite{Dong2014}
\begin{equation}
c_i = \frac{\rho_i}{\tilde{\rho}_i}, \ 1\leqslant i\leqslant N; \quad
\sum_{i=1}^N c_i = 1; \quad
\rho = \sum_{i=1}^N \rho_i.
\label{equ:volfrac_expr}
\end{equation}

The mass balance of the $N$ individual fluid
phases is described by the $(N-1)$ independent
mass balance equations \cite{Dong2014}
\begin{equation}
\frac{\partial}{\partial t}\left(\rho_i - \rho_N  \right)
+ \mathbf{u}\cdot\nabla \left(\rho_i - \rho_N  \right)
= -\nabla\cdot\mathbf{J}_{ai},
\quad 1\leqslant i\leqslant N-1,
\label{equ:mass_balance}
\end{equation}
where $\rho_i$ ($1\leqslant i\leqslant N$) is
the density of fluid $i$ {\em within the  mixture},
and $\mathbf{u}$ is the {\em volume averaged mixture velocity}
and can be rigorously shown to be
divergence free \cite{Dong2014}
\begin{equation}
\nabla\cdot\mathbf{u} = 0.
\label{equ:continuity}
\end{equation}
$\mathbf{J}_{ai}$ ($1\leqslant i\leqslant N-1$)
are ($N-1$) diffusive fluxes whose forms are
to be determined based on the second law of 
thermodynamics. 

The momentum balance of the
system is described by \cite{Dong2014}
\begin{equation}
\rho\left(\frac{\partial\mathbf{u}}{\partial t} 
+\mathbf{u}\cdot\nabla\mathbf{u} \right)
+ \tilde{\mathbf{J}}\cdot\nabla\mathbf{u}
= -\nabla p + \nabla\cdot\mathbf{S},
\label{equ:momentum_balance}
\end{equation}
where $p$ is the pressure, and $\mathbf{S}$
is a stress tensor whose form is
to be determined based on the second law of thermodynamics.
The density of fluid $i$ within the mixture $\rho_i$, 
the volume fraction $c_i$,
and the mixture density $\rho$, 
are given by 
\begin{equation}
\left\{
\begin{split}
&
\rho_i(\vec{\phi}) = 
\frac{1}{\Gamma} + \sum_{j=1}^{N-1}\left(
  \delta_{ij} - \frac{\tilde{\gamma}_j}{\Gamma}
\right)\varphi_j(\vec{\phi}), 
\quad 1\leqslant i\leqslant N, \\
%\left\{
%\begin{array}{ll}
%\frac{1}{\Gamma}[1-\sum_{j=1}^{N-1} \tilde{\gamma}_j\varphi_j(\vec{\phi})], &
%i=N, \\
%\rho_N(\vec{\phi}) + \varphi_i(\vec{\phi}), & 1\leqslant i\leqslant N-1;
%\end{array}
%\right.
&
c_i(\vec{\phi}) = \tilde{\gamma}_i\rho_i(\vec{\phi}) = 
\frac{\tilde{\gamma}_i}{\Gamma} + \sum_{j=1}^{N-1}\left(
  \tilde{\gamma}_i\delta_{ij} - \frac{\tilde{\gamma}_i\tilde{\gamma}_j}{\Gamma}
\right)\varphi_j(\vec{\phi}), 
\quad 1\leqslant i\leqslant N, \\
&
\rho(\vec{\phi})=\sum_{i=1}^N \rho_i = 
\frac{N}{\Gamma} + \sum_{i=1}^{N-1}\left(1 - \frac{N}{\Gamma}\tilde{\gamma}_i \right)\varphi_i(\vec{\phi}),
\end{split}
\right.
\label{equ:density_expr}
\end{equation}
where $\delta_{ij}$ is the Kronecker delta, and
\begin{equation}
\varphi_i(\vec{\phi}) \equiv \rho_i - \rho_N, \quad 1\leqslant i\leqslant N-1.
\label{equ:varphi_expr}
\end{equation}
The flux $\tilde{\mathbf{J}}$ is given by
\begin{equation}
\tilde{\mathbf{J}} = \sum_{i=1}^{N-1}\left(1-\frac{N}{\Gamma}\tilde{\gamma}_i \right)
\mathbf{J}_{ai}.
\end{equation}

We introduce a free energy density 
function $W(\vec{\phi},\nabla\vec{\phi})$
in the spirit of the phase field approach
to account for the interfacial energy for
the diffuse interfaces.
%
% 2nd law
%
The appropriate forms for
$\mathbf{J}_{ai}$ in \eqref{equ:mass_balance}
and $\mathbf{S}$ in \eqref{equ:momentum_balance}
are determined by invoking
the second law of thermodynamics, which for isothermal systems
requires that the following inequality should hold,
\begin{equation}
\frac{d}{d t}\int_{\Omega(t)} e(\mathbf{u},\vec{\phi},\nabla\vec{\phi})
\leqslant P_c
\end{equation}
where $\Omega(t)$ is an arbitrary volume that is
transported
with the mixture velocity $\mathbf{u}$,
the total energy density function
$e(\mathbf{u},\vec{\phi},\nabla\vec{\phi})$ 
is given by
$
e(\mathbf{u},\vec{\phi},\nabla\vec{\phi})
=\frac{1}{2}\rho(\vec{\phi})|\mathbf{u}|^2
+ W(\vec{\phi},\nabla\vec{\phi}),
$
and $P_c$ is the conventional power
expended on $\Omega(t)$.
This inequality is eventually reduced to (see \cite{Dong2014} for details)
\begin{equation}
%-\mathcal{D} = 
-\left(
  \mathbf{S} + \sum_{i=1}^{N-1} \nabla\phi_i\otimes
  \frac{\partial W}{\partial\nabla\phi_i}
\right): \frac{1}{2}\mathbf{D}(\mathbf{u})
+ \sum_{i=1}^{N-1} \nabla \mathcal{C}_i\cdot\mathbf{J}_{ai}
\leqslant 0
\label{equ:2nd_law}
\end{equation}
where 
$\mathbf{D}(\mathbf{u})=\nabla\mathbf{u}+\nabla\mathbf{u}^T$,
and $\mathcal{C}_i$ ($1\leqslant i\leqslant N-1$)
are ($N-1$) effective chemical potentials given by
the linear system
\begin{equation}
\sum_{j=1}^{N-1}\frac{\partial\varphi_j}{\partial\phi_i} \mathcal{C}_j
= \frac{\partial W}{\partial\phi_i}
-\nabla\cdot\frac{\partial W}{\partial(\nabla\phi_i)}.
\quad 1\leqslant i\leqslant N-1.
\label{equ:chempot}
\end{equation}

In \cite{Dong2014}
the following constitutive relations are chosen to ensure the inequality \eqref{equ:2nd_law}
\begin{subequations}
\begin{align}
&
\mathbf{J}_{ai} = -\tilde{m}_i(\vec{\phi})\nabla \mathcal{C}_i, \quad 1\leqslant i\leqslant N-1
\label{equ:const_relation_flux} \\
&
\mathbf{S} + \sum_{i=1}^{N-1} \nabla\phi_i\otimes
  \frac{\partial W}{\partial\nabla\phi_i}
= \mu(\vec{\phi})\mathbf{D}(\mathbf{u}),
\end{align}
\end{subequations}
where $\tilde{m}_i(\vec{\phi})\geqslant 0$ and
$\mu(\vec{\phi})\geqslant 0$ are non-negative
quantities playing the roles of mobility and viscosity,
respectively.
With these constitutive relations, 
the momentum equation \eqref{equ:momentum_balance} and the mass balance
equations \eqref{equ:mass_balance} are transformed into
\begin{equation}
\rho\left(
  \frac{\partial\mathbf{u}}{\partial t}
  + \mathbf{u}\cdot\nabla\mathbf{u}
\right)
+ \tilde{\mathbf{J}}\cdot\nabla\mathbf{u}
=
-\nabla p
+ \nabla\cdot\left[
  \mu(\vec{\phi}) \mathbf{D}(\mathbf{u})
\right]
- \sum_{i=1}^{N-1} \nabla\cdot\left(
  \nabla\phi_i \otimes \frac{\partial W}{\partial\nabla\phi_i}
\right),
\label{equ:nse_original}
\end{equation}
\begin{equation}
\sum_{j=1}^{N-1}\frac{\partial\varphi_i}{\partial\phi_j}\left(
  \frac{\partial\phi_j}{\partial t} + \mathbf{u}\cdot\nabla\phi_j
\right)
=
\nabla\cdot\left[
  \tilde{m}_i(\vec{\phi}) \nabla \mathcal{C}_i
\right],
\qquad 1 \leqslant i \leqslant N-1.
\label{equ:CH_original}
\end{equation}

The general phase field model of \cite{Dong2014} for the N-phase
system consists of equations \eqref{equ:nse_original},
\eqref{equ:continuity} and \eqref{equ:CH_original},
in which $\mathcal{C}_i$ ($1\leqslant i\leqslant N-1$) are given by the linear
system \eqref{equ:chempot}.
The order parameters $\vec{\phi}$ are  defined 
through equation \eqref{equ:varphi_expr}
(see \cite{Dong2015} for a family of order parameters),
and the form of the free energy density function
$W(\vec{\phi},\nabla\vec{\phi})$ is yet to be specified.
Once the functions $\varphi_i(\vec{\phi})$ ($1\leqslant i\leqslant N-1$)
and the free energy density function $W(\vec{\phi},\nabla\vec{\phi})$
are specified, essentially all the other quantities
in the model can be computed.

%\section{N-Phase Flow Formulation and Contact-Angle Boundary Condition}

% what are the contributions?
% (1) N-phase formulation with mobility matrix
%     (i) difference in general N-phase model: constitutive relation
%     (ii) how to determine mobility matrix m_ij: consistency
%     (iii) N-phase potential energy: proof of consistency with 2 phases
% (2) N-phase contact-angle BC
%     reduction/consistency with 2-phase contact-angle BC
%

\section{Modified N-Phase Model and Physical Formulation}
\label{sec:method}
\label{sec:formulation}

\subsection{A Modified General Phase Field Model for N-phase Mixture}

%% New general N-phase model

The point of departure of the current work is 
in the choice of the constitutive relation \eqref{equ:const_relation_flux}
for the fluxes $\mathbf{J}_{ai}$ involved in the mass balance equations.
In the present paper we consider a modified constitutive relation,
\begin{equation}
\mathbf{J}_{ai} = -\sum_{j=1}^{N-1}\tilde{m}_{ij}(\vec{\phi}) \nabla \mathcal{C}_j,
\quad 1\leqslant i\leqslant N-1,
\label{equ:const_relation_flux_curr}
\end{equation}
where $\tilde{m}_{ij}(\vec{\phi})$ ($1\leqslant i,j\leqslant N-1$)
are coefficients. We require that the matrix formed by these
coefficients
\begin{equation}
\tilde{\mathbf{m}} = \begin{bmatrix} \tilde{m}_{ij} \end{bmatrix}_{(N-1)\times (N-1)}
\end{equation}
be symmetric positive definite (SPD)
to ensure the non-positivity of the second term on
the left hand side of the inequality \eqref{equ:2nd_law}.
Note that this constitutive relation is more general than \eqref{equ:const_relation_flux}.
By requiring that $\tilde{\mathbf{m}}$ be a diagonal matrix,
one can reduce \eqref{equ:const_relation_flux_curr} to \eqref{equ:const_relation_flux}.
With the constitutive relation \eqref{equ:const_relation_flux_curr}
the mass balance equations \eqref{equ:mass_balance}
 are  transformed into
\begin{equation}
\sum_{j=1}^{N-1}\frac{\partial\varphi_i}{\partial\phi_j}\left(
  \frac{\partial\phi_j}{\partial t} + \mathbf{u}\cdot\nabla\phi_j
\right)
=
\sum_{j=1}^{N-1} \nabla\cdot\left[
  \tilde{m}_{ij}(\vec{\phi}) \nabla \mathcal{C}_j
\right],
\qquad 1 \leqslant i \leqslant N-1.
\label{equ:CH_general}
\end{equation}

The equations \eqref{equ:nse_original},
\eqref{equ:continuity}, and \eqref{equ:CH_general}
constitute a modified phase field model
for the N-phase system that is  more general than the one from \cite{Dong2014}.
This model similarly satisfies the thermodynamic principles:
the conservations of mass
and momentum, the second law of thermodynamics, and
Galilean invariance.
This new N-phase model is the basis for the 
developments in the current work.

In this new model, $\tilde{\mathbf{J}}$ in
the momentum equation \eqref{equ:nse_original}
is given by
\begin{equation}
\tilde{\mathbf{J}} = \sum_{i=1}^{N-1}\left(1-\frac{N}{\Gamma}\tilde{\gamma}_i \right)\mathbf{J}_{ai}
= -\sum_{i,j=1}^{N-1}\left(1-\frac{N}{\Gamma}\tilde{\gamma}_i \right)
  \tilde{m}_{ij} \nabla \mathcal{C}_j.
\label{equ:J_expr}
\end{equation}
Once the functions $W(\vec{\phi},\nabla\vec{\phi})$
and $\varphi_i(\vec{\phi})$ ($1\leqslant i\leqslant N-1$)
are known, the chemical potentials $\mathcal{C}_i$ ($1\leqslant i\leqslant N-1$)
are computed from the linear system \eqref{equ:chempot},
and the densities $\rho_i(\vec{\phi})$ ($1\leqslant i\leqslant N$),
volume fractions $c_i(\vec{\phi})$,
and $\rho(\vec{\phi})$ are computed based on \eqref{equ:density_expr}.
%The volume fractions $c_i(\vec{\phi})$ ($1\leqslant i\leqslant N$)
%are computed based on \eqref{equ:volfrac_expr}.
%
We assume that the mixture dynamic viscosity $\mu(\vec{\phi})$
is given by an analogous expression  to that for the
mixture density $\rho(\vec{\phi})$ as follows,
\begin{equation}
\mu(\vec{\phi}) = \sum_{i=1}^N \tilde{\mu}_i c_i(\vec{\phi})
= \frac{\Gamma_{\mu}}{\Gamma} + \sum_{i=1}^{N-1}\left(
  \tilde{\mu}_i - \frac{\Gamma_{\mu}}{\Gamma}
\right) \tilde{\gamma}_i \varphi_i(\vec{\phi}).
\label{equ:mu_expr}
\end{equation}

% energy law for new model

The density $\rho(\vec{\phi})$ given by \eqref{equ:density_expr}
and the flux $\tilde{\mathbf{J}}$ given by 
\eqref{equ:J_expr} satisfy the relation
\begin{equation}
\frac{\partial\rho}{\partial t} + \mathbf{u}\cdot\nabla \rho
= -\nabla\cdot\tilde{\mathbf{J}}
\end{equation}
thanks to equation \eqref{equ:CH_general}.
By using this relation and assuming that all flux terms 
vanish on the domain boundary, it can be shown that 
the model consisting of equations \eqref{equ:nse_original},
\eqref{equ:continuity} and \eqref{equ:CH_general}
admits the following energy law
\begin{equation}
\frac{\partial}{\partial t}\int_{\Omega}\left[
  \frac{1}{2}\rho|\mathbf{u}|^2
  + W(\vec{\phi},\nabla\vec{\phi})
\right]
= -\int_{\Omega} \frac{\mu}{2}\|\mathbf{D}(\mathbf{u}) \|^2
-\int_{\Omega}\sum_{i,j=1}^{N-1} \tilde{m}_{ij}\nabla \mathcal{C}_i\cdot\nabla \mathcal{C}_j
\end{equation}
where $\Omega$ denotes the flow domain
and $\partial\Omega$ denotes its boundary.

% order parameters

The above N-phase model is a general phase field model.
To arrive at a specific physical formulation suitable
for numerical simulations, the model further requires 
\begin{itemize}

\item
the specification of a set of order 
parameters $\phi_i$ ($1\leqslant i\leqslant N-1$)
through equation \eqref{equ:varphi_expr};

\item
the specification of the form of the free energy density function
$W(\vec{\phi},\nabla\vec{\phi})$;

\item
the determination of the coefficients $\tilde{m}_{ij}$.

\end{itemize}
  
The order parameters for the N-phase system
have been discussed extensively
in \cite{Dong2015}, and a family of
order parameters has been introduced therein.
This family allows the use of many commonly-used physical
variables (e.g. volume fractions, volume fractions differences,
densities, density differences)
to formulate the system.
In the present paper we will employ the family of
order parameters introduced in \cite{Dong2015}
to the formulate the N-phase system.
More specifically, we define the order parameters 
$\phi_i$ ($1\leqslant i\leqslant N-1$) according 
to \cite{Dong2015}
\begin{equation}
\varphi_i \equiv \rho_i - \rho_N 
= \sum_{j=1}^N a_{ij} \phi_j + b_i, 
\quad 1\leqslant i\leqslant N-1,
\label{equ:gop_def}
\end{equation}
where $a_{ij}$ and $b_i$ ($1\leqslant i,j\leqslant N-1$)
are prescribed constants such that the matrix
\begin{equation}
\mathbf{A}_1 = \begin{bmatrix} a_{ij} \end{bmatrix}_{(N-1)\times(N-1)}
\label{equ:A1_expr}
\end{equation}
must be non-singular.
By prescribing a set of $a_{ij}$ and $b_i$, one will 
define a specific set of order parameters 
$\phi_i$ ($1\leqslant N-1$).
The coefficients $a_{ij}$ and $b_i$ for several
most commonly-used formulations
have been provided in \cite{Dong2015},
which correspond to  employing several commonly-encountered physical
variables  as the order parameters $\phi_i$, 
In the following, the development of the physical formulation  will be presented in terms of 
a general set of order parameters $\phi_i$ defined by \eqref{equ:gop_def}.
However,
in the numerical simulations of this paper, we will use a specific formulation,
which corresponds to 
using the volume fractions $c_i$ ($1\leqslant i\leqslant N-1$)
as the order parameters, i.e.
\begin{equation}
\phi_i = c_i, \quad 
\phi_i \in [0,1],
\quad 1\leqslant i\leqslant N-1.
\label{equ:gop_volfrac}
\end{equation}
The coefficients $a_{ij}$ and $b_i$ for this formulation 
are given by (see \cite{Dong2015})
\begin{equation}
a_{ij} = \tilde{\rho}_i \delta_{ij} + \tilde{\rho}_N, \quad
b_i = -\tilde{\rho}_N, \quad
1\leqslant i, j \leqslant N-1,
\label{equ:volfrac_aij}
\end{equation}
where $\delta_{ij}$ is the Kronecker delta.

The specification of the free energy density
function $W(\vec{\phi},\nabla\vec{\phi})$
and the determination of the coefficients
$\tilde{m}_{ij}$ call for considerations of
the consistency conditions ($\mathscr{C}$1)--($\mathscr{C}$4).
%For simplicity we assume in the current paper that
%$\tilde{m}_{ij}$ ($1\leqslant i,j\leqslant N-1$) are 
%all constants. 
We focus on $W(\vec{\phi},\nabla\vec{\phi})$
and $\tilde{m}_{ij}$
in the subsequent sections.

\subsection{N-Phase Free Energy Density Function}
\label{sec:free_energy}

To determine an appropriate form for the free energy
density function $W(\vec{\phi},\nabla\vec{\phi})$ 
and the coefficients $\tilde{m}_{ij}$,
we consider the set of consistency conditions ($\mathscr{C}$1)--($\mathscr{C}$4)
and insist that the N-phase formulation should honor these
conditions.
In~\cite{Dong2014,Dong2015} we have considered the consistency
of the free energy density function between N-phase and 
two-phase systems, and derived an explicit form for the mixing energy density
coefficients involved in the free energy density function therein.
The consistency conditions considered in \cite{Dong2014,Dong2015} are weaker;
Those conditions are equivalent to a combination of ($\mathscr{C}$1) and a subset of the
consistency condition ($\mathscr{C}$2) corresponding to $M=2$.

Following \cite{Dong2014}, we assume
the following form for the N-phase free energy density function
\begin{equation}
W(\vec{\phi},\nabla\vec{\phi}) =
\sum_{i,j=1}^{N-1} \frac{\lambda_{ij}}{2}\nabla\phi_i\cdot\nabla\phi_j
+ H(\vec{\phi}),
\label{equ:free_energy}
\end{equation}
where $\phi_i$ ($1\leqslant i\leqslant N-1$) are
the ($N-1$) independent order parameters and the coefficients
$\lambda_{ij}$ ($1\leqslant i,j\leqslant N-1$) are
referred to as the mixing energy density coefficients. Because
the term involving $\lambda_{ij}$ is a quadratic form,
the coefficients $\lambda_{ij}$ can always chosen to be symmetric.
We further require that
the matrix formed by these coefficients
\begin{equation}
\mathbf{A} = \begin{bmatrix}\lambda_{ij}  \end{bmatrix}_{(N-1)\times (N-1)}
\label{equ:lambda_matrix}
\end{equation}
be symmetric positive definite (SPD) to ensure
the positivity of the first term on the right hand side (RHS)
of \eqref{equ:free_energy}.
We assume that $\lambda_{ij}$ ($1\leqslant i,j\leqslant N-1$) are all
constants in the present paper.
$H(\vec{\phi})$ is referred to as
the potential free energy density function.
We deal with the $\lambda_{ij}$ term in 
\eqref{equ:free_energy} in this subsection, and
will defer the discussions about $H(\vec{\phi})$
to a later one (Section \ref{sec:potential_energy}).
Whenever needed in discussions of this subsection, we will make the following
assumption about $H(\vec{\phi})$:
\begin{enumerate}[($\mathscr{A}$1):]

\item
The potential free energy density function
$H(\vec{\phi})$ in \eqref{equ:free_energy} satisfies 
the consistency condition  ($\mathscr{C}$2).

\end{enumerate}
We will verify this property about
$H(\vec{\phi})$ later in Section \ref{sec:potential_energy}.

We require that the N-phase free energy density function \eqref{equ:free_energy}
satisfy the consistency property ($\mathscr{C}$2). 
It then follows from ($\mathscr{C}$2) and the assumption ($\mathscr{A}$1)
that the $\sum_{i,j=1}^{N-1}\frac{\lambda_{ij}}{2}\nabla\phi_i\cdot\nabla\phi_j$
term in \eqref{equ:free_energy} also satisfies the consistency property ($\mathscr{C}$2).

In the rest of this subsection 
we determine $\lambda_{ij}$ ($1\leqslant i,j\leqslant N-1$)
based on the consistency condition ($\mathscr{C}$2). 
We proceed in two steps.
We first use a subset of ($\mathscr{C}$2) that corresponds to $M=2$, i.e.~the consistency
 between N-phase and two-phase systems 
as discussed in \cite{Dong2014,Dong2015}, to determine uniquely
the values of $\lambda_{ij}$;
Then we show that with these $\lambda_{ij}$
values the free energy density function \eqref{equ:free_energy}
satisfies ($\mathscr{C}$2) for any $M$ ($2\leqslant M\leqslant N-1$)
under the assumption ($\mathscr{A}$1) for $H(\vec{\phi})$.

In \cite{Dong2015} 
we have derived a relation
between $\lambda_{ij}$ for
 a general set of 
order parameters $\phi_i$ defined by 
\eqref{equ:gop_def} 
and that for a special set using the volume fractions $c_i$ ($1\leqslant i\leqslant N-1$) as the order parameters 
as defined by \eqref{equ:gop_volfrac}--\eqref{equ:volfrac_aij}.
This relation is given by
\begin{equation}
\mathbf{A} 
%= \mathbf{Y}^T \bm{\Lambda}\mathbf{Y}
= (\mathbf{ZA}_1)^T \bm{\Lambda} (\mathbf{ZA}_1)
\label{equ:lambda_mat_gop}
\end{equation}
where $\mathbf{A}$ and $\mathbf{A}_1$ are defined respectively by 
\eqref{equ:lambda_matrix} and \eqref{equ:A1_expr}
corresponding to
a general set of order parameters $\phi_i$ ($1\leqslant i\leqslant N-1$)
defined by \eqref{equ:gop_def}, and
\begin{equation}
\bm{\Lambda} = \begin{bmatrix} \Lambda_{ij}  \end{bmatrix}_{(N-1)\times(N-1)}
\label{equ:lambda_mat_volfrac}
\end{equation} 
where $\Lambda_{ij}$ ($1\leqslant i,j\leqslant N-1$)
are the $\lambda_{ij}$  
values corresponding to the special set using the volume fractions 
as the order parameters as defined by \eqref{equ:gop_volfrac}--\eqref{equ:volfrac_aij}.
The matrix $\mathbf{Z}$ is given by
\begin{equation}
\mathbf{Z} = \begin{bmatrix} e_{ij}  \end{bmatrix}_{(N-1)\times(N-1)},
\quad \text{where} \
e_{ij} = \tilde{\gamma}_i\delta_{ij} 
   - \frac{\tilde{\gamma}_i\tilde{\gamma}_j}{\Gamma}, \
1\leqslant i\leqslant N, \ 1\leqslant j\leqslant N-1.
\label{equ:Z_mat_expr}
\end{equation}
In light of \eqref{equ:lambda_mat_gop},
we only need to determine the 
$\Lambda_{ij}$, i.e.~the $\lambda_{ij}$
values corresponding to the volume fractions 
as the order parameters defined by \eqref{equ:gop_volfrac}
and \eqref{equ:volfrac_aij}.

Let us now determine the values
for $\Lambda_{ij}$ ($1\leqslant i,j\leqslant N-1$)
based on the consistency property ($\mathscr{C}$2) restricted to $M=2$,
i.e.~the consistency between N-phase and 
two-phase systems.
%The process for deriving $\Lambda_{ij}$ below is similar to that
%of \cite{Dong2015}, 
%but the resultant $\Lambda_{ij}$ are very different from
%that given in \cite{Dong2015}, leading to
%a different N-phase free energy density function
%with improved properties.
%We impose the requirement that, if only a pair of
%two fluids is present in the N-phase system (while all
%the other fluids are absent),
%the N-phase free energy density function 
%should reduce to the corresponding two-phase free energy
%density function  for these two fluids \cite{Dong2014,Dong2015}.

%For a two-phase system it is well known that
%the mixing energy density coefficient can be related to
% the surface tension associated with
%the interface between the two fluids.
For a two-phase system, the free energy density function 
for a commonly-used 
formulation \cite{YueFLS2004,Dong2015} is as follows,
\begin{equation}
W(\phi_1,\nabla\phi_1) = 
\frac{\lambda_{11}}{2}\nabla\phi_1\cdot\nabla\phi_1
+\frac{\lambda_{11}}{4\eta^2}(1-\phi_1^2)^2 = 
\frac{1}{2}\left(\frac{6}{\sqrt{2}}\sigma_{12}\eta\right) \nabla c_1\cdot\nabla c_1
+ \frac{6}{\sqrt{2}}\frac{\sigma_{12}}{\eta} c_1^2(1-c_1)^2,
\label{equ:2p_energy}
\end{equation}
where 
$c_1$ and $c_2$ are volume fractions of 
the two fluids, 
$\eta$ is the characteristic interfacial thickness,
$\sigma_{12}$ is the surface tension between the
two fluids, and the phase field variable $\phi_1$ is
given by
\begin{equation}
\phi_1 = c_1 - c_2, \quad
c_1 = \frac{1}{2}(1+\phi_1), \quad
c_2 = \frac{1}{2}(1-\phi_1).
\label{equ:2p_order_param}
\end{equation}
The two-phase mixing energy density coefficient $\lambda_{11}$
is given by \cite{YueFLS2004}
\begin{equation}
\lambda_{11} = \frac{3}{2\sqrt{2}}\sigma_{12}\eta,
\label{equ:2p_lambda}
\end{equation}
which is derived by requiring in a one-dimensional setting that
at equilibrium the integral of the two-phase free energy
density function across the interface should match the 
surface tension.

In order to determine $\Lambda_{ij}$
we consider an N-phase system in which 
fluids $k$ and $l$ ($1\leqslant k<l\leqslant N$)
are the only fluids present therein, i.e.
\begin{equation}
c_i \equiv 0, \quad \rho_i \equiv 0, \quad
\text{if} \ i \neq k \ \text{and} \ i\neq l, \
\text{for} \ 1\leqslant i\leqslant N.
\label{equ:nphase_spec_config}
\end{equation}
Let $\sigma_{ij}$ ($1\leqslant i,j\leqslant N$)
denote the  surface tension associated with
the interface formed between fluids $i$ and $j$, with
the property
\begin{equation}
\sigma_{ij} = \sigma_{ji}, \ 1\leqslant i,j\leqslant N; \quad
\sigma_{ii} = 0, \ 1\leqslant i\leqslant N; \quad
\sigma_{ij}>0, \ \text{if} \ i\neq j.
\label{equ:surften_property}
\end{equation}
We consider the special set using the volume fractions $c_i$ ($1\leqslant i\leqslant N-1$)
as the order parameters as defined by 
\eqref{equ:gop_volfrac} and \eqref{equ:volfrac_aij}.
In light of the form of two-phase potential free energy density
function in \eqref{equ:2p_energy},
we make another assumption about  $H(\vec{\phi})$ as follows,
\begin{enumerate}[($\mathscr{A}$1):]
\addtocounter{enumi}{1}

\item
For the N-phase system characterized by \eqref{equ:nphase_spec_config},
the potential free energy $H(\vec{\phi})$ should reduce to 
\begin{equation}
H(\vec{\phi}) = (\beta \sigma_{kl}) c_k^2(1-c_k)^2, 
% \ \text{where} \ \beta=\frac{6}{\sqrt{2}}\frac{1}{\eta}
\label{equ:potential_energy_reduction}
\end{equation}
where 
\begin{equation}
\beta = \frac{6}{\sqrt{2}}\frac{1}{\eta}.
\label{equ:beta_expr}
\end{equation}

\end{enumerate}
This property will be verified in Section \ref{sec:potential_energy}.

Let us look into the reduction of the N-phase
free energy function \eqref{equ:free_energy}
for the N-phase system characterized by \eqref{equ:nphase_spec_config}.
We differentiate two cases:
(i) $1\leqslant k<l=N$,
and (ii) $1\leqslant k<l\leqslant N-1$.
In the first case the free energy density is transformed into
\begin{equation}
W(\vec{\phi},\nabla\vec{\phi}) = 
\sum_{i,j=1}^{N-1}\frac{\Lambda_{ij}}{2}\nabla\phi_i\cdot\nabla\phi_j
+ H(\vec{\phi})
= \frac{\Lambda_{kk}}{2}\nabla c_k\cdot\nabla c_k
+ (\beta\sigma_{kN}) c_k^2(1-c_k)^2,
\end{equation}
where equations \eqref{equ:gop_volfrac}
and \eqref{equ:potential_energy_reduction}
and the conditions \eqref{equ:nphase_spec_config}
have been used.
Comparing the above form with
the two-phase free energy density function \eqref{equ:2p_energy},
we have the relations
\begin{equation}
\Lambda_{kk} = \frac{6}{\sqrt{2}} \eta \sigma_{kN}, 
\quad 1\leqslant k \leqslant N-1.
\label{equ:lambda_kk_expr}
\end{equation}
In the second case ($1\leqslant k<l\leqslant N-1$)
the N-phase free energy density function
is reduced to
\begin{equation}
\begin{split}
W(\vec{\phi},\nabla\vec{\phi})& =
\sum_{i,j=1}^{N-1}\frac{\Lambda_{ij}}{2}\nabla\phi_i\cdot\nabla\phi_j
+ H(\vec{\phi}) \\
& = \frac{\Lambda_{kk}}{2}\nabla c_k\cdot\nabla c_k
+ \Lambda_{kl}\nabla c_k\cdot\nabla c_l
+ \frac{\Lambda_{ll}}{2}\nabla c_l\cdot\nabla c_l
+ (\beta\sigma_{kl}) c_k^2(1-c_k)^2 \\
& = \frac{1}{2}(\Lambda_{kk}+\Lambda_{ll} - 2\Lambda_{kl})\nabla c_k\cdot\nabla c_k + (\beta\sigma_{kl}) c_k^2(1-c_k)^2,
\end{split}
\end{equation}
where we have used \eqref{equ:gop_volfrac},
\eqref{equ:potential_energy_reduction},
the relation $\sum_{i=1}^N c_i=1$ in \eqref{equ:volfrac_expr},
and the conditions \eqref{equ:nphase_spec_config}.
Comparing the above form with
the two-phase free energy density function \eqref{equ:2p_energy},
we have the relation.
\begin{equation}
\Lambda_{kk}+\Lambda_{ll} - 2\Lambda_{kl} = \frac{6}{\sqrt{2}}\eta\sigma_{kl},
\quad
1\leqslant k<l\leqslant N-1.
\label{equ:lambda_kl_relation}
\end{equation}
By combining  \eqref{equ:lambda_kk_expr} and
\eqref{equ:lambda_kl_relation} and 
noting the symmetry $\Lambda_{kl}=\Lambda_{lk}$,
we obtain
\begin{equation}
\Lambda_{kl} = \frac{3}{\sqrt{2}}\eta(\sigma_{kN} + \sigma_{lN} - \sigma_{kl}),
\quad
1\leqslant k, l \leqslant N-1.
\label{equ:lambda_ij_volfrac}
\end{equation}
These are the mixing energy density coefficients
with the volume fractions as the order parameters.

On can note that the mixing energy density coefficients
obtained here are  different from those of
\cite{Dong2015}. This is due to the reduction property
\eqref{equ:potential_energy_reduction} about
the potential free energy density $H(\vec{\phi})$ assumed here.
%and the form of the two-phase free energy density function
%\eqref{equ:2p_energy} employed here.
While both the current free energy density function
and that of \cite{Dong2015} 
can consistently reduce to the two-phase free energy
density,
the mixing energy density coefficients in \cite{Dong2015} lead to
different interfacial thicknesses for the 
interfaces formed between different pairs of fluids. 
More precisely, the characteristic
 thickness of a fluid interface resulting from the free energy form
of \cite{Dong2015} is proportional to
the surface tension associated with that interface.
In contrast, with the free energy density coefficients
obtained here different fluid interfaces have 
the same interfacial thickness, characterized
by the constant $\eta$ in \eqref{equ:lambda_ij_volfrac}.

The mixing energy density coefficients $\lambda_{ij}$
for a general set of order parameters defined
by \eqref{equ:gop_def} can be computed based on 
\eqref{equ:lambda_mat_gop}, where 
$\Lambda_{ij}$ are given by \eqref{equ:lambda_ij_volfrac}.

% comments on Lambda_ij, compare with results in Dong(2015)
% comment on how to compute lambda_ij for general order parameters
% commen on SPD of lambda_ij matrix
% prove that it satisfies consistency condition (D1)
% 

Having determined $\lambda_{ij}$, let us now look into
the consistency property ($\mathscr{C}$2) with any
$2\leqslant M\leqslant N-1$. Specifically, we have the following theorem:
\begin{theorem}
\label{thm:thm_1}

Under assumption ($\mathscr{A}$1), 
the N-phase free energy density function \eqref{equ:free_energy}
satisfies the consistency property ($\mathscr{C}$2),
with $\lambda_{ij}$ ($1\leqslant i,j\leqslant N-1$) given by
\eqref{equ:lambda_mat_gop} and $\Lambda_{ij}$ ($1\leqslant i,j\leqslant N-1$)
given by \eqref{equ:lambda_ij_volfrac}.
%provided that $H(\vec{\phi})$
%satisfies the assumption ($\mathscr{A}$1).

\end{theorem}
\noindent A proof of this theorem is provided in Appendix A.

% comment on SPD-ness of lambda_ij as computed above

Let us make a comment on the symmetric positive definiteness
of the matrix $\mathbf{A}$ as computed
using \eqref{equ:lambda_mat_gop}.
Note first that this matrix will be symmetric.
Because the matrices $\mathbf{Z}$ and $\mathbf{A}_1$
in \eqref{equ:lambda_mat_gop}
are both non-singular \cite{Dong2015}, 
the matrices $\mathbf{A}$
and $\bm{\Lambda}$ will have the same positive definiteness.
This positive definiteness is determined only by the pairwise
surface tensions $\sigma_{ij}$ because of
 equation \eqref{equ:lambda_ij_volfrac}.
%The conditions on $\sigma_{ij}$ that ensure
%the positive definiteness of the matrix $\bm{\Lambda}$
%are an open question at the moment (see also \cite{Dong2015}). 
Therefore, in this paper we will require that the pairwise
surface tensions $\sigma_{ij}$ ($1\leqslant i\neq j\leqslant N$) among the $N$ fluids
are such that the matrix $\bm{\Lambda}$ defined by \eqref{equ:lambda_mat_volfrac}, with $\Lambda_{ij}$
given by \eqref{equ:lambda_ij_volfrac},
is positive definite.

% what else to discuss here? 

\subsection{Determination of  $\tilde{m}_{ij}$}
\label{sec:mij}

% what is the logic for determining m_ij?
% (1) equation for rho_i
% (2) determine m_ij based on condition for K term:
%     if fluid k is absent at t=0, then fluid k will not be
%        generated at t>0
% (3) show that with the m_ij computed, the governing equations
%     satisfy consistency condition (D2), if H term satisfy
%     certain assumption
%
% what does consistency condition (D2) mean?
% if fluid k is absent from system, then N-phase governing equations
%    must reduce to (N-1)-phase governing equations
% (1) what are the N-phase governing equations?
%     what are the (N-1)-phase governing equations?
% (2) what are the conditions in order to reduce to each other?

We next consider the consistency property ($\mathscr{C}$3)
and determine the coefficients
$\tilde{m}_{ij}$ ($1\leqslant i,j\leqslant N-1$)
in \eqref{equ:CH_general} 
based on this property.
We assume that
$\tilde{m}_{ij}$ are all constants
in this paper.

% first determine m_ij

Combining equations \eqref{equ:density_expr}
and \eqref{equ:CH_general}, we obtain
\begin{equation}
\frac{\partial c_i}{\partial t}
+\mathbf{u}\cdot\nabla c_i = 
\sum_{k,j=1}^{N-1}
  \left(\tilde{\gamma}_i\delta_{ij} - \frac{\tilde{\gamma}_i\tilde{\gamma}_j}{\Gamma} \right)
\tilde{m}_{jk} \nabla^2 \mathcal{C}_k
= \sum_{k,j=1}^{N-1}
  e_{ij}\tilde{m}_{jk} \nabla^2 \mathcal{C}_k,
\quad 1\leqslant i\leqslant N,
\label{equ:mass_balance_volfrac}
\end{equation}
where $e_{ij}$ ($1\leqslant i\leqslant N$, $1\leqslant j\leqslant N-1$)
is defined in \eqref{equ:Z_mat_expr}.
%In light of \eqref{equ:volfrac_expr}, the above equation can be rewritten
%in terms of the volume fractions $c_i$
%\begin{equation}
%\frac{\partial c_i}{\partial t}
%+\mathbf{u}\cdot\nabla c_i 
%= \sum_{k,j=1}^{N-1}
%  e_{ij}\tilde{m}_{jk} \nabla^2 \mathcal{C}_k,
%\quad 1\leqslant i\leqslant N.
%\label{equ:mass_balance_volfrac}
%\end{equation}

Define
\begin{equation}
\begin{split}
&
\vec{c} = \begin{bmatrix}
c_i
\end{bmatrix}_{N\times 1},
\quad
\mathbf{Z}_N = \begin{bmatrix}
e_{Nj}
\end{bmatrix}_{1\times (N-1)},
\quad
\overline{\mathbf{Z}} = \begin{bmatrix}
\begin{array}{l}
\mathbf{Z} \\
\mathbf{Z}_N
\end{array}
\end{bmatrix}_{N\times(N-1)}, 
%\quad
%
%\tilde{\mathbf{m}} = \begin{bmatrix}
%\tilde{m}_{ij}
%\end{bmatrix}_{(N-1)\times(N-1)} \\
%
\quad
\bm{\Phi} = \begin{bmatrix}\phi_i  \end{bmatrix}_{(N-1)\times 1},
\\
&
\frac{\partial H}{\partial\bm{\Phi}}=
\begin{bmatrix}\frac{\partial H}{\partial\phi_i}  \end{bmatrix}_{(N-1)\times 1},
\quad
\vec{\mathcal{C}} = \begin{bmatrix}
 \mathcal{C}_i
\end{bmatrix}_{(N-1)\times 1},
\end{split}
\label{equ:c_vec_def}
\end{equation}
where $\mathbf{Z}$ is defined in \eqref{equ:Z_mat_expr}.
Then equation \eqref{equ:mass_balance_volfrac}
can be expressed in matrix form as
\begin{equation}
\frac{\partial\vec{c}}{\partial t}
+ \mathbf{u}\cdot\nabla\vec{c}
= \overline{\mathbf{Z}}\tilde{\mathbf{m}}\nabla^2\vec{\mathcal{C}}.
\label{equ:mass_balance_volfrac_mat}
\end{equation}
With respect to a general set of order parameters
defined by \eqref{equ:gop_def}, the linear algebraic
system \eqref{equ:chempot} for the chemical potentials
is transformed into
\begin{equation}
\sum_{j=1}^{N-1} a_{ji}\mathcal{C}_j = 
\frac{\partial H}{\partial\phi_i} - \sum_{j=1}^{N-1}\lambda_{ij}\nabla^2\phi_i,
\quad 1\leqslant i\leqslant N-1,
\end{equation}
where we have used \eqref{equ:free_energy}.
In matrix form, this equation can be written as
\begin{equation}
\mathbf{A}_1^T\vec{\mathcal{C}} = 
\frac{\partial H}{\partial\bm{\Phi}} 
- \mathbf{A}\nabla^2\bm{\Phi},
\quad
\vec{\mathcal{C}} = \mathbf{A}_1^{-T}\frac{\partial H}{\partial\bm{\Phi}} 
- \mathbf{A}_1^{-T}\mathbf{A}\nabla^2\bm{\Phi},
\label{equ:chempot_2}
\end{equation}
where $\mathbf{A}$ and $\mathbf{A}_1$ are defined
in \eqref{equ:lambda_matrix} and \eqref{equ:A1_expr}
respectively. 

Consequently, equation \eqref{equ:mass_balance_volfrac_mat}
becomes
\begin{equation}
\frac{\partial\vec{c}}{\partial t}
+ \mathbf{u}\cdot\nabla\vec{c}
= \nabla^2\left[
 -\left(\overline{\mathbf{Z}}\tilde{\mathbf{m}}\mathbf{A}_1^{-T}\mathbf{A}\right)\nabla^2\bm{\Phi}
+ \left(\overline{\mathbf{Z}}\tilde{\mathbf{m}}\mathbf{A}_1^{-T}\right)\frac{\partial H}{\partial\bm{\Phi}}
\right]
\label{equ:mass_balance_volfrac_mat_2}
\end{equation}
Let
\begin{equation}
\overline{\mathbf{Z}}\tilde{\mathbf{m}}\mathbf{A}_1^{-T}\mathbf{A}
= \begin{bmatrix}\tilde{N}_{ij}  \end{bmatrix}_{N\times(N-1)},
\quad
\overline{\mathbf{Z}}\tilde{\mathbf{m}}\mathbf{A}_1^{-T}
= \begin{bmatrix}\tilde{S}_{ij}  \end{bmatrix}_{N\times(N-1)}.
\label{equ:mat_N_def}
\end{equation}
Then equation \eqref{equ:mass_balance_volfrac_mat_2} can be written
in terms of the components as
\begin{equation}
\frac{\partial c_i}{\partial t}
+ \mathbf{u}\cdot\nabla c_i = 
\nabla^2\left[
- \sum_{j=1}^{N-1}\tilde{N}_{ij}\nabla^2\phi_j
+ \sum_{j=1}^{N-1}\tilde{S}_{ij}\frac{\partial H}{\partial \phi_j}
\right],
\quad 1\leqslant i\leqslant N,
\label{equ:mass_balance_volfrac_3}
\end{equation}
where $c_i$ ($1\leqslant i\leqslant N$) are 
the volume fractions and $\phi_i$ ($1\leqslant i\leqslant N-1$)
are the order parameters defined by \eqref{equ:gop_def}.

% Now how to choose m_ij

To satisfy the consistency property ($\mathscr{C}$3) it suffices to
 consider the reduction of the N-phase system to
the ($N-1$)-phase system. According to  ($\mathscr{C}$3),
if any one fluid phase is absent from the N-phase system, then it should remain
absent over time.
Suppose fluid $k$ is absent from the N-phase system,
i.e.
\begin{equation}
c_k \equiv 0, \ \text{for some} \ k,  \ 1\leqslant k\leqslant N.
\label{equ:condition_F1}
\end{equation}
In light of \eqref{equ:mass_balance_volfrac_3},
the consistency condition ($\mathscr{C}$3)  requires
\begin{equation}
\nabla^2\left[
- \sum_{j=1}^{N-1}\tilde{N}_{kj}\nabla^2\phi_j
+ \sum_{j=1}^{N-1}\tilde{S}_{kj}\frac{\partial H}{\partial \phi_j}
\right] \equiv 0, 
\ \text{if} \ c_k \equiv 0,
\ \text{for arbitrary} \ \phi_i.
\label{equ:condition_E1}
\end{equation}
A sufficient condition to ensure \eqref{equ:condition_E1} is
\begin{equation}
\sum_{j=1}^{N-1}\tilde{N}_{kj}\nabla^2\phi_j = 0, % \text{\em const.}
\quad \text{if} \ c_k\equiv 0, \ \text{for any} \ 1\leqslant k\leqslant N \ \text{and arbitrary} \ \phi_i,
\label{equ:condition_E2}
\end{equation}
with 
\begin{equation}
\sum_{j=1}^{N-1}\tilde{S}_{kj}\frac{\partial H}{\partial \phi_j} = 0, % \text{\em const.}
\quad \text{if} \ c_k\equiv 0, \ \text{for any} \ 1\leqslant k\leqslant N \ \text{and arbitrary} \ \phi_i.
\label{equ:condition_E3}
\end{equation}
We next use \eqref{equ:condition_E2} to determine
$\tilde{m}_{ij}$, and equation
\eqref{equ:condition_E3} is a condition on
the potential free energy density function
$H(\vec{\phi})$ to ensure consistency.

In light of equations  \eqref{equ:density_expr}
and \eqref{equ:gop_def},
the condition \eqref{equ:condition_F1} can be 
transformed into
\begin{equation}
0 \equiv c_k %= \tilde{\gamma}_k\rho_k
= \frac{\gamma_k}{\Gamma}
+ \sum_{i=1}^{N-1}e_{ki}\left(
  \sum_{j=1}^{N-1} a_{ij}\phi_j + b_i
\right).
\end{equation}
It follows that
\begin{equation}
\sum_{j=1}^{N-1}\left(\sum_{i=1}^{N-1}e_{ki}a_{ij}  \right)\nabla^2\phi_j = 0
\label{equ:condition_F2}
\end{equation}
for arbitrary $\phi_i$.
Comparing equations \eqref{equ:condition_E2} and 
\eqref{equ:condition_F2}, we conclude that if
\begin{equation}
\tilde{N}_{kj} = d_k \sum_{i=1}^{N-1} e_{ki} a_{ij},
\quad 1\leqslant k\leqslant N, \ 
1\leqslant j\leqslant N-1,
\label{equ:condition_G1}
\end{equation}
for constants $d_k\neq 0$ ($1\leqslant k\leqslant N$),
then the condition \eqref{equ:condition_E2}
will be satisfied.
Let 
\begin{equation}
\mathbf{D} = \text{diag}\left(d_1,d_2,\dots,d_{N-1}  \right).
\end{equation}
In light of \eqref{equ:mat_N_def} and \eqref{equ:c_vec_def},
equation \eqref{equ:condition_G1} leads to
\begin{subequations}
\begin{align}
&
\mathbf{Z}\tilde{\mathbf{m}}\mathbf{A}_1^{-T}\mathbf{A} 
= \mathbf{DZA}_1, \label{equ:mij_equ_1}
\\
&
\mathbf{Z}_N \tilde{\mathbf{m}}\mathbf{A}_1^{-T}\mathbf{A}
= d_N \mathbf{Z}_N \mathbf{A}_1. \label{equ:mij_equ_2}
\end{align}
\end{subequations}
Equation \eqref{equ:mij_equ_1} can be written as
\begin{equation}
\mathbf{Z}\tilde{\mathbf{m}}\mathbf{Z}^T = 
\mathbf{D}\left(\mathbf{ZA}_1 \right) \mathbf{A}^{-1}
\left(\mathbf{ZA}_1 \right)^T.
\end{equation}
Noting that $\mathbf{A}$ and $\tilde{\mathbf{m}}$
are both required to be general SPD matrices,
and that $\mathbf{Z}$ and $\mathbf{A}_1$
are  non-singular, 
we conclude that 
\begin{equation}
\mathbf{D} = m_0 \mathbf{I}, 
\quad m_0>0 \ \text{being a constant},
\end{equation}
where  $\mathbf{I}$ is the 
identity matrix. This leads to 
\begin{equation}
\begin{bmatrix} \tilde{m}_{ij}  \end{bmatrix}_{(N-1)\times(N-1)}
= \tilde{\mathbf{m}} 
= m_0\mathbf{A}_1\mathbf{A}^{-1}\mathbf{A}_1^T,
\label{equ:mij_expr}
\end{equation}
which provides an explicit expression for
the coefficients $\tilde{m}_{ij}$.
Substitution of this expression into \eqref{equ:mij_equ_2} results in
\begin{equation}
d_N = m_0.
\end{equation}

% implications of this m_ij

With $\tilde{m}_{ij}$ given by \eqref{equ:mij_expr},
the volume fraction equation \eqref{equ:mass_balance_volfrac_mat_2}
is transformed into
\begin{equation}
\frac{\partial\vec{c}}{\partial t}
+ \mathbf{u}\cdot\nabla\vec{c}
= m_0 \nabla^2\left[
 -\left(\overline{\mathbf{Z}}\mathbf{A}_1\right) \nabla^2\bm{\Phi}
+ \left(\overline{\mathbf{Z}}\mathbf{A}_1\mathbf{A}^{-1}\right)\frac{\partial H}{\partial\bm{\Phi}}
\right].
\label{equ:volfrac_equ}
\end{equation}
The phase field equation \eqref{equ:CH_general}
for the N-phase system
is transformed into, in terms of the general set of order parameters
$\phi_i$ defined by \eqref{equ:gop_def},
\begin{subequations}
\begin{align}
&
\frac{\partial\bm{\Phi}}{\partial t}
+ \mathbf{u}\cdot\nabla\bm{\Phi}
= m_0\nabla^2\left[
-\nabla^2\bm{\Phi}
+ \mathbf{A}^{-1}\frac{\partial H}{\partial\bm{\Phi}}
\right], \label{equ:CH_matform}
\\
&
\frac{\partial\phi_i}{\partial t} + 
\mathbf{u}\cdot\nabla\phi_i
= m_0\nabla^2\left[
-\nabla^2\phi_i + \sum_{j=1}^{N-1}\zeta_{ij} \frac{\partial H}{\partial \phi_j}
\right],
\quad 1\leqslant i\leqslant N-1, 
\label{equ:CH_final}
\end{align}
\end{subequations}
where
\begin{equation}
\mathbf{A}^{-1} = \begin{bmatrix}\zeta_{ij}  \end{bmatrix}_{(N-1)\times(N-1)},
\label{equ:theta_expr}
\end{equation}
and $\mathbf{A}$ is given by \eqref{equ:lambda_mat_gop},
and we have used 
\eqref{equ:chempot_2} and \eqref{equ:mij_expr}.
%Hereafter we will refer to $m_0$ as the mobility of
%the N-phase system.

We conclude that, with $\tilde{m}_{ij}$ given by \eqref{equ:mij_expr}
and if $H(\vec{\phi})$ satisfies the condition \eqref{equ:condition_E3},
then the N-phase formulation satisfies the consistency property
($\mathscr{C}$3).

It can be noted that
the coefficients $\tilde{m}_{ij}$ given by
\eqref{equ:mij_expr} are independent of
the choice of the set of order parameters. Let
$\bm{\Lambda}_1$ denote the matrix $\mathbf{A}_1$
(see \eqref{equ:A1_expr}) corresponding to
the set with volume fractions as the order parameters as defined by
\eqref{equ:gop_volfrac}, i.e.
\begin{equation}
\bm{\Lambda}_1 = \begin{bmatrix} a_{ij} \end{bmatrix}_{(N-1)\times(N-1)},
\ \ a_{ij} \ \text{given by equation \eqref{equ:volfrac_aij}}.
\end{equation}
It is straightforward to verify that 
\begin{equation}
\bm{\Lambda}_1 = \mathbf{Z}^{-1},
\end{equation}
where $\mathbf{Z}$ is given by \eqref{equ:Z_mat_expr}.
Then equation \eqref{equ:mij_expr} can be transformed
into
\begin{equation}
\tilde{\mathbf{m}} = m_0\mathbf{A}_1\mathbf{A}^{-1}\mathbf{A}_1^T
= m_0\bm{\Lambda}_1\bm{\Lambda}^{-1}\bm{\Lambda}_1^T,
\end{equation}
where we have used \eqref{equ:lambda_mat_gop},
and the matrix $\bm{\Lambda}$ is given 
by \eqref{equ:lambda_mat_volfrac} and \eqref{equ:lambda_ij_volfrac}.

\subsection{Implications of Consistency Property ($\mathscr{C}$4)}

% now that we have m_ij and the phase field equations
% let us look at the implications of consistency condition (D2)
% focus on volume fractions as order parameters
% (D2) is equivalent to: if fluid k is from N-phase system
%   then N-phase governing equations should reduce to those
%   for the corresponding (N-1)-phase system
% 

Having determined the coefficients $\tilde{m}_{ij}$, let
us explore the implications of the consistency property
($\mathscr{C}$4). 
Substitution of the free energy form \eqref{equ:free_energy} into
the momentum equation \eqref{equ:nse_original} leads to
\begin{equation}
\rho\left(
  \frac{\partial\mathbf{u}}{\partial t}
  + \mathbf{u}\cdot\nabla\mathbf{u}
\right)
+ \tilde{\mathbf{J}}\cdot\nabla\mathbf{u}
=
-\nabla p
+ \nabla\cdot\left[
  \mu \mathbf{D}(\mathbf{u})
\right]
- \sum_{i,j=1}^{N-1} \nabla\cdot\left(\lambda_{ij}
  \nabla\phi_i \otimes \nabla\phi_j
\right)
\label{equ:nse_trans_1}
\end{equation}
where $\rho(\vec{\phi})$ and $\mu(\vec{\phi})$
are given by \eqref{equ:density_expr}
and \eqref{equ:mu_expr} respectively, $\lambda_{ij}$
($1\leqslant i,j\leqslant N-1$) are given by \eqref{equ:lambda_mat_gop},
and
\begin{equation}
\tilde{\mathbf{J}}(\vec{\phi},\nabla\vec{\phi})
= -m_0\sum_{i,j=1}^{N-1}\left(1-\frac{N}{\Gamma}\tilde{\gamma}_i \right)
a_{ij} \nabla\left(
  -\nabla^2\phi_j + \sum_{k=1}^{N-1} \zeta_{jk}\frac{\partial H}{\partial\phi_k}
\right).
\label{equ:J_expr_1}
\end{equation}
Equations \eqref{equ:nse_trans_1}, \eqref{equ:continuity}
and \eqref{equ:CH_final} constitute the governing equations
for the N-phase system, with the potential free energy
density function $H(\vec{\phi})$ to be specified.

% focus on volume fractions as order parameters
% then discuss conditions for reduction of governing equations

\begin{remark}

Consider the family of order parameters
defined by \eqref{equ:gop_def}.
It can be shown that the values for the terms $\rho(\vec{\phi})$,
$\mu(\vec{\phi})$, $\tilde{\mathbf{J}}(\vec{\phi},\nabla\vec{\phi})$ and
$
\sum_{i,j=1}^{N-1}\nabla\cdot\left[  
\lambda_{ij}\nabla\phi_i\otimes\nabla\phi_j
\right]
$
are independent of the choice of the set of 
order parameters from this family.
Therefore the momentum equations \eqref{equ:nse_trans_1}
and \eqref{equ:continuity} are invariant 
with respect to a different choice of
the order parameters from this family.
On the other hand, with the choice of a different set of order
parameters,
the phase field equations \eqref{equ:CH_final}
 transforms accordingly  such that its
form will remain the same under the new set of order
parameters.
The phase field equations
\eqref{equ:CH_final} formulated in terms of different sets of order parameters
are equivalent.   

\end{remark}

In light of the above property of the governing equations, 
in the rest of this subsection we will focus on the
governing equations formulated in terms of the 
volume fractions as the order parameters,
as defined by \eqref{equ:gop_volfrac} and
\eqref{equ:volfrac_aij}.
In subsequent discussions $\phi_i$ 
($1\leqslant i\leqslant N-1$) are understood to be
the volume fractions $c_i$ ($1\leqslant i\leqslant N-1$)
in the governing equations, and 
the model parameters (e.g.~$\lambda_{ij}$, 
$a_{ij}$) correspond to those with
volume fractions as the order parameters.
To satisfy the consistency property ($\mathscr{C}$4),
it suffices to consider the reduction of the N-phase governing equations 
when a fluid phase $k$ (for any $1\leqslant k\leqslant N$)
is absent from the system.

Consider first the phase field equations \eqref{equ:CH_final}
in light of the consistency property ($\mathscr{C}$4).
We re-write them as
\begin{equation}
\frac{\partial c_i^{(N)}}{\partial t}
+ \mathbf{u}\cdot\nabla c_i^{(N)}
= m_0\nabla^2\left[
  -\nabla^2 c_i^{(N)} 
  + \sum_{j=1}^{N-1} \Theta_{ij}^{(N)}\frac{\partial H^{(N)}}{\partial c_j^{(N)}}
\right],
\quad 1\leqslant i\leqslant N-1,
\label{equ:volfrac_equ_i}
\end{equation}
where the superscript ${N}$ in $(\cdot)^{(N)}$ 
stresses that the variable is with
respect to the N-phase system,
and $\Theta_{ij}$ ($1\leqslant i,j\leqslant N-1$)
denote $\zeta_{ij}$ corresponding to
the volume fractions as the order parameters, i.e.
\begin{equation}
\begin{bmatrix}\Theta_{ij}\end{bmatrix}_{(N-1)\times(N-1)}=
\bm{\Lambda}^{-1} = 
\begin{bmatrix}\Lambda_{ij} \end{bmatrix}^{-1}_{(N-1)\times(N-1)},
\quad \Lambda_{ij} \ \text{given by} \ \eqref{equ:lambda_ij_volfrac}.
\label{equ:theta_ij_volfrac}
\end{equation}
These equations imply the following equation
for $c_N$,
\begin{equation}
\frac{\partial c_N^{(N)}}{\partial t}
+ \mathbf{u}\cdot\nabla c_N^{(N)}
=-\sum_{i=1}^{N-1}\left( \frac{\partial c_i^{(N)}}{\partial t}
+ \mathbf{u}\cdot\nabla c_i^{(N)} \right)
= m_0\nabla^2\left[
  -\nabla^2 c_N^{(N)} 
  - \sum_{i,j=1}^{N-1} \Theta_{ij}^{(N)}\frac{\partial H^{(N)}}{\partial c_j^{(N)}}
\right]
\label{equ:volfrac_equ_N}
\end{equation}
The equations \eqref{equ:volfrac_equ_i}
and \eqref{equ:volfrac_equ_N} together
are equivalent to 
the set of $N$ equations \eqref{equ:volfrac_equ}
in matrix form for the volume fractions.
Define $L_i^{(N)}$ ($1\leqslant i\leqslant N$, $N\geqslant 2$),
\begin{equation}
L_i^{(N)} = \sum_{j=1}^{N-1}\Theta_{ij}^{(N)}\frac{\partial H^{(N)}}{\partial c_j^{(N)}}, \ 1\leqslant i\leqslant N-1; \quad
L_N^{(N)} = -\sum_{i=1}^{N-1}L_i^{(N)}.
\label{equ:Li_def}
\end{equation}
Then the equations \eqref{equ:volfrac_equ_i}
and \eqref{equ:volfrac_equ_N} can be written
in a unified form
\begin{equation}
\frac{\partial c_i^{(N)}}{\partial t}
+ \mathbf{u}\cdot\nabla c_i^{(N)}
= m_0\nabla^2\left[
  -\nabla^2 c_i^{(N)} 
  + L_i^{(N)}
\right],
\quad 1\leqslant i\leqslant N.
\label{equ:volfrac_equ_3}
\end{equation}

Suppose that fluid $k$ (for some $1\leqslant k\leqslant N$) 
is absent from the N-phase system,
namely, $c_k^{(N)}\equiv 0$.
%Then the correspondence 
%between $c_i^{(N-1)}$ ($1\leqslant i\leqslant N-1$)
%and $c_i^{(N)}$ ($1\leqslant i\leqslant N$) is given by \eqref{equ:relation_cN_cN1}.
Then 
\begin{equation*}
c_i^{(N-1)} = \left\{
\begin{array}{ll}
c_i^{(N)} & 1\leqslant i\leqslant k-1 \\
c_{i+1}^{(N)} & k\leqslant i\leqslant N-1.
\end{array}
\right.
\end{equation*}
The consistency condition ($\mathscr{C}$4) requires that
\begin{subequations}
\begin{align}
&
\nabla^2 L_k^{(N)} = 0, \\
&
\nabla^2 L_i^{(N-1)} = \nabla^2 L_i^{(N)}, \quad 1\leqslant i\leqslant k-1, \\
&
\nabla^2 L_i^{(N-1)} = \nabla^2 L_{i+1}^{(N)}, \quad k\leqslant i\leqslant N-1.
\end{align}
\end{subequations}
A sufficient condition to ensure
the above is
\begin{subequations}
\begin{align}
&
L_k^{(N)} = 0,  \ \ \text{if} \ c_k^{(N)}\equiv 0, \ \text{for any} \ 1\leqslant k\leqslant N,
\label{equ:condition_H1}
\\
&
L_i^{(N-1)} = L_i^{(N)}, \quad 1\leqslant i\leqslant k-1, 
\ \ \text{if} \ c_k^{(N)}\equiv 0, \ \text{for any} \ 1\leqslant k\leqslant N,
\label{equ:condition_H2} \\
&
L_i^{(N-1)} = L_{i+1}^{(N)}, \quad k\leqslant i\leqslant N-1,
\ \ \text{if} \ c_k^{(N)}\equiv 0, \ \text{for any} \ 1\leqslant k\leqslant N.
\label{equ:condition_H3}
\end{align}
\end{subequations}
Note that the condition \eqref{equ:condition_H1}
is equivalent to that given in \eqref{equ:condition_E3}.
These are the conditions that the potential free energy
density function $H(\vec{\phi})$ should satisfy in
order to ensure consistency.

Let us next consider the momentum equation \eqref{equ:nse_trans_1}
in light of the consistency condition ($\mathscr{C}4$).
If fluid $k$ (for any $1\leqslant k\leqslant N$) is absent from the N-phase system,
($\mathscr{C}$4) requires that
\begin{align}
&
\rho^{(N)}(\vec{c}^{(N)}) = \rho^{(N-1)}(\vec{c}^{(N-1)}), \label{equ:rho_consist}
\\
&
\mu^{(N)}(\vec{c}^{(N)}) = \mu^{(N-1)}(\vec{c}^{(N-1)}), \label{equ:mu_consist}
 \\
&
\sum_{i,j=1}^{N-1} \nabla\cdot\left[
\Lambda_{ij}^{(N)}\nabla c_i^{(N)}\otimes c_j^{(N)}
\right]
= \sum_{i,j=1}^{N-2} \nabla\cdot\left[
\Lambda_{ij}^{(N-1)}\nabla c_i^{(N-1)}\otimes c_j^{(N-1)}
\right], \label{equ:surften_consist}
\\
&
\tilde{\mathbf{J}}^{(N)}(\vec{c}^{(N)},\nabla\vec{c}^{(N)}) = \tilde{\mathbf{J}}^{(N-1)}(\vec{c}^{(N-1)},\nabla\vec{c}^{(N-1)}).
\label{equ:J_consist}
\end{align}
The following theorem confirms the above relations, provided that the conditions
\eqref{equ:condition_H1}--\eqref{equ:condition_H3} are satisfied.
\begin{theorem}
\label{thm:thm_2}

If any one fluid phase is absent from the N-phase system, then 
the relations given by \eqref{equ:rho_consist}--\eqref{equ:surften_consist}
hold. If further the potential free energy density function $H(\vec{\phi})$
satisfies the conditions \eqref{equ:condition_H1}--\eqref{equ:condition_H3},
then the relation \eqref{equ:J_consist} holds.

\end{theorem}
\noindent A proof of the above theorem is provided in Appendix B.

% summary of results
% what are the conditions to ensure (D2)?

In summary, the N-phase governing equations
consist of the equations \eqref{equ:nse_trans_1},
\eqref{equ:continuity} and \eqref{equ:CH_final},
in which $\lambda_{ij}$ ($1\leqslant i,j\leqslant N-1$)
are given by \eqref{equ:lambda_mat_gop}, \eqref{equ:lambda_mat_volfrac}
and \eqref{equ:lambda_ij_volfrac},
 $\zeta_{ij}$ ($1\leqslant i,j\leqslant N-1$)
are given by \eqref{equ:theta_expr},
$\tilde{\mathbf{J}}$ is given by \eqref{equ:J_expr_1},
and $\rho$ and $\mu$ are given by \eqref{equ:density_expr},
\eqref{equ:mu_expr} and \eqref{equ:gop_def}.
If the potential free energy density function $H(\vec{\phi})$
satisfies the conditions \eqref{equ:condition_H1}--\eqref{equ:condition_H3},
then the N-phase formulation satisfies the consistency
properties ($\mathscr{C}$3) and ($\mathscr{C}$4).
In other words, if only $M$ ($2\leqslant M\leqslant N-1$)
fluid phases are present in the N-phase system,
then the N-phase governing equations will reduce
to those for the corresponding $M$-phase system,
and initially absent fluid phases will remain absent over time. 

% what else to discuss here?

\subsection{N-Phase Potential Energy Density Function}
\label{sec:potential_energy}

% what are the conditions H should meet to ensure consistency?
% how to construct such H functions? is it an open question?
%   does such function exist?
% employ a form suggested by Boyer. 
% this formulation with Boyer potential energy satisfies a weaker
%    condition than (D2) with this N-phase model 
% show that if only two fluids are present, then N-phase formulation
%    will satisfy (D1) and (D2)

Let us now look into the potential free energy density function
$H(\vec{\phi})$. 
To ensure the full reduction consistency of the N-phase formulation,
$H(\vec{\phi})$ should satisfy: 
(i) assumption ($\mathscr{A}$1),
(ii) assumption ($\mathscr{A}$2),
and (iii) the reduction relations \eqref{equ:condition_H1}--\eqref{equ:condition_H3}.
In addition, $H(\vec{\phi})$ should be invariant with different choices
of the set of 
the order parameters $\phi_i$, be non-negative (or
bounded from below), and be multi-welled.
% (reaching a minimum 
%at $c_i = 0$ or $c_i=1$ for $1\leqslant i\leqslant N$).

The construction of a potential energy density function that satisfies the above
properties, in particular the conditions \eqref{equ:condition_H1}--\eqref{equ:condition_H3},
is a highly non-trivial and challenging matter.
How to construct such a fully consistent potential energy density function 
is still an open problem. 
A set of  function forms with certain interesting properties that are conducive to
the potential energy construction have been suggested in \cite{BoyerM2014}.
We consider a  function form
suggested by \cite{BoyerM2014} as follows,
\begin{equation}
H = \frac{\beta}{2}\sum_{i,j=1}^N \frac{\sigma_{ij}}{2}
\left[
  f(c_i) + f(c_j) - f(c_i+c_j)
\right], \quad
\text{with} \ f(c) = c^2(1-c)^2,
\label{equ:potential_energy}
\end{equation}
where the constant $\beta$ is given by \eqref{equ:beta_expr}, 
$\sigma_{ij}$ ($1\leqslant i,j\leqslant N$) are
the pairwise surface tensions satisfying the property \eqref{equ:surften_property},
and $c_i$ ($1\leqslant i\leqslant N$)
are the volume fractions.

In this paper we will employ \eqref{equ:potential_energy} for the potential
energy density function.
It is straightforward to verify that this function satisfies the assumptions ($\mathscr{A}$1)
and ($\mathscr{A}$2), by noticing that
if a fluid phase $k$ (for any $1\leqslant k\leqslant N$)
is absent then the function is reduced to
$ %\begin{equation}
H = \frac{\beta}{2}\sum^N_{ 
\substack{
i,j=1 \\
i,j \neq k }
}
\frac{\sigma_{ij}}{2}
\left[
  f(c_i) + f(c_j) - f(c_i+c_j)
\right].
$ %\end{equation}
Therefore, the N-phase free energy function $W(\vec{\phi},\nabla\vec{\phi})$ 
defined in \eqref{equ:free_energy},
with $H(\vec{\phi})$ given by \eqref{equ:potential_energy}, 
satisfies the consistency property ($\mathscr{C}$2). 
Evidently it also satisfies the consistency property ($\mathscr{C}$1).

This potential energy density function
does not, however, guarantee the general reduction
relations \eqref{equ:condition_H1}--\eqref{equ:condition_H3}.
So the N-phase formulation consisting of \eqref{equ:nse_trans_1},
\eqref{equ:continuity} and \eqref{equ:CH_final},
with $H(\vec{\phi})$ given by \eqref{equ:potential_energy},
does not satisfy the general consistency conditions ($\mathscr{C}$3) and ($\mathscr{C}$4),
if $M$ is an arbitrary number with $2\leqslant M\leqslant N-1$
therein.

However, with the N-phase formulation
given by equations \eqref{equ:nse_trans_1},
\eqref{equ:continuity} and \eqref{equ:CH_final},
the potential energy density 
$H(\vec{\phi})$ given by \eqref{equ:potential_energy} 
does satisfy an important subset of the consistency 
conditions ($\mathscr{C}$3) and ($\mathscr{C}$4),
when only a pair of two fluids (for any pair) is present in the system. 
More specifically, we have the following theorem
\begin{theorem}
\label{thm:thm_3}

If only a pair of fluids, fluid $k$ and fluid $l$ ($1\leqslant k<l\leqslant N$),
are present in the N-phase system, while all the other fluids are absent,
i.e.~the system is characterized by \eqref{equ:nphase_spec_config},
then with $H(\vec{\phi})$ given by \eqref{equ:potential_energy}
\begin{subequations}
\begin{align}
&
L_i^{(N)} = 0, \quad \text{if} \ i\neq k \ \text{and} \ i\neq l, \ \text{for} \ 1\leqslant i\leqslant N,
\label{equ:condition_2p_1}
\\
&
L_k^{(N)} = L_1^{(2)}, 
\label{equ:condition_2p_2}
\\
&
L_l^{(N)} = L_2^{(2)},
\label{equ:condition_2p_3}
\\
&
\tilde{\mathbf{J}}^{(N)} = \tilde{\mathbf{J}}^{(2)},
\label{equ:condition_2p_4}
\end{align}
\end{subequations}
where $L_i^{(N)}$ ($1\leqslant i\leqslant N$) are
defined in \eqref{equ:Li_def}.

\end{theorem}
\noindent A proof of this theorem is provided in Appendix C.

Therefore, the current N-phase formulation
given by \eqref{equ:nse_trans_1}, \eqref{equ:continuity} and \eqref{equ:CH_final},
with the potential energy density function given by \eqref{equ:potential_energy},
satisfies the consistency conditions ($\mathscr{C}$3) and ($\mathscr{C}$4) with $M=2$.
In other words,
if only a pair of two fluid phases are present in
the system, the N-phase governing
equations will fully reduce to those for the corresponding 
two-phase system consisting of these two fluids.
This N-phase formulation is fully consistent with two-phase formulations.

\begin{comment}
As noted by \cite{BoyerM2014},
due to the negative sign involved therein,
it is unknown whether the potential energy density
function \eqref{equ:potential_energy}
is bounded from below.
Numerous numerical experiments indicate that
this energy density function produces 
qualitatively reasonable and
quantitatively accurate results.
Some of the simulation results will be 
presented in Section \ref{sec:tests}.
\end{comment}

% what else to discuss here?
% comments on full consistency with M phases?
% call for action

\begin{comment}
The set of conditions,
\eqref{equ:condition_H1}--\eqref{equ:condition_H3},
on the potential free energy density function
ensures the full reduction consistency
of the governing equations between the N-phase
and the M-phase ($2\leqslant M\leqslant N-1$) 
systems.
How to construct energy density functions
that satisfy these conditions is an important,
and mathematically interesting problem.
This is still an open problem. 
The potential energy density function given by
\eqref{equ:potential_energy} is but an intermediate,
albeit crucial, step toward this goal.
Solving this problem requires much 
future efforts from the community.
\end{comment}

\section{An N-Phase Contact-Angle Boundary Condition}
\label{sec:cabc}

% what is the logic?
% (1) proposed form of N-phase contact-angle BC
% (2) reduction property
% (3) two-phase contact-angle BC
% (4) how to determine coefficients
% (5) prove that BC is fully reduction consistent
%
% consider only static contact angles
% only (N-1) independent contact angles due to Young's relations
% 4-th order equations, need two independent BCs on each boundary

In this section we employ the consistency condition ($\mathscr{C}$5)
to devise a boundary condition
to account for  the multitude
of contact angles
on solid-wall surfaces.
We impose the requirement that the contact-angle boundary condition should satisfy
 ($\mathscr{C}$5).
Only static (or equilibrium)
contact angles will be considered in the current paper.

% contact-angle BC

Let $\partial\Omega$ denote the solid wall boundary.
We propose the following form of  boundary
condition to account for the contact angles,
\begin{equation}
\mathbf{n}\cdot\nabla c_i = \sum_{j=1}^N \xi_{ij} c_i c_j, 
\quad 1\leqslant i\leqslant N,
\quad \text{on} \ \partial\Omega,
\label{equ:cabc}
\end{equation}
where $\mathbf{n}$ is the outward-pointing unit
vector normal to the wall boundary, 
$c_i$ ($1\leqslant i\leqslant N$) are the 
volume fractions, and $\xi_{ij}$ ($1\leqslant i,j\leqslant N$)
are constant coefficients to be determined.
Taking into account the constraint on the volume fractions in
equation \eqref{equ:volfrac_expr}, we have
\begin{equation}
0 = \sum_{i=1}^N \mathbf{n}\cdot\nabla c_i
= \sum_{i,j=1}^N \xi_{ij} c_i c_j
= \sum_{i,j=1}^N \frac{1}{2}\left(\xi_{ij} + \xi_{ji} \right) c_i c_j,
\quad \text{on} \ \partial\Omega
\end{equation}
for arbitrary $c_i$ satisfying \eqref{equ:volfrac_expr}.
Consequently,
\begin{equation}
\xi_{ij} = -\xi_{ji}, \ 1\leqslant i\neq j\leqslant N;
\quad
\xi_{ii} = 0, \ 1\leqslant i\leqslant N.
\label{equ:xi_ij_property}
\end{equation}
Because of the above property,
only ($N-1$) boundary conditions are independent among the
$N$ conditions in \eqref{equ:cabc}.

We impose the requirement that the boundary condition
\eqref{equ:cabc} satisfy the the consistency property ($\mathscr{C}$5)
for any $2\leqslant M\leqslant N-1$.
We will first determine the coefficients $\xi_{ij}$ using 
the condition ($\mathscr{C}$5) with $M=2$, 
i.e.~by requiring consistency with two-phase contact-angle boundary conditions.
We then show that with the computed $\xi_{ij}$ values
the  boundary condition \eqref{equ:cabc}
satisfies ($\mathscr{C}$5) for arbitrary $2\leqslant M\leqslant N-1$.

Let us define
$\theta_{ij}$ ($1\leqslant i\neq j\leqslant N$)
as the static (equilibrium) contact angle between the wall and 
the fluid interface formed by fluids $i$ and $j$,
measured on the side of fluid $i$.
Then
\begin{equation}
\theta_{ij} = \pi - \theta_{ji}, \quad
\cos \theta_{ij} = -\cos\theta_{ji}, \quad
1\leqslant i,j\leqslant N, \ i\neq j.
\label{equ:ca_def}
\end{equation}
Let $\sigma_{w,i}$ ($1\leqslant i\leqslant N$)
denote the interfacial tension between fluid $i$
and the solid wall. The Young's relation provides
\begin{equation}
\sigma_{w,i} - \sigma_{w,j} = -\sigma_{ij}\cos\theta_{ij},
\quad
1\leqslant i\neq j\leqslant N.
\label{equ:young_relation}
\end{equation}
Based on this relation we have
\begin{equation}
\left\{
\begin{split}
&
-\sigma_{ij}\cos\theta_{ij} 
= (\sigma_{w,i}-\sigma_{w,N}) - (\sigma_{w,j} -\sigma_{w,N})
= -\sigma_{iN}\cos\theta_{iN} + \sigma_{jN}\cos\theta_{jN}, \\
&
\cos\theta_{ij} 
= \frac{\sigma_{iN}}{\sigma_{ij}}\cos\theta_{iN}
- \frac{\sigma_{jN}}{\sigma_{ij}}\cos\theta_{jN},
\quad 
1\leqslant i\neq j \leqslant N-1.
\end{split}
\right.
\label{equ:ca_relation}
\end{equation}
Therefore, all the contact angles 
$\theta_{ij}$ ($1\leqslant i\neq j\leqslant N$)
in the N-phase system can be expressed in terms of
the angles
$\theta_{iN}$ ($1\leqslant i\leqslant N-1$).
We will use $\theta_{iN}$ ($1\leqslant i\leqslant N-1$)
as the ($N-1$) independent contact 
angles in the current work.

Let us now consider the consistency of the contact-angle
boundary conditions between N-phase and two-phase
systems, and determine the coefficients $\xi_{ij}$. 
Two-phase contact-angle boundary conditions have been
investigated in a number of works (see e.g.~\cite{Jacqmin2000,YueZF2010,Dong2012} among others).
Here we employ the form given by \cite{Dong2012},
\begin{equation}
\mathbf{n}\cdot\nabla\phi_1 
= -\frac{1}{\lambda_{11}}f^\prime_{w}(\phi_1),
\label{equ:2p_cabc}
\end{equation}
where $\phi_1$ is the two-phase phase field variable
defined by \eqref{equ:2p_order_param}, and
$\lambda_{11}$ is the two-phase mixing
energy density coefficient given by \eqref{equ:2p_lambda}.
The function $f^\prime_w(\phi_1)$ is given by~\cite{Dong2012}
\begin{equation}
f_w^\prime(\phi_1) = 
-\frac{3}{4}\sigma_{12}\cos\theta_{12} (1-\phi_1^2),
\label{equ:2p_wall_energy_deriv}
\end{equation}
where $\sigma_{12}$ is the surface tension between the
two fluids, and $\theta_{12}$ is the static 
contact angle between the interface and the wall
measured on the side of the {\em first} fluid.
In light of \eqref{equ:2p_order_param}, \eqref{equ:2p_lambda}
and \eqref{equ:2p_wall_energy_deriv},
the two-phase contact-angle condition \eqref{equ:2p_cabc} can be
transformed into
\begin{equation}
\mathbf{n}\cdot\nabla c_1 
= \left(\frac{\sqrt{2}}{\eta} \cos\theta_{12}\right) c_1 c_2,
\label{equ:2p_cabc_1}
\end{equation}
where $\eta$ is the characteristic interfacial thickness.

Let us look into the reduction of the N-phase
contact-angle boundary condition \eqref{equ:cabc}
when only two fluid phases are present in the N-phase system.
Suppose fluids $k$ and $l$ ($1\leqslant k<l\leqslant N$)
are the only fluids present in the system (all the other
fluids are absent), that is, the system is characterized by
 \eqref{equ:nphase_spec_config}.
The boundary condition \eqref{equ:cabc} is
transformed into
\begin{align}
&
\mathbf{n}\cdot\nabla c_i = c_i\sum_{j=1}^N \xi_{ij}c_j = 0,
\quad \text{if} \ i\neq k \ \text{and} \ i\neq l, 
\ \text{for} \ 1\leqslant i\leqslant N, \\
&
\mathbf{n}\cdot\nabla c_k = c_k\sum_{j=1}^N \xi_{kj}c_j
= \xi_{kk}c_k^2 + \xi_{kl}c_k c_l
= \xi_{kl}c_k c_l \label{equ:np_cabc_reduce_2}
\\
&
\mathbf{n}\cdot\nabla c_l = c_l\sum_{j=1}^N \xi_{lj}c_j
= \xi_{lk}c_l c_k + \xi_{ll}c_l^2
= -\xi_{kl}c_k c_l 
= -\mathbf{n}\cdot\nabla c_k,
\end{align}
where we have used \eqref{equ:nphase_spec_config} and
\eqref{equ:xi_ij_property}.
A comparison between \eqref{equ:np_cabc_reduce_2}
and the two-phase contact-angle boundary condition
\eqref{equ:2p_cabc_1} leads to
\begin{equation}
\xi_{kl} = \frac{\sqrt{2}}{\eta}\cos\theta_{kl}, \quad
1\leqslant k<l \leqslant N.
\end{equation}
In light of \eqref{equ:ca_relation},
 $\xi_{ij}$ can be expressed as
\begin{equation}
\left\{
\begin{split}
&
\xi_{ij} = \frac{\sqrt{2}}{\eta}\left(
  \frac{\sigma_{iN}}{\sigma_{ij}}\cos\theta_{iN}
 - \frac{\sigma_{jN}}{\sigma_{ij}}\cos\theta_{jN}
\right), \quad
1\leqslant i\neq j\leqslant N, \\
&
\xi_{ii} = 0, \quad 1\leqslant i\leqslant N.
\end{split}
\right.
\label{equ:xi_ij_expr}
\end{equation}

Having determined $\xi_{ij}$, let us now show that
the N-phase contact-angle boundary condition
\eqref{equ:cabc}, with $\xi_{ij}$
given by \eqref{equ:xi_ij_expr},
satisfies the consistency property ($\mathscr{C}$5)
between the N-phase and M-phase systems  for any $2\leqslant M\leqslant N-1$.
It suffices to show that ($\mathscr{C}$5) is satisfied 
for $M=N-1$, that is, 
if one fluid phase is
absent from the system  the boundary
condition \eqref{equ:cabc} will reduce to
that for the corresponding ($N-1$)-phase
system.
%This is because if this property holds the condition
%(R1) will follow for arbitrary $M$ ($2\leqslant M\leqslant N-1$)
%by repeated application of this property.

Suppose fluid $k$ is absent from the N-phase system,
i.e.~$c_k\equiv 0$, for some $k$ ($1\leqslant k\leqslant N$).
We distinguish two cases:
(i) $1\leqslant k\leqslant N-1$, and
(ii) $k=N$.
In the first case, we have the correspondence
relations \eqref{equ:relation_cN_cN1},
\eqref{equ:surften_N_N1}, and
\begin{equation}
\theta_{i,N-1}^{(N-1)} = \left\{
\begin{array}{ll}
\theta_{iN}^{(N)}, & 1\leqslant i\leqslant k-1, \\
\theta_{i+1,N}^{(N)}, & k\leqslant i\leqslant N-1,
\end{array}
\right.
\label{equ:ca_relation_N_N1}
\end{equation}
where the superscript $(N)$ again highlights that
the quantity is for the N-phase system.
Based on \eqref{equ:surften_N_N1}, 
\eqref{equ:ca_relation_N_N1} and 
\eqref{equ:xi_ij_expr}, we further
have the relation
\begin{equation}
\xi_{ij}^{(N-1)} = \left\{
\begin{array}{ll}
\xi_{ij}^{(N)}, & 1\leqslant i\leqslant k-1, \ 1\leqslant j\leqslant k-1, \\
\xi_{i,j+1}^{(N)}, & 1\leqslant i\leqslant k-1, \ k\leqslant j\leqslant N-1,\\
\xi_{i+1,j}^{(N)}, & k\leqslant i\leqslant N-1, \ 1\leqslant j\leqslant k-1, \\
\xi_{i+1,j+1}^{(N)}, & k\leqslant i\leqslant N-1, \ k\leqslant j\leqslant N-1.
\end{array}
\right.
\label{equ:xi_relation_N_N1}
\end{equation}
Therefore, for $1\leqslant i\leqslant k-1$,
\begin{equation*}
\begin{split}
\mathbf{n}\cdot\nabla c_i^{(N)}
&= \sum_{j=1}^N \xi_{ij}^{(N)} c_i^{(N)} c_j^{(N)}
= \left(
  \sum_{j=1}^{k-1} + \sum_{j=k+1}^N
\right) \xi_{ij}^{(N)} c_i^{(N)} c_j^{(N)} \\
&
= \sum_{j=1}^{k-1}\xi_{ij}^{(N)} c_i^{(N)} c_j^{(N)}
+ \sum_{j=k}^{N-1}\xi_{i,j+1}^{(N)} c_i^{(N)} c_{j+1}^{(N)}
= \left(
  \sum_{j=1}^{k-1} + \sum_{j=k}^{N-1}
\right) \xi_{ij}^{(N-1)} c_i^{(N-1)} c_j^{(N-1)} \\
&
= \sum_{j=1}^{N-1} \xi_{ij}^{(N-1)} c_i^{(N-1)} c_j^{(N-1)} \\
 \Longrightarrow \ \ & \ \
\mathbf{n}\cdot\nabla c_i^{(N-1)}
= \sum_{j=1}^{N-1} \xi_{ij}^{(N-1)} c_i^{(N-1)} c_j^{(N-1)},
\quad 1\leqslant i\leqslant k-1,
\end{split} 
\end{equation*}
where we have used \eqref{equ:relation_cN_cN1}
and \eqref{equ:xi_relation_N_N1}.
For $k\leqslant i\leqslant N-1$,
\begin{equation*}
\begin{split}
\mathbf{n}\cdot\nabla c_{i+1}^{(N)}
& = \sum_{j=1}^N \xi_{i+1,j}^{(N)} c_{i+1}^{(N)} c_j^{(N)}
= \left(
  \sum_{j=1}^{k-1} + \sum_{j=k+1}^N
\right) \xi_{i+1,j}^{(N)} c_{i+1}^{(N)} c_j^{(N)} \\
& 
= \sum_{j=1}^{k-1} \xi_{i+1,j}^{(N)} c_{i+1}^{(N)} c_j^{(N)}
+ \sum_{j=k}^{N-1} \xi_{i+1,j+1}^{(N)} c_{i+1}^{(N)} c_{j+1}^{(N)}
= \left(
  \sum_{j=1}^{k-1} + \sum_{j=k}^{N-1}
\right) \xi_{ij}^{(N-1)} c_{i}^{(N-1)} c_j^{(N-1)} \\
&
= \sum_{j=1}^{N-1} \xi_{ij}^{(N-1)} c_{i}^{(N-1)} c_j^{(N-1)} \\
\Longrightarrow \ \ & \ \
\mathbf{n}\cdot\nabla c_i^{(N-1)}
= \sum_{j=1}^{N-1} \xi_{ij}^{(N-1)} c_i^{(N-1)} c_j^{(N-1)},
\quad k\leqslant i\leqslant N-1,
\end{split}
\end{equation*}
where we again have used \eqref{equ:relation_cN_cN1}
and \eqref{equ:xi_relation_N_N1}.
One also notes in this case that
\begin{equation*}
\mathbf{n}\cdot\nabla c_k^{(N)} 
= \sum_{j=1}^N \xi_{kj}^{(N)} c_k^{(N)} c_j^{(N)}
= 0.
\end{equation*}

In the second case ($k=N$, and $c_N\equiv 0$)
we have the correspondence relations
given by \eqref{equ:relation_cN_1} and \eqref{equ:relation_sigma_N}
and
\begin{equation}
\left\{
\begin{split}
&
\theta_{i,N-1}^{(N-1)} = \theta_{i,N-1}^{(N)}, \quad
1\leqslant i\leqslant N-1; \\
&
\xi_{ij}^{(N-1)} = \xi_{ij}^{(N)}, \quad
1\leqslant i,j\leqslant N-1.
\end{split}
\right.
\label{equ:relation_xi_N_2}
\end{equation}
Therefore, for $1\leqslant i\leqslant N-1$,
\begin{equation*}
\begin{split}
\mathbf{n}\cdot\nabla c_i^{(N)}
&= \sum_{j=1}^N \xi_{ij}^{(N)} c_i^{(N)} c_j^{(N)}
= \sum_{j=1}^{N-1} \xi_{ij}^{(N)} c_i^{(N)} c_j^{(N)}
= \sum_{j=1}^{N-1} \xi_{ij}^{(N-1)} c_i^{(N-1)} c_j^{(N-1)}, \\
\Longrightarrow \ \ & \ \
\mathbf{n}\cdot\nabla c_i^{(N-1)}
= \sum_{j=1}^{N-1} \xi_{ij}^{(N-1)} c_i^{(N-1)} c_j^{(N-1)},
\quad 1\leqslant i\leqslant N-1,
\end{split}
\end{equation*}
where we have used \eqref{equ:relation_cN_1}
and \eqref{equ:relation_xi_N_2}.
Note also that in this case
\begin{equation}
\mathbf{n}\cdot\nabla c_N^{(N)} = 0.
\end{equation}

Combining the above discussions, we conclude
that the contact-angle boundary condition
\eqref{equ:cabc}, with $\xi_{ij}$ given
by \eqref{equ:xi_ij_expr},
reduces to that for the ($N-1$)-phase system
if any one fluid phase is absent from the N-phase system.
Therefore, this contac-angle boundary condition
satisfies the consistency property ($\mathscr{C}$5)
between N-phase and M-phase systems for any $2\leqslant M\leqslant N-1$.

% express BC using general order parameters

In light of %\eqref{equ:volfrac_expr} and
\eqref{equ:density_expr}, 
and noting that only ($N-1$) among
the $N$ contact-angle boundary conditions in \eqref{equ:cabc}
are independent,
we can re-write these boundary conditions 
(for $1\leqslant i\leqslant N-1$)
in terms of a general set of
 order parameters $\phi_i$ defined
by \eqref{equ:gop_def} as follows,
\begin{equation}
\mathbf{n}\cdot\nabla\phi_i = 
\sum_{j=1}^{N-1}\sum_{s=1}^{N} y_{ij}\xi_{js}c_j(\vec{\phi})c_s(\vec{\phi}),
\quad 1\leqslant i\leqslant N-1, 
\ \text{on} \ \partial\Omega,
\label{equ:cabc_1}
\end{equation}
where 
\begin{equation}
\begin{bmatrix} y_{ij}  \end{bmatrix}_{(N-1)\times(N-1)}
= \left(\mathbf{ZA}_1  \right)^{-1},
\quad
\mathbf{Z} \ \text{defined in} \ \eqref{equ:Z_mat_expr},
\ \mathbf{A}_1 \ \text{defined in} \ \eqref{equ:A1_expr},
\end{equation}
and 
\begin{equation}
c_i(\vec{\phi}) = \frac{\tilde{\gamma}_i}{\Gamma}
+ \sum_{j=1}^{N-1}\left(
  \tilde{\gamma}_i\delta_{ij} - \frac{\tilde{\gamma}_i\tilde{\gamma}_j}{\Gamma}
\right) \left(
  \sum_{s=1}^{N-1}a_{js}\phi_s + b_j
\right), 
\quad 1\leqslant i\leqslant N.
\label{equ:volfrac_expr_1}
\end{equation}

Since the phase field equations \eqref{equ:CH_final}
are of fourth spatial order, two independent
boundary conditions are needed on each domain boundary.
In addition to the contact-angle boundary condition 
\eqref{equ:cabc_1} on wall boundaries,
for the other boundary condition we will impose the zero flux
of the chemical potentials, i.e.
\begin{equation}
\mathbf{n}\cdot\nabla\left[
  -\nabla^2\phi_i 
  + \sum_{j=1}^{N-1}\zeta_{ij}\frac{\partial H}{\partial\phi_j}
\right] = 0,
\quad 1\leqslant i\leqslant N-1,
\quad \text{on} \ \partial\Omega.
\label{equ:bc_noflux_chempot}
\end{equation}
This condition can equivalently be expressed in terms of the volume fractions
as follows
\begin{equation}
\mathbf{n}\cdot\nabla\left[
  -\nabla^2 c_i 
  + \sum_{j=1}^{N-1}\Theta_{ij}\frac{\partial H}{\partial c_j}
\right] = 0,
\quad 1\leqslant i\leqslant N-1,
\quad \text{on} \ \partial\Omega.
\label{equ:bc_noflux_chempot_1}
\end{equation}

Based on discussions in this and the previous sections, we arrive at the following
theorem:
\begin{theorem}

The contact-angle boundary condition \eqref{equ:cabc} satisfies 
the consistency property ($\mathscr{C}$5). 
The boundary condition \eqref{equ:bc_noflux_chempot} satisfies
the consistency property
($\mathscr{C}$5) if the potential energy function $H(\vec{\phi})$
satisfies the conditions \eqref{equ:condition_H1}--\eqref{equ:condition_H3}.

\end{theorem}

% what else to discuss here?

\section{Numerical Algorithm and Implementation}
\label{sec:algorithm}

% what is the logic?
% (1) governing equations
% (2) boundary conditions
% (3) algorithm formulation
%     mainly for phase field equations
% (4) weak formulation, 
%     implementation with spectral elements
% (5) some comments on the algorithm

Let $\Omega$ denote the flow domain and
$\partial\Omega$ denote the domain boundary, and
we assume that the domain boundary consists of all
solid walls with certain wetting property.
The N-phase system is described by the equations
\eqref{equ:nse_trans_1}, \eqref{equ:continuity}
and \eqref{equ:CH_final},
together with the boundary conditions
\eqref{equ:cabc_1} and \eqref{equ:bc_noflux_chempot}
for the phase field variables and
the following Dirichlet condition for
the velocity,
\begin{equation}
\mathbf{u} = \mathbf{w}(\mathbf{x},t), \quad \text{on} \ \partial\Omega,
\label{equ:dbc}
\end{equation}
where $\mathbf{w}(\mathbf{x},t)$
is the boundary velocity.

To facilitate subsequent development we
re-write the momentum equation \eqref{equ:nse_trans_1} into 
\begin{equation}
  \frac{\partial\mathbf{u}}{\partial t}
  + \mathbf{u}\cdot\nabla\mathbf{u}
+ \frac{1}{\rho}\tilde{\mathbf{J}}\cdot\nabla\mathbf{u}
=
-\frac{1}{\rho}\nabla P
+ \frac{\mu}{\rho}\nabla^2\mathbf{u}
+ \frac{1}{\rho}\nabla\mu\cdot\mathbf{D}(\mathbf{u})
- \frac{1}{\rho}\sum_{i,j=1}^{N-1} \lambda_{ij}\nabla^2\phi_j \nabla\phi_i
+ \frac{1}{\rho}\mathbf{f}(\mathbf{x},t),
\label{equ:nse_final}
\end{equation} 
where we have added an external body force $\mathbf{f}$, and
$P$ is an auxiliary pressure,
\begin{equation}
P = p + 
\sum_{i,j=1}^{N-1}\frac{\lambda_{ij}}{2}\nabla\phi_i\cdot\nabla\phi_j,
\end{equation}
which will also be loosely referred to as the pressure
where no confusion arises.

We re-write the phase field equations \eqref{equ:CH_final}
into
\begin{equation}
\frac{\partial\phi_i}{\partial t} + 
\mathbf{u}\cdot\nabla\phi_i
= m_0\nabla^2\left[
-\nabla^2\phi_i + \sum_{j=1}^{N-1}\zeta_{ij} h_j(\vec{\phi})
\right] + g_i(\mathbf{x},t),
\quad 1\leqslant i\leqslant N-1, 
\label{equ:CH_final_1}
\end{equation}
where
\begin{equation}
  h_i(\vec{\phi}) = \frac{\partial H}{\partial \phi_j},
  \quad 1\leqslant i\leqslant N-1,
  \label{equ:hi_expr}
\end{equation}
and we have added a 
source term $g_i$ ($1\leqslant i\leqslant N-1$)
to each of these equations.
The source terms $g_i$ are prescribed functions
for the purpose of numerical
testing only, and will be set to $g_i=0$
in actual simulations.

Similarly, we also modify the boundary conditions
\eqref{equ:bc_noflux_chempot} and
\eqref{equ:cabc_1} 
by adding certain source terms as follows,
\begin{align}
&
\mathbf{n}\cdot\nabla\left[
  -\nabla^2\phi_i 
  + \sum_{j=1}^{N-1}\zeta_{ij} h_j(\vec{\phi})
\right] = g_{ai}(\mathbf{x},t),
\quad 1\leqslant i\leqslant N-1,
\quad \text{on} \ \partial\Omega.
\label{equ:bc_chempot_2} \\
&
\mathbf{n}\cdot\nabla\phi_i = 
\sum_{j=1}^{N-1}\sum_{s=1}^{N} y_{ij}\xi_{js}c_j(\vec{\phi})c_s(\vec{\phi})
+ g_{bi}(\mathbf{x},t),
\quad 1\leqslant i\leqslant N-1, 
\ \text{on} \ \partial\Omega.
\label{equ:cabc_2} 
\end{align}
The source terms $g_{ai}$ and $g_{bi}$ ($1\leqslant i\leqslant N-1$)
in the above modified boundary conditions
are all prescribed functions 
for numerical testing only,
and will be set to $g_{ai}=0$ and
$g_{bi}=0$ in actual simulations.

% ready for description of algorithm

Let us consider below  how to numerically solve  the system consisting
of \eqref{equ:nse_final}, \eqref{equ:continuity}
and \eqref{equ:CH_final_1}, together with the
boundary conditions \eqref{equ:dbc} for the
velocity $\mathbf{u}$ and
\eqref{equ:bc_chempot_2} and \eqref{equ:cabc_2}
for the phase field variables $\phi_i$.
Because the momentum equations \eqref{equ:nse_final} and \eqref{equ:continuity}
have the same form as
those from \cite{Dong2015,Dong2014,Dong2012}, they can be numerically treated
in the same way.
In the current work we solve the momentum equations using the algorithm
we developed in \cite{Dong2015}, and we have provided a summary of this
 algorithm 
%for the momentum
% equations \eqref{equ:nse_final} and \eqref{equ:continuity}
in the Appendix D of this paper.
The phase field equations \eqref{equ:CH_final_1}
are different from, and are simpler in form than,
those of the N-phase formulation of \cite{Dong2014,Dong2015}.
Each of these equations bears a similarity to
the two-phase phase field equation (see \cite{DongS2012}).
A strategy similar to that for two-phase
flows can be used to numerically treat the N-phase
phase field equations \eqref{equ:CH_final_1}.

In the following we present an algorithm for solving
the N-phase phase field equations \eqref{equ:CH_final_1}.
Let $n\geqslant 0$ denote the time step index, and
$(\cdot)^n$ denote the quantity $(\cdot)$ at time step $n$.
Then, given $(\mathbf{u}^n, \phi_i^n)$, we solve for
the phase field variables $\phi_i^{n+1}$
as follows: 
%\underline{For $\phi_i^{n+1}$:}
\begin{subequations}
  \begin{multline}
    \frac{\gamma_0\phi_i^{n+1}-\hat{\phi}_i}{\Delta t}
    + \mathbf{u}^{*,n+1}\cdot\nabla\phi_i^{*,n+1} \\
    = m_0\nabla^2\left[
      -\nabla^2\phi_i^{n+1}
      + \frac{S}{\eta^2}\left(\phi_i^{n+1}-\phi_i^{*,n+1}  \right)
      + \sum_{j=1}^{N-1}\zeta_{ij}h_j(\vec{\phi}^{*,n+1})
      \right]
    + g_i^{n+1},
    \quad 1\leqslant i\leqslant N-1.
    \label{equ:phase_1}
  \end{multline}
  \begin{equation}
    \mathbf{n}\cdot\nabla\left[-\nabla^2\phi_i^{n+1}
      + \frac{S}{\eta^2}\left(\phi_i^{n+1} - \phi_i^{*,n+1}  \right)
      + \sum_{j=1}^{N-1}\zeta_{ij}h_j(\vec{\phi}^{*,n+1})  \right] = g_{ai}^{n+1},
    \quad 1\leqslant i\leqslant N-1,
    \ \ \text{on} \ \partial\Omega,
    \label{equ:phase_2}
  \end{equation}
  \begin{equation}
    \mathbf{n}\cdot\nabla\phi_i^{n+1}
    = \sum_{j=1}^{N-1}\sum_{s=1}^N y_{ij}\xi_{js}c_j(\vec{\phi}^{*,n+1})c_s(\vec{\phi}^{*,n+1})
    + g_{bi}^{n+1},
    \quad 1\leqslant i\leqslant N-1,
    \ \ \text{on} \ \partial\Omega.
    \label{equ:phase_3}
  \end{equation}
\end{subequations}
%
% explain symbols
In the above equations,
$\Delta t$ is the time step size,
and $S$ is a chosen constant to be determined later.
Let $J$ ($J=1$ or $2$) denote the temporal order of accuracy of
the algorithm.
If $\chi$ denotes an arbitrary variable, then
$\chi^{*,n+1}$ is a $J$-th order explicit approximation
of $\chi^{n+1}$ in the above equations, given by
\begin{equation}
  \chi^{*,n+1} = \left\{
  \begin{array}{ll}
    \chi^n, & J=1, \\
    2\chi^n - \chi^{n-1}, & J=2.
  \end{array}
  \right.
  \label{equ:var_star_def}
\end{equation}
$\hat{\chi}$ and $\gamma_0$ are such that
$
\frac{1}{\Delta t}(\gamma_0 \chi^{n+1}-\hat{\chi})
$
is an approximation of
$
\left. \frac{\partial\chi}{\partial t}\right|^{n+1}
$
using the $J$-th order backward differentiation formulation,
specifically given by
\begin{equation}
  \hat{\chi} = \left\{
  \begin{array}{ll}
    \chi^n, & J=1, \\
    2\chi^n - \frac{1}{2}\chi^{n-1}, & J=2,
  \end{array}
  \right.
  \quad
  \gamma_0 = \left\{
  \begin{array}{ll}
    1, & J=1, \\
    3/2, & J=2.
  \end{array}
  \right.
  \label{equ:var_hat_def}
\end{equation}

The key construction in the above scheme is
the extra term,
$
\frac{S}{\eta^2}\left(\phi_i^{n+1}-\phi_i^{*,n+1}  \right),
$
in equations \eqref{equ:phase_1} and \eqref{equ:phase_2}, 
which is equal to zero to the $J$-th order accuracy.
This extra term is inspired by the algorithm of \cite{DongS2012}
for two-phase flows.
This term, together with the explicit treatment of
the $\sum_{j=1}^{N-1}\zeta_{ij}h_j(\vec{\phi})$ term,
allows the reformulation of the ($N-1$) coupled
fourth-order phase field equations into
$2(N-1)$ Helmholtz-type equations that are de-coupled
from one another.

We employ $C^0$-continuous
high-order spectral elements \cite{KarniadakisS2005,ZhengD2011}
for spatial discretizations in the current work.
To facilitate the implementation of
the algorithm \eqref{equ:phase_1}--\eqref{equ:phase_3}
using spectral elements, we first reformulate
each of the fourth-order equations in \eqref{equ:phase_1}
into two de-coupled Helmholtz type equations,
and then obtain their weak forms.

Equation \eqref{equ:phase_1} can be re-written as
\begin{equation}
  \nabla^2\left[
    \nabla^2\phi_i^{n+1} - \frac{S}{\eta^2}\phi_i^{n+1}
    \right]
  + \frac{\gamma_0}{m_0\Delta t}\phi_i^{n+1}
  = Q_i + \nabla^2 R_i,
  \quad 1\leqslant i\leqslant N-1,
  \label{equ:phase_trans_1}
\end{equation}
where
\begin{equation}
  \left\{
  \begin{split}
    &
    Q_i = \frac{1}{m_0}\left(
    g_i^{n+1} + \frac{\hat{\phi}_i}{\Delta t}
    -\mathbf{u}^{*,n+1}\cdot\nabla\phi_i^{*,n+1}
    \right), \\
    &
    R_i = \sum_{j=1}^{N-1}\zeta_{ij} h_j(\vec{\phi}^{*,n+1})
    - \frac{S}{\eta^2}\phi_i^{*,n+1}.
  \end{split}
  \right.
  \label{equ:Qi_expr}
\end{equation}
Each of these fourth-order equations for $\phi_i^{n+1}$ is similar in form
to that encountered for two-phase flows. Therefore,
one can use an idea well-known for two-phase flows
 to re-formulate
each of the equations \eqref{equ:phase_trans_1} into two
de-coupled Helmholtz type equations. 
We refer to \cite{DongS2012} for the details
of this re-formulation, and will directly provide
the re-formulated forms as follows,
\begin{subequations}
  \begin{align}
    &
    \nabla\psi_i^{n+1} - \left(\alpha + \frac{S}{\eta^2}\right)\psi_i^{n+1}
    = Q_i + \nabla^2 R_i,
    \quad 1\leqslant i \leqslant N-1, \label{equ:CH_psi} \\
    &
    \nabla^2\phi_i^{n+1} + \alpha\phi_i^{n+1} = \psi_i^{n+1},
    \quad 1\leqslant i\leqslant N-1, \label{equ:CH_phi}
  \end{align}  
\end{subequations}
where $\psi_i^{n+1}$ is an auxiliary variable defined by
\eqref{equ:CH_phi}, $\alpha$ is a constant given by
\begin{equation}
  \alpha = \frac{S}{2\eta^2}\left[
    -1 + \sqrt{1 - \frac{4\gamma_0}{m_0\Delta t}\left(\frac{\eta^2}{S} \right)^2 }
    \right],
  \label{equ:alpha_expr}
\end{equation}
and the chosen constant $S$ must satisfy the
condition
\begin{equation}
  S \geqslant \eta^2\sqrt{\frac{4\gamma_0}{m_0\Delta t} }.
  \label{equ:S_expr}
\end{equation}
Note that the condition \eqref{equ:S_expr} implies that
$\alpha < 0$ and $\alpha+\frac{S}{\eta^2}>0$.
%The combination of the two equations \eqref{equ:CH_psi} and
%\eqref{equ:CH_phi} is equivalent
%to the single fourth-order equation \eqref{equ:phase_trans_1}.
To compute $\phi_i^{n+1}$, we first solve
equation \eqref{equ:CH_psi} for $\psi_i^{n+1}$,
and then solve equation \eqref{equ:CH_phi}
for $\phi_i^{n+1}$. It can be
noted that these two equations are  de-coupled.

% transform of BCs

In light of \eqref{equ:CH_phi}, the boundary condition
\eqref{equ:phase_2} can be transformed into
\begin{equation}
  \begin{split}
  \mathbf{n}\cdot\nabla\psi_i^{n+1} =&
  \left(\alpha + \frac{S}{\eta^2}  \right)\mathbf{n}\cdot\nabla\phi_i^{n+1}
  - g_{ai}^{n+1}
  + \mathbf{n}\cdot\nabla R_i \\
  =&
  \left(\alpha + \frac{S}{\eta^2}\right) \left[
    \sum_{j=1}^{N-1}\sum_{s=1}^N y_{ij}\xi_{js}c_j(\vec{\phi}^{*,n+1})c_s(\vec{\phi}^{*,n+1})
    + g_{bi}^{n+1}
    \right]
  - g_{ai}^{n+1}
  + \mathbf{n}\cdot\nabla R_i, 
  \end{split}
  \label{equ:phase_2_trans}
\end{equation}
where $R_i$ is defined in \eqref{equ:Qi_expr}, and 
we have used \eqref{equ:phase_3}.

% weak forms

We next derive the weak forms for the equations \eqref{equ:CH_psi}
and \eqref{equ:CH_phi}.
Let $\varphi\in H^1(\Omega)$ denote the test function.
Take the $L^2$ inner product between $\varphi$ and equation \eqref{equ:CH_psi},
and we get
\begin{multline}
  \int_{\Omega} \nabla\psi_i^{n+1}\cdot\nabla\varphi
  + \left(\alpha + \frac{S}{\eta^2}  \right)\int_{\Omega}\psi_i^{n+1}\varphi \\
  = -\int_{\Omega} Q_i\varphi
  + \int_{\Omega}\nabla R_i\cdot\nabla\varphi
  + \int_{\partial\Omega}\left(
  \mathbf{n}\cdot\nabla\psi_i^{n+1}
  -\mathbf{n}\cdot\nabla R_i
  \right) \varphi, \ \ \forall \varphi\in H^1(\Omega),
  \ \ 1\leqslant i\leqslant N-1,
\end{multline}
where we have used integration by part.
Substitution of $\mathbf{n}\cdot\nabla\psi_i^{n+1}$
in \eqref{equ:phase_2_trans} into the above equation
leads to the final weak form about $\psi_i^{n+1}$,
\begin{equation}
  \begin{split}
  \int_{\Omega} \nabla\psi_i^{n+1}\cdot\nabla\varphi
 & + \left(\alpha + \frac{S}{\eta^2}  \right)\int_{\Omega}\psi_i^{n+1}\varphi 
  = -\int_{\Omega} Q_i\varphi
  + \int_{\Omega}\nabla R_i\cdot\nabla\varphi
    \\
 & + \left(\alpha + \frac{S}{\eta^2}  \right)\int_{\partial\Omega}
  \left[
    \sum_{j=1}^{N-1}\sum_{s=1}^N y_{ij}\xi_{js}c_j(\vec{\phi}^{*,n+1})c_s(\vec{\phi}^{*,n+1}) + g_{bi}^{n+1}
    \right] \varphi
  -\int_{\partial\Omega}g_{ai}^{n+1}\varphi, \\
  &
  \ \forall \varphi\in H^1(\Omega),
  \ 1\leqslant i\leqslant N-1.
  \end{split}
  \label{equ:psi_weakform}
\end{equation}

Taking the $L^2$ inner product between
the test function $\varphi$ and equation \eqref{equ:CH_phi},
we obtain the weak form about $\phi_i^{n+1}$,
\begin{multline}
  \int_{\Omega}\nabla\phi_i^{n+1}\cdot\nabla \varphi
  -\alpha\int_{\Omega} \phi_i^{n+1}\varphi
  = -\int_{\Omega}\psi_i^{n+1}\varphi \\
  + \int_{\partial\Omega} \left[
    \sum_{j=1}^{N-1}\sum_{s=1}^N y_{ij}\xi_{js}c_j(\vec{\phi}^{*,n+1})c_s(\vec{\phi}^{*,n+1}) + g_{bi}^{n+1}
    \right] \varphi,
  \ \ \forall \varphi\in H^1(\Omega),
  \ \ 1\leqslant i\leqslant N-1
  \label{equ:phi_weakform}
\end{multline}
where we have used integration by part
and equation \eqref{equ:phase_3}.

Equations \eqref{equ:psi_weakform} and
\eqref{equ:phi_weakform} are in weak forms, and
all the terms on the right hand sides can be
computed directly using $C^0$ spectral elements
(or finite elements). These weak forms
can be discretized using the standard procedure
for spectral elements \cite{KarniadakisS2005}.

% summary of procedure for computing phase field variables
% laplace(phi_i)

Combining the above discussions, we arrive at
a method for computing the phase field variables numerically.
During each time step,
given $(\phi_i^n,\mathbf{u}^n)$,
we compute
$(\phi_i^{n+1},\nabla^2\phi_i^{n+1},\tilde{\mathbf{J}}^{n+1},\rho^{n+1},\mu^{n+1})$
using the following procedure (referred to as the {\bf Advance-Phase} 
hereafter), \\
\underline{{\bf Advance-Phase} procedure:}
\begin{enumerate}

\item
  Compute $Q_i$ and $R_i$ ($1\leqslant i\leqslant N-1$) based on
  \eqref{equ:Qi_expr}.

\item
  Solve equation \eqref{equ:psi_weakform}
  for $\psi_i^{n+1}$ ($1\leqslant i\leqslant N-1$).

\item
  Solve equation \eqref{equ:phi_weakform}
  for $\phi_i^{n+1}$ ($1\leqslant i\leqslant N-1$).

\item
  Compute $\nabla^2\phi_i^{n+1}$ ($1\leqslant i\leqslant N-1$) based on
  \eqref{equ:CH_phi} as follows:
  \begin{equation}
    \nabla^2\phi_i^{n+1} = \psi_i^{n+1} - \alpha\phi_i^{n+1},
    \quad 1\leqslant i\leqslant N-1.
    \label{equ:lap_phi}
  \end{equation}

\item
  Compute
  \begin{equation}
    \tilde{\mathbf{J}}^{n+1} = \tilde{\mathbf{J}}(\vec{\phi}^{n+1},\nabla\vec{\phi})
    \label{equ:J_expr_3}
  \end{equation}
  based on equation \eqref{equ:J_expr_1},
  where $\nabla^2\phi_i^{n+1}$ ($1\leqslant i\leqslant N-1$)
  are computed based on \eqref{equ:lap_phi}.

\item
  Compute
  \begin{equation}
    \rho^{n+1} = \rho(\vec{\phi}^{n+1}), \quad
    \mu^{n+1}  = \mu(\vec{\phi}^{n+1})
    \label{equ:rho_mu}
  \end{equation}
  based on equations \eqref{equ:density_expr} and
  \eqref{equ:mu_expr}, where $\varphi_i(\vec{\phi})$
  are given by \eqref{equ:gop_def}.
  When the maximum density ratio among
  the N fluids is large (beyond about $10^2$),
  we follow \cite{Dong2015} and further clamp the values
  of $\rho^{n+1}$ and $\mu^{n+1}$ as follows,
  \begin{equation}
    \rho^{n+1} = \left\{
    \begin{array}{ll}
      \rho^{n+1}, & \text{if} \ \rho^{n+1}\in[\tilde{\rho}_{\min},\tilde{\rho}_{\max}] \\
      \tilde{\rho}_{\max}, & \text{if} \ \rho^{n+1} > \tilde{\rho}_{\max} \\
      \tilde{\rho}_{\min}, & \text{if} \ \rho^{n+1} < \tilde{\rho}_{\min},
    \end{array}
    \right.
    \quad
    \mu^{n+1} = \left\{
    \begin{array}{ll}
      \mu^{n+1}, & \text{if} \ \mu^{n+1}\in[\tilde{\mu}_{\min},\tilde{\mu}_{\max}] \\
      \tilde{\mu}_{\max}, & \text{if} \ \mu^{n+1} > \tilde{\mu}_{\max} \\
      \tilde{\mu}_{\min}, & \text{if} \ \mu^{n+1} < \tilde{\mu}_{\min},
    \end{array}
    \right.
    \label{equ:rho_mu_clamp}
  \end{equation}
  where
  $\tilde{\rho}_{\max} = \max_{1\leqslant i\leqslant N}\{\tilde{\rho}_i\}$,
  $\tilde{\rho}_{\min} = \min_{1\leqslant i\leqslant N}\{\tilde{\rho}_i\}$,
  $\tilde{\mu}_{\max} = \max_{1\leqslant i\leqslant N}\{\tilde{\mu}_i\}$,
  and $\tilde{\mu}_{\min} = \min_{1\leqslant i\leqslant N}\{\tilde{\mu}_i\}$.
  
\end{enumerate}

% overall algorithm for N-phase flow simulations

We combine the algorithm for the phase field equations
of this section and the algorithm for the momentum equations
discussed in the Appendix D to obtain an overall method
for N-phase flow simulations. Given
$(\mathbf{u}^n,P^n,\phi_i^n)$, our method consists of
the following de-coupled steps for computing
$(\mathbf{u}^{n+1},P^{n+1},\phi_i^{n+1})$:
\begin{itemize}

\item
  Using the {\bf Advance-Phase} procedure discussed above
  to compute $\phi_i^{n+1}$, $\nabla^2\phi_i^{n+1}$,
  $\tilde{\mathbf{J}}^{n+1}$, $\rho^{n+1}$, and $\mu^{n+1}$.

\item
  Solve equation \eqref{equ:p_weakform} in  Appendix D for $P^{n+1}$.

\item
  Solve equation \eqref{equ:vel_weakform} in the Appendix D for
  $\mathbf{u}^{n+1}$.
  
\end{itemize}

\section{Representative Numerical Tests}
\label{sec:tests}

% what will be discussed in this section?

In this section we test the method presented in previous sections
by considering
several two-dimensional flow  problems involving multiple
immiscible incompressible fluid components
and partially wettable solid-wall surfaces.
These multiphase problems involve large density
ratios and large viscosity ratios among
the fluids, and the wettability of the wall surface
has a significant influence on the behavior of the system.
We will compare numerical simulation results with
theory to demonstrate the physical accuracy
of the method developed herein.

\begin{table}
\begin{center}
\begin{tabular*}{1.0\textwidth}{@{\extracolsep{\fill}} l c | l c}
\hline
variables/parameters & normalization constant &
variables/parameters & normalization constant \\
$\mathbf{x}$, $\eta$ & $L$ 
& $t$, $\Delta t$ & $L/U_0$ \\
$\mathbf{u}$, $\tilde{\mathbf{u}}$, $\mathbf{w}$ & $U_0$
& $p$, $P$, $W(\vec{\phi},\nabla\vec{\phi})$, $H(\vec{\phi})$ & $\varrho_dU_0^2$ \\
$\lambda_{ij}$, $\Lambda_{ij}$ & $\varrho_dU_0^2L^2$
& $\rho$, $\rho_i$, $\tilde{\rho}_i$, $\rho_0$, $\varphi_i$, $a_{ij}$, $b_i$  & $\varrho_d$ \\
$\mu$, $\tilde{\mu}_i$ & $\varrho_dU_0L$ 
& $m_0$ & $U_0L^3$ \\
$\zeta_{ij}$, $\Theta_{ij}$ & $\frac{1}{\varrho_dU_0^2L^2}$
& $\tilde{\mathbf{J}}$, $\mathbf{J}_{ai}$ & $\varrho_dU_0$ \\
$\sigma_{ij}$ & $\varrho_dU_0^2L$
& $g_i$ & $U_0/L$ \\
$\mathbf{f}$ & $\varrho_dU_0^2/L$
& $g_{ai}$ & $1/L^3$ \\
$\phi_i$, $c_i$, $S$, $y_{ij}$, $\gamma_0$ & $1$
& $\xi_{ij}$, $g_{bi}$  & $1/L$ \\
$e_{ij}$, $\tilde{\gamma}_i$, $\Gamma$ & $1/\varrho_d$
& $\Gamma_{\mu}$, $\nu_0$ & $U_0L$ \\
$\alpha$, $\psi_i$ & $1/L^2$ 
& $\tilde{m}_{ij}$ & $\varrho_dL/U_0$ \\
$\mathcal{C}_i$ & $U_0^2$
& $g_r$ (gravity) & $U_0^2/L$ \\
\hline
\end{tabular*}
\caption{
Normalization of flow variables and physical parameters.
$L$ is a length scale. $U_0$ is a velocity scale.
$\varrho_d$ is a density scale.
}
\label{tab:normalization}
\end{center}
\end{table}

% normalization of variables and parameters

We briefly mention the normalization of physical
variables and parameters. 
As discussed in detail in \cite{Dong2014},
as long as the variables are normalized consistently,
the non-dimensionalized problem
(governing equations, boundary/initial conditions)
will have the same form as the dimensional problem.
Let $L$ denote a length scale, $U_0$ denote a
velocity scale, and ${\varrho}_d$ denote a density scale.
In Table \ref{tab:normalization} we list the normalization
constants for all the physical variables and parameters.
For example, the non-dimensional pairwise surface tension is 
$\frac{\sigma_{ij}}{{\varrho}_dU_0^2L}$ according to
this table. In subsequent discussions
all variables are assumed to be in non-dimensional forms unless otherwise
specified, and the normalization is conducted
based on Table \ref{tab:normalization}.

\subsection{Convergence Rates}

\begin{figure}
\centerline{
\includegraphics[height=2.5in]{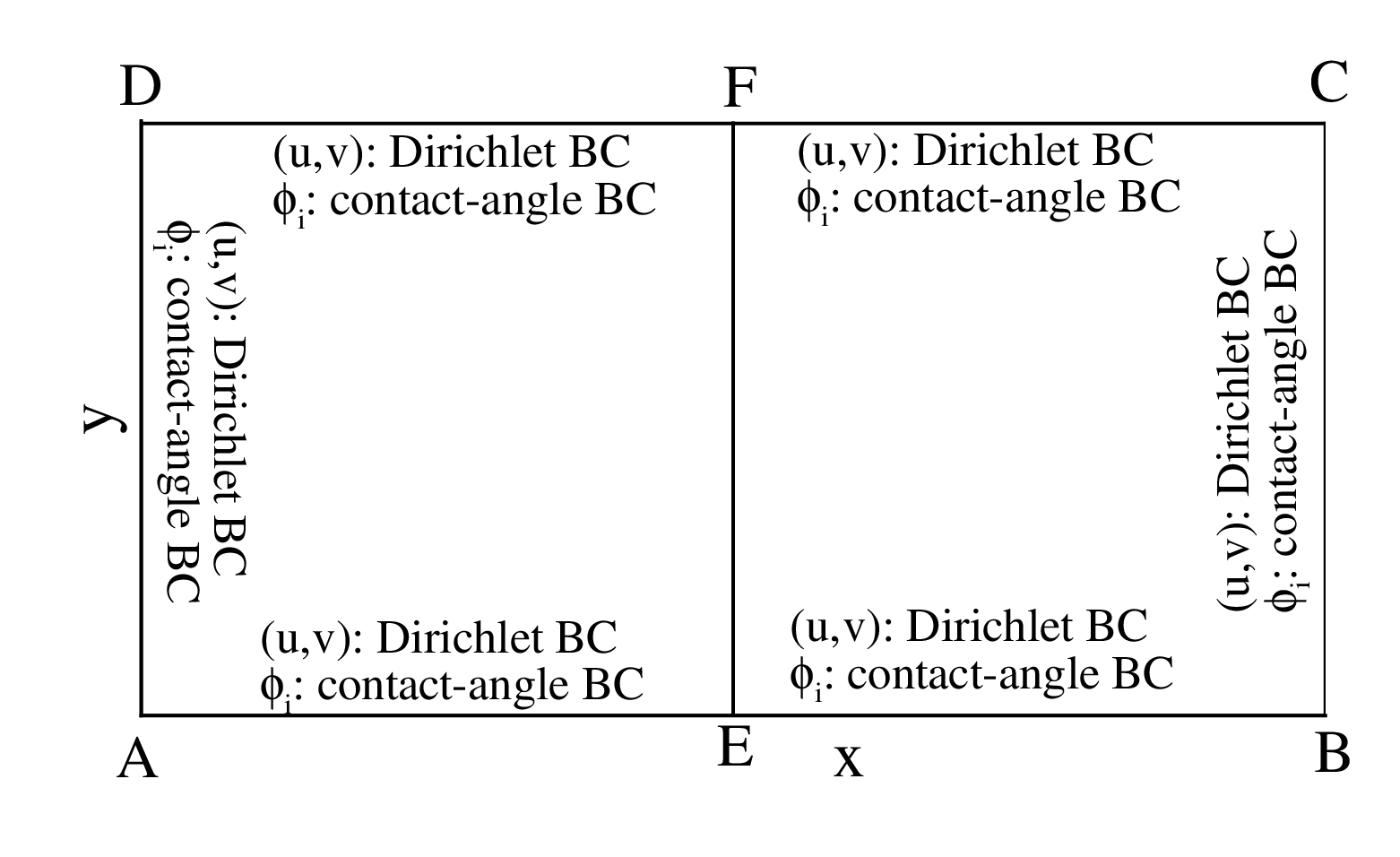}(a)
}
\centerline{
\includegraphics[width=3in]{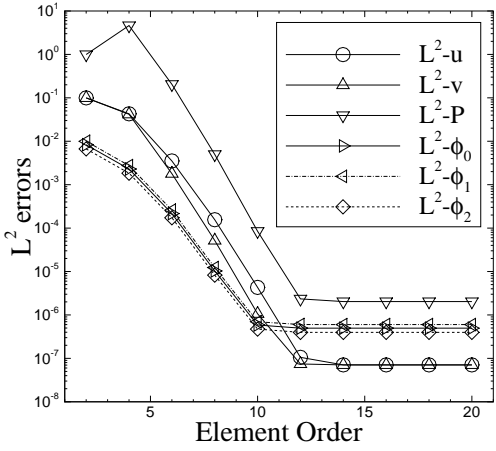}(b)
\includegraphics[width=3in]{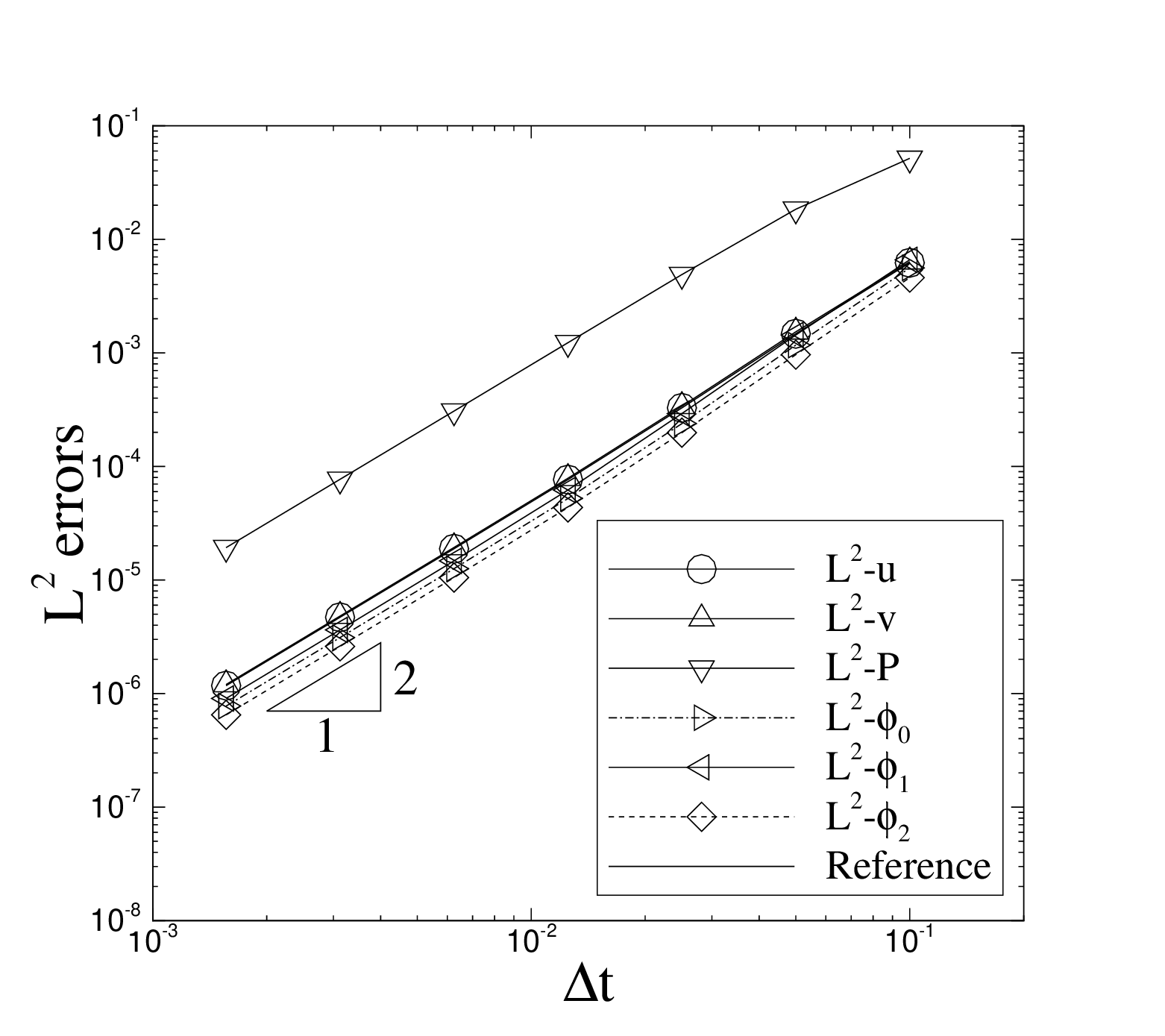}(c)
}
\caption{
Spatial/temporal convergence rates:
(a) Problem configuration;
(b) $L^2$ errors of flow variables versus element order
(fixed $\Delta t=0.001$ and $t_f=0.1$);
(c) $L^2$ errors of flow variables versus time step 
size $\Delta t$ (fixed element order $16$ and $t_f=1.0$).
}
\label{fig:conv}
\end{figure}

The goal of this section is to demonstrate 
the spatial and temporal convergence rates
of the method developed herein
using a contrived analytic solution under
non-trivial contact angles.

The setting of this problem is 
illustrated in Figure \ref{fig:conv}(a).
We consider a rectangular domain $\overline{ABCD}$,
$0\leqslant x\leqslant 2$ and $-1\leqslant y\leqslant 1$,
and a four-fluid mixture (i.e. $N=4$) contained in this domain.
Assume field distributions for the velocity, pressure
and phase field functions given by 
the following expressions
\begin{equation}
\left\{
\begin{split}
&
u = A_0 \sin (ax) \cos(\pi y)\sin(\omega_0 t) \\
&
v = -(A_0 a/\pi) \cos (ax)\sin(\pi y) \sin(\omega_0 t) \\
&
P = A_0\sin(ax)\sin(\pi y) \cos(\omega_0 t) \\
&
\phi_1 = \frac{1}{6}\left[1 + A_1\cos(a_1x)\cos(\bar{b}_1y)\sin(\omega_1 t)  \right] \\
&
\phi_2 = \frac{1}{6}\left[1 + A_2\cos(a_2x)\cos(\bar{b}_2y)\sin(\omega_2 t)  \right] \\
&
\phi_3 = \frac{1}{6}\left[1 + A_3\cos(a_3x)\cos(\bar{b}_3y)\sin(\omega_3 t)  \right], \\
\end{split}
\right.
\label{equ:anal_expr}
\end{equation}
where ($u$, $v$) are the $x$ and $y$ components
of the velocity $\mathbf{u}$.
$A_i$, $\omega_i$ ($0\leqslant i\leqslant 3$), 
$a$, $a_i$ and $\bar{b}_i$ ($1\leqslant i\leqslant 3$)
are constant parameters to be specified later.
The above  expressions for the velocity
satisfy equation \eqref{equ:continuity}.
We choose the $\mathbf{f}$ term in equation \eqref{equ:nse_final}
such that the analytic expressions given in \eqref{equ:anal_expr}
satisfy equation \eqref{equ:nse_final}.
We choose the $g_i$ ($1\leqslant i\leqslant 3$)
terms in equations \eqref{equ:CH_final_1} such that
the expressions from \eqref{equ:anal_expr}
satisfy each of the ($N-1$) equations \eqref{equ:CH_final_1}.

% BCs and ICs

For the boundary conditions, we impose velocity Dirichlet condition
\eqref{equ:dbc} on all domain boundaries,
where the boundary velocity $\mathbf{w}$ are chosen according
to the velocity analytic expressions in \eqref{equ:anal_expr}.
For the phase field variables we impose 
the contact-angle conditions 
\eqref{equ:bc_chempot_2}--\eqref{equ:cabc_2}
on the domain boundaries,
where the source terms $g_{ai}$ and $g_{bi}$ ($1\leqslant i\leqslant 3$)
are chosen such that the analytic expressions for $\phi_i$
from \eqref{equ:anal_expr} satisfy the equations
\eqref{equ:bc_chempot_2}--\eqref{equ:cabc_2}
on the boundaries with contact angles
$(\theta_{14},\theta_{24},\theta_{34})=(120^0,30^0,135^0)$.
For the initial conditions, we choose the initial
velocity and initial phase field distributions
according to the analytic expressions of \eqref{equ:anal_expr}
by setting $t=0$.

\begin{table}
\begin{center}
\begin{tabular}{ l c | l c}
\hline
Parameters & Values & Parameters & Values \\ \hline
$A_0$ & $2.0$ 
& $A_1$, $A_2$, $A_3$ & $1.0$ \\
$a$, $a_1$, $a_2$, $a_3$ & $\pi$
& $\bar{b}_1$, $\bar{b}_2$, $\bar{b}_3$ & $\pi$ \\
$\omega_0$, $\omega_1$ & $1.0$
& $\omega_2$ & $1.2$ \\
$\omega_3$ & $0.8$ 
& $\eta$ & $0.1$ \\
$\tilde{\rho}_1$ & $1.0$
& $\tilde{\rho}_2$ & $3.0$ \\
$\tilde{\rho}_3$ & $2.0$
& $\tilde{\rho}_4$ & $4.0$ \\
$\tilde{\mu}_1$ & $0.01$
& $\tilde{\mu}_2$ & $0.02$ \\
$\tilde{\mu}_3$ & $0.03$
& $\tilde{\mu}_4$ & $0.04$ \\
$\sigma_{12}$ & $6.236E-3$
& $\sigma_{13}$ & $7.265E-3$ \\
$\sigma_{14}$ & $3.727E-3$
& $\sigma_{23}$ & $8.165E-3$ \\
$\sigma_{24}$ & $5.270E-3$
& $\sigma_{34}$ & $6.455E-3$ \\
$\theta_{14}$ & $120^0$
& $\theta_{24}$ & $30^0$ \\
$\theta_{34}$ & $135^0$
& $m_0$ & $1.0E-5$ \\
$\rho_0$ & $\min(\tilde{\rho}_1,\dots,\tilde{\rho}_4)$ 
& $\nu_0$ & $\max\left(\frac{\tilde{\mu}_1}{\tilde{\rho}_1},\dots,\frac{\tilde{\mu}_4}{\tilde{\rho}_4}\right)$ \\
 $J$ (temporal order) & $2$ 
& $t_f$ & $0.1$ or $1.0$  \\
 $\Delta t$ & (varied) 
& Number of elements & $2$ \\
 Element order & (varied) \\
\hline
\end{tabular}
\caption{Parameter values for the convergence-rate tests.}
\label{tab:conv}
\end{center}
\end{table}

% how to simulate?

To simulate the problem we discretize
the domain using two quadrilateral spectral elements
of the same size, as shown in Figure \ref{fig:conv}(a).
The volume fractions are chosen as the order parameters,
as defined in \eqref{equ:gop_volfrac}.
We integrate in time, from $t=0$ to $t=t_f$ ($t_f$ to
be specified later),
the governing equations for this four-phase
system using the algorithm developed in Section \ref{sec:algorithm}.
Then we compare the numerical solution and the exact
solution given in \eqref{equ:anal_expr} at $t=t_f$
to quantify the numerical errors for different
flow variables.
The values for the physical and numerical parameters
of this problem are listed in Table \ref{tab:conv}.

Two group of tests have been performed. In the first group,
we fix the integration time at $t_f=0.1$ and 
the time step size at $\Delta t=0.001$ ($100$ time steps),
and vary the element order systematically
between $2$ and $20$. The same element order has been used
for all elements.
Figure \ref{fig:conv}(b) plots the numerical errors at $t=t_f$
in $L^2$ norm for different flow variables
as a function of the element order.
It is evident that within a certain range of
the element order (below about $12$) 
the errors decrease exponentially with increasing element order,
exhibiting an exponential convergence rate in space.
Beyond the element order of about $12$, 
the error curves level off as the element order
further increases, exhibiting a saturation caused
by the temporal truncation error.

In the second group of tests, we fix the integration
time at $t_f=1.0$ and the element order at a large value $16$,
and vary the time step size systematically
between $\Delta t=0.0015625$ and $\Delta t=0.1$.
Figure \ref{fig:conv}(c) shows the numerical errors at $t=t_f$
in $L^2$ norm for different variables as a function
of $\Delta t$ in logarithmic scales.
A reference line for a temporal second-order convergence 
rate has also been shown in the plot.
It is evident that the numerical errors exhibit a second-order
convergence rate in time.  

The above results indicate that the method developed
herein has a spatial exponential convergence rate
and a temporal second-order convergence rate
with multiple fluid components and different
contact angles in the system.

\subsection{Equilibrium Liquid Drops on  Partially Wettable Wall --
Comparison with de Gennes Theory}
\label{sec:3p_compare}

In this section we study the equilibrium configurations of
two liquid drops -- a water drop and an oil drop in ambient air --
that are sufficiently far apart from each other,
resting on a partially-wettable horizontal wall. 
This is a three-phase problem ($N=3$). However,
because the two liquid drops are far apart,
their interactions are weak, and 
the equilibrium shape of each drop will
essentially be the same as the shape of that drop alone 
in the air.
This allows us to compare qualitatively and quantitatively
the numerical results of three-phase simulations on the contact angle
effects
with the de Gennes theory~\cite{deGennesBQ2003} on
the equilibrium drop shape for two-phase problems.

% zero gravity

\begin{figure}
\centering
\includegraphics[width=6in]{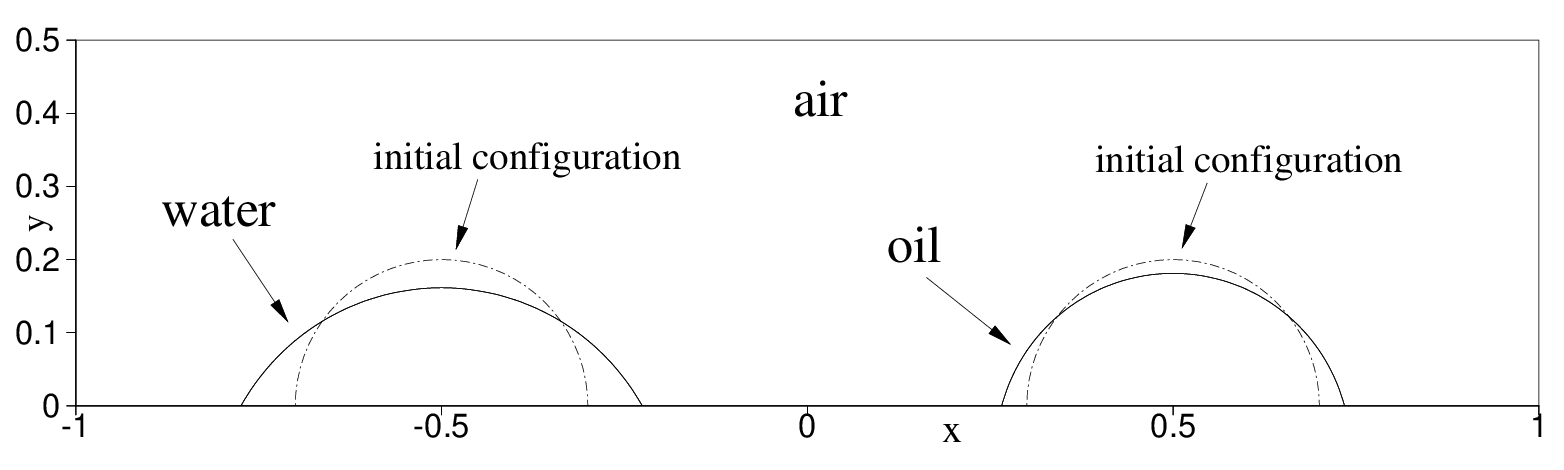}(a)
\includegraphics[width=6in]{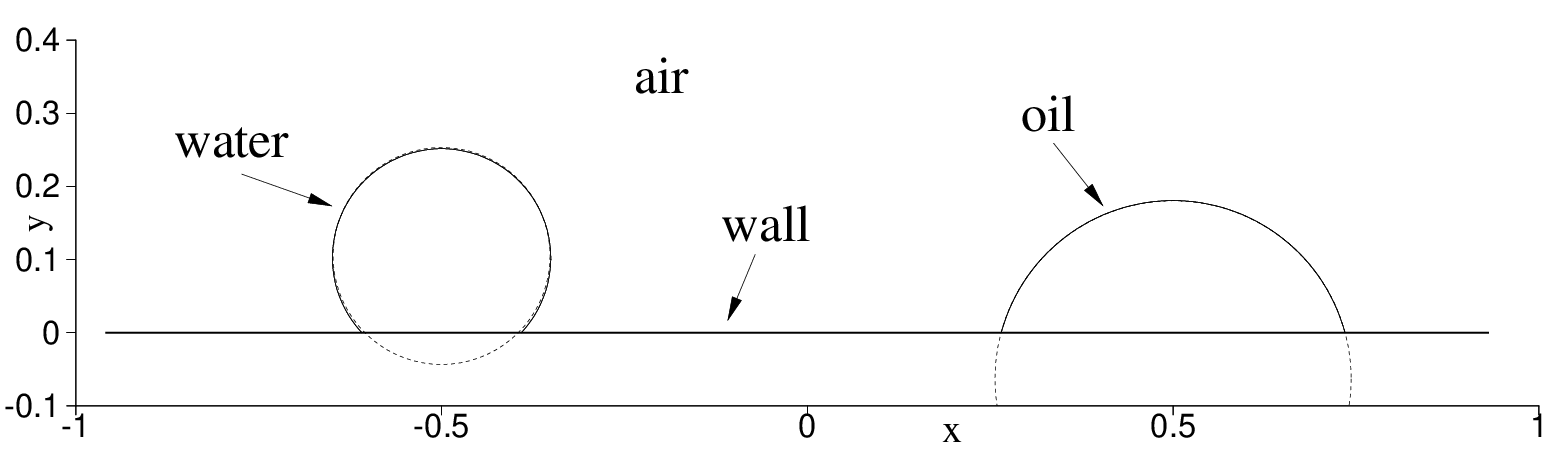}(b)
\caption{
Three-phase flows (zero gravity), 
equilibrium configurations of a water drop and an oil
drop on a partially-wettable horizontal wall surface with contact angles:
(a) $(\theta_{aw},\theta_{ao})=(60^0, 75^0)$;
(b) $(\theta_{aw},\theta_{ao})=(135^0, 75^0)$.
Solid curves denote the drop profiles from the simulations,
visualized by the contour levels $c_i=\frac{1}{2}$ ($i=1,2,3$).
In (a) the dashed-dot curves show the initial configurations (half circles)
of the liquid drops.
In (b) the dashed curves show the theoretical equilibrium drop profiles
for the corresponding contact angles.
}
\label{fig:zero_G_config}
\end{figure}

We consider a rectangular domain,
$-L\leqslant x\leqslant L$ and $0\leqslant y\leqslant L/2$,
where $L=4$cm; see Figure \ref{fig:zero_G_config}(a).
The bottom and top sides of the domain are solid walls,
with certain wettability properties.
In the horizontal direction the domain is assumed
to be periodic at $x=\pm L$.
The domain is filled with air.
Two liquid drops, a water drop and an oil drop,
both initially semi-circular with a radius $R_0=L/5$,
are held at rest on the bottom wall.
Initially the center of the water drop
is at $\mathbf{X}_w=(x_{0w},y_{0w})=(-L/2,0)$,
and the center of the oil drop is at
$\mathbf{X}_o=(x_{0o},y_{0o})=(L/2,0)$.
The gravity is in the vertical direction, pointing
downward.
We use $\theta_{aw}$ and $\theta_{ao}$ to
denote the static contact angles of the air-water interface
and the air-oil interface on the wall, respectively.
Note that these are the angles measured on the side of
the liquid by convention.
At $t=0$ the system is released and starts to evolve,
eventually reaching an equilibrium state.
Our goal is to study the equilibrium configuration
of this three-phase system.

\begin{table}
\begin{center}
\begin{tabular}{l l l l}
\hline
Density $[kg/m^3]$: & Air -- $1.2041$ & Water -- $998.207$ & Oil -- $870$ \\
Dynamic viscosity $[kg/(m\cdot s)]$: &  Air -- $1.78\times 10^{-4}$ & Water -- $1.002\times 10^{-3}$
& Oil -- $9.15\times 10^{-2}$ \\
Surface tension $[kg/s^2]$: & Air/water -- $0.0728$ &
Air/oil -- $0.055$ (or varied) &
Oil/water -- $0.04$ \\
Gravitational acceleration $[m/s^2]$: & $0$ or $9.8$ (or varied) \\
\hline
\end{tabular}
\caption{Physical parameter values for the air/water/oil
three-phase system.}
\label{tab:air_water_param}
\end{center}
\end{table}

% physical parameter values, 
% and normalization

The physical parameters for this problem include
the densities and dynamic viscosities of 
the three fluids (air, water and oil),
as well as the three pairwise
surface tensions among them, together with
the gravitational acceleration.
The values for these parameters employed in
this paper are listed in 
Table \ref{tab:air_water_param}.
We use $L$ as the length scale and
the air density as the density scale $\varrho_d$,
and choose the velocity scale as
$U_0 = \sqrt{g_{r0}L}$,
where $g_{r0}=1m/s^2$. 
Then the problem is normalized
based on  Table \ref{tab:normalization}.

% how do you simulate the problem?
% mesh and BCs, ICs

In the simulations we assign the water, oil and air 
as the first, second, and third fluids,
respectively. The two independent contact angles 
for this three-phase system are therefore
$\theta_{13}$ (air-water contact angle $\theta_{aw}$) and
$\theta_{23}$ (air-oil contact angle $\theta_ao$).
We employ  the 
volume fractions $c_i$ ($1\leqslant i\leqslant N-1$)
as the order parameters; see equations
\eqref{equ:gop_volfrac} and \eqref{equ:volfrac_aij}.

To simulate this problem we partition 
the domain using $100$ quadrilateral spectral elements
of the same size, with $20$ elements along
the $x$ direction and $5$ elements along 
the $y$ direction. We employ an element order of
$14$ within each element. 
%
% BCs and ICs, numerical parameter values
%
On the top and bottom walls ($y=0,L/2$),
we impose the Dirichlet condition \eqref{equ:dbc}
with $\mathbf{w}=0$ for the velocity,
and the contact-angle boundary conditions
\eqref{equ:bc_chempot_2}--\eqref{equ:cabc_2}
with $g_{ai}=0$ and $g_{bi}=0$ for
the phase field variables $\phi_i$ ($1\leqslant i\leqslant N-1$).
In the horizontal direction,
periodic boundary conditions are imposed
for all flow variables.

% numerical parameter values

\begin{table}
\begin{center}
\begin{tabular}{l l}
\hline
Parameters & Values \\
$\zeta_{ij}$ & Computed based on \eqref{equ:theta_expr} and \eqref{equ:lambda_mat_gop} \\
$\eta/L$ & $0.01$ \\
$m_0/(U_0L^3)$ & $10^{-9}$ \\
$\rho_0$ & $\min(\tilde{\rho}_1, \tilde{\rho}_2, \tilde{\rho}_3)$ \\
$\nu_0$ & $5\max\left(\frac{\tilde{\mu}_1}{\tilde{\rho}_1},\frac{\tilde{\mu}_2}{\tilde{\rho}_2}, \frac{\tilde{\mu}_3}{\tilde{\rho}_3}\right)$ \\
$S$ & $\eta^2\sqrt{\frac{4\gamma_0}{m_0\Delta t} }$ \\
$\alpha$ & Computed based on \eqref{equ:alpha_expr} \\
$U_0\Delta t/L$ & $2.0\times 10^{-6}$ \\
$\theta_{13}$, $\theta_{23}$ & ranging between $15^0$ and $165^0$ \\
$J$ (temporal order) & $2$ \\
Number of elements & $100$ \\
Element order & $14$ \\
\hline
\end{tabular}
\caption{Simulation parameter values for air/water/oil three-phase
problem.}
\label{tab:param_comp}
\end{center}
\end{table}

The governing equations \eqref{equ:nse_final}, \eqref{equ:continuity},
and \eqref{equ:CH_final_1} with $g_i=0$,
together with the above boundary conditions are
solved using the algorithm presented in Section
\ref{sec:algorithm}. 
The initial velocity is set to $\mathbf{u}=0$,
and the initial phase field distributions are set
to
\begin{equation}
\left\{
\begin{split}
&
\phi_0 = \frac{1}{2}\left(1 - \tanh\frac{|\mathbf{x}-\mathbf{X}_w|-R_0}{\sqrt{2}\eta}  \right) \\
&
\phi_1 = \frac{1}{2}\left(1 - \tanh\frac{|\mathbf{x}-\mathbf{X}_o|-R_0}{\sqrt{2}\eta}  \right)
\end{split}
\right.
\label{equ:init_phase}
\end{equation}
where $\mathbf{X}_w$ and $\mathbf{X}_o$
are the initial center coordinates of the 
water and oil drops, respectively.
Table \ref{tab:param_comp}
summaries the values of the numerical parameters
employed in the simulations.

% what else to discuss about simulation settings?

% sketch of equilibrium drop shape

\begin{figure}
\centerline{
\includegraphics[width=2.8in]{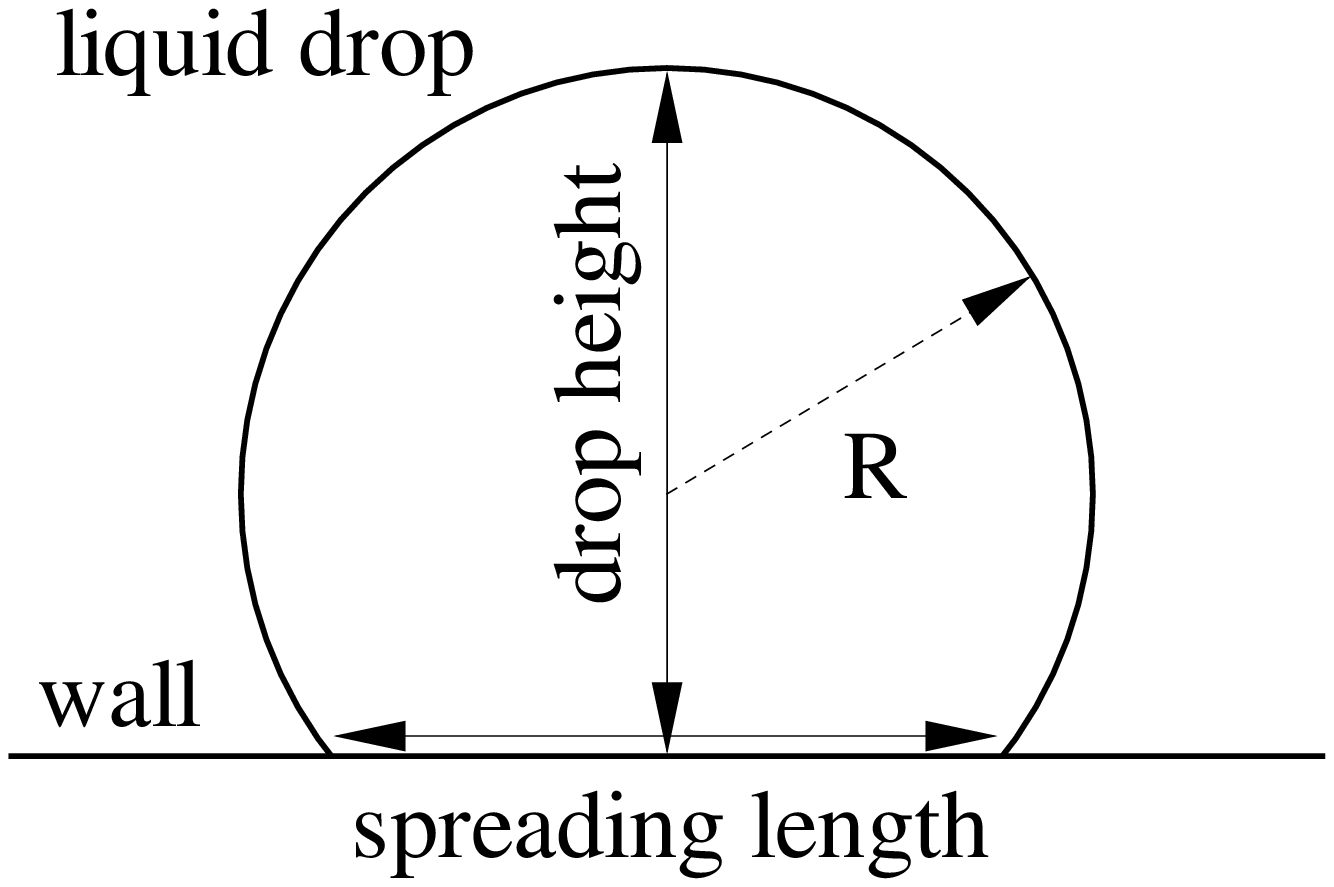}(a)
\includegraphics[width=3.1in]{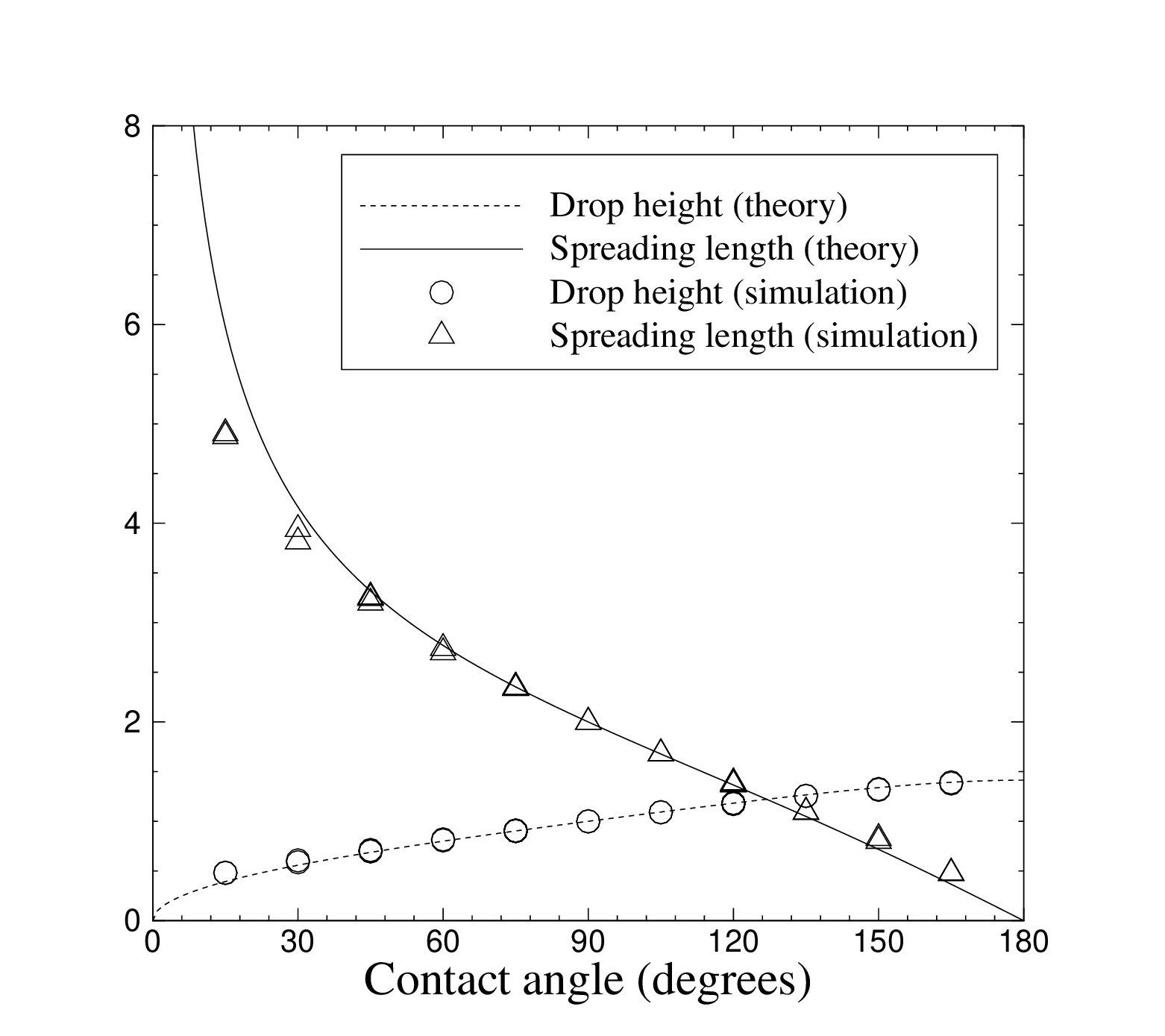}(b)
}
\caption{
Three-phase flows (zero gravity):
(a) Sketch of the equilibrium liquid-drop shape.
(b) Comparison of the equilibrium drop height and spreading length as
a function of the contact angle between the simulations and 
the de Gennes theory~\cite{deGennesBQ2003}.
Symbols include simulation results for both the water and oil
drops.
Multiple data points corresponding to a contact-angle value are results from
different simulation cases with different 
$(\theta_{aw},\theta_{ao})$ combinations. 
For example, two combinations with 
$(\theta_{aw},\theta_{ao})=(60^0, 75^0)$ and 
$(\theta_{aw},\theta_{ao})=(120^0, 60^0)$ both 
contribute to the data for contact angle $60^0$.
}
\label{fig:zero_G_param}
\end{figure}

\subsubsection{Zero Gravity}

We first focus on the case with no gravity, i.e.
$g_r=0$, where $g_r$ denotes the magnitude of
gravitational acceleration.

Since the two drops are sufficiently far apart on the wall,
their influence on each other is small. 
In case of zero gravity the surface tensions are the
only forces that come into play in the system.
The equilibrium profile of
each drop will be a circular cap (or a spherical cap
in three dimensions), which intersects 
the wall surface at the prescribed 
contact angle~\cite{deGennesBQ2003}.

% qualitative comparison

In Figure \ref{fig:zero_G_config} we show the
equilibrium configurations of the system
from the simulations corresponding to
two sets of contact-angle values,
$(\theta_{aw},\theta_{ao})=(60^0,75^0)$
for Figure \ref{fig:zero_G_config}(a)
and $(\theta_{aw},\theta_{ao})=(135^0,75^0)$
for Figure \ref{fig:zero_G_config}(b).
The solid curves correspond to the contour levels
$c_i = \frac{1}{2}$ ($1\leqslant i\leqslant 3$).
In Figure \ref{fig:zero_G_config}(a) 
the dashed-dot curves correspond to the 
initial profiles (semi-circular)
of the water and oil drops.
In Figure \ref{fig:zero_G_config}(b)
we have also shown two dashed circles as references.
The intersecting angles of these circles at the wall
are exactly $135^0$ (in the water region)
and $75^0$ (in the oil region).
In addition, the caps
formed between these dashed circles and the wall
have exactly the same area as
the initial semi-circular shapes
of the water and oil drops (i.e. $\frac{1}{2}\pi R_0^2$).
The water-drop and oil-drop 
profiles obtained from the simulations (solid curves)
almost exactly overlap with those of 
the dashed circular caps in Figure \ref{fig:zero_G_config}(b).
This indicates that the simulation has
produced results that are qualitatively consistent
with the theory~\cite{deGennesBQ2003}.

% quantitative comparison

To provide a quantitative comparison, we focus on
the parameters spreading length $L_s$ and drop height $H_d$,
as defined in Figure \ref{fig:zero_G_param}(a),
of the equilibrium drop profile at zero gravity.
Let $R$ denote the radius of the circle at equilibrium,
and $\theta_E$ denote the contact angle.
Then based on the volume conservation 
of the liquid drop we can obtain the following
relations~\cite{deGennesBQ2003,Dong2012},
\begin{equation}
\left\{
\begin{split}
&
R = R_0\sqrt{\frac{\pi/2}{\theta_E - \sin\theta_E\cos\theta_E}} \\
&
H_d = R(1-\cos\theta_E) \\
&
L_s = 2R\sin\theta_E,
\end{split}
\right.
\label{equ:3p_param_exact}
\end{equation}
where the initial drop profile is assumed to
be semi-circular with a radius $R_0$.
These theoretical expressions for the equilibrium drop parameters
allow for quantitative
comparisons with numerical simulations.

We have performed a series of numerical simulations of
this three-phase problem with various combinations
for the contact angles $(\theta_{aw},\theta_{ao})$,
in particular, with a fixed $\theta_{aw}=120^0$
and $\theta_{ao}$ varied systematically 
between $15^0$ and $165^0$, and with a fixed
$\theta_{ao}=75^0$ and $\theta_{aw}$ varied systematically
between $15^0$ and $165^0$.
For each pair of contact angles $(\theta_{aw},\theta_{ao})$,
we have conducted simulations of this three-phase problem, and
obtained the spreading length and the drop height
from the equilibrium profiles of the water and oil drops.
In Figure \ref{fig:zero_G_param}(b)
we plot the spreading length and the drop height (symbols)
as a function of the contact angle for both the water
and the oil drops. 
For comparison we have also included in this plot the theoretical
relations given by \eqref{equ:3p_param_exact}
(see the solid/dashed curves).
Note that at zero gravity the theoretical relations
\eqref{equ:3p_param_exact} apply to both water and oil drops.
Accordingly, we have not differentiated the water 
and oil drops when plotting the numerical results
in Figure \ref{fig:zero_G_param}(b), and the symbols
represent results for both the water drops and the oil drops
from the simulations.
We observe that the numerical results for the equilibrium drop
height are in good agreement with the theoretical
 results for the whole range of contact-angle values ($[15^0,165^0]$)
consider here.
For the spreading length, the numerical results also agree
quite well with the theoretical results in the bulk
range of contact angles ($[30^0,150^0]$).
However, at very small (or very large) contact angles
(e.g. $15^0$ and $165^0$)
we observe a larger discrepancy between
the numerically obtained values and the theoretical
values for the spreading length.
 
The results of this subsection indicate that
our simulation results compare favorably with
the de Gennes theory~\cite{deGennesBQ2003}
both qualitatively and quantitatively at zero gravity
with multiple fluid components and multiple types
of contact angles.

% what else to discuss here?

\subsubsection{Effects of Gravity and Surface Tension}

In this subsection we consider how the 
gravity influences the equilibrium profiles of
the water and oil drops.
We will also look into the effect of surface tensions
on the system.

% gravity effects
% 3 fluid phases and 4 fluid phases

\begin{figure}
\centering
\includegraphics[width=5in]{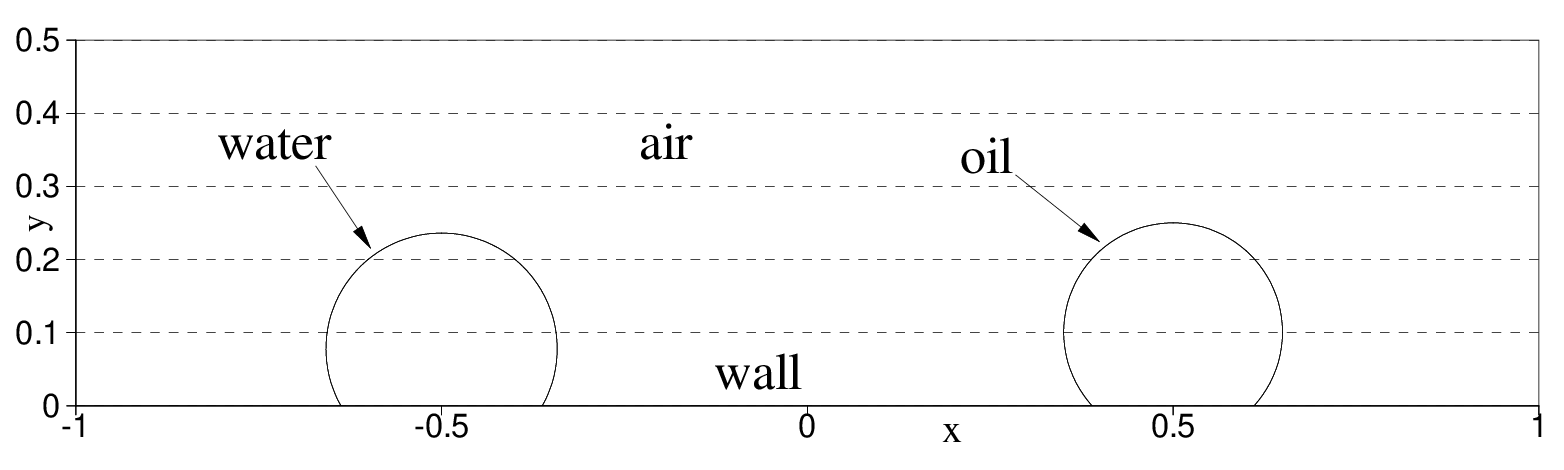}(a)
\includegraphics[width=5in]{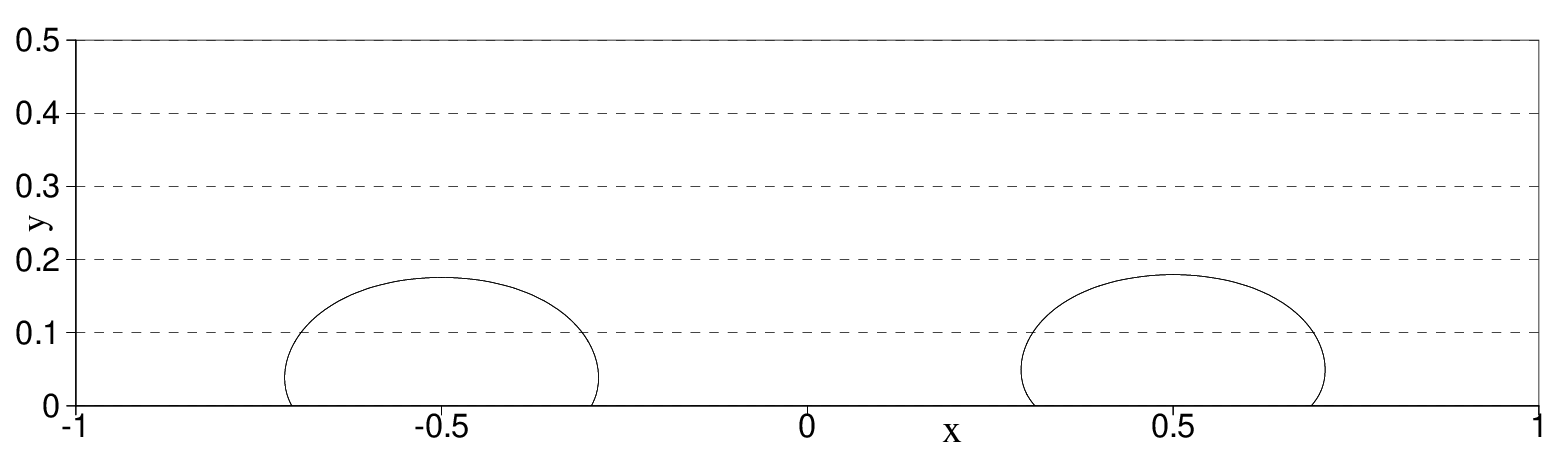}(b)
\includegraphics[width=5in]{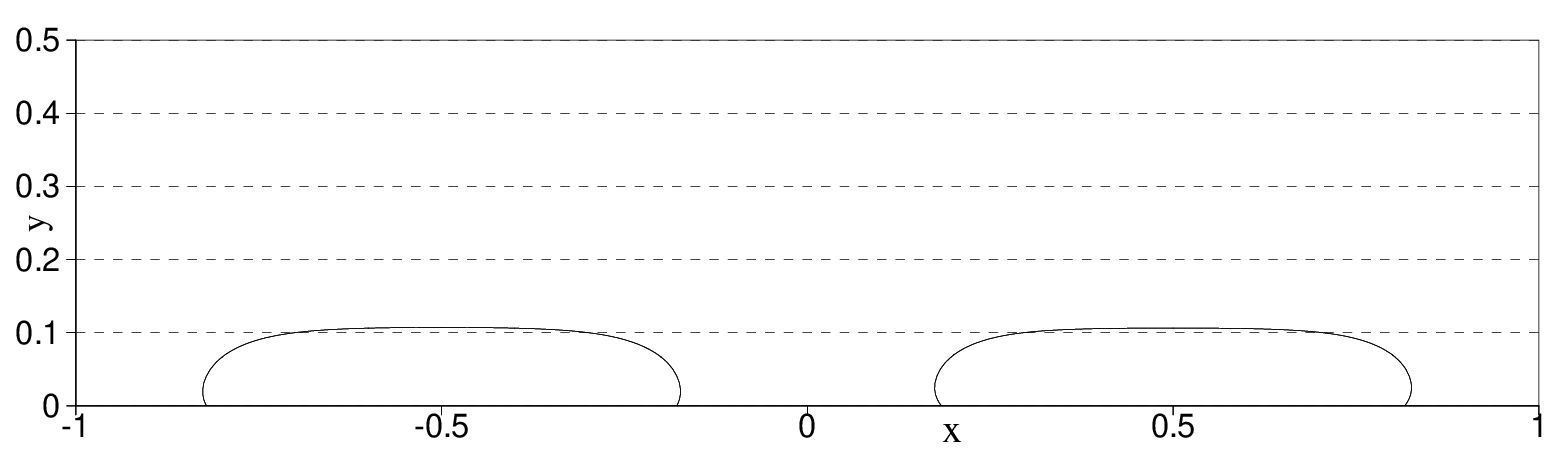}(c)
\caption{
Three-phase flows,
effect of gravity on equilibrium profiles of
water/oil drops with contact angles
$(\theta_{aw},\theta_{ao})=(120^0, 135^0)$:
(a) zero gravity,
(b) gravity $g_r = 3m/s^2$,
(c) gravity $g_r = 12 m/s^2$.
}
\label{fig:3p_prof_G}
\end{figure}

In the presence of gravity, the equilibrium profile
of the liquid drop is determined by
the balance of three effects.
These effects are associated with
(i) the gravity, which tends to spread the drop
on the wall;
(ii) the surface tension, which tends to restore
the drop to a circular cap;
(iii) the contact angle, which the drop profile
must respect at the wall.
More specifically, one can define a capillary length associated 
with the liquid-air interface~\cite{deGennesBQ2003},
$\kappa^{-1}=\sqrt{\frac{\sigma_{la}}{\rho_l g_r}}$,
where $\sigma_{la}$ is the surface tension associated
with the interface, $\rho_l$ is the liquid density
and $g_r$ is the magnitude of the gravitational acceleration.
If the drop size is much smaller than $\kappa^{-1}$,
then the surface tension is dominant and the
liquid drop forms a circular cap at equilibrium.
If the drop size is much larger than $\kappa^{-1}$,
then the gravity is dominant and the drop
forms a puddle (or pancake-like shape) at equilibrium, with a flat liquid
surface~\cite{deGennesBQ2003}.
Moreover, if the gravity is dominant (liquid forming a puddle),
by considering the force and the Young's relation
one can obtain the following expression for the puddle
thickness (height) in terms of other physical parameters (see \cite{deGennesBQ2003})
\begin{equation}
  H_{\infty} = 2\kappa^{-1}\sin\left(\frac{\theta_E}{2}  \right) = 2 \sqrt{\frac{\sigma_{la}}{\rho_l g_r}} \sin\left(\frac{\theta_E}{2}  \right)
  \label{equ:puddle_thickness}
\end{equation}
where $H_{\infty}$ denotes the asymptotic puddle thickness,
and $\theta_E$ is the equilibrium contact angle at the wall.

% drop profiles with gravity

We have performed two groups of numerical experiments to study
the effects of the gravity and the surface tension, respectively.

In the first group of experiments, we fix all the other physical parameters
at those values given in Table \ref{tab:air_water_param}
(air-oil surface tension is fixed at $0.055 kg/s^2$),
and vary only the magnitude of the gravitational acceleration $g_r$ systematically.
We have conducted a series of simulations of this three-phase system
corresponding to these gravity values.
Figure \ref{fig:3p_prof_G} shows the equilibrium profiles
of the water and oil drops corresponding to three
gravity values. The drop profiles
again are visualized by the volume-fraction contour
levels $c_i=\frac{1}{2}$ ($1\leqslant i\leqslant 3$).
These results are for an air-water
contact angle of $120^0$ and an air-oil contact angle of $135^0$.
With zero gravity, the drops form circular caps (Figure \ref{fig:3p_prof_G}(a)).
With a large gravity $g_r=12m/s^2$, both drops form a puddle
on the wall, with a flattened top surface (Figure \ref{fig:3p_prof_G}(c)).
With an intermediate gravity magnitude $g_r=3m/s^2$,
the  drops form oval caps (or elongated circular caps)
on the wall (Figure \ref{fig:3p_prof_G}(b)).
These observations are consistent with the
theory of \cite{deGennesBQ2003}.

\begin{figure}
\centerline{
\includegraphics[height=3in]{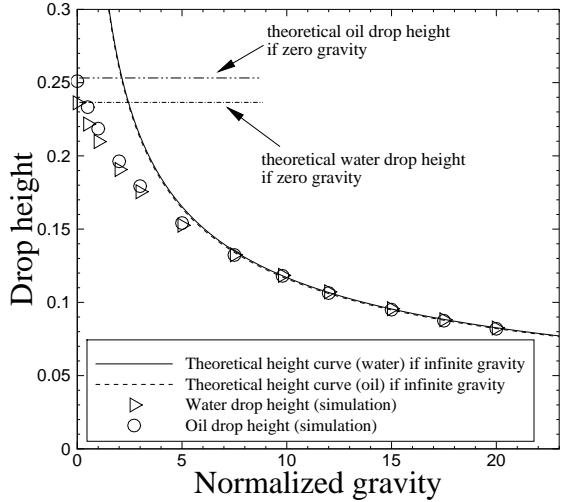}
}
\caption{
Effect of gravity: equilibrium water/oil drop heights
as a function of gravity, with contact angles
$(\theta_{aw},\theta_{ao})=(120^0, 135^0)$.
Solid/dashed curves show the theoretical asymptotic drop-height curves
if gravity is dominant. The theoretical water/oil drop heights
at zero gravity are also marked in the figure.
}
\label{fig:gravity}
\end{figure}

% quantitative comparison

We have computed the
heights of the water/oil drops from their equilibrium profiles
corresponding to each gravity value.
In Figure \ref{fig:gravity} we plot the heights of the water drop and
the oil drop  as a function of the gravity obtained from our simulations
(see the symbols) corresponding to a set of fixed contact angles
$(\theta_{aw},\theta_{ao})=(120^0,135^0)$. 
For comparison we have included in this plot the theoretical drop-height (puddle thickness)
for water and for oil if the gravity is dominant,
computed based on equation \eqref{equ:puddle_thickness},
as a function of the gravity for the same set of contact angles;
see the solid and dashed curves.
Note that these theoretical height curves are  valid only for sufficiently
large gravity values. They are invalid if the gravity is small.
In addition, we have also included in this figure the
theoretical heights for the water drop and the oil drop
at zero gravity for  $(\theta_{aw},\theta_{ao})=(120^0,135^0)$,
which are computed based on equation
\eqref{equ:3p_param_exact}.
It can be observed that the drop-height values from numerical
simulations agree with the theoretical values very well when
the gravity becomes large (beyond about $7.5kg/m^2$).
At zero gravity the simulation results are also in
good agreement with the theoretical results.
For intermediate gravity values, the simulation results
exhibit a transition between the results of these two
extreme cases.

\begin{figure}
\centering
\includegraphics[width=5in]{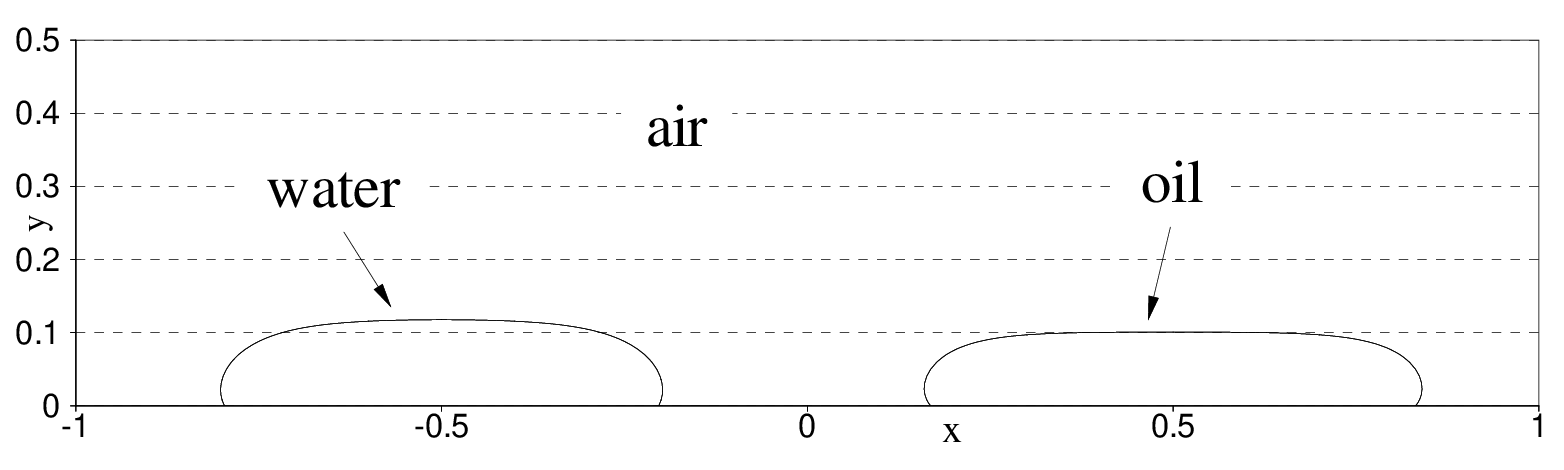}(a)
\includegraphics[width=5in]{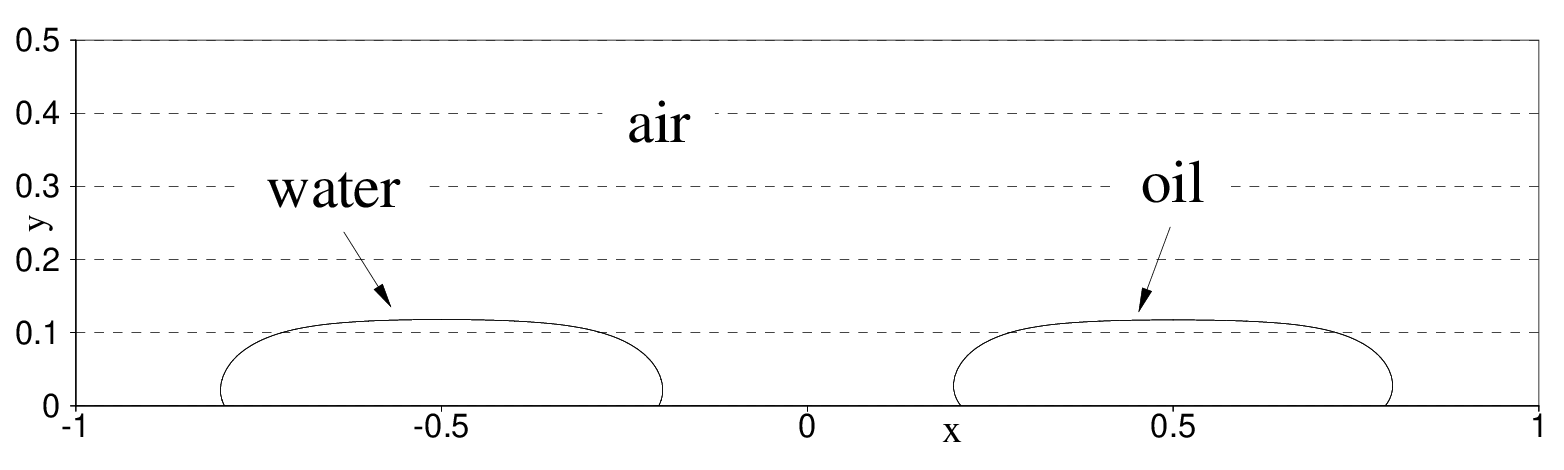}(b)
\caption{
Three-phase flows, effect of air-oil surface tension 
on equilibrium drop profiles:
(a) $\sigma_{ao}=0.04 kg/s^2$,
(b) $\sigma_{ao}=0.055 kg/s^2$.
Contact angles are 
$(\theta_{aw},\theta_{ao})=(120^0, 135^0)$.
All other physical parameters are fixed.
}
\label{fig:3p_config_surften}
\end{figure}

\begin{figure}
\centerline{
\includegraphics[height=3in]{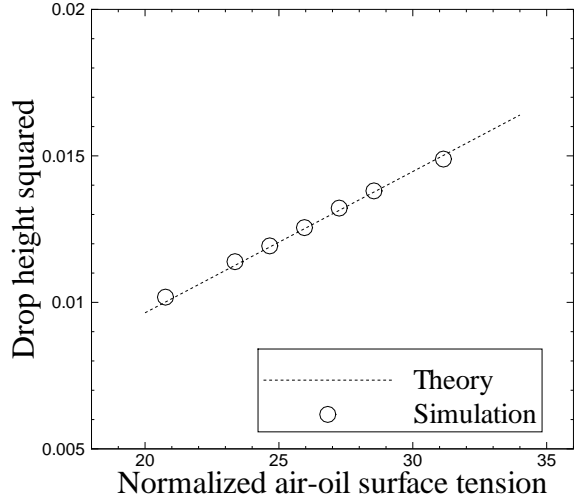}
}
\caption{
Three-phase flows, comparison of the oil-puddle thickness squared as a function
of the normalized air/oil surface tension between simulation
and the de Gennes theory.
Contact angles are fixed at $(\theta_{aw},\theta_{ao})=(120^0, 135^0)$.
}
\label{fig:3p_thickness_surften}
\end{figure}

In the second group of experiments we investigate
the effect of the surface tension on the liquid drop heights
when the gravity is dominant (i.e. liquid forming a puddle).
In these tests we vary the air-oil surface tension
systematically over a range of values while fixing all the other
physical parameters. 
We use the normal gravitational acceleration $g_r=9.8 m/s^2$,
and the air-water and air-oil contact angles are
fixed at $(\theta_{aw},\theta_{ao})=(120^0,135^0)$.
The values for the rest of the physical parameters (excluding the air-oil surface
tension) are given in Table \ref{tab:air_water_param}.
Figure \ref{fig:3p_config_surften}
shows the equilibrium configurations of this three-phase system
corresponding to two air-oil surface tensions
$\sigma_{ao}=0.04 kg/s^2$ and $0.055kg/s^2$.
One can observe that both the water and oil drops form a puddle on
the wall in this case, and the thickness of the oil puddle
is notably influenced by the air-oil surface tension.
We have computed the oil-puddle thickness
from the equilibrium configurations corresponding to
each air-oil surface tension value.
In Figure \ref{fig:3p_thickness_surften} we compare
the oil-puddle thickness squared as a function of
the air-oil surface tension between the simulations
and the de-Gennes theory (see equation \eqref{equ:puddle_thickness}).
The simulation results agree quite well with
the theoretical relation.

The results of this section and the comparisons
with the de Gennes theory \cite{deGennesBQ2003}
indicate that, for multiphase problems involving solid walls
and multiple types of fluid interfaces and  contact angles,
the method developed herein 
produces physically accurate results.

\subsection{Compound Drops of Multiple Fluids on Horizontal Wall Surfaces -- Effect of Contact Angles}

We study the equilibrium configurations of compound drops 
formed by multiple fluids on a wall surface in this section, and how
the various contact angles influence the drop configurations.
Two multiphase systems will be considered, consisting of three and four
fluid components, respectively. 
Because the interactions among the fluids are strong,
in certain cases the profile of the compound drop can be dramatically 
modified with a small change in the contact angles.

\subsubsection{Three Fluid Components}

% water/oil drops touching each other

\begin{figure}
\centering
\includegraphics[width=5in]{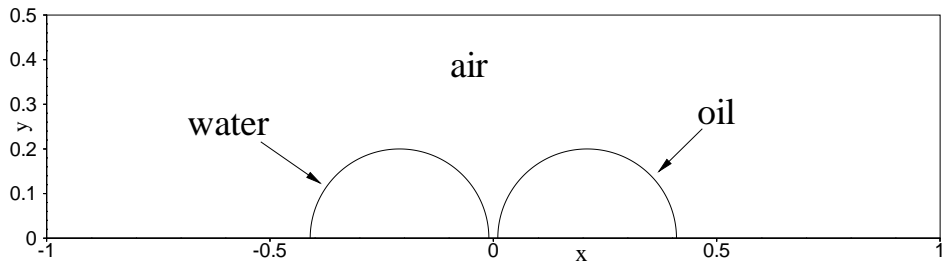}
\caption{
Compound liquid drop: initial profiles of water and oil drops.
}
\label{fig:3p_config_touching}
\end{figure}

First let us consider a three-phase flow problem consisting of air, water and oil, with
a setting similar to that of Section \ref{sec:3p_compare},
More specifically, we consider the rectangular domain as shown in
Figure \ref{fig:3p_config_touching}, $-L\leqslant x\leqslant L$ and $0\leqslant y\leqslant 0.5L$
($L=4cm$), where the top and bottom are solid walls and in the horizontal direction it is periodic.
A water drop and an oil drop, both semi-circular initially with radius $R_0=0.2L$,
are in ambient air and  held at rest on the bottom wall.
The two drops are placed next to and almost touching each other.
The water-drop center is located at $\mathbf{X}_w=(x_{0w},y_{0w})=(-0.21L, 0)$,
and the oil-drop center is at $\mathbf{X}_o=(x_{0o},y_{0o})=(0.21L, 0)$.
The gravity is in the $-y$ direction.
Let $\theta_{aw}$ denote the contact angle between
the air-water interface and the wall  when measured
on the water side, and $\theta_{ao}$ denote the contact angle between the air-oil interface
and the wall when measured on the oil side.
At $t=0$ the system is released, and evolves to equilibrium
eventually. Because the two liquid drops are very close to each
other, they merge and form a compound drop on the wall. Our goal is to study
how the profile of the compound drop at equilibrium 
is affected by the wall wettability,
i.e.~the contact angles $\theta_{aw}$ and $\theta_{ao}$.

% to discuss physical parameters, simulation parameters, mesh etc.

In the present simulations
we employ the values listed in Table \ref{tab:air_water_param} for
the physical parameters about the air, water and oil
and the interfaces formed by these fluids.
Water, oil and air are assigned as the first, second and third fluids,
respectively. Similar to in Section \ref{sec:3p_compare},
the problem is normalized by 
choosing $L$ as the length scale, the air density as
the density scale ${\varrho}_d$, and 
$\sqrt{g_{r0}L}$ (where $g_{r0}=1m/s^2$) as the velocity scale $U_0$.

\begin{table}
\begin{center}
\begin{tabular}{l l}
\hline
Parameters & Values \\
$\phi_i$ & defined by \eqref{equ:gop_volfrac}, volume fractions as order parameters \\
$\zeta_{ij}$ & Computed based on \eqref{equ:theta_expr} and \eqref{equ:lambda_mat_gop} \\
$\eta/L$ & $0.01$ \\
$m_0/(U_0L^3)$ & $10^{-7}$ \\
$\rho_0$ & $\min(\tilde{\rho}_1, \tilde{\rho}_2, \tilde{\rho}_3)$ \\
$\nu_0$ & $5\max\left(\frac{\tilde{\mu}_1}{\tilde{\rho}_1},\frac{\tilde{\mu}_2}{\tilde{\rho}_2}, \frac{\tilde{\mu}_3}{\tilde{\rho}_3}\right)$ \\
$S$ & $\eta^2\sqrt{\frac{4\gamma_0}{m_0\Delta t} }$ \\
$\alpha$ & Computed based on \eqref{equ:alpha_expr} \\
$U_0\Delta t/L$ & $2.0\times 10^{-6}$ \\
$\theta_{13}$, $\theta_{23}$ & ranging between $45^0$ and $135^0$ \\
$J$ (temporal order) & $2$ \\
Number of elements & $100$ \\
Element order & $14$ \\
\hline
\end{tabular}
\caption{Simulation parameter values for air/water/oil three-phase
problem.}
\label{tab:3p_touching_param}
\end{center}
\end{table}

\begin{figure}
\centering
\includegraphics[width=5in]{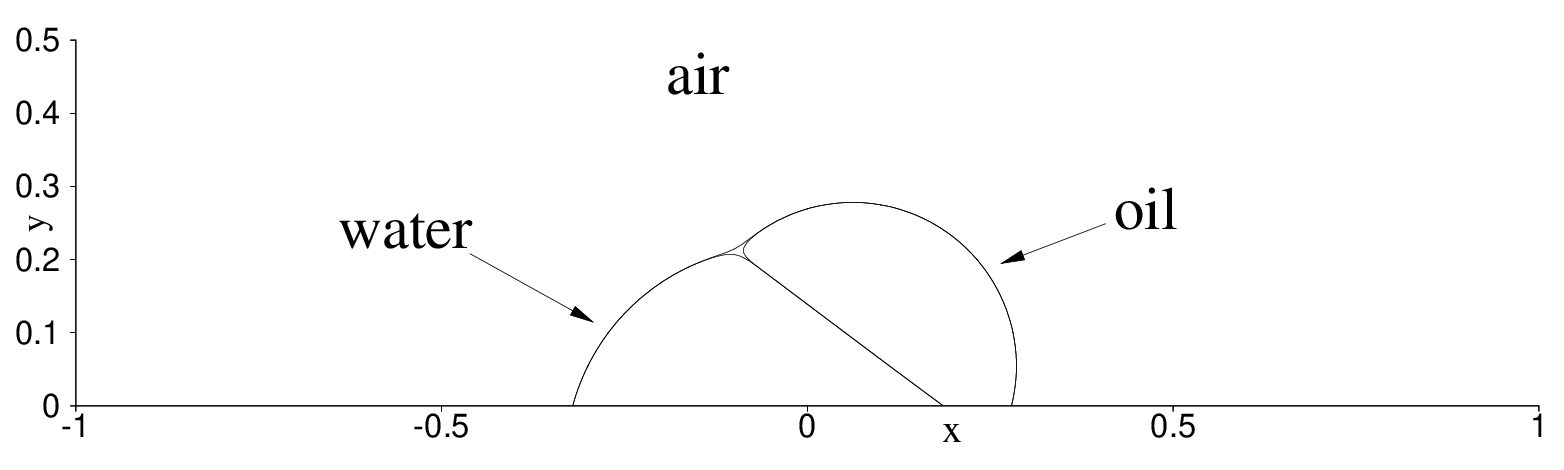}(a)
\includegraphics[width=5in]{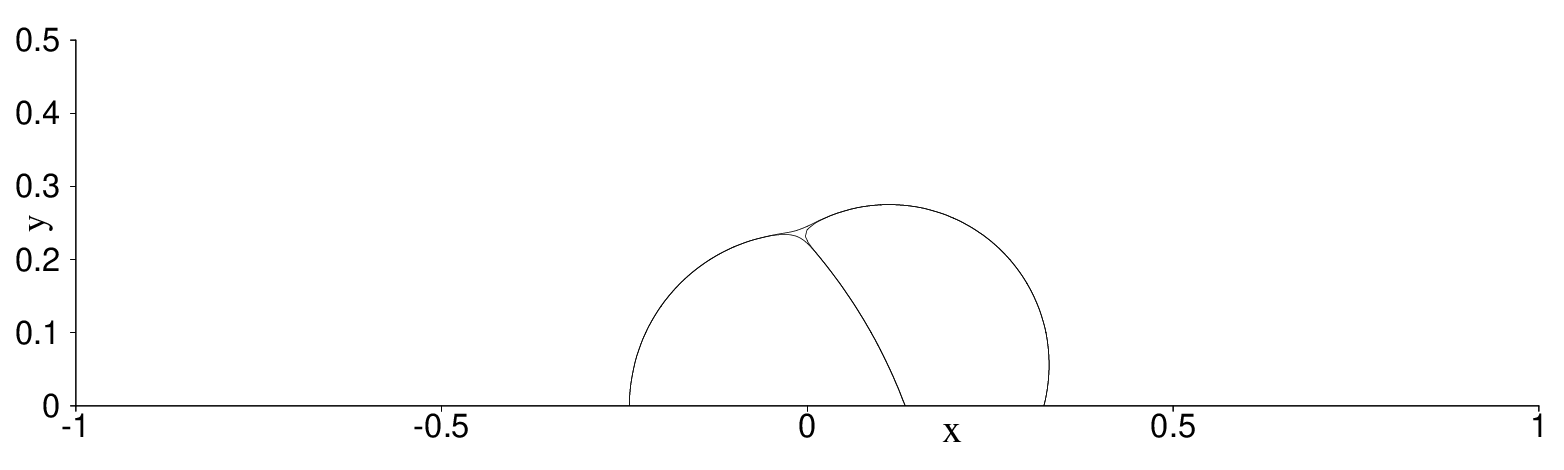}(b)
\includegraphics[width=5in]{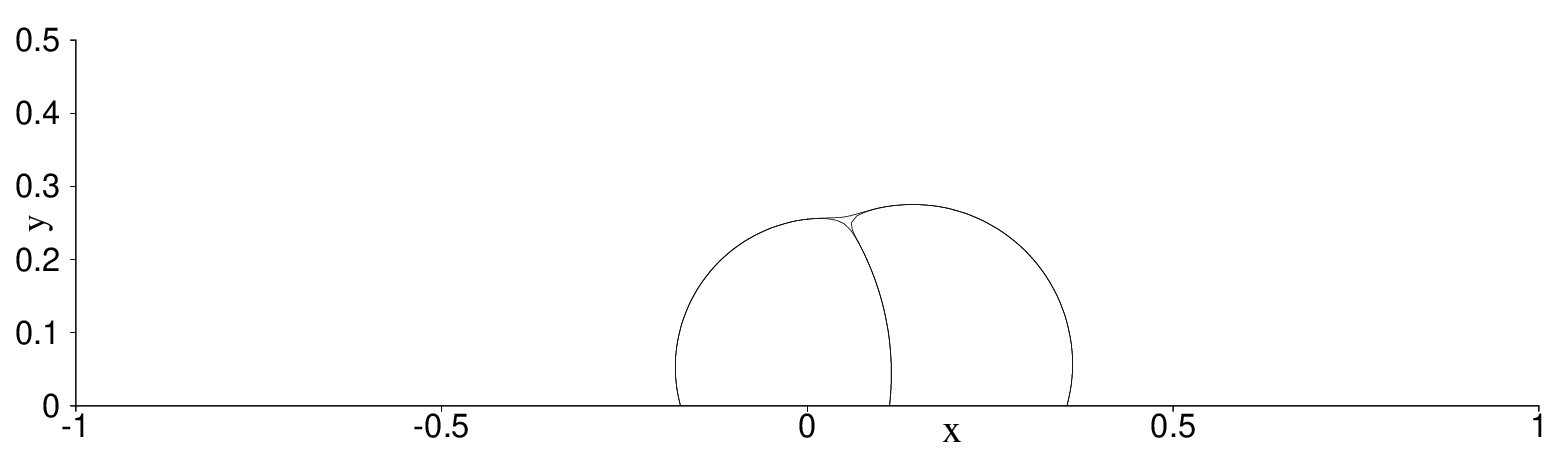}(c)
\includegraphics[width=5in]{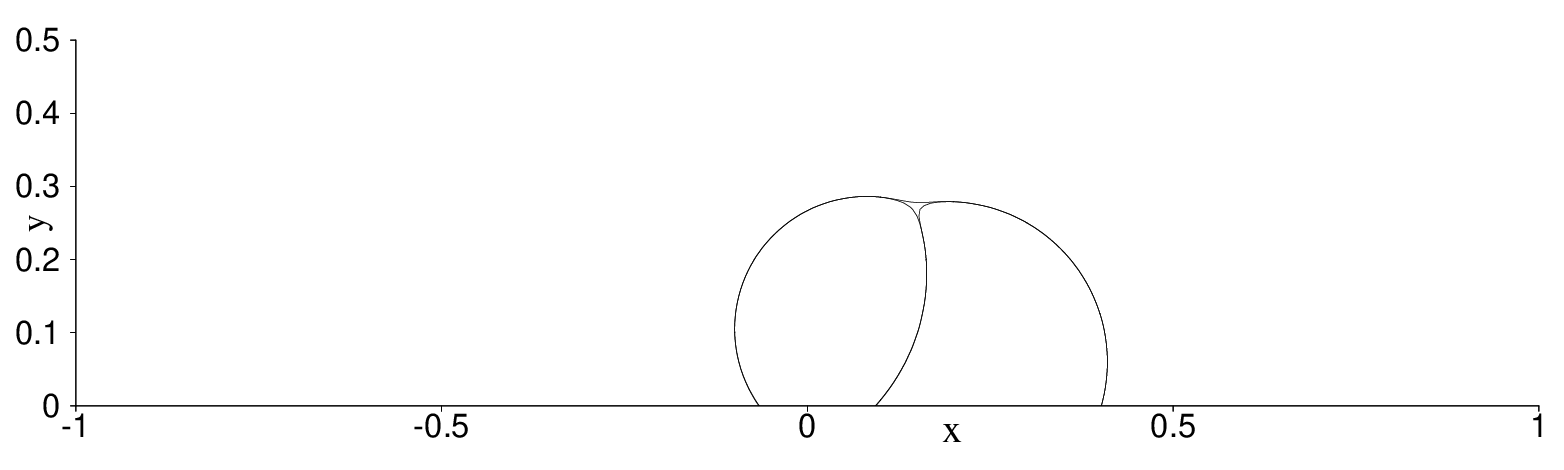}(d)
\includegraphics[width=5in]{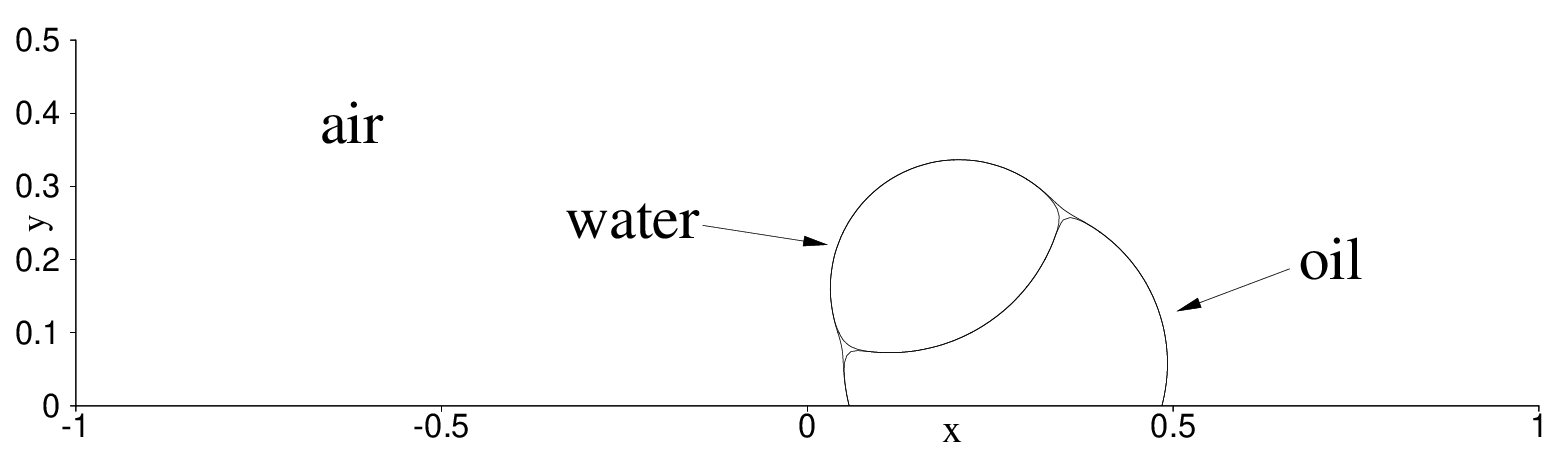}(e)
\caption{
Profiles of the compound liquid drop formed by water and oil, with
the air/oil contact angle fixed at $\theta_{ao}=105^0$
and the air/water contact angle varied:
(a) $\theta_{aw}=75^0$,
(b) $\theta_{aw}=90^0$,
(c) $\theta_{aw}=105^0$,
(d) $\theta_{aw}=125^0$,
(e) $\theta_{aw}=135^0$.
}
\label{fig:3p_angle_aw_effect}
\end{figure}

The domain is partitioned using $100$ spectral elements (with $20$ and 
$5$ elements in the $x$ and $y$ directions, respectively),
and an element order $14$ is employed in the simulations.
On the top and bottom walls, no slip condition is imposed
for the velocity, and for the phase field functions
the contact-angle conditions \eqref{equ:bc_chempot_2}--\eqref{equ:cabc_2}
are imposed with  $g_{ai}=0$ and $g_{bi}=0$. Periodic conditions
are imposed for all flow variables in the horizontal direction.
The initial velocity is assumed to be zero, and the initial
distributions of the phase field functions are
given by the expressions \eqref{equ:init_phase} by noting
that $x_{0w}=-0.21L$ and $x_{0o}=0.21L$ in this case.
Table \ref{tab:3p_touching_param} lists the simulation parameters
employed for this test problem.

% effect of theta_{aw}

We observe that the wettability of the wall,
i.e.~the contact angles of the various fluids forming
the compound drop, can considerably affect the equilibrium configurations
of the drop.
To demonstrate this point, let us assume that the gravity is absent,
and the system is influenced only by the surface tensions among
the air, water and oil.
The results shown in Figure \ref{fig:3p_angle_aw_effect} demonstrate
the effect of the air-water contact angle on the equilibrium shape
of the water/oil compound drops. Plotted here are the profiles of the 
fluid interfaces, visualized by the contour levels
of the volume fractions $c_i(\vec{\phi})=\frac{1}{2}$ ($1\leqslant i\leqslant 3$)
for the three fluids.
In this group of tests, the contact angle of the air-oil interface  has been
fixed at $\theta_{ao}=105^0$, while the contact angle of the air-water interface
$\theta_{aw}$ is varied in a range of values, from $75^0$ to $135^0$.
Around the three-phase line where the three fluid components intersect
a small star-shaped region can be observed. 
As pointed out in \cite{Dong2014,Dong2015}, such a region is formed
by the contour levels because
no fluid has a volume fraction larger than $\frac{1}{2}$ in that region.
%
% what are the observations?
%  forming compound drop, configuration change, contact angle between oil/water
We can observe that the water and oil form a compound drop.
The contact angle of the water-oil interface and
the overall profile of the compound drop are affected by
the air-water contact angle remarkably.
With air-water contact angle $\theta_{aw}=75^0$ 
the water partially goes underneath the oil at 
equilibrium (Figure \ref{fig:3p_angle_aw_effect}),
with the contact angle of the water-oil interface (measured on the water side)
being about $\theta_{ow}\approx 34^0$ according to 
equation \eqref{equ:ca_relation}.
%
% other angles
As the air-water contact angle increases 
the contact angle of the water-oil interface increases
more rapidly, and the region occupied by the water
becomes more ``plump'' within the compound drop
 (Figures \ref{fig:3p_angle_aw_effect}(b)-(d)).
Based on equation \eqref{equ:ca_relation},
as the air-water contact angle increases to about
$\theta_{aw}= 138.16^0$ the water-oil contact angle
will reach $\theta_{ow}=180^0$. 
In practice we have observed from numerical experiments 
that, when the air-water 
contact angle is below but close to this value, 
the water tends to move away from the wall at
equilibrium because of the large contact angle of
the water-oil interface.
For example, with an air-water
contact angle $\theta_{aw}=135^0$
the water forms a drop on the shoulder of
the oil region, and the water drop is not in contact
with the wall any more under current simulation conditions;
see Figure \ref{fig:3p_angle_aw_effect}(e).

\begin{figure}
\centering
\includegraphics[width=5in]{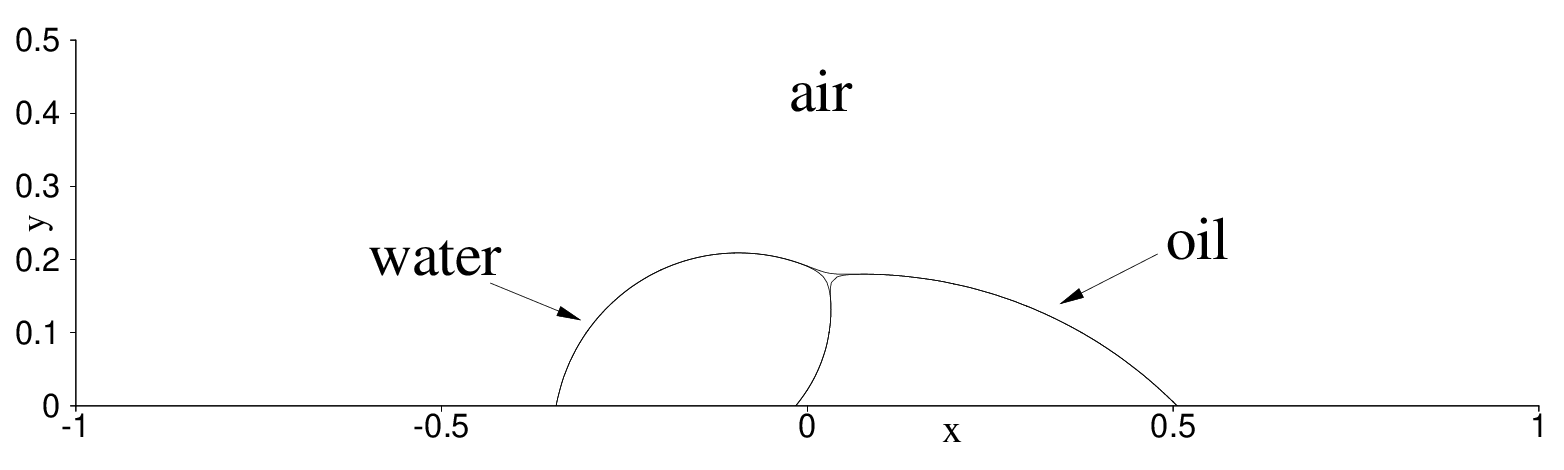}(a)
\includegraphics[width=5in]{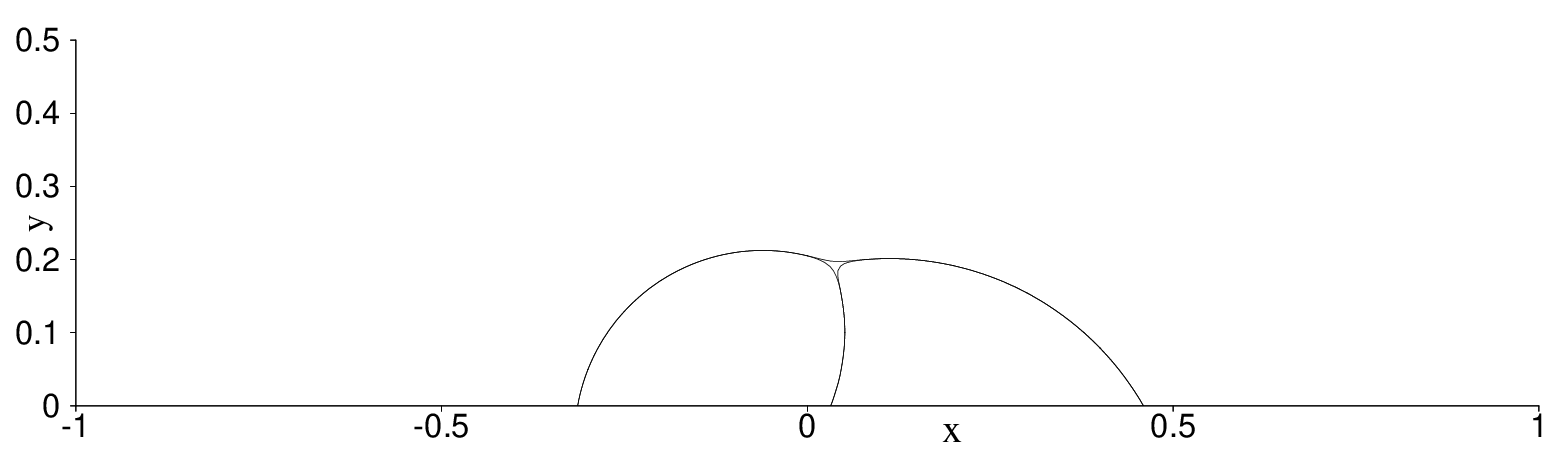}(b)
\includegraphics[width=5in]{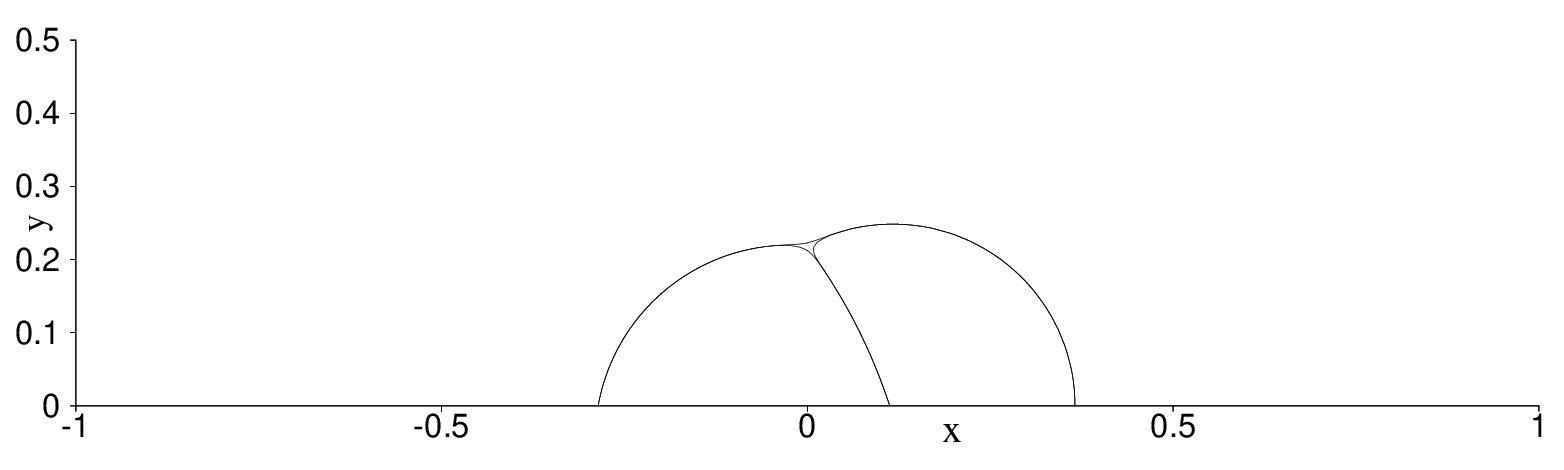}(c)
\includegraphics[width=5in]{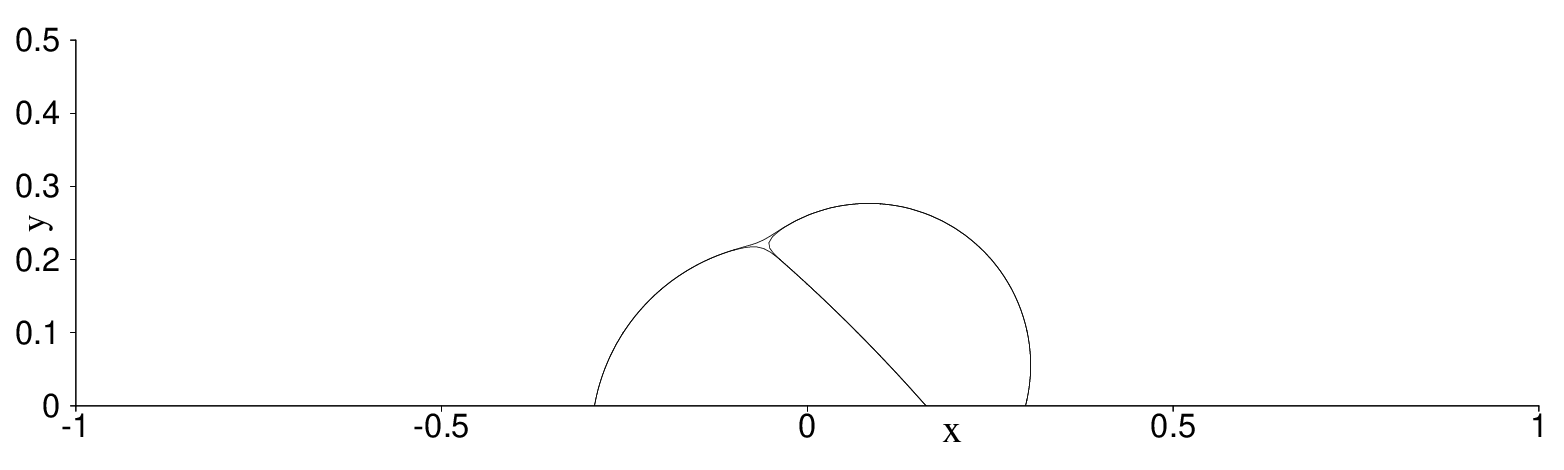}(d)
\caption{
Profiles of the compound liquid drop formed by water and oil, with
the air/water contact angle fixed at $\theta_{aw}=80^0$
and the air/oil contact angle varied:
(a) $\theta_{ao}=45^0$,
(b) $\theta_{ao}=60^0$,
(c) $\theta_{ao}=90^0$,
(d) $\theta_{ao}=105^0$.
}
\label{fig:3p_angle_ao_effect}
\end{figure}

Figure \ref{fig:3p_angle_ao_effect} shows the equilibrium configurations 
of the compound water-oil drop corresponding to several values of
the air-oil contact angle. 
In this group of tests the air-water contact angle is fixed
at $\theta_{aw}=80^0$, and the air-oil contact angle
is varied in a range of values between $\theta_{ao}=45^0$ and $\theta_{ao}=105^0$.
According to equation \eqref{equ:ca_relation},
the contact angle of the water-oil interface (measured on the water side) 
varies between $\theta_{ow}\approx 131^0$ (Figure \ref{fig:3p_angle_ao_effect}(a))
and $\theta_{ow}\approx 48^0$ (Figure \ref{fig:3p_angle_ao_effect}(d)).
It is evident that the overall profile of 
the compound drop, and the profiles of the water and oil regions
within the drop, have been dramatically influenced by
the change in the air-oil contact angle.

\begin{figure}
\centering
\includegraphics[width=5in]{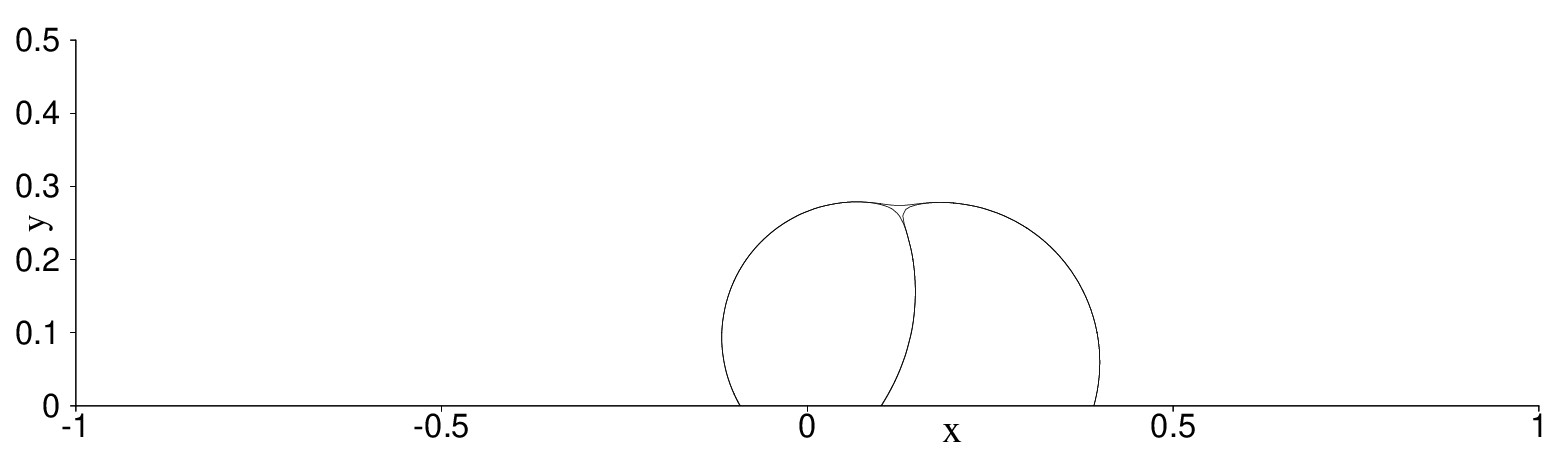}(a)
\includegraphics[width=5in]{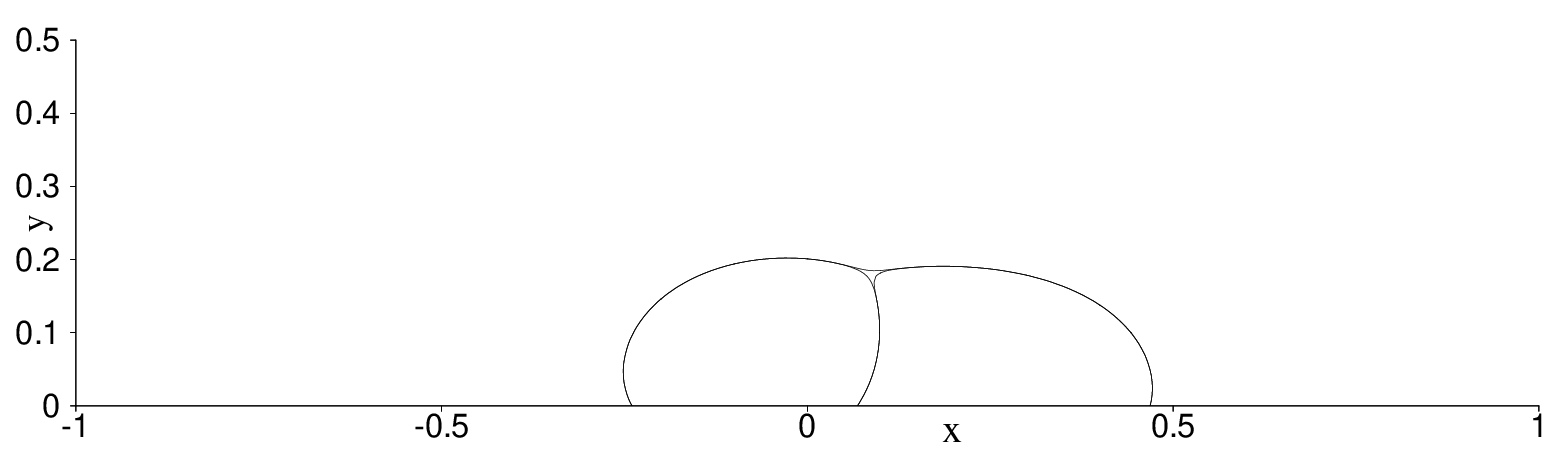}(b)
\includegraphics[width=5in]{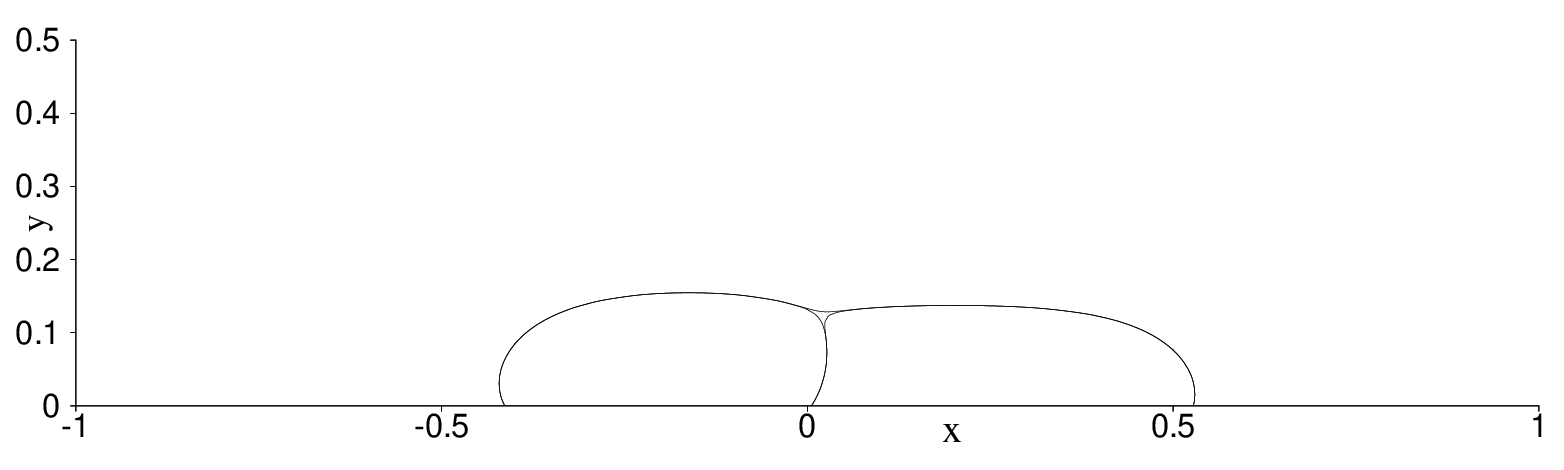}(c)
\caption{
Effect of gravity on profiles of compound liquid drop
formed by water and oil with contact angles
$(\theta_{aw},\theta_{ao})=(120^0,105^0)$:
(a) zero gravity,
(b) $g_r=2m/s^2$,
(c) $g_r=5m/s^2$.
}
\label{fig:3p_gravity_effect}
\end{figure}

It is also observed that the gravity can affect
the equilibrium configuration of the compound drop significantly.
Figure \ref{fig:3p_gravity_effect} shows the equilibrium
configurations of the compound drop of water and oil 
corresponding to three values of the gravitational acceleration:
$g_r=0$, $2m/s^2$ and $5m/s^2$.
These results correspond to the air-water and air-oil
contact angles $(\theta_{aw},\theta_{ao})=(120^0,105^0)$.
The increase in the gravity tends to spread the compound drop 
onto the wall, reducing the drop height. The drop becomes very 
stretched along the horizontal direction at large gravity values.

% what else to discuss here?

\subsubsection{Four Fluid Components}

\begin{figure}
\centerline{
\includegraphics[width=5in]{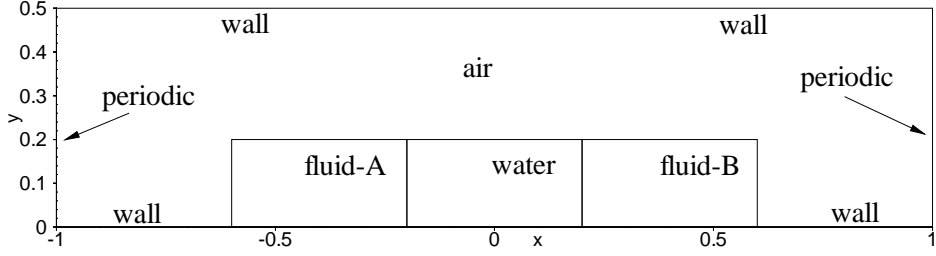}
}
\caption{
Initial configuration of a compound liquid drop
consisting of water, fluid-A and fluid-B (in ambient air) on a horizontal
wall.
}
\label{fig:4p_setting}
\end{figure}

We next study a four-phase problem, and consider a compound
liquid drop consisting of three liquids  (in ambient air)
on a solid wall surface. We will look into 
the effects of various contact angles.  The multitude of
independent contact angles much complicates the interactions among
different fluids.

Specifically, we consider the problem as sketched in Figure \ref{fig:4p_setting}.
A rectangular domain, of dimensions $-L\leqslant x\leqslant L$ and
$0\leqslant y\leqslant L/2$ (where $L=4cm$) and with solid walls on the top and bottom
sides, is filled with air and
contains three liquids (water, fluid-A and fluid-B) within. 
The initial region occupied by these liquids are shown in Figure \ref{fig:4p_setting}.
The three liquid regions all have an initial height $0.2L$ (i.e.~$0\leqslant y\leqslant 0.2L$).
In the horizontal direction water occupies
the region $-0.2L\leqslant x\leqslant 0.2L$, fluid-A occupies $-0.6L\leqslant x\leqslant -0.2L$,
and fluid-B occupies $0.2L\leqslant x\leqslant 0.6L$.
All these fluids are assumed to be incompressible and immiscible
with one another. In the horizontal direction
the domain is assumed to be periodic at $x=\pm L$.
The gravity is ignored for this problem.
All the four fluids are held at rest initially.
Then at $t=0$ the system is released and starts to evolve
under the six pairwise surface tensions among these fluids,
eventually reaching an equilibrium configuration.
Our goal is to investigate the effects of various contact angles
 on the equilibrium configuration
of this  system.
We employ the physical parameter values as listed
in Table \ref{tab:4p_param} for this problem.

\begin{table}
\begin{center}
\begin{tabular}{l l}
\hline 
density [$kg/m^3$]: & air -- $1.2041$,  water -- $998.207$,   fluid-A -- $870$,  fluid-B -- $400$ \\
dynamic viscosity [$kg/m\cdot s$]: & air -- $1.78E-4$,   water -- $1.002E-3$, 
   fluid-A -- $0.0915$,   fluid-B -- $0.02$ \\
surface tension [$kg/s^2$]: & air/water -- $0.0728$,   air/fluid-A -- $0.055$, 
   air/fluid-B -- $0.06$, \\
 &  water/fluid-A -- $0.044$,   water/fluid-B -- $0.045$,   fluid-A/fluid-B -- $0.048$ \\
\hline
\end{tabular}
\end{center}
\caption{
Physical parameter values for the four-phase flow problem with
air, water, fluid-A and fluid-B.
}
\label{tab:4p_param}
\end{table}

% how to simulate the problem?

We assign the water, fluid-A, fluid-B and air as the first, second, third
and fourth fluids respectively in the simulations.
Therefore the contact angles of
the air-water interface ($\theta_{aw}$), the air/fluid-A interface ($\theta_{aA}$)
and the air/fluid-B interface ($\theta_{aB}$) are chosen 
as the independent contact angles of this four-phase system.
The normalization proceeds according to Table \ref{tab:normalization}
by choosing $L$ as the length scale, the air density as
the density scale $\varrho_d$, and $\sqrt{g_{r0}L}$ (where $g_{r0}=1m/s^2$) as
the velocity scale $U_0$.
We employ the volume fractions as the order parameters in this problem;
see equations \eqref{equ:gop_volfrac} and \eqref{equ:volfrac_aij}.

To simulate the problem we discretize the domain using $256$ quadrilateral
spectral elements, with $32$ and $8$ elements along the $x$ and
$y$ directions respectively.
We use an element order $12$ for each element in the simulations.
On the top and bottom walls we impose the Dirichlet boundary
condition \eqref{equ:dbc} with $\mathbf{w}=0$ for the velocity,
and the contact-angle boundary 
conditions \eqref{equ:bc_chempot_2}--\eqref{equ:cabc_2} with $g_{ai}=0$ and
$g_{bi}=0$ for
the phase field functions.
Periodic conditions are imposed for all the flow variables
in the horizontal direction.
%
% initial conditions
The initial velocity is zero. The initial phase field distributions
are given by,
\begin{equation}
\left\{
\begin{split}
&
\phi_{10} = \frac{1}{8}\left(1+\tanh\frac{x+0.2L}{\sqrt{2}\eta}  \right)
  \left(1-\tanh\frac{x-0.2L}{\sqrt{2}\eta}  \right)
  \left(1-\tanh\frac{y-0.2L}{\sqrt{2}\eta}  \right) \\
&
\phi_{20} = \frac{1}{8}\left(1+\tanh\frac{x+0.6L}{\sqrt{2}\eta}  \right)
  \left(1-\tanh\frac{x+0.2L}{\sqrt{2}\eta}  \right)
  \left(1-\tanh\frac{y-0.2L}{\sqrt{2}\eta}  \right) \\
&
\phi_{30} = \frac{1}{8}\left(1+\tanh\frac{x-0.2L}{\sqrt{2}\eta}  \right)
  \left(1-\tanh\frac{x-0.6L}{\sqrt{2}\eta}  \right)
  \left(1-\tanh\frac{y-0.2L}{\sqrt{2}\eta}  \right) \\
&
\phi_{40} = 1-\phi_{10} - \phi_{20} - \phi_{30}.
\end{split}
\right.
\end{equation}
Table \ref{tab:4p_param_simul} lists the values of the simulation parameters 
for this problem.

\begin{table}
\begin{center}
\begin{tabular}{ll}
\hline
Parameters & Values \\
$\phi_i$ & defined by \eqref{equ:gop_volfrac} \\
$\zeta_{ij}$ & Computed based on \eqref{equ:theta_expr} and \eqref{equ:lambda_mat_gop} \\
$\eta/L$ & $0.01$ \\
$m_0/(U_0L^3)$ & $10^{-9}$ \\
$\rho_0$ & $\min(\tilde{\rho}_1, \tilde{\rho}_2, \tilde{\rho}_3)$ \\
$\nu_0$ & $5\max\left(\frac{\tilde{\mu}_1}{\tilde{\rho}_1},\frac{\tilde{\mu}_2}{\tilde{\rho}_2}, \frac{\tilde{\mu}_3}{\tilde{\rho}_3}\right)$ \\
$S$ & $\eta^2\sqrt{\frac{4\gamma_0}{m_0\Delta t} }$ \\
$\alpha$ & Computed based on \eqref{equ:alpha_expr} \\
$U_0\Delta t/L$ & $2.5\times 10^{-6}$ \\
$\theta_{14}$, $\theta_{24}$, $\theta_{34}$ & ranging between $60^0$ and $120^0$ \\
$J$ (temporal order) & $2$ \\
Number of elements & $256$ \\
Element order & $12$ \\
\hline
\end{tabular}
\end{center}
\caption{Simulation parameter values for the four-phase air/water/fluid-A/fluid-B problem.}
\label{tab:4p_param_simul}
\end{table}

\begin{figure}
\centering
\includegraphics[width=5in]{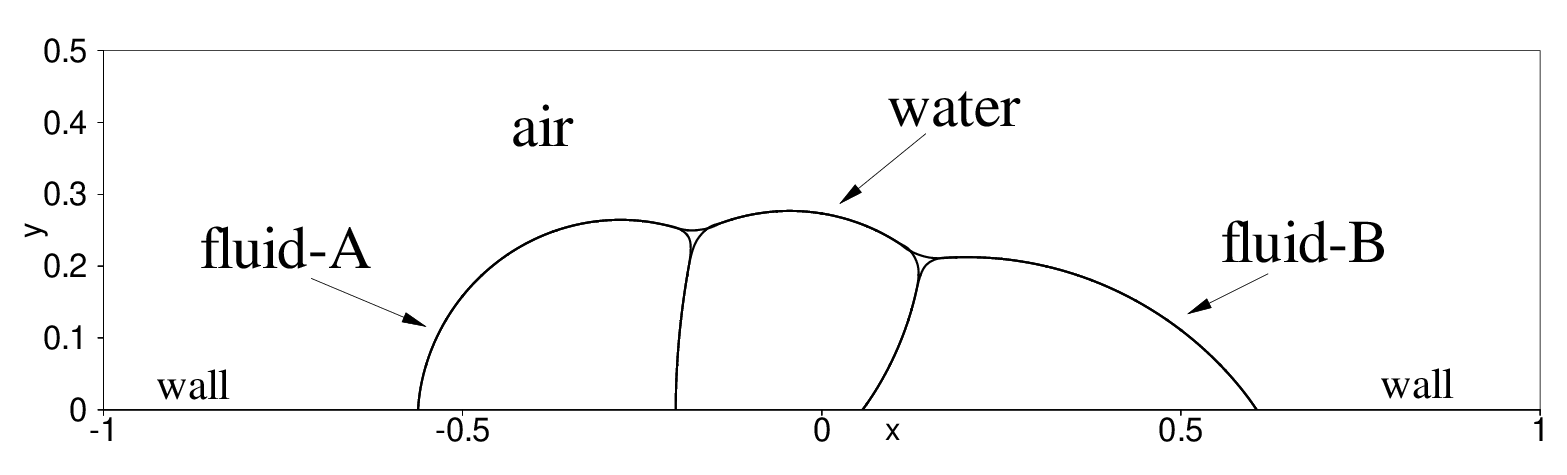}(a)
\includegraphics[width=5in]{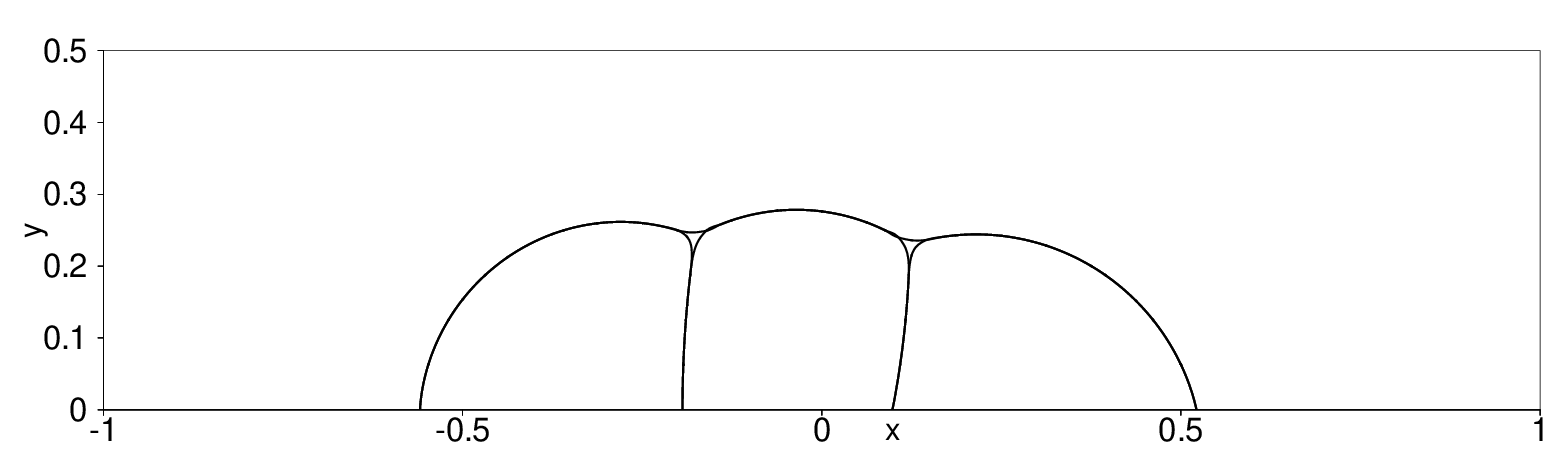}(b)
\includegraphics[width=5in]{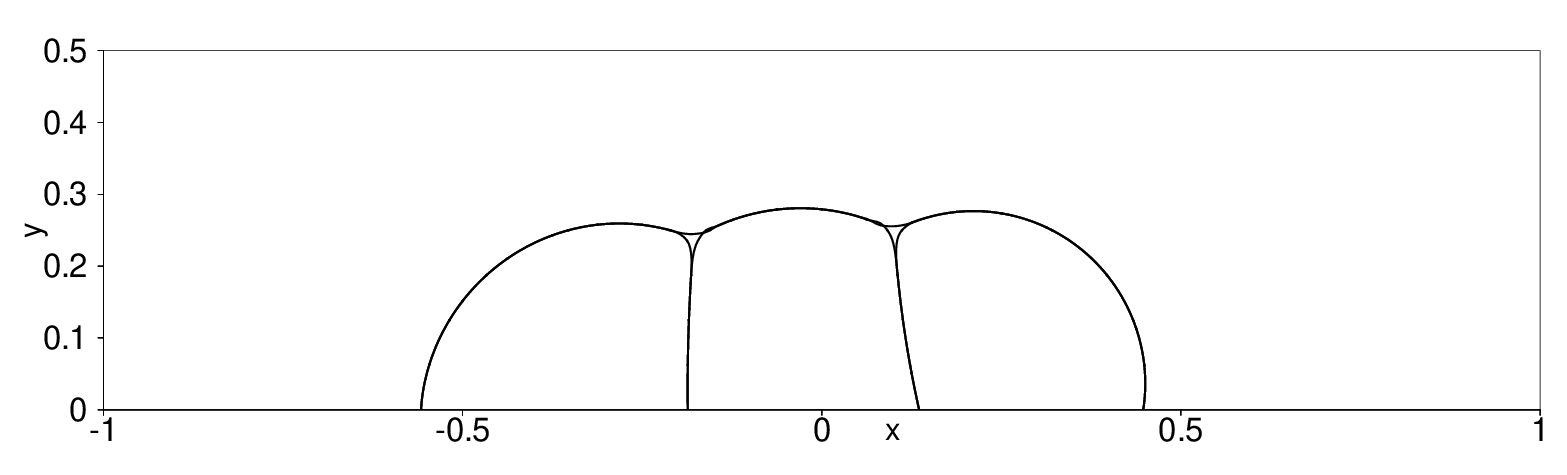}(c)
\includegraphics[width=5in]{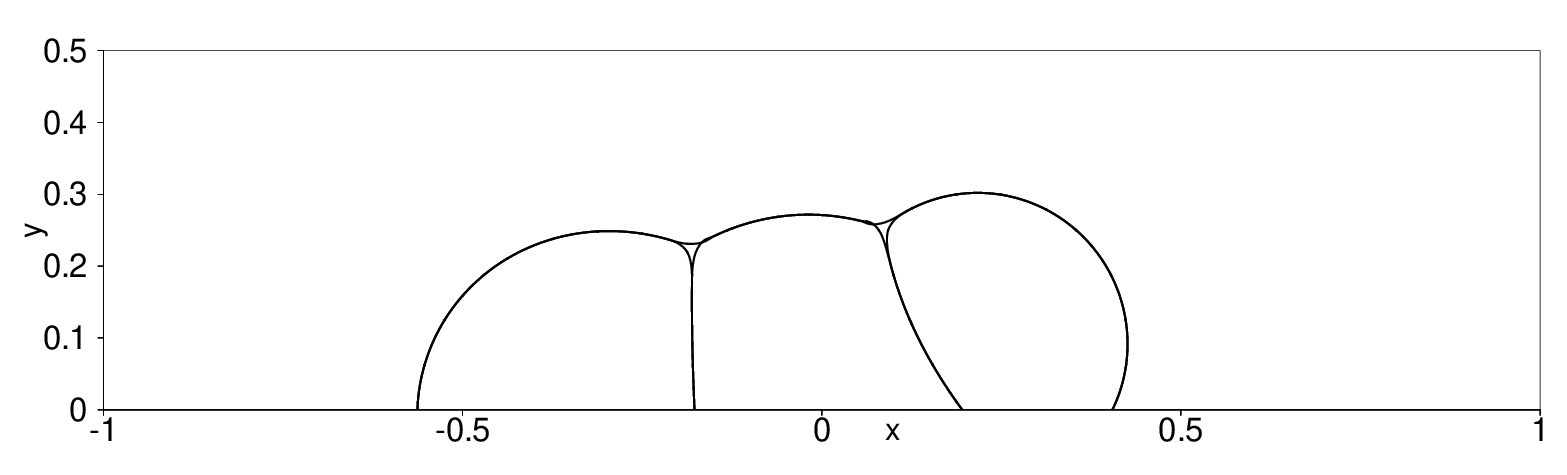}(d)
\caption{
Profiles of a compound liquid drop formed by water, fluid-A and
fluid-B (ambient air) on a wall, with air/fluid-A and air/water contact
angles fixed at $\theta_{aA}=\theta_{aw}=90^0$ and
the air/fluid-B contact angle varied as:
(a) $\theta_{aB}=60^0$,
(b) $\theta_{aB}=80^0$,
(c) $\theta_{aB}=100^0$,
(d) $\theta_{aB}=120^0$.
}
\label{fig:4p_effect_aB}
\end{figure}

Let us first look into the effect of the air/fluid-B contact angle on
the equilibrium configuration of the system.
In this set of tests we fix the contact angles of 
the air/water interface and the air/fluid-A interface at
$\theta_{aw}=\theta_{aA}=90^0$, and vary the air/fluid-B
contact angle in a range of values.
Figure \eqref{fig:4p_effect_aB} shows the configurations of
this system corresponding to four air/fluid-B
contact angles:
$\theta_{aB}=60^0$, $80^0$, $100^0$ and $120^0$.
The profiles of the fluid regions
are visualized by the volume-fraction contour levels
$c_i=\frac{1}{2}$ ($1\leqslant i\leqslant 4$).
One can observe that the change in the 
air/fluid-B contact angle not only impacts
the fluid-B region, but also significantly affects
the water region.
Let $\theta_{wA}$ denote the contact angle
of the water/fluid-A interface measured on the water side,
and $\theta_{wB}$ denote the contact angle of
the water/fluid-B interface measured on the water side.
Then according to equation \eqref{equ:ca_relation}
$\theta_{wA}=90^0$  under the current conditions.
On the other hand, as the $\theta_{aB}$ varies
between $60^0$ and $120^0$,
the water/fluid-B contact angle
changes between $\theta_{wB}\approx 132^0$ and
$\theta_{wB}\approx 48^0$ according to equation
\eqref{equ:ca_relation}.
The results shown in Figures \ref{fig:4p_effect_aB}(a)--(d)
are qualitatively consistent with these
theoretical results.

\begin{figure}
\centering
\includegraphics[width=5in]{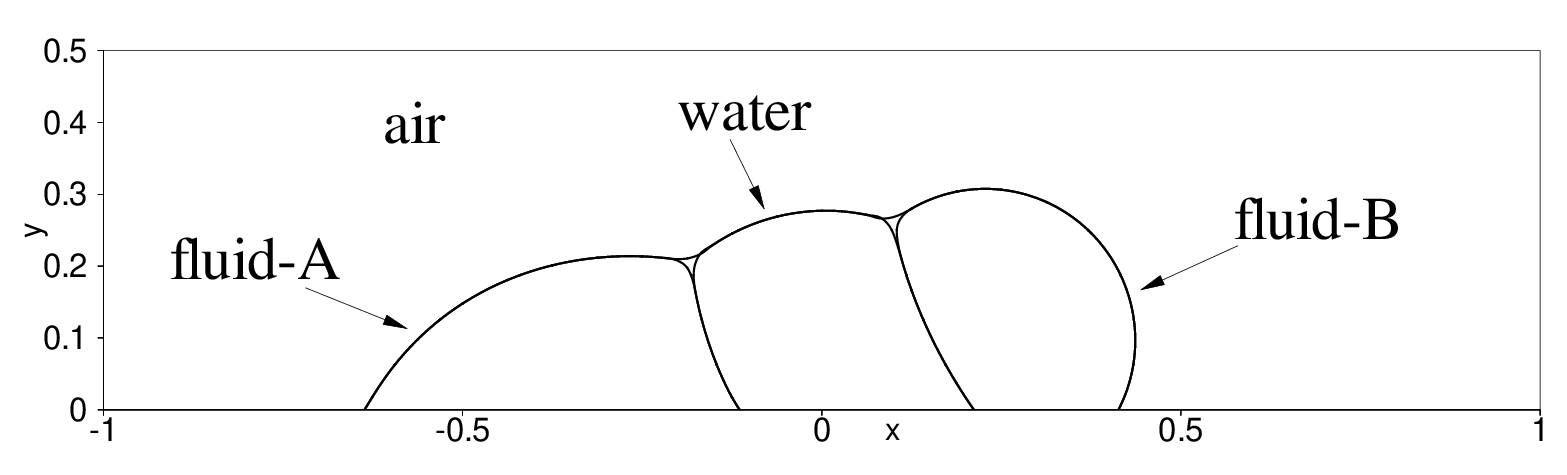}(a)
\includegraphics[width=5in]{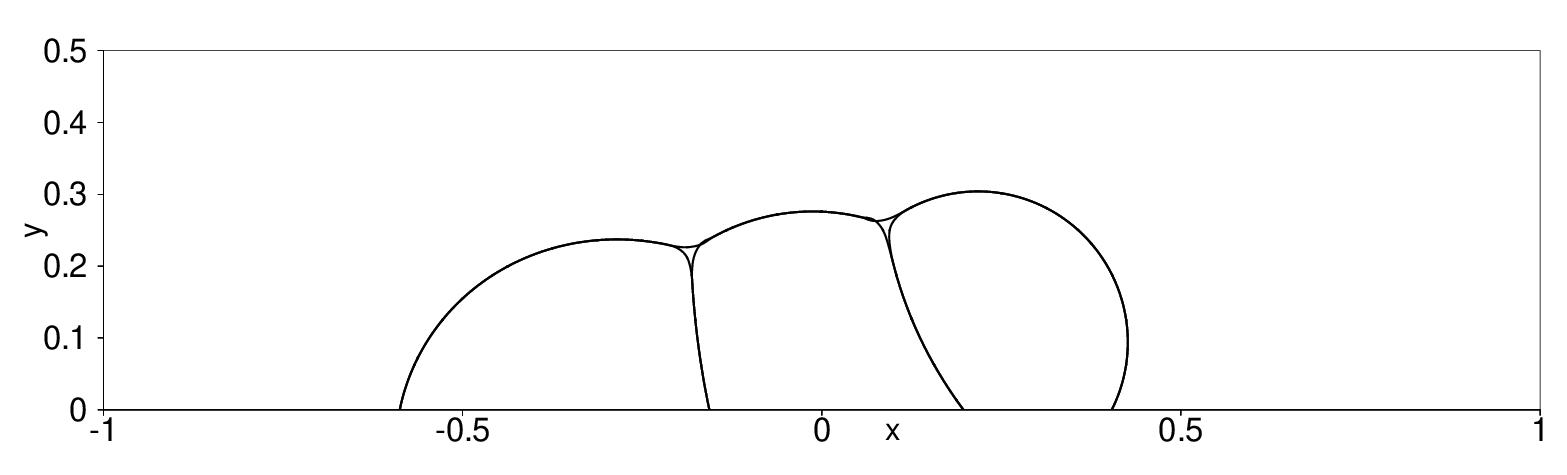}(b)
\includegraphics[width=5in]{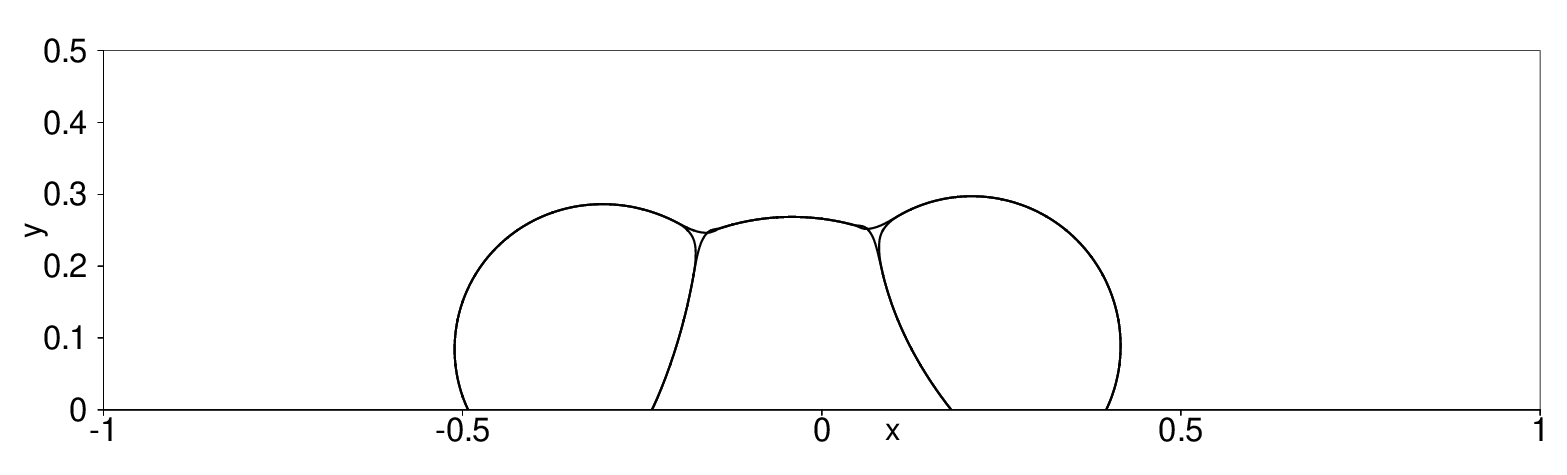}(c)
\caption{
Profiles of a compound liquid drop formed by water, fluid-A and
fluid-B (ambient air) on a wall, with air/water and air/fluid-B contact
angles fixed at $\theta_{aw}=90^0$ and $\theta_{aB}=120^0$ and
the air/fluid-A contact angle varied as:
(a) $\theta_{aA}=60^0$,
(b) $\theta_{aB}=80^0$,
(c) $\theta_{aB}=120^0$.
}
\label{fig:4p_effect_aA}
\end{figure}

Figure \ref{fig:4p_effect_aA} demonstrates the effect of the air/fluid-A
contact angle on the equilibrium configuration of
the system. In this set of tests the air/water contact angle
is fixed at $\theta_{aw}=90^0$ and
the air/fluid-B contact angle is fixed at
$\theta_{aB}=120^0$,
while the air/fluid-A contact angle $\theta_{aA}$ is varied
in a range of values.
The variation in $\theta_{aA}$ alters the profiles
of the fluid-A region and the water region noticeably, while
the fluid-B region seems little affected.
As $\theta_{aA}$ increases from $60^0$ to $120^0$ the region occupied
by the fluid-A becomes more compact,
and the water/fluid-A contact angle is varied between
$\theta_{wA}\approx 129^0$ and $\theta_{wA}\approx 51^0$
according to equation \eqref{equ:ca_relation}.
Note that the water/fluid-B contact angle is $\theta_{wB}\approx 48^0$
based on \eqref{equ:ca_relation} under the current conditions.

\begin{figure}
\centering
\includegraphics[width=5in]{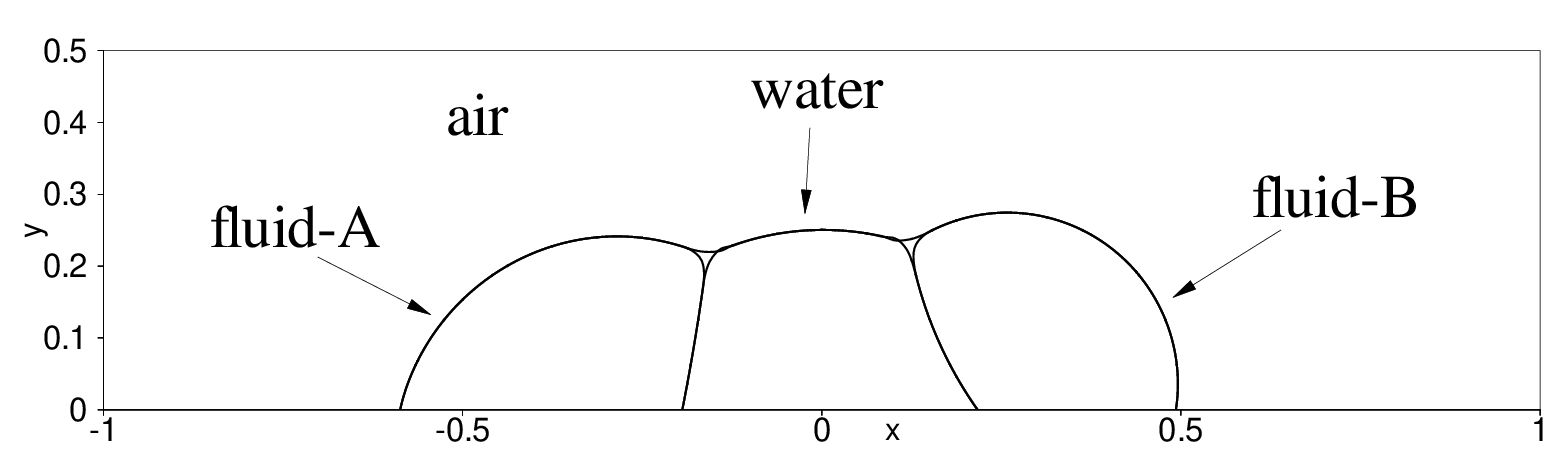}(a)
\includegraphics[width=5in]{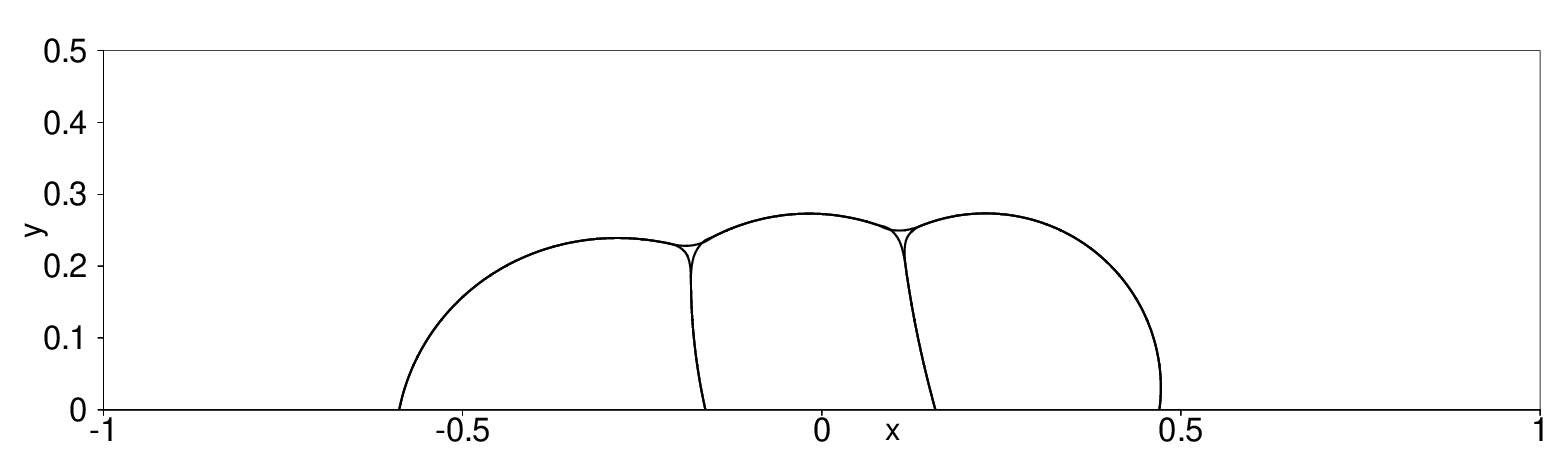}(b)
\includegraphics[width=5in]{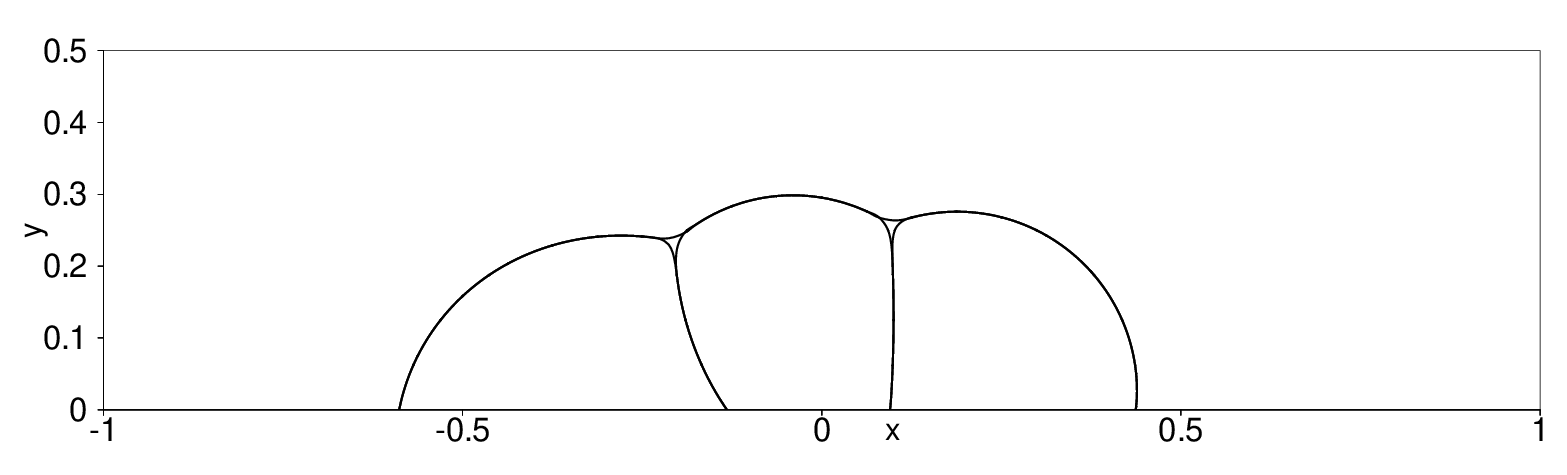}(c)
\caption{
Profiles of a compound liquid drop formed by water, fluid-A and
fluid-B (ambient air) on a wall, with air/fluid-A and air/fluid-B contact
angles fixed at $\theta_{aA}=80^0$ and $\theta_{aB}=100^0$ and
the air/water contact angle varied as:
(a) $\theta_{aw}=75^0$,
(b) $\theta_{aw}=90^0$,
(c) $\theta_{aw}=105^0$.
}
\label{fig:4p_effect_aw}
\end{figure}

Figure \ref{fig:4p_effect_aw} demonstrates the effect of the air-water
contact angle $\theta_{aw}$ on the configuration of the compound drop.
In this group of tests the air/fluid-A and air/fluid-B contact
angles are fixed at $\theta_{aA}=80^0$ and $\theta_{aB}=100^0$ respectively,
and the air/water contact angle is varied.
Note that even though the air/water interface is not in direct contact with the wall
in this problem, the air/water contact angle $\theta_{aw}$ 
influences the contact angles of the other fluid interfaces
in the system
because of the Young's relation \eqref{equ:young_relation}.
The profiles of the regions occupied by
the water, fluid-A and fluid-B have all been modified
by the variation in the air/water contact angle.
The water region appears to become more rounded with 
increasing air/water contact angle.
As the air/water contact angle increases from 
$\theta_{aw}=75^0$ to $\theta_{aw}=105^0$,
the water/fluid-A contact angle (measured on the water side)
increases from $\theta_{wA}\approx 78^0$ 
to $\theta_{wA}\approx 130^0$ according to \eqref{equ:ca_relation},
and the water/fluid-B contact angle (measured on
the water side) increases
from $\theta_{wB}\approx 49^0$ to $\theta_{wB}\approx 101^0$.
The results in Figure \ref{fig:4p_effect_aw} are qualitatively
consistent with these theoretical predictions.

% summary and comments on contact angle effects
% when any contact-angle increases, the region of that fluid tend to become
%    more plump and rounded. The adjacent regions will adjust themselves accordingly.

\subsection{Impact of Liquid Drops on Partially Wettable Walls}

The goal of this section is to demonstrate the performance of the method
developed herein for studying the dynamics of multiphase flows involving
the wall wettability. We will consider the impact of a water drop and an oil drop
on a partially wettable horizontal wall.

\begin{figure}
\centerline{
\includegraphics[width=2in]{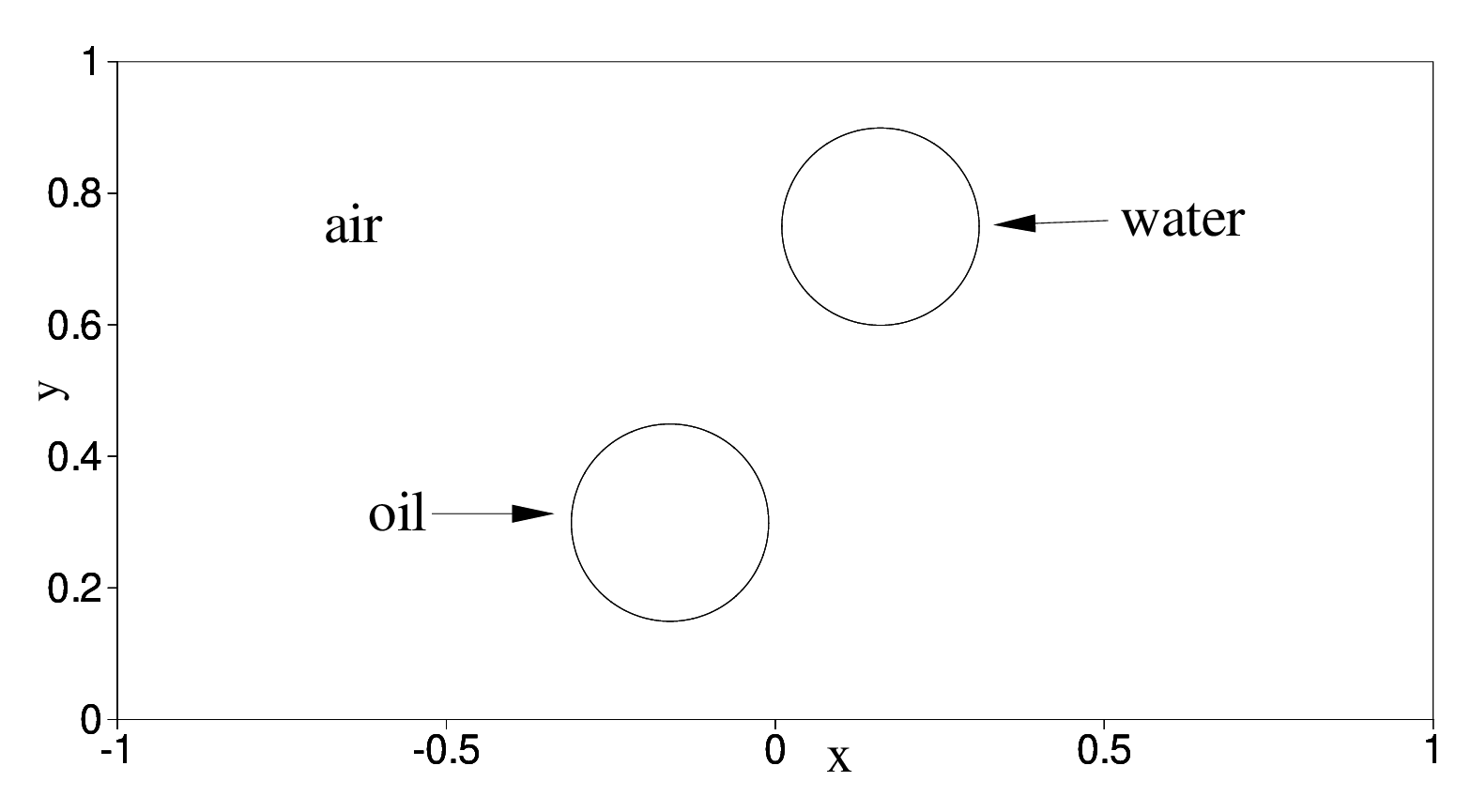}(a)
\includegraphics[width=2in]{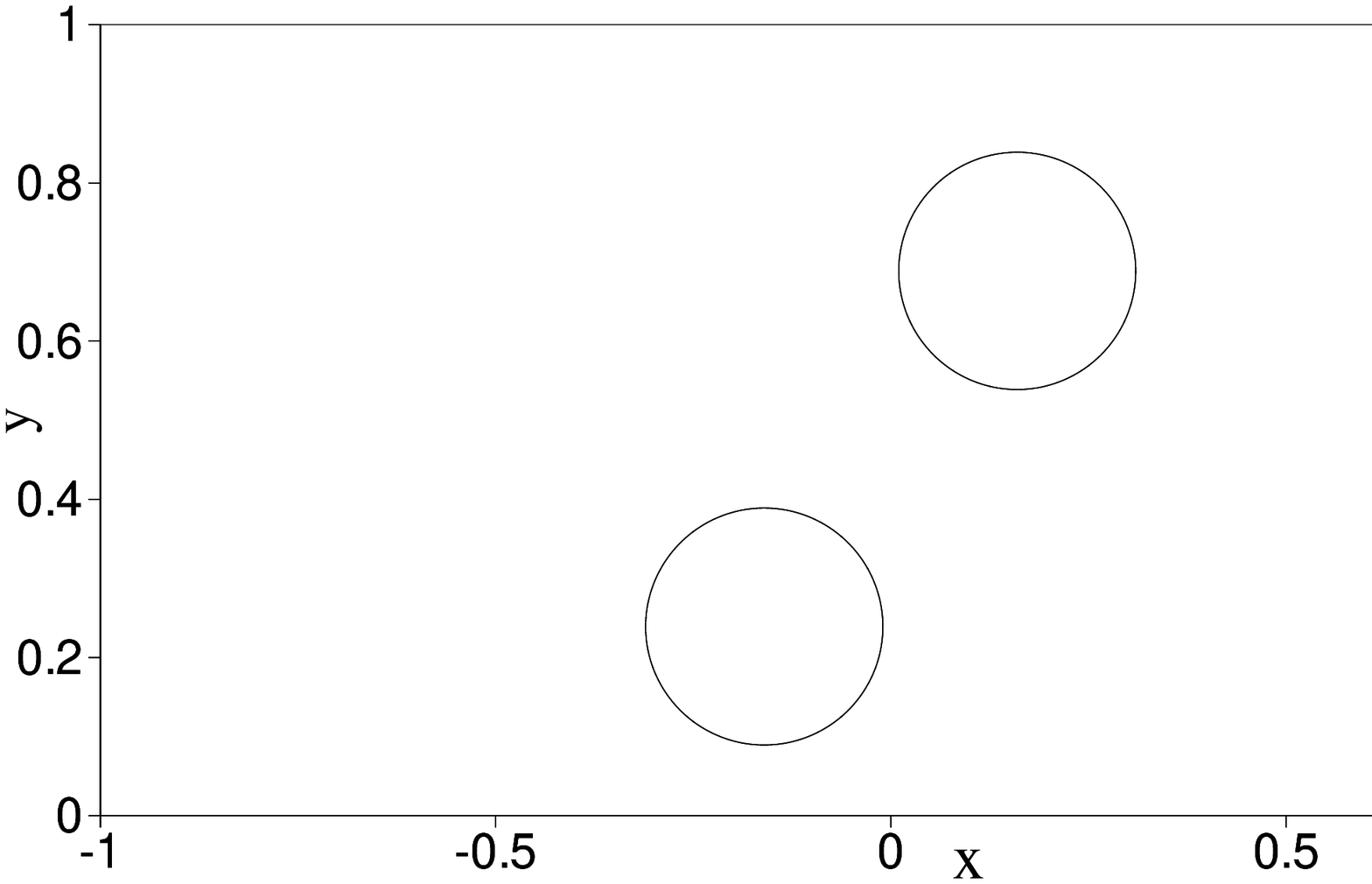}(b)
\includegraphics[width=2in]{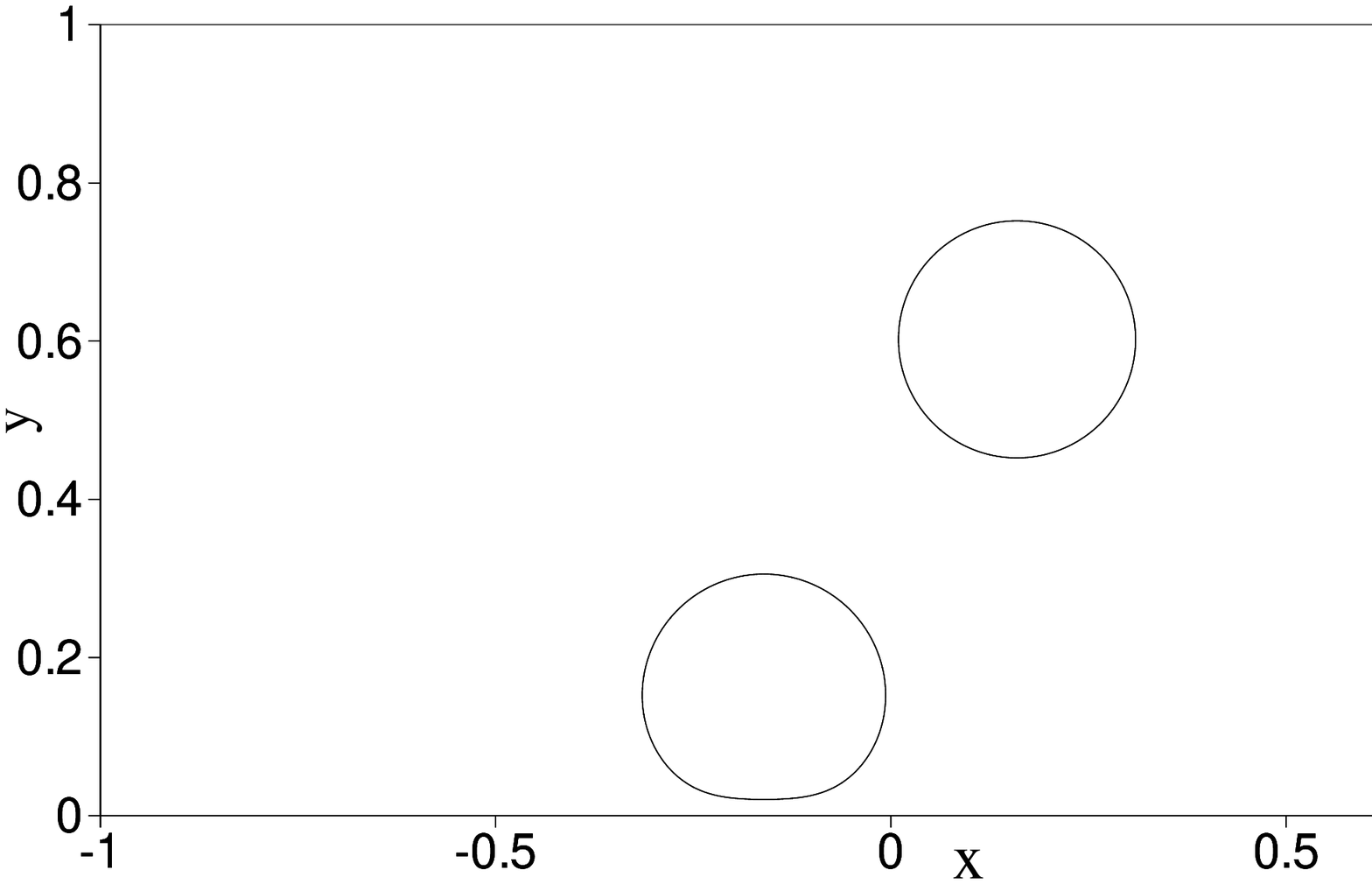}(c)
}
\centerline{
\includegraphics[width=2in]{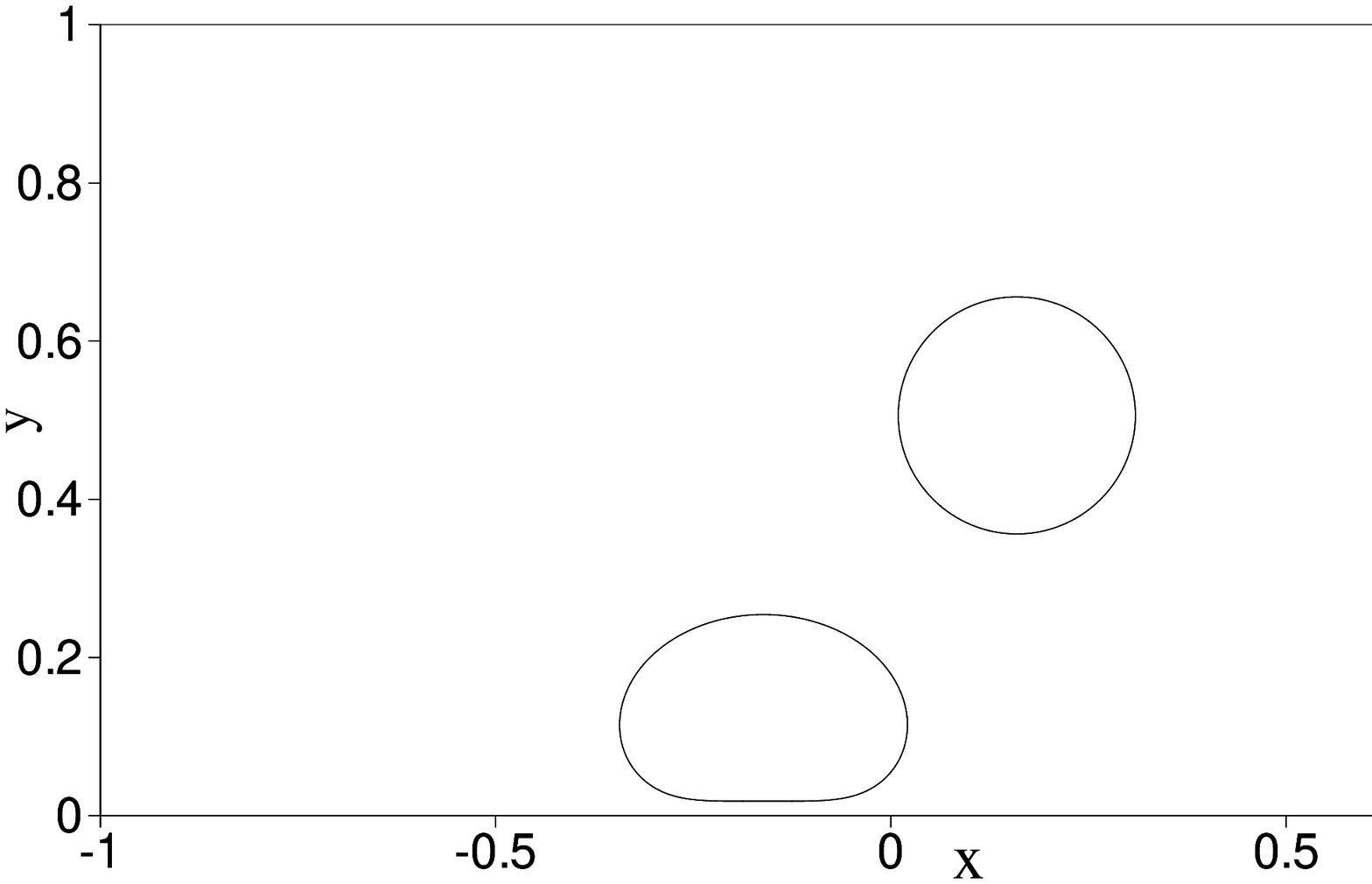}(d)
\includegraphics[width=2in]{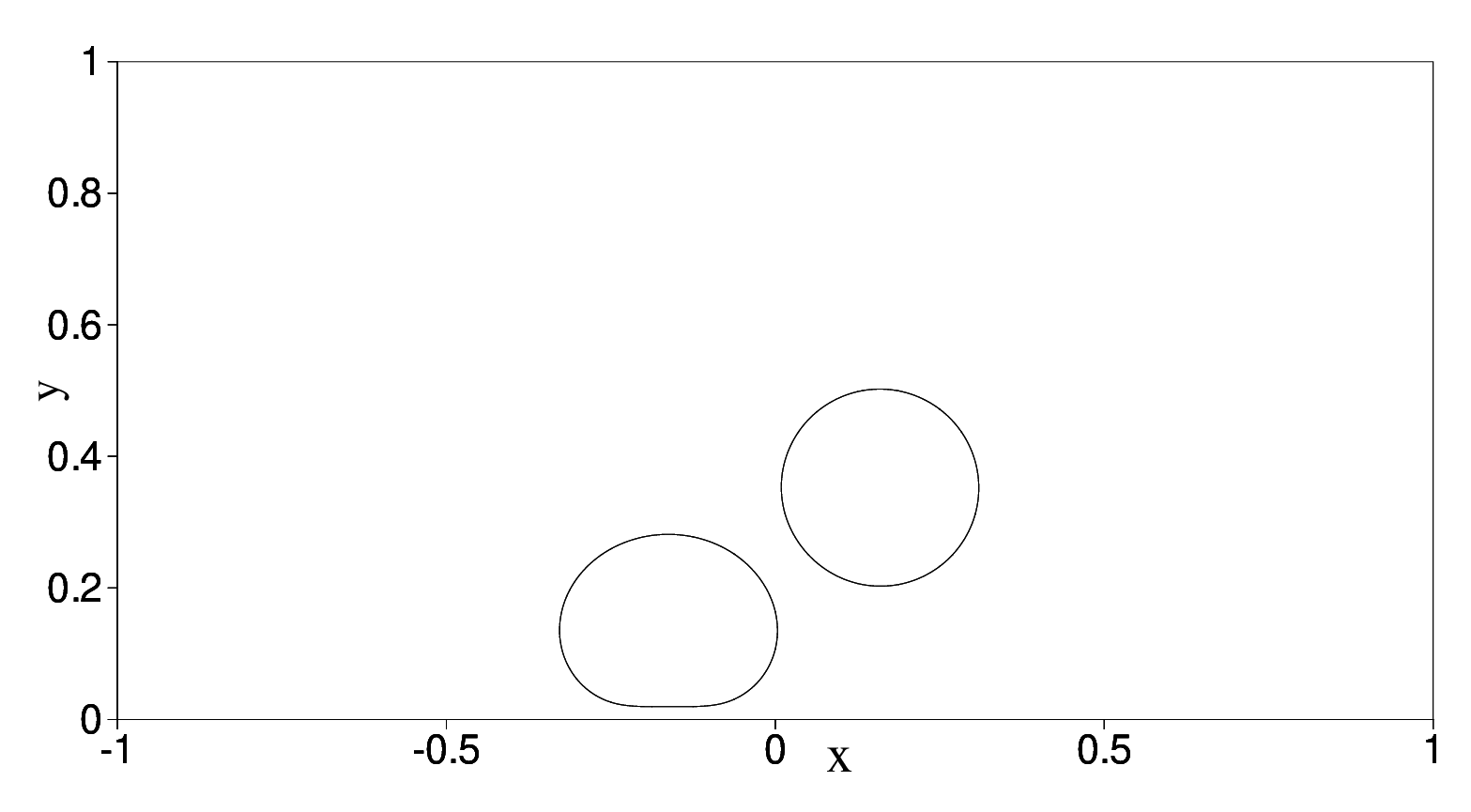}(e)
\includegraphics[width=2in]{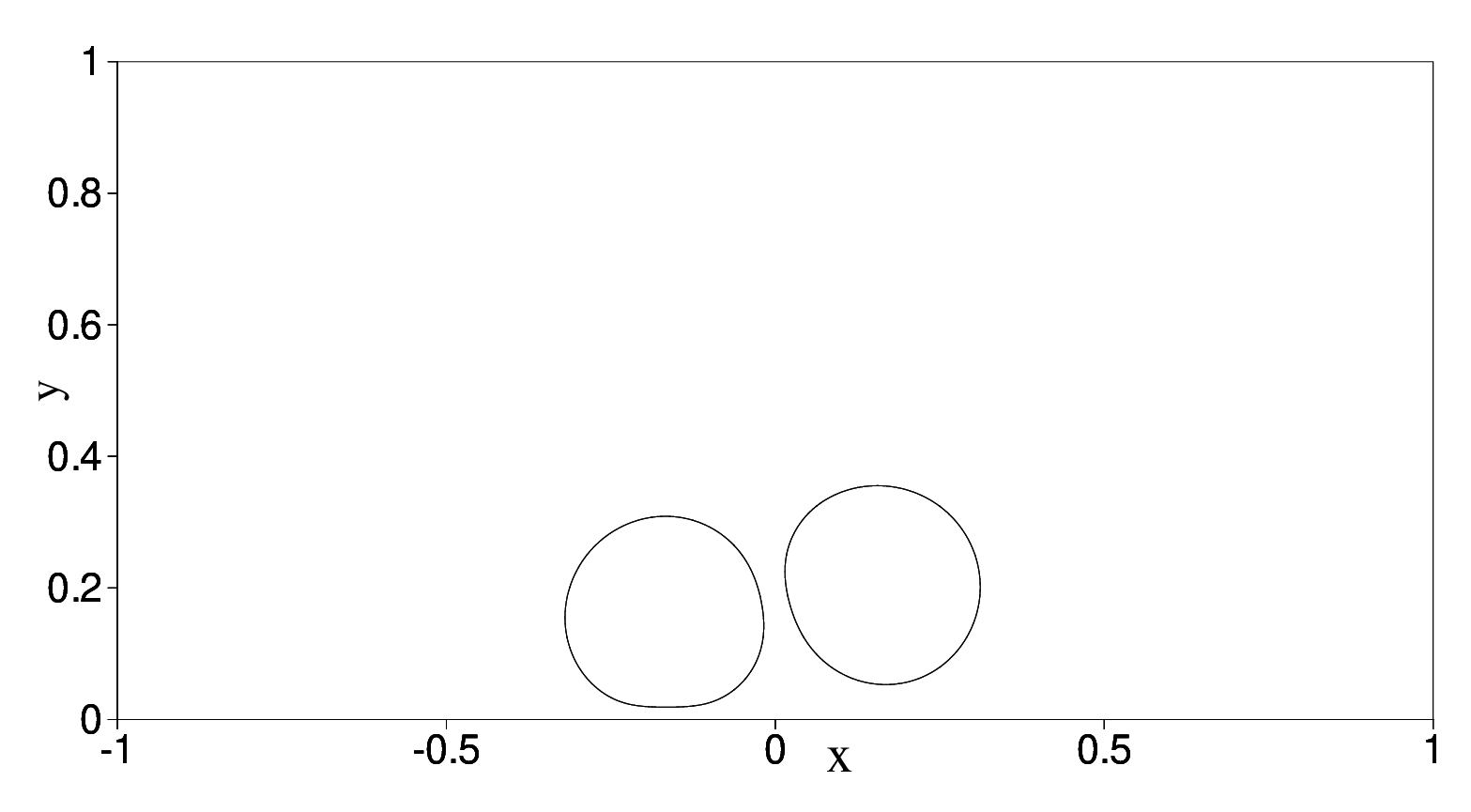}(f)
}
\centerline{
\includegraphics[width=2in]{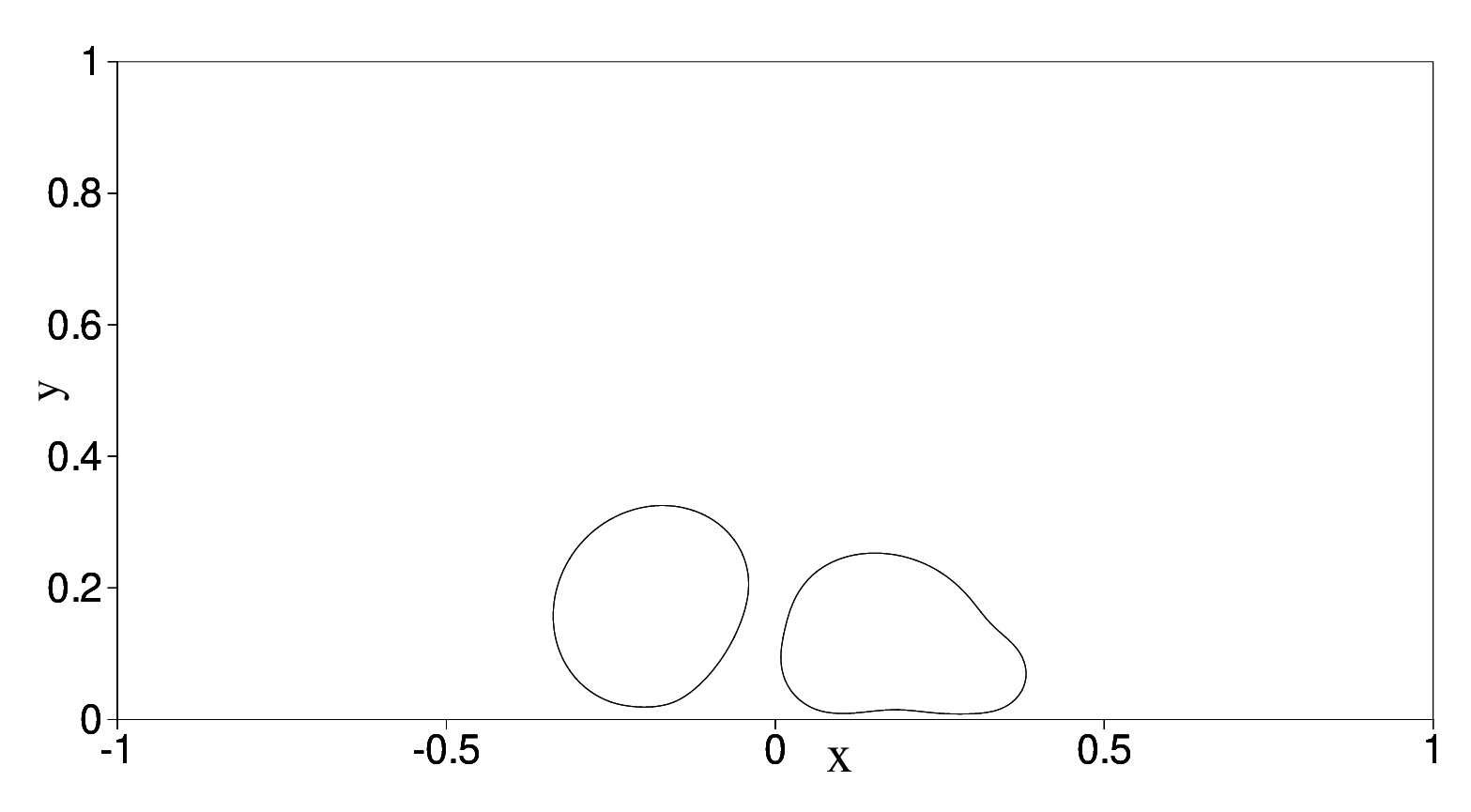}(g)
\includegraphics[width=2in]{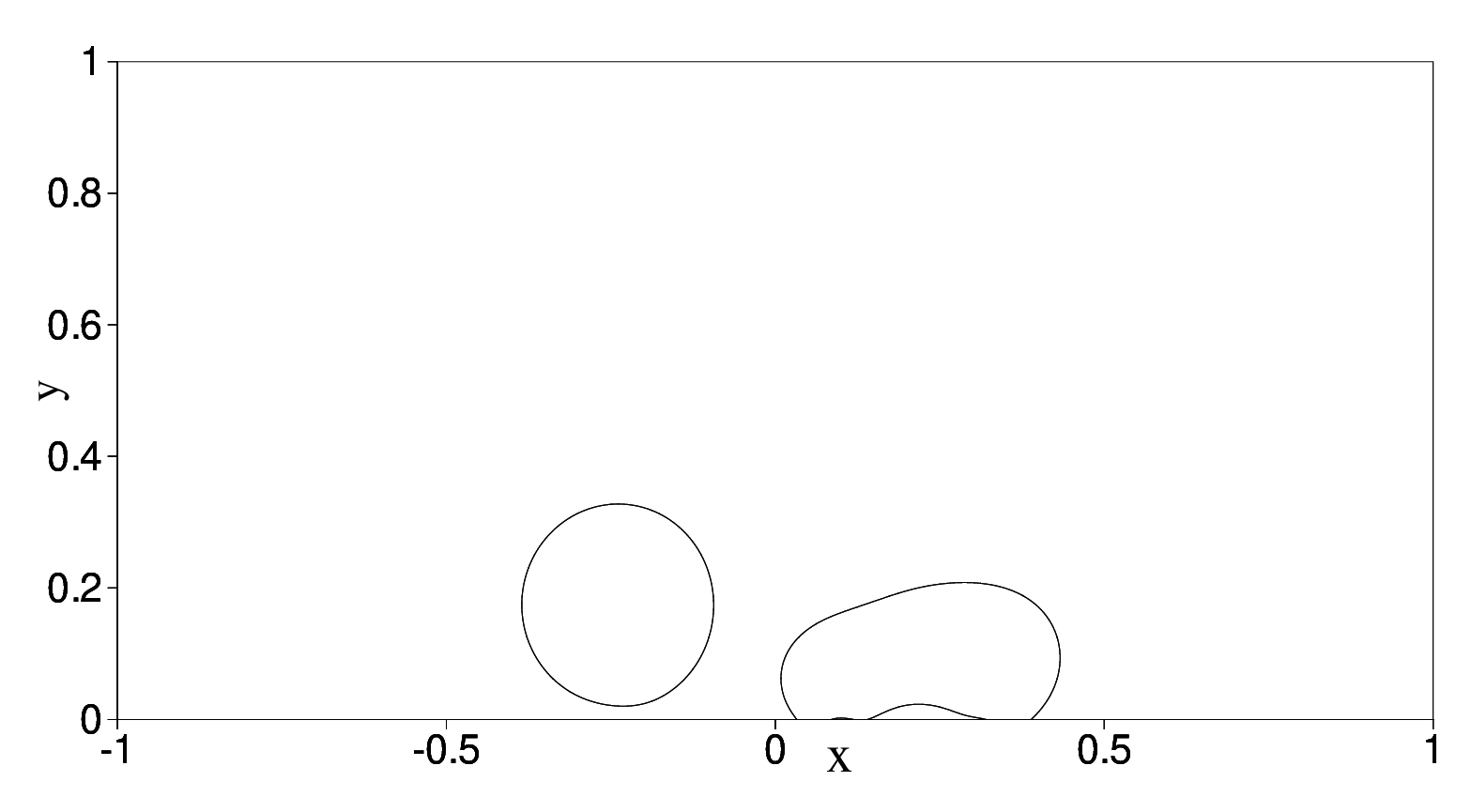}(h)
\includegraphics[width=2in]{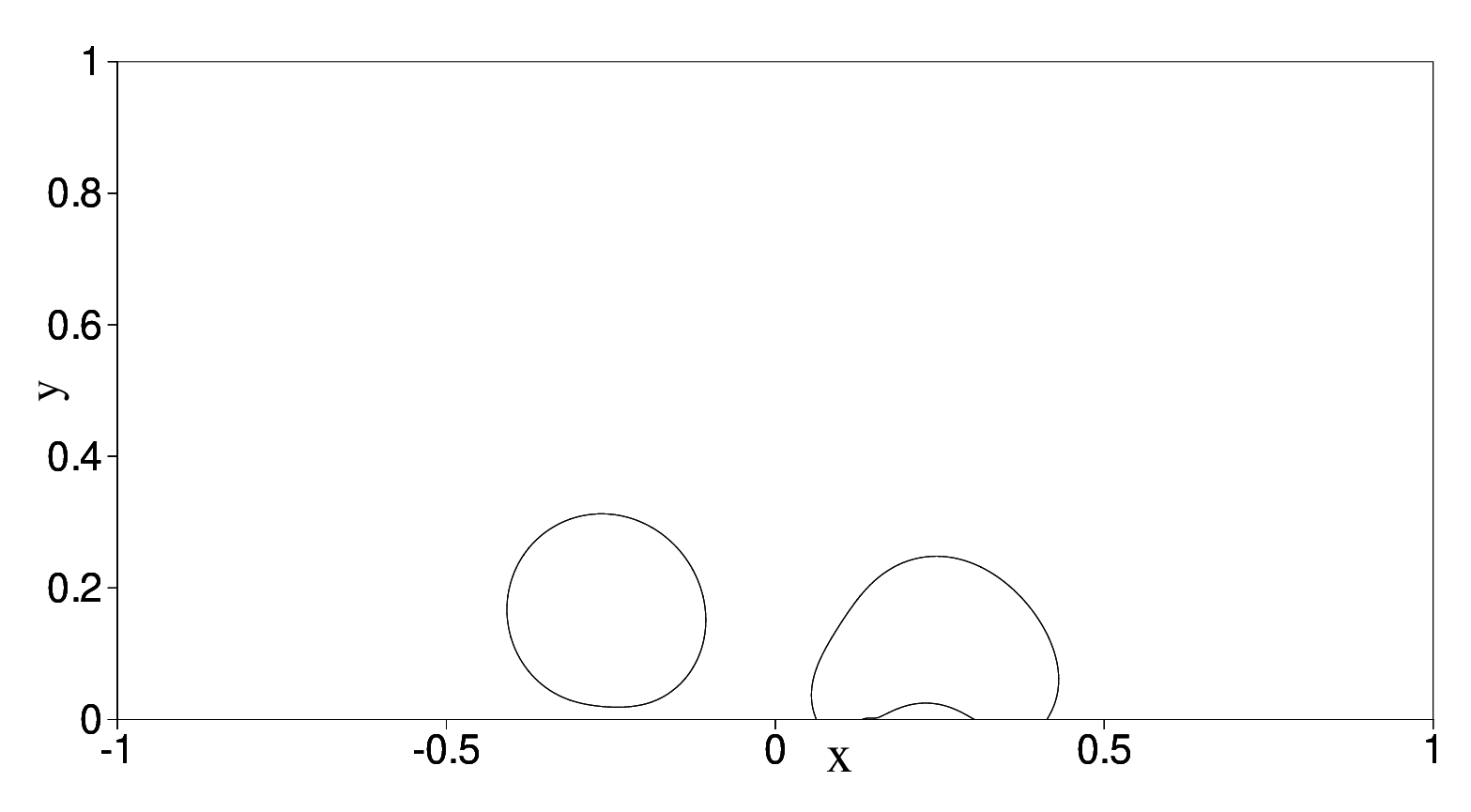}(i)
}
\centerline{
\includegraphics[width=2in]{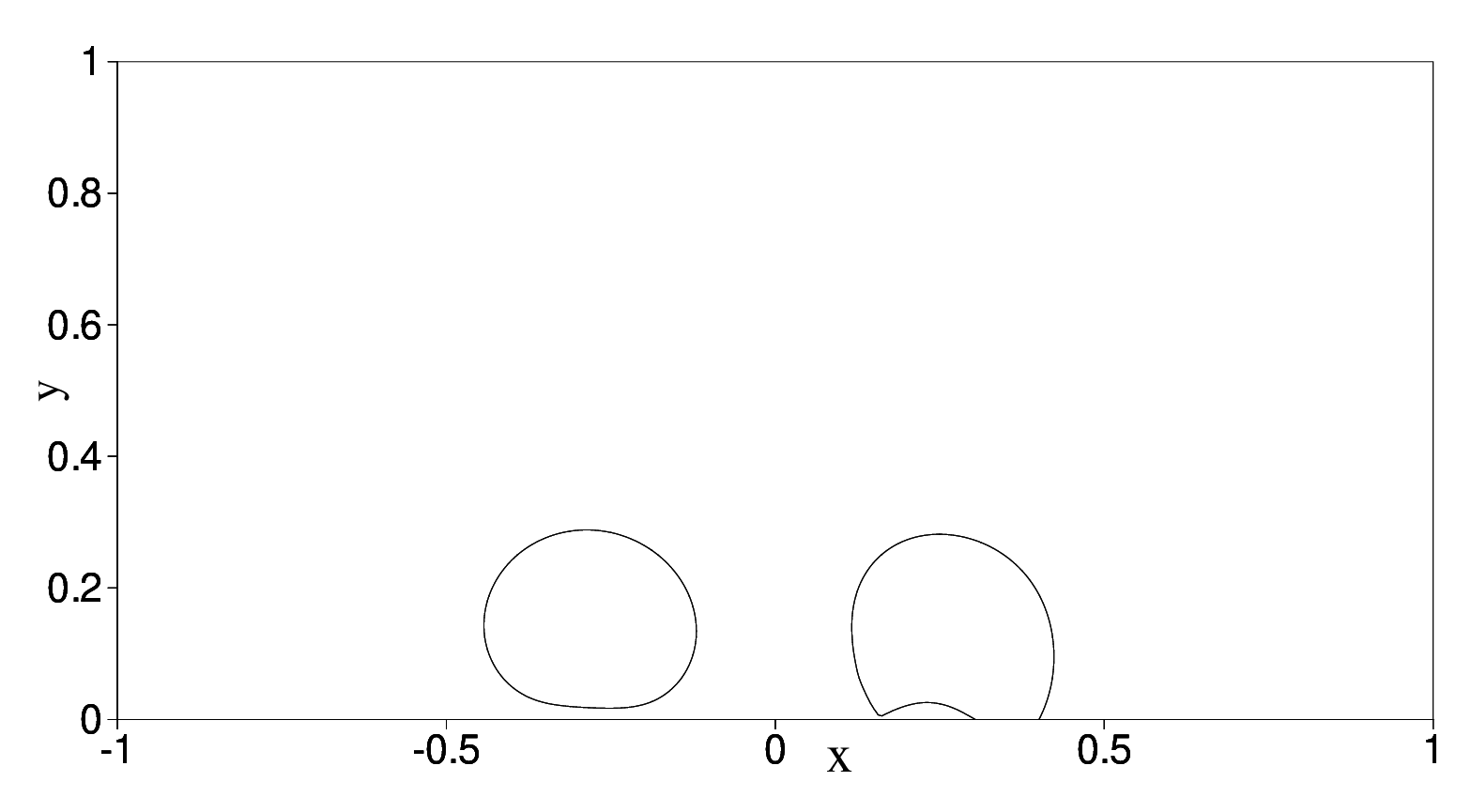}(j)
\includegraphics[width=2in]{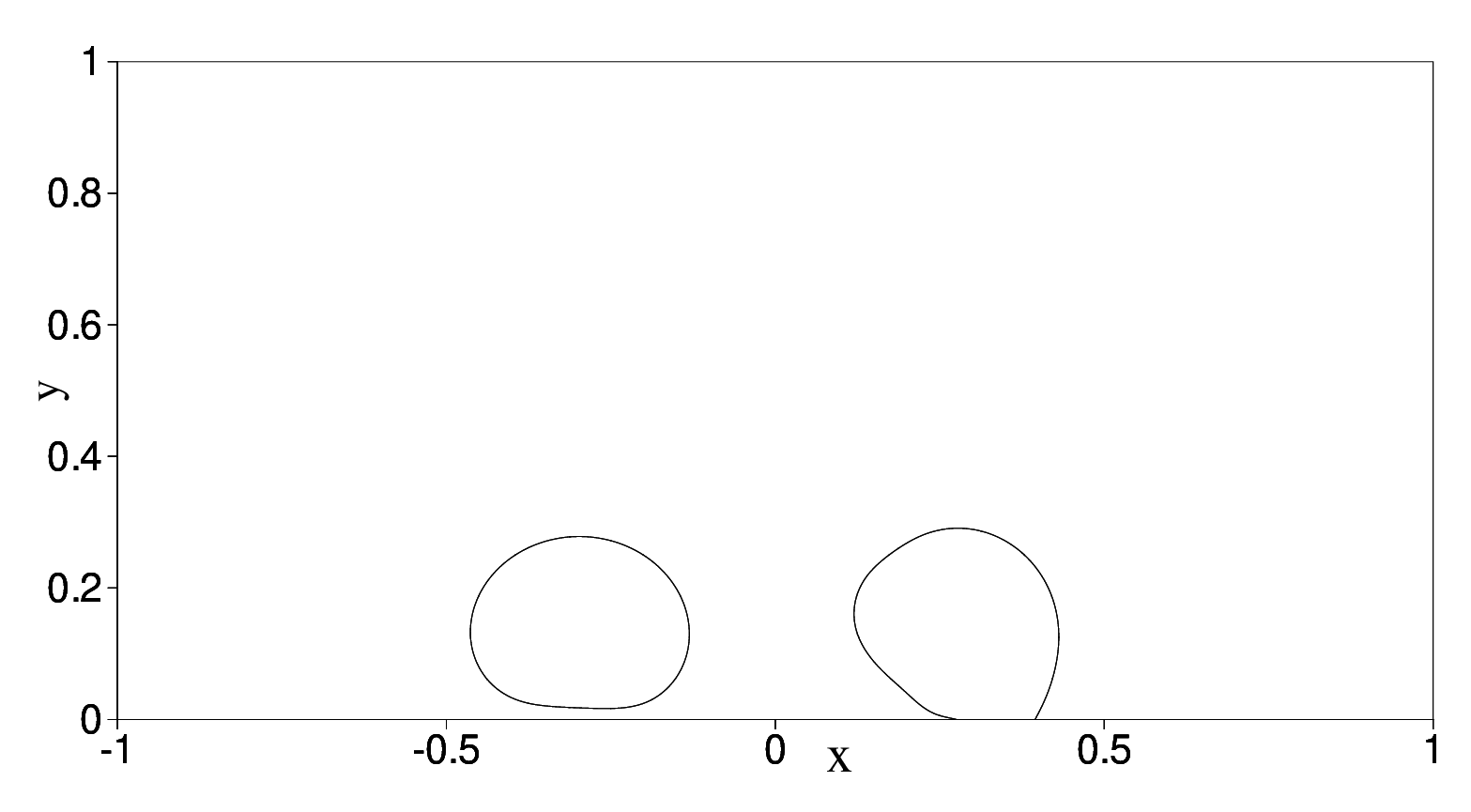}(k)
\includegraphics[width=2in]{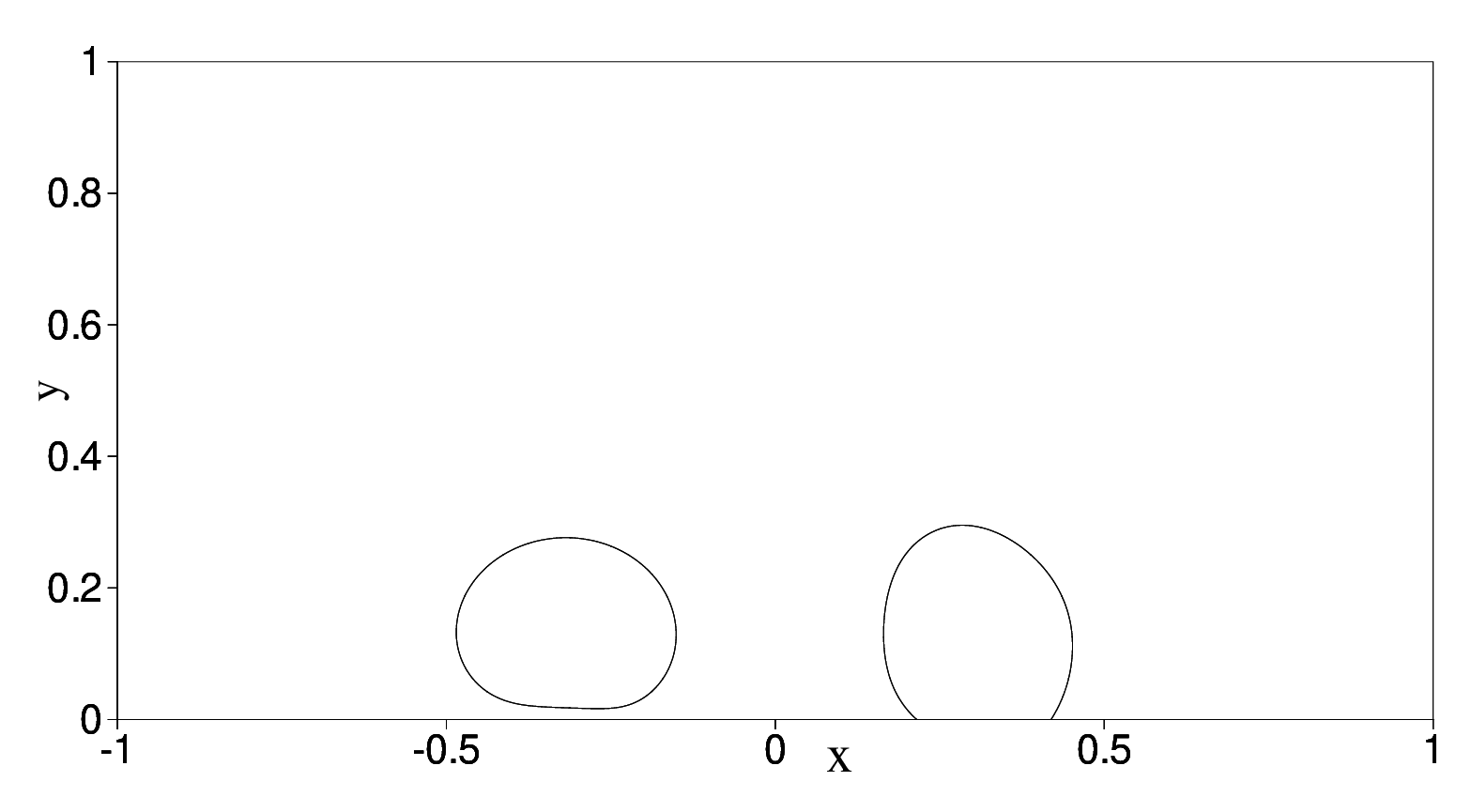}(l)
}
\centerline{
\includegraphics[width=2in]{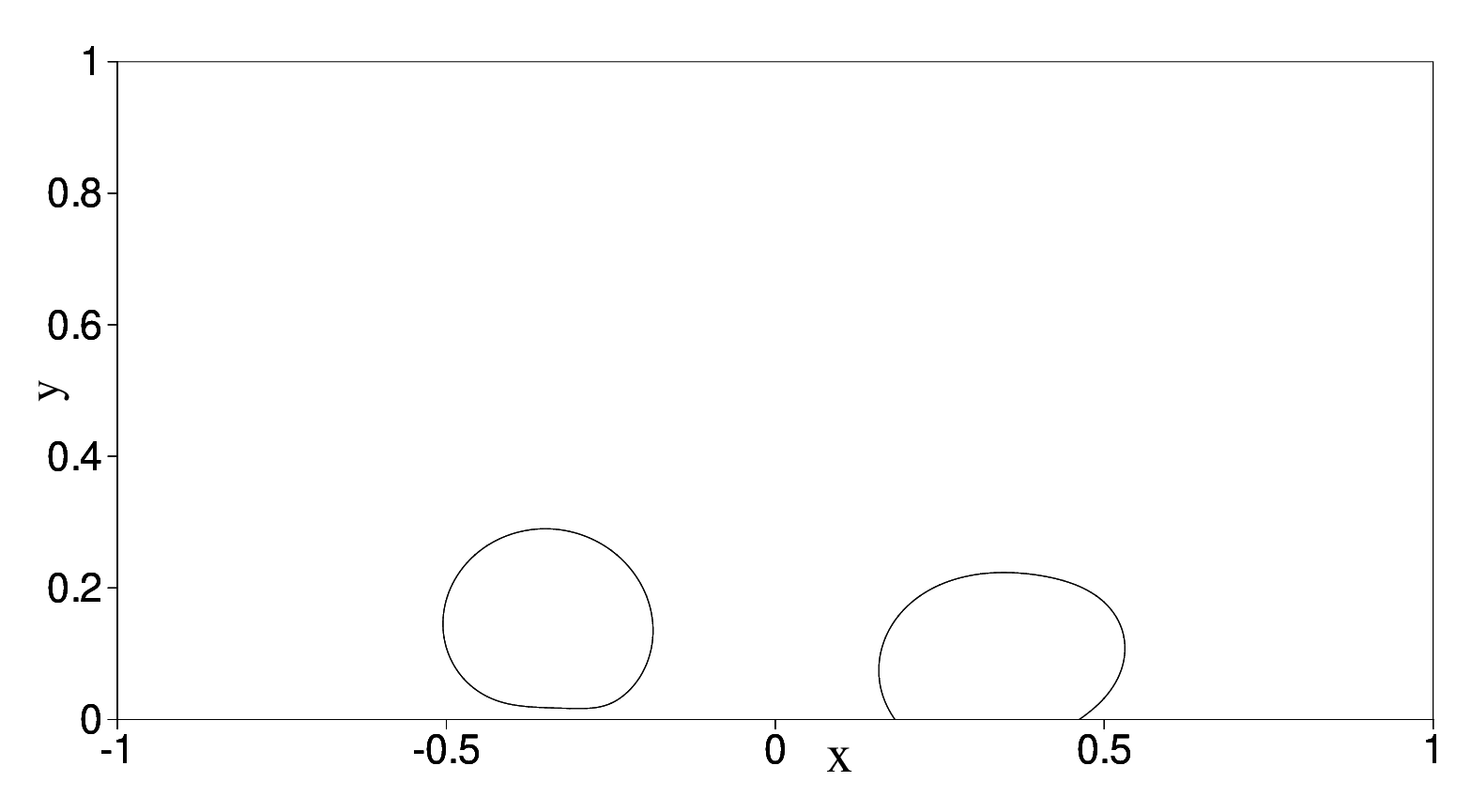}(m)
\includegraphics[width=2in]{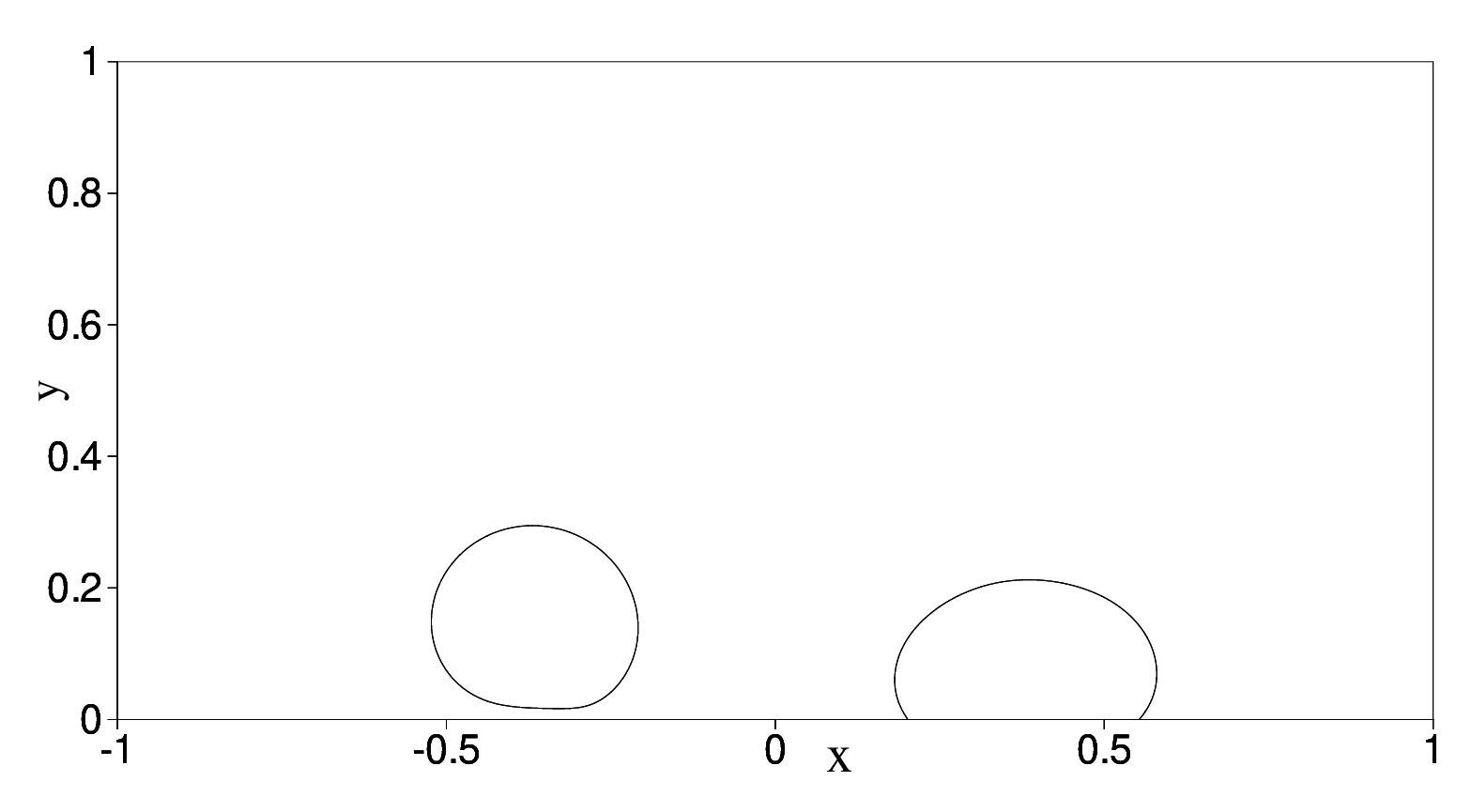}(n)
\includegraphics[width=2in]{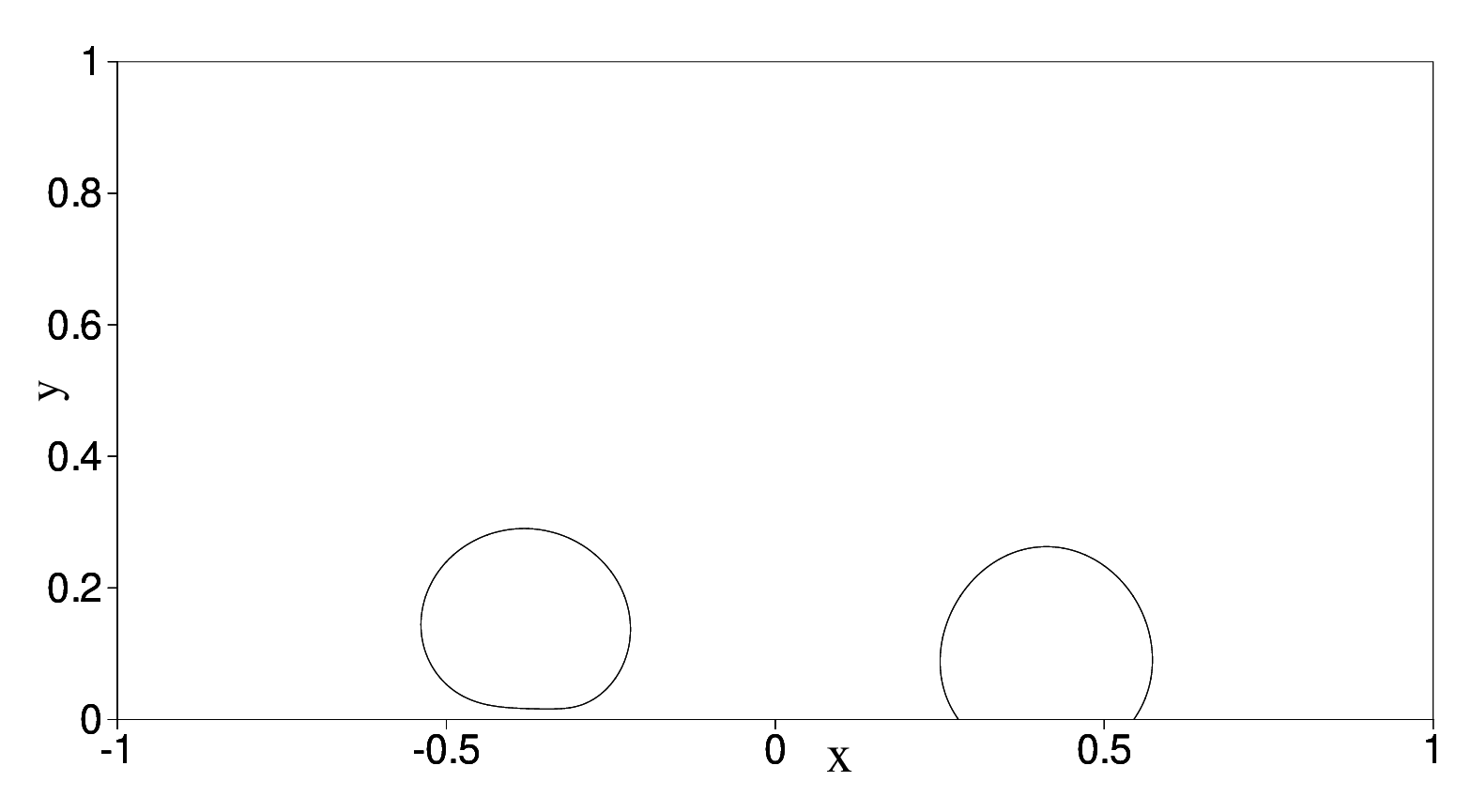}(o)
}
\caption{
Impact of water/oil drops on horizontal wall with contact angles
($\theta_{ao},\theta_{aw})=(120^0,120^0)$: temporal sequence of snapshots
of drop profiles at time instants
(a) $t=0.0125$,
(b) $t=0.1125$,
(c) $t=0.175$,
(d) $t=0.225$,
(e) $t=0.2875$,
(f) $t=0.3375$,
(g) $t=0.375$,
(h) $t=0.4375$,
(i) $t=0.4625$,
(j) $t=0.5$,
(k) $t=0.525$,
(l) $t=0.5625$,
(m) $t=0.625$,
(n) $t=0.6875$,
(o) $t=0.75$.
}
\label{fig:phase_profile_120_120}

\end{figure}

More specifically, the problem configuration is illustrated in
Figure \ref{fig:phase_profile_120_120}(a).
We consider a rectangular domain of dimensions
$-L\leqslant x\leqslant L$ and $0\leqslant y\leqslant L$,
where $L=8mm$. The top and bottom of the domain are two solid
walls. In the horizontal direction
the domain is periodic at $x=\pm L$.
The domain is filled with air. A drop of wall and a drop of
oil, both initially circular with a radius $R_0=0.15L$, are suspended in the air and
held at rest. The center of the water drop is located 
at $\mathbf{X}_w=(x_w,y_w)=(0.16L,0.75L)$, and
the oil-drop center is located at
$\mathbf{X}_o=(x_o,y_o)=(-0.16L,0.3L)$.
The gravity is assumed to be in the $-y$ direction.
At $t=0$, the system is released, and the water and oil drops
fall through the air. The oil drop impacts the wall first, followed
by the water drop. Let $\theta_{aw}$ denote the 
contact angle between the air-water interface and the wall (measured
on the water side), and 
$\theta_{ao}$ denote the contact angle between the air-oil interface 
and the wall (measured on the oil side).
We look into the effect of the wall wettability on 
the dynamic behavior of this three-phase system.

We use the values listed in Table \ref{tab:air_water_param} for the physical parameters
of air, water and oil in this problem.
The gravitational acceleration is $9.8m/s^2$ in this problem.
We take $L$ as the length scale, $U_0=\sqrt{g_{r0}L}$ (where $g_{r0}=1m/s^2$)
as the velocity scale, and the air density as the density scale
$\varrho_d$. The variables and parameters are then
normalized according to Table \ref{tab:normalization}.
We assign water, oil and air as the first, second, and third fluids,
respectively. The formulation with the volume fractions $c_i$ ($1\leqslant i\leqslant N-1$)
 as the order parameters is employed in the simulations.

\begin{table}
\begin{center}
\begin{tabular}{l l}
\hline
Parameters & Values \\
$\phi_i$ & defined by \eqref{equ:gop_volfrac}, volume fractions as order parameters \\
$\zeta_{ij}$ & Computed based on \eqref{equ:theta_expr} and \eqref{equ:lambda_mat_gop} \\
$\eta/L$ & $0.01$ \\
$m_0/(U_0L^3)$ & $10^{-8}$ \\
$\rho_0$ & $\min(\tilde{\rho}_1, \tilde{\rho}_2, \tilde{\rho}_3)$ \\
$\nu_0$ & $2\max\left(\frac{\tilde{\mu}_1}{\tilde{\rho}_1},\frac{\tilde{\mu}_2}{\tilde{\rho}_2}, \frac{\tilde{\mu}_3}{\tilde{\rho}_3}\right)$ \\
$S$ & $\eta^2\sqrt{\frac{4\gamma_0}{m_0\Delta t} }$ \\
$\alpha$ & Computed based on \eqref{equ:alpha_expr} \\
$U_0\Delta t/L$ & $2.5\times 10^{-6}$ \\
$\theta_{13}$ ($\theta_{aw}$)  &  $120^0$, $150^0$ \\
$\theta_{23}$ ($\theta_{ao}$) & $120^0$ \\
$J$ (temporal order) & $2$ \\
Number of elements & $512$ \\
Element order & $12$ \\
\hline
\end{tabular}
\caption{Simulation parameter values for the dynamic problem of
water and oil drops impacting the wall.
}
\label{tab:dyn_3p}
\end{center}
\end{table}

% discretizations, BCs, ICs, and simulation parameters

In order to simulate the problem, we discretize the domain using
$512$ equal-sized quadrilateral spectral elements,
with $32$ elements in the $x$ direction and
$16$ elements in the $y$ direction.
An element order $12$ is employed for each element in
the simulations. 
%
% BCs
On the top and bottom walls we impose the Dirichlet condition \eqref{equ:dbc}
with $\mathbf{w}=0$ for the velocity, and 
the contac-angle conditions \eqref{equ:bc_chempot_2}--\eqref{equ:cabc_2}
with $g_{ai}=0$ and $g_{bi}=0$ for the phase field variables.
On the horizontal boundaries $x=\pm L$ periodic conditions
are imposed on all the flow variables.
%
% ICs
The initial velocity is set to $\mathbf{u}=0$. Ihe initial
phase field distributions are given by equation \eqref{equ:init_phase},
in which the drop coordinates and radii are replaced by values 
for this problem.
Table \ref{tab:dyn_3p} lists the values of  simulation parameters
for this problem.

% contact angles and result discussions

We have considered two sets of contact angles to demonstrate 
the effect of the wall wettability on the dynamics of the system.
The air-oil contact angle is $\theta_{ao}=120^0$ in both sets,
while the air-water contact angle is $\theta_{aw}=120^0$ in one
set and $\theta_{aw}=150^0$ in the other.

We first look into the dynamic behavior of the liquid drops
for the set of contact angles
$(\theta_{aw},\theta_{ao})=(120^0,120^0)$.
Figure \ref{fig:phase_profile_120_120} is a temporal sequence of snapshots 
 of the fluid interfaces of this system, visualized by
the contours of volume fractions $c_i=\frac{1}{2}$ ($1\leqslant i\leqslant 3$).
As the system is released, the water and oil drops fall
through the air (Figure \ref{fig:phase_profile_120_120}(a)-(c)). 
The oil drop impacts the wall first (Figure \ref{fig:phase_profile_120_120}(c)). 
The impact has caused the oil drop to deform (Figure \ref{fig:phase_profile_120_120}(d)).
Subsequently, the falling water drop rushes by the oil drop (Figure \ref{fig:phase_profile_120_120}(e)-(f)).
The hydrodynamic interactions between the water and air and between
the air and oil push the oil drop laterally along the 
wall away from the water drop (Figure \ref{fig:phase_profile_120_120}(g)-(m)).
These interactions also cause 
the water drop to move sideways after it impacts the 
bottom wall (Figure \ref{fig:phase_profile_120_120}(h)-(o)).
The impact onto the wall has caused the water drop to deform
significantly. The water drop appears to have trapped a small pocket of
air underneath (Figure \ref{fig:phase_profile_120_120}(h)-(i)), 
which escapes from the underside of the water drop subsequently 
as the water drop moves sideways (Figure \ref{fig:phase_profile_120_120}(j)-(k)).

\begin{figure}
\centerline{
\includegraphics[width=2in]{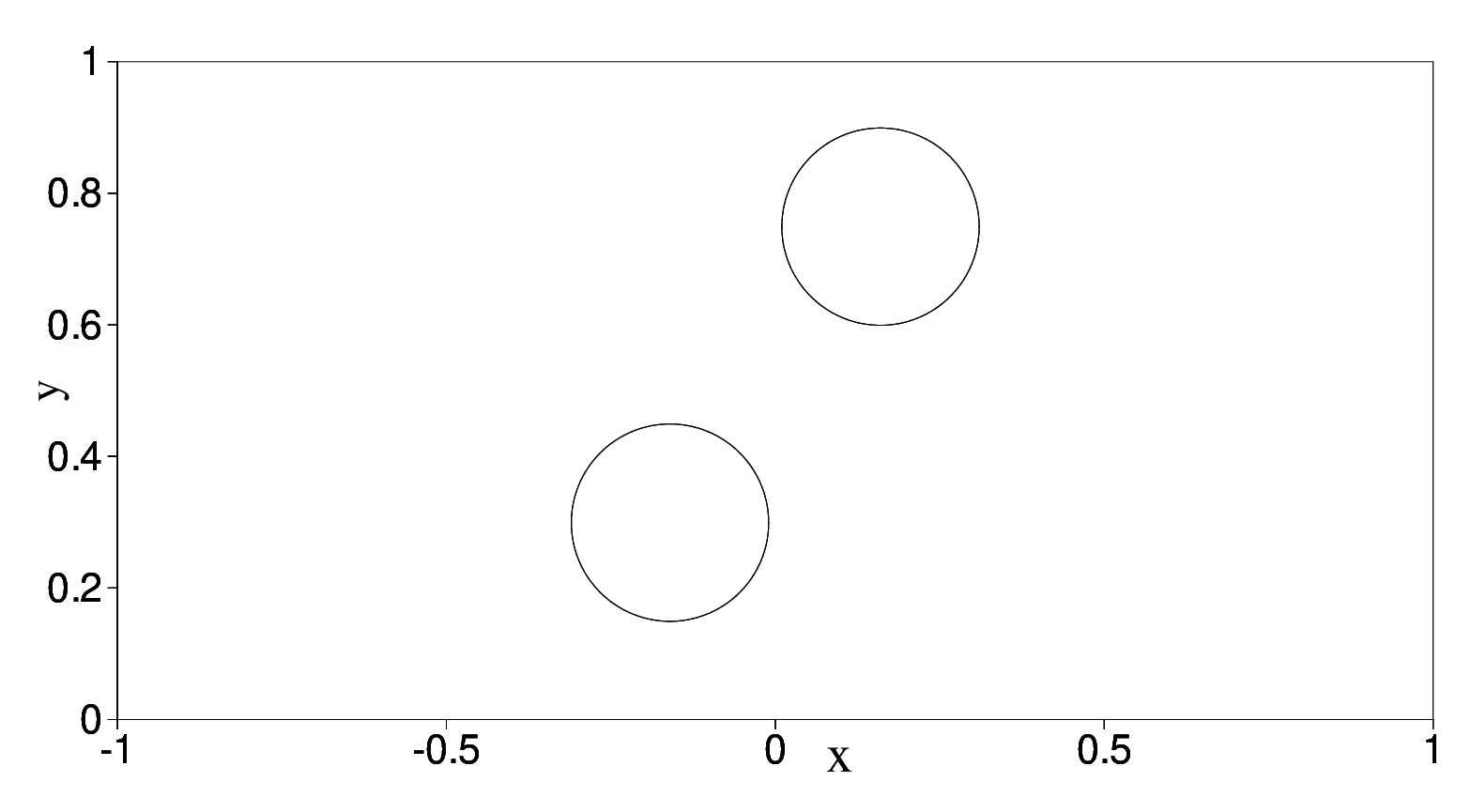}(a)
\includegraphics[width=2in]{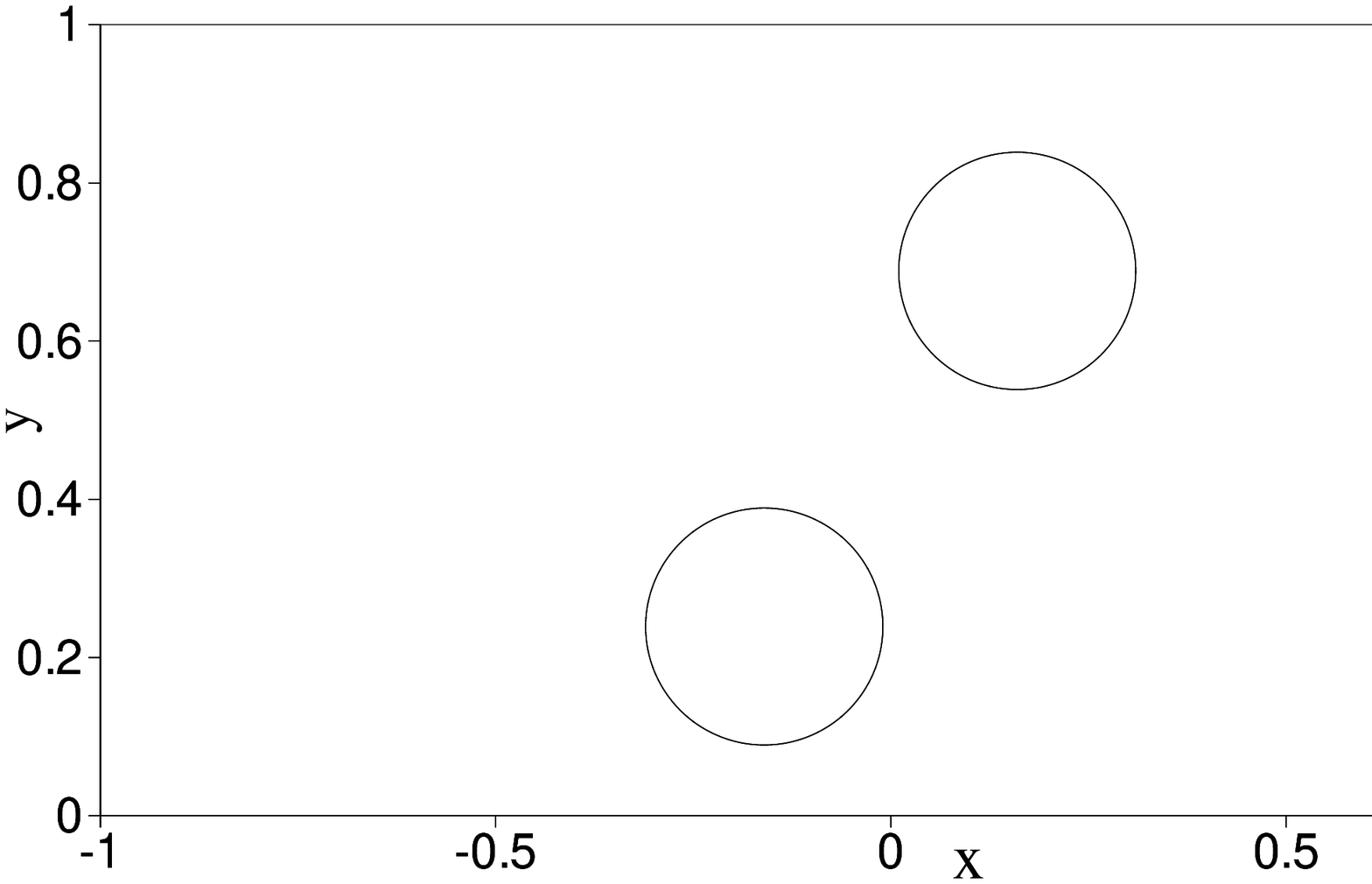}(b)
\includegraphics[width=2in]{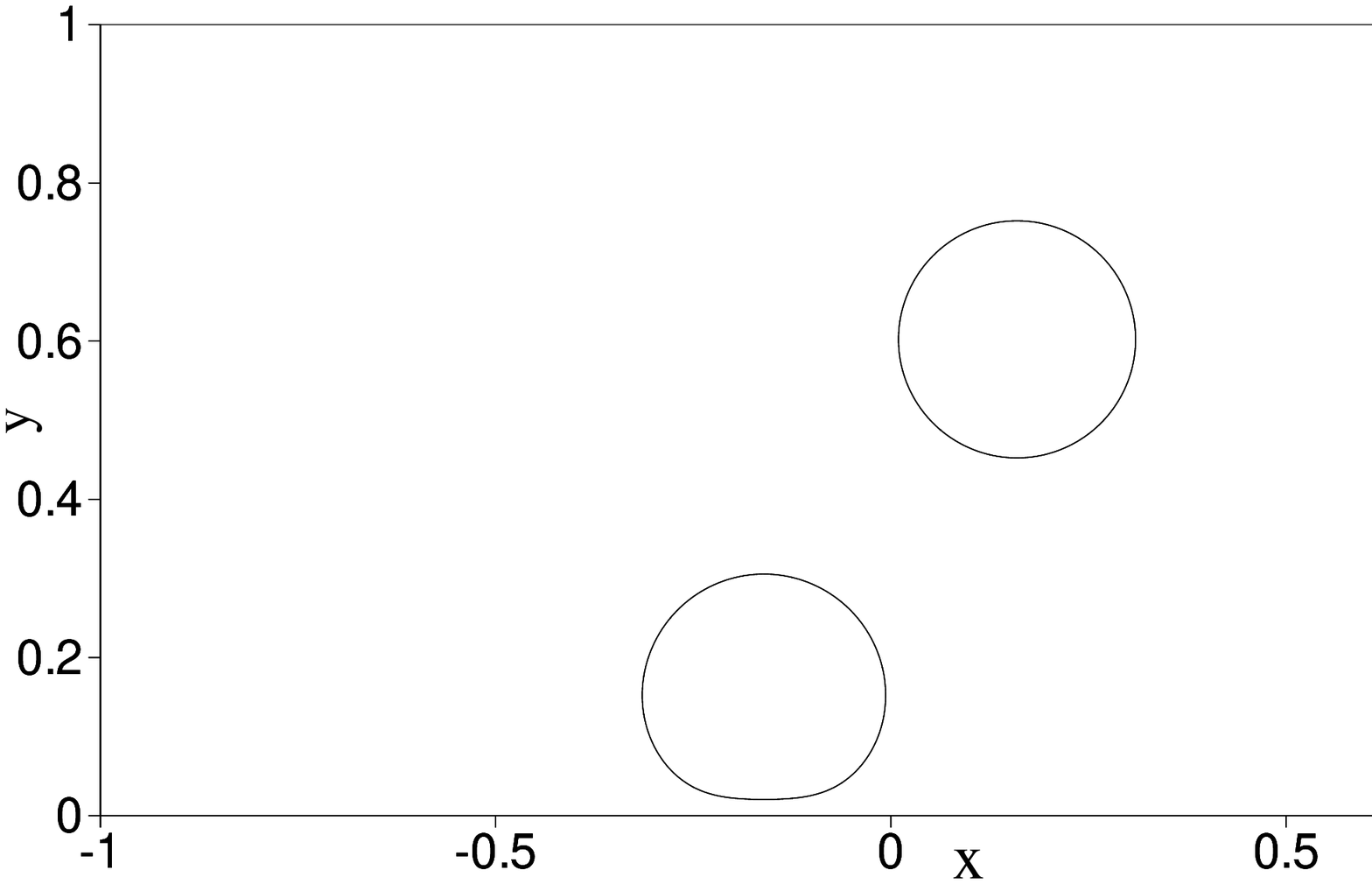}(c)
}
\centerline{
\includegraphics[width=2in]{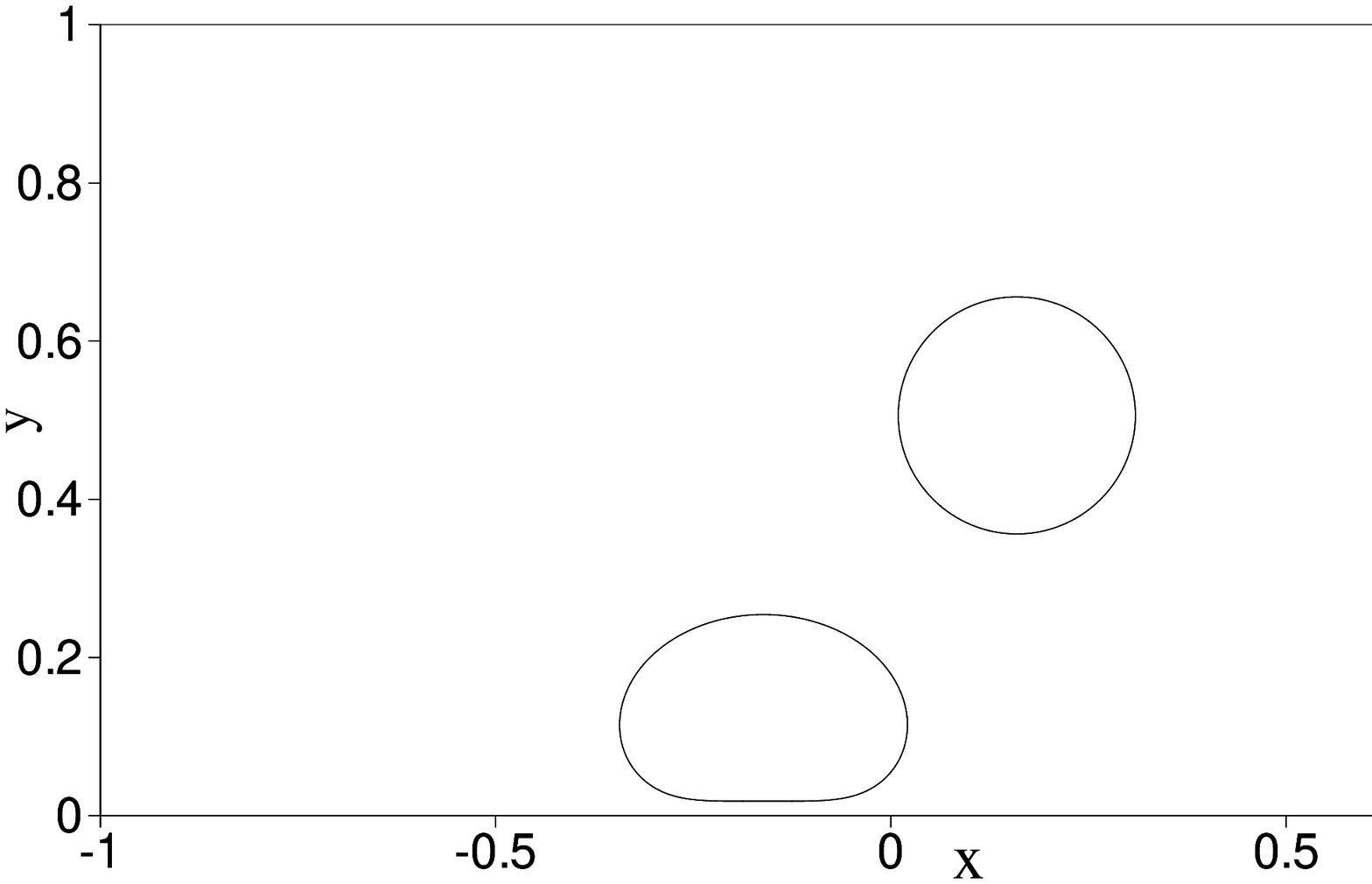}(d)
\includegraphics[width=2in]{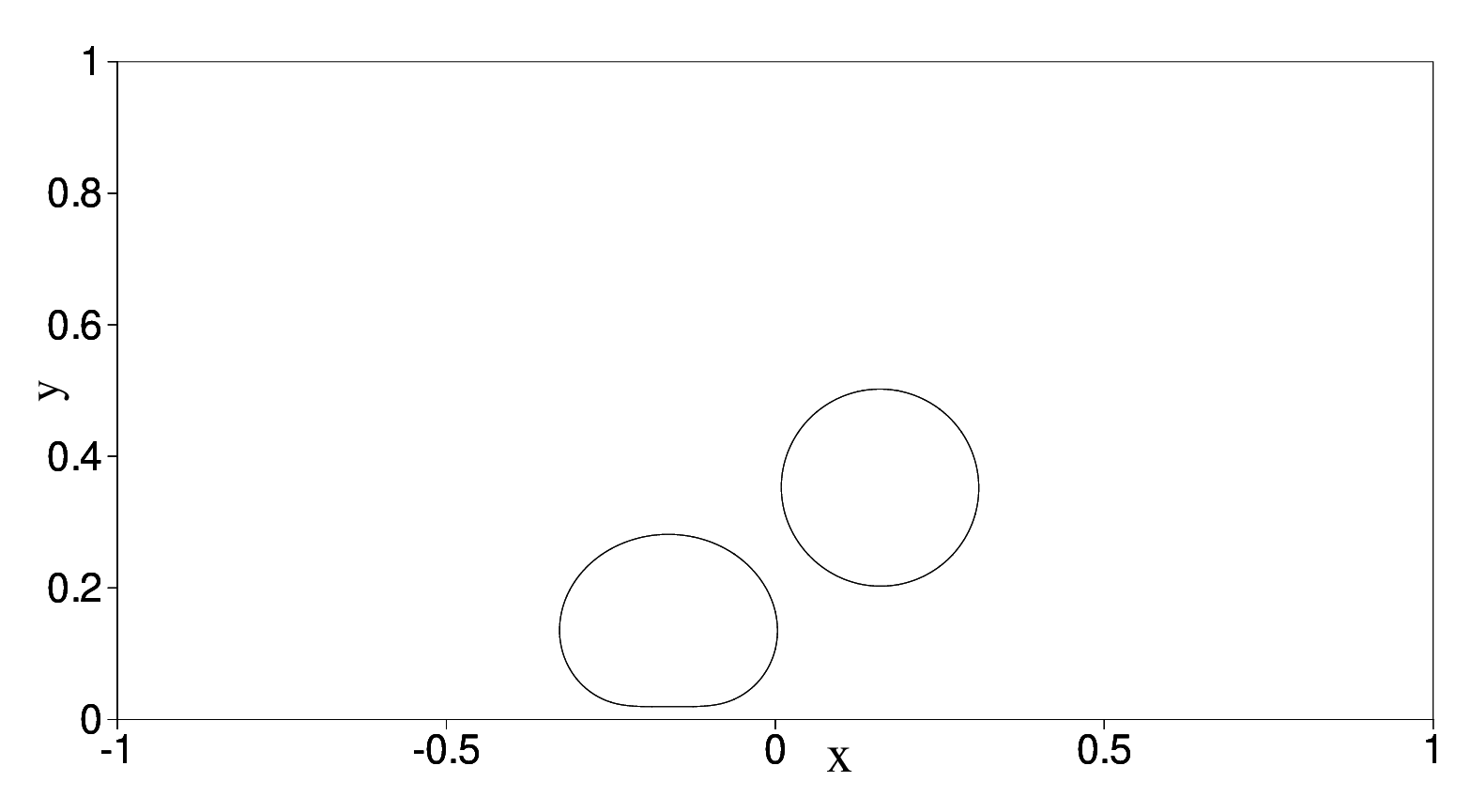}(e)
\includegraphics[width=2in]{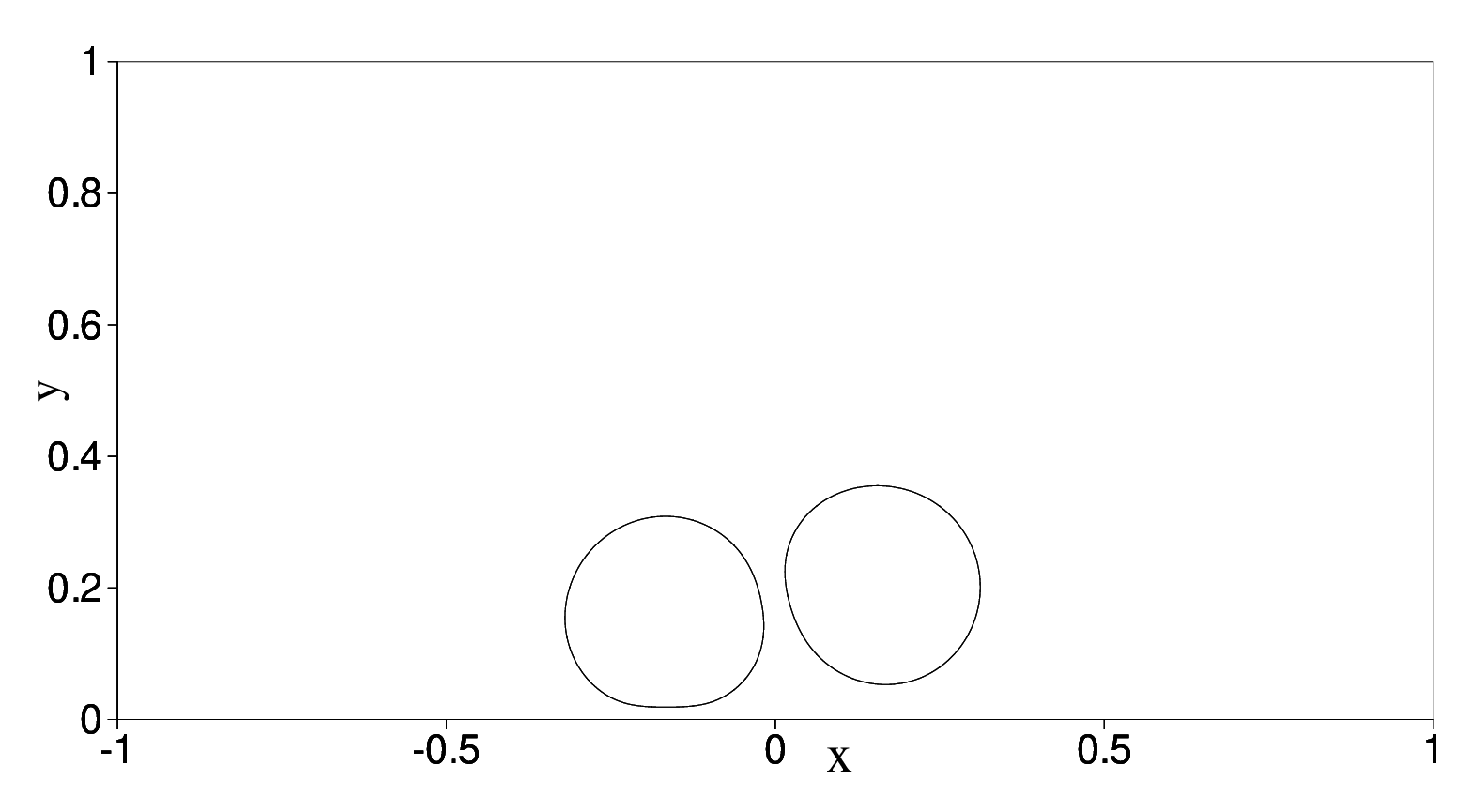}(f)
}
\centerline{
\includegraphics[width=2in]{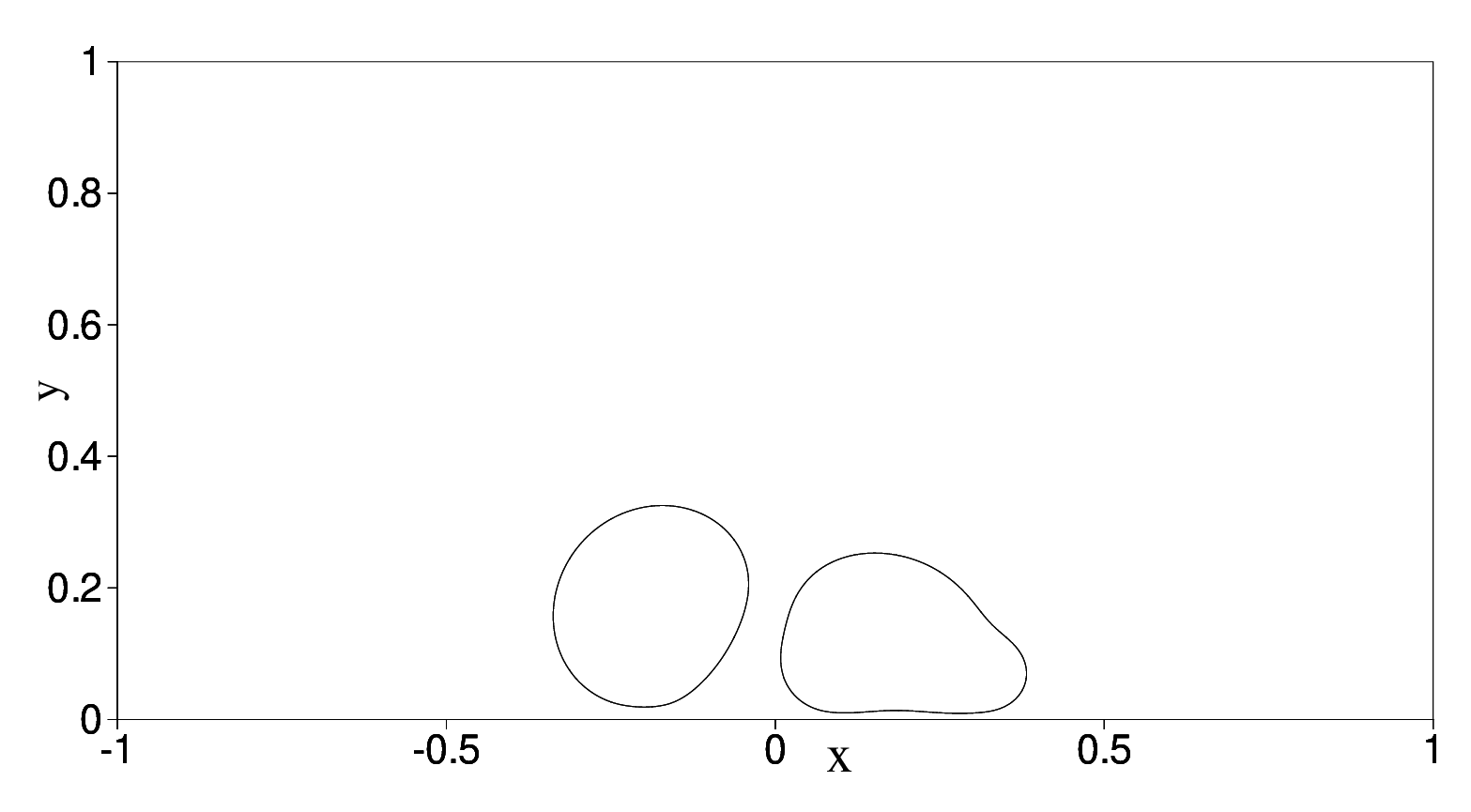}(g)
\includegraphics[width=2in]{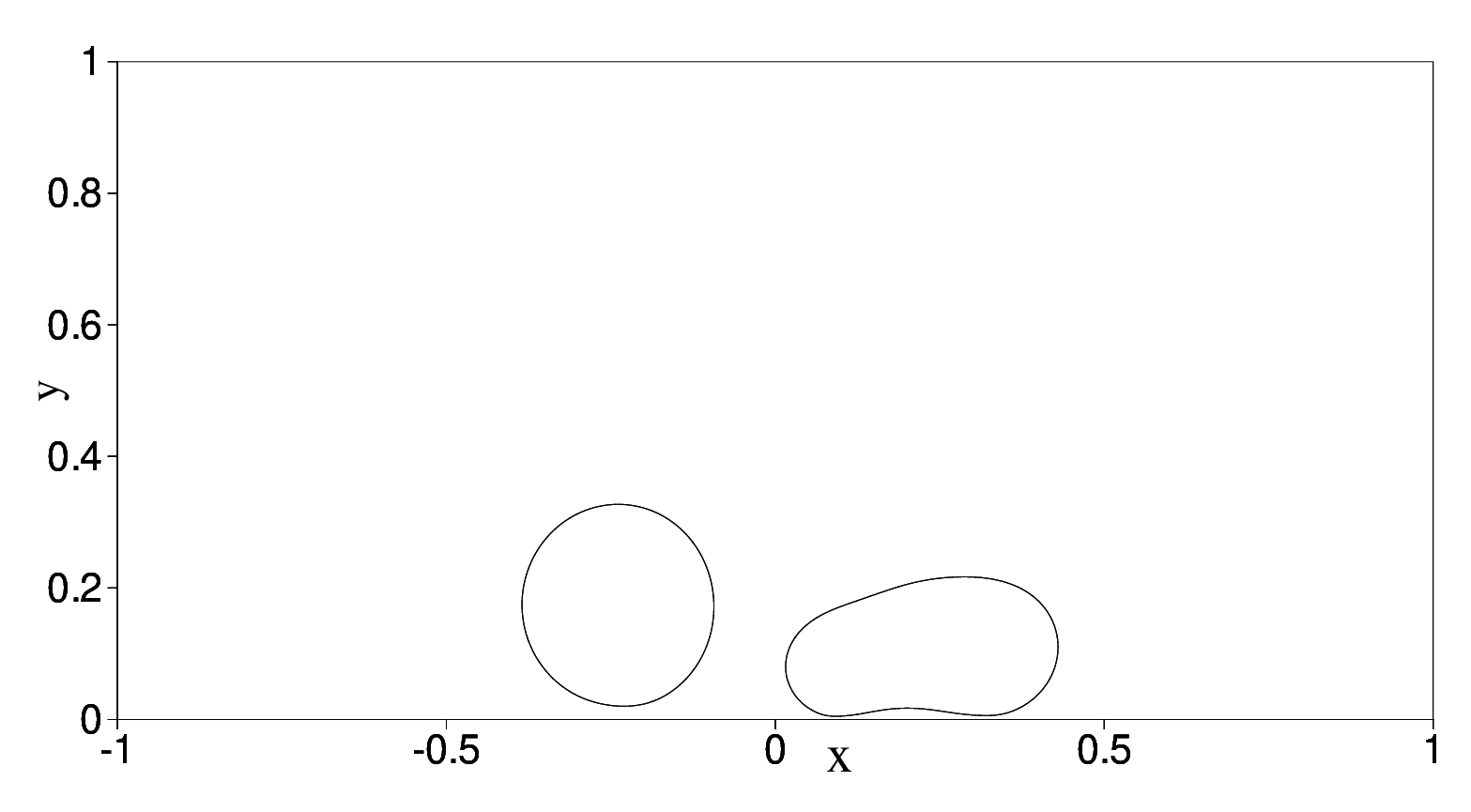}(h)
\includegraphics[width=2in]{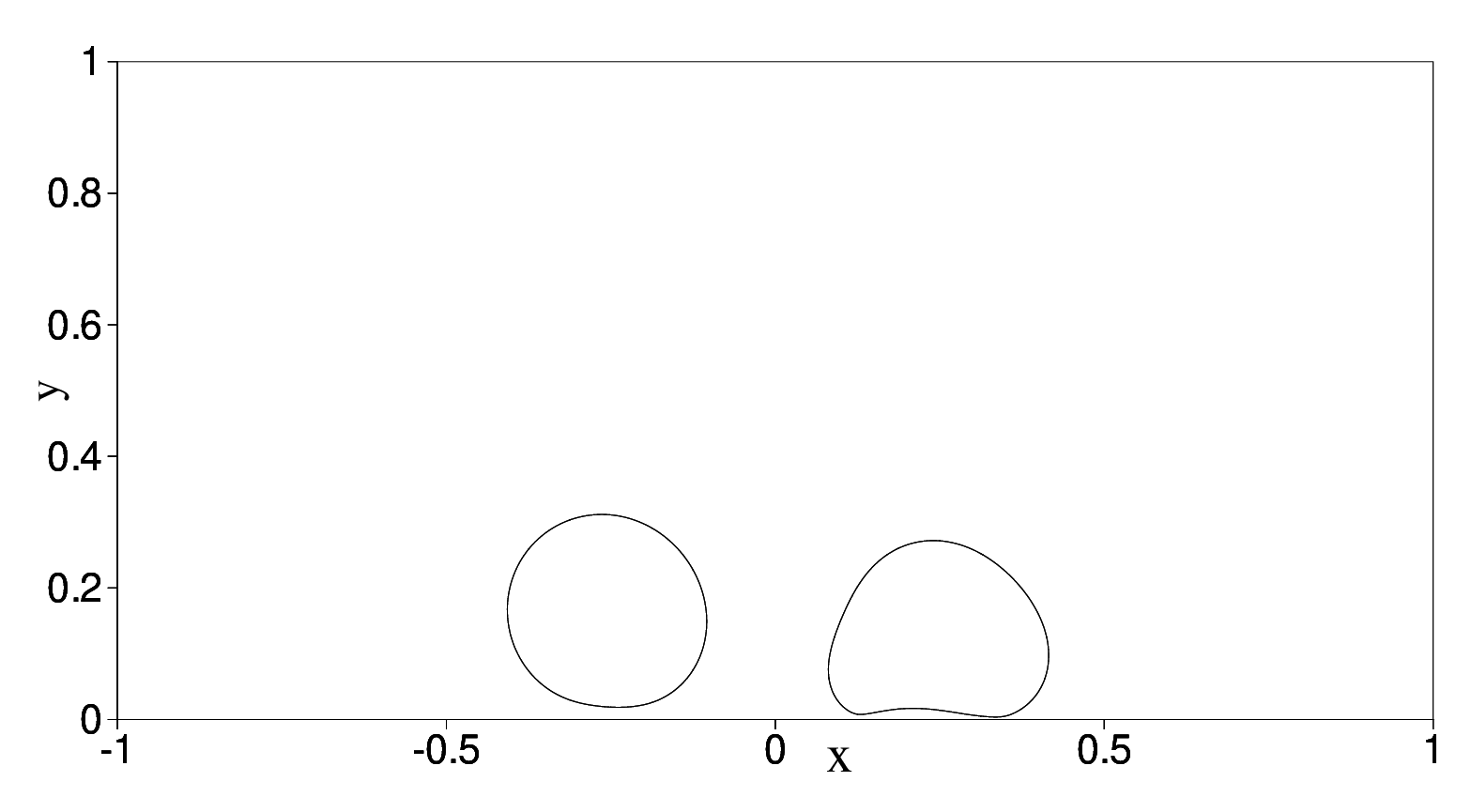}(i)
}
\centerline{
\includegraphics[width=2in]{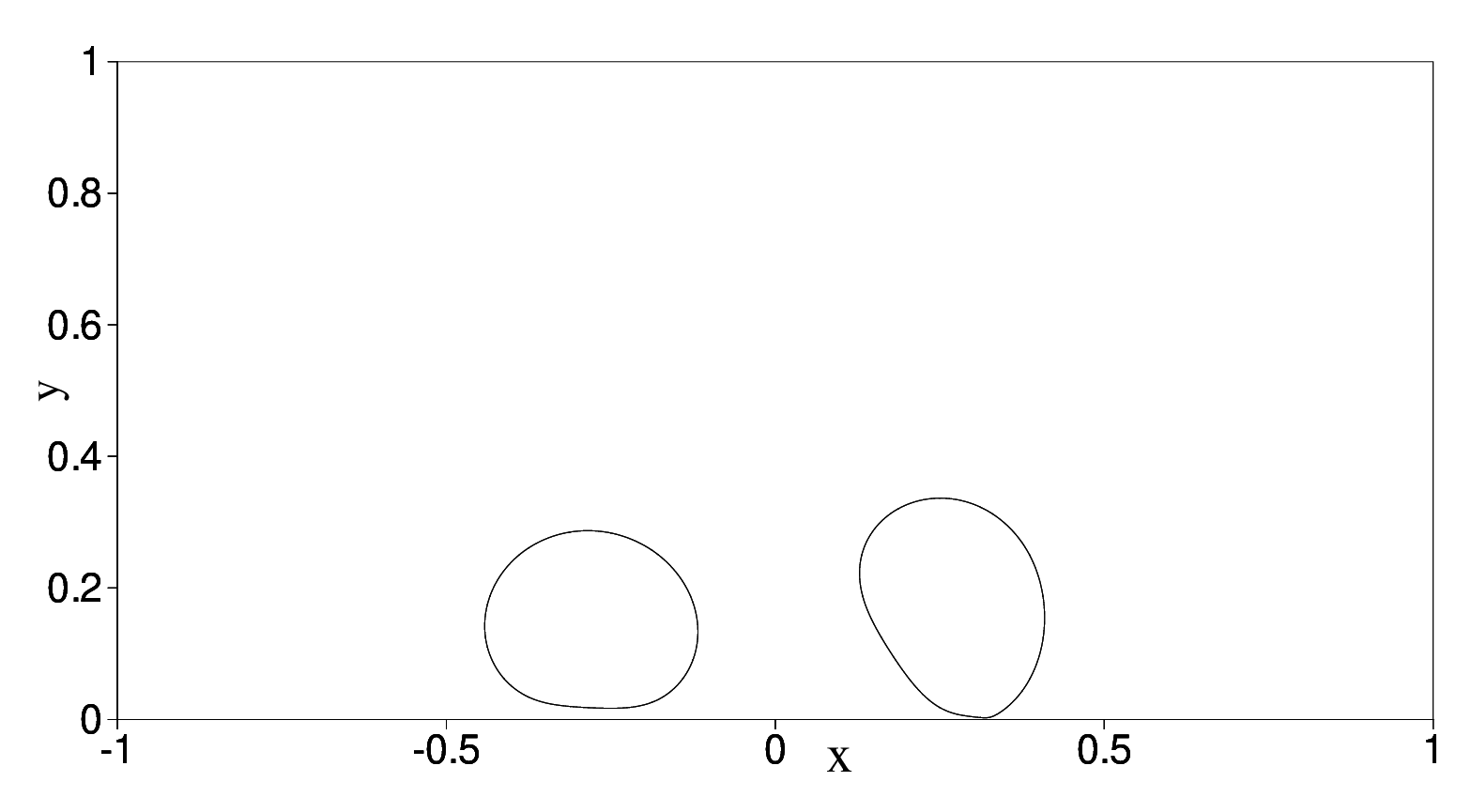}(j)
\includegraphics[width=2in]{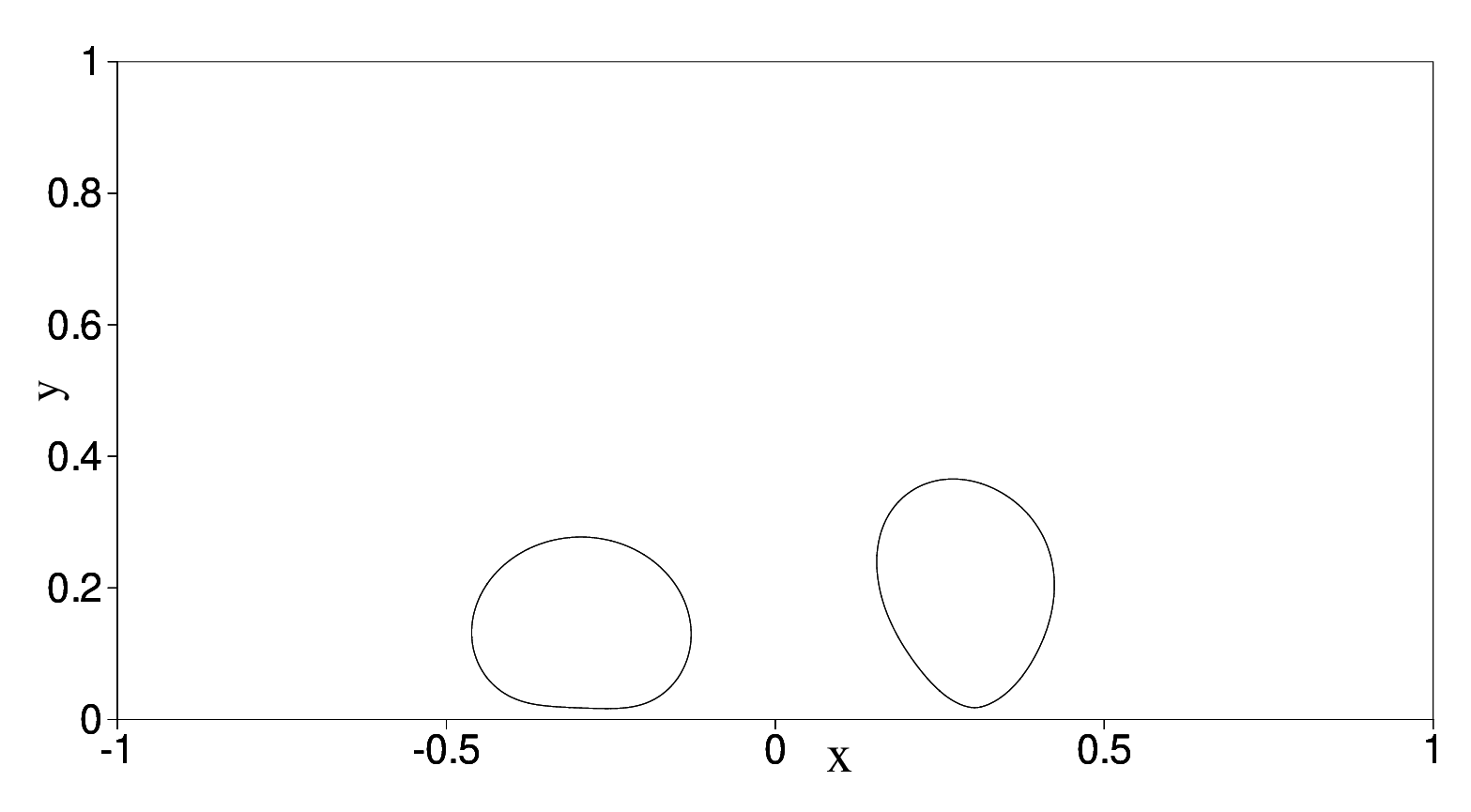}(k)
\includegraphics[width=2in]{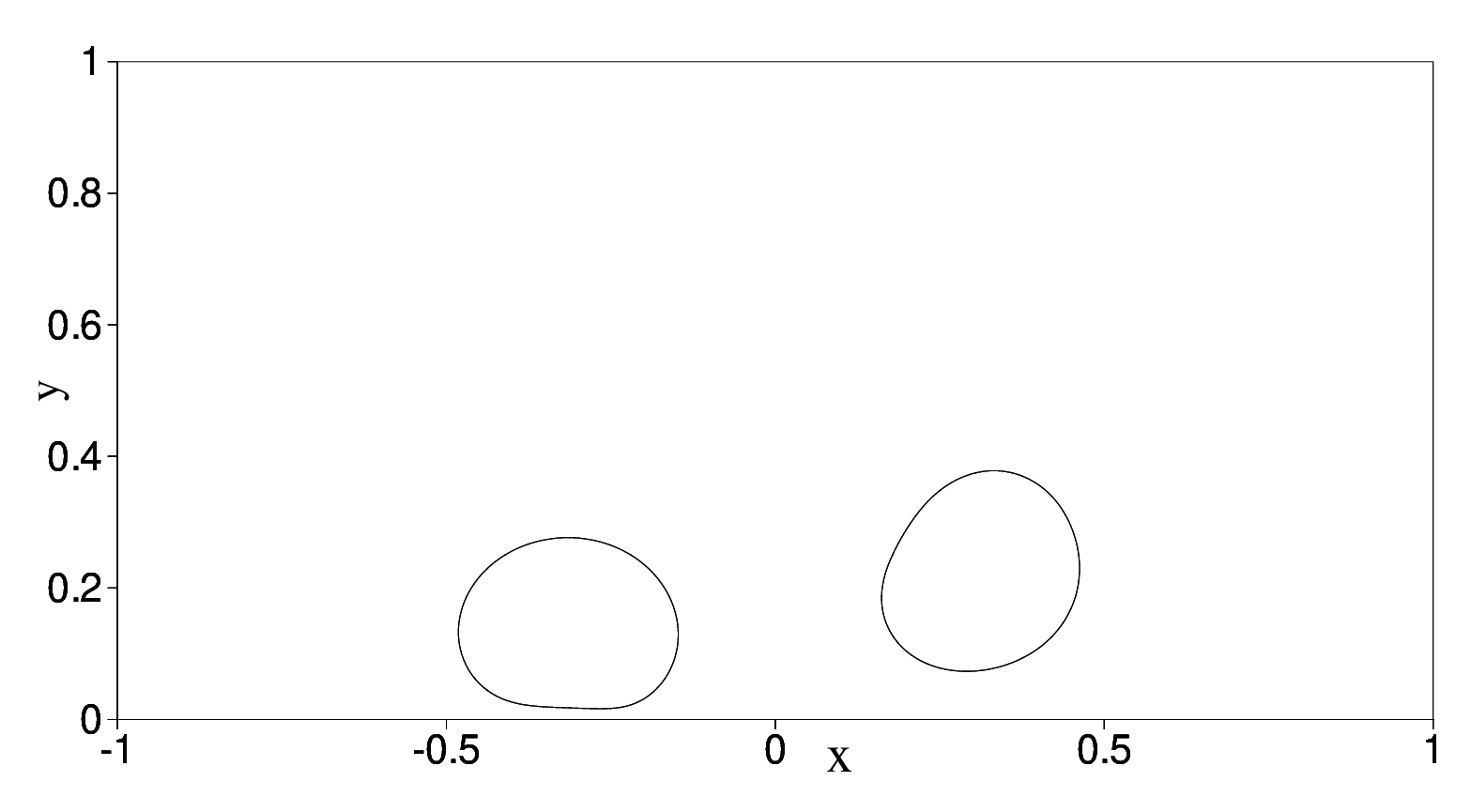}(l)
}
\centerline{
\includegraphics[width=2in]{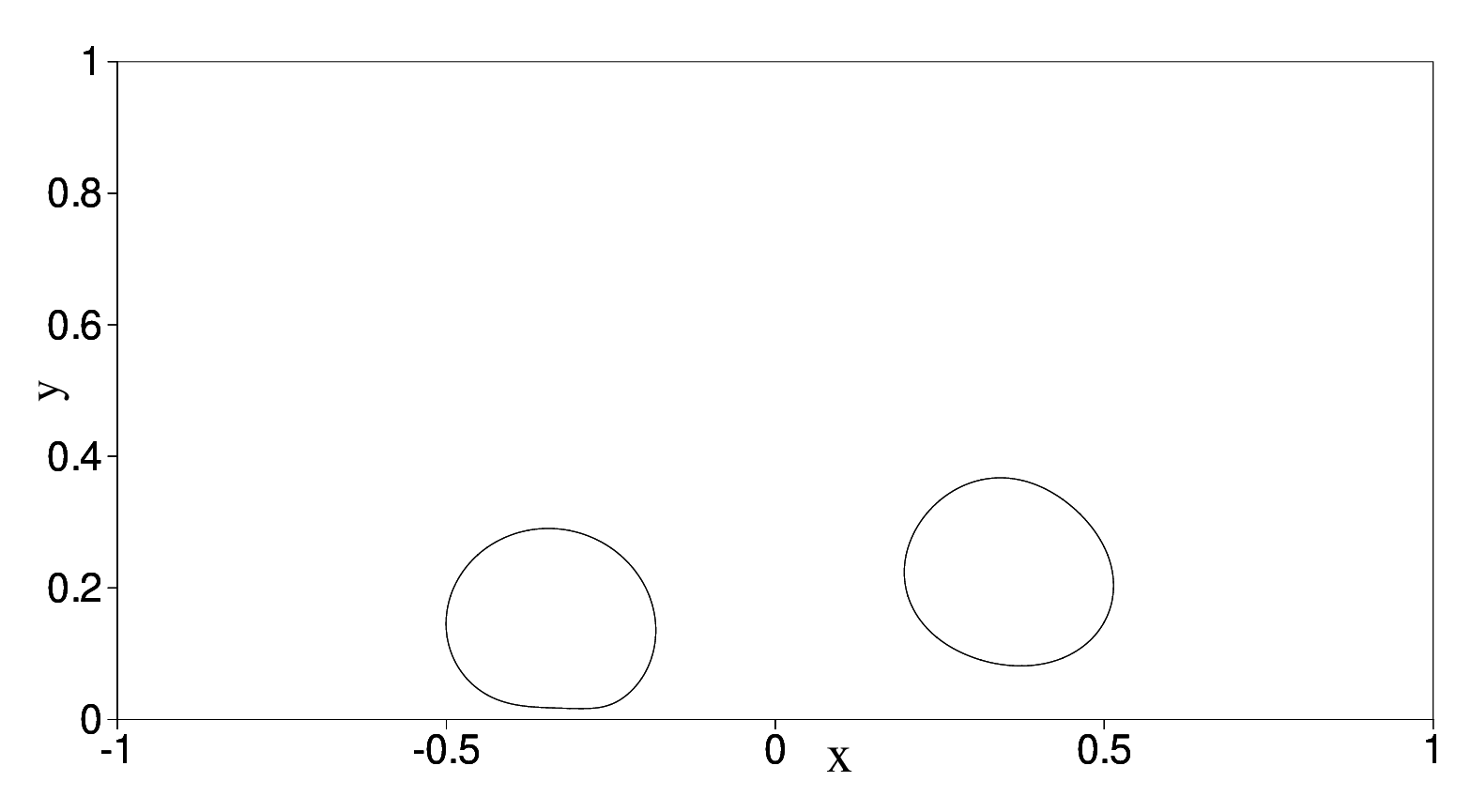}(m)
\includegraphics[width=2in]{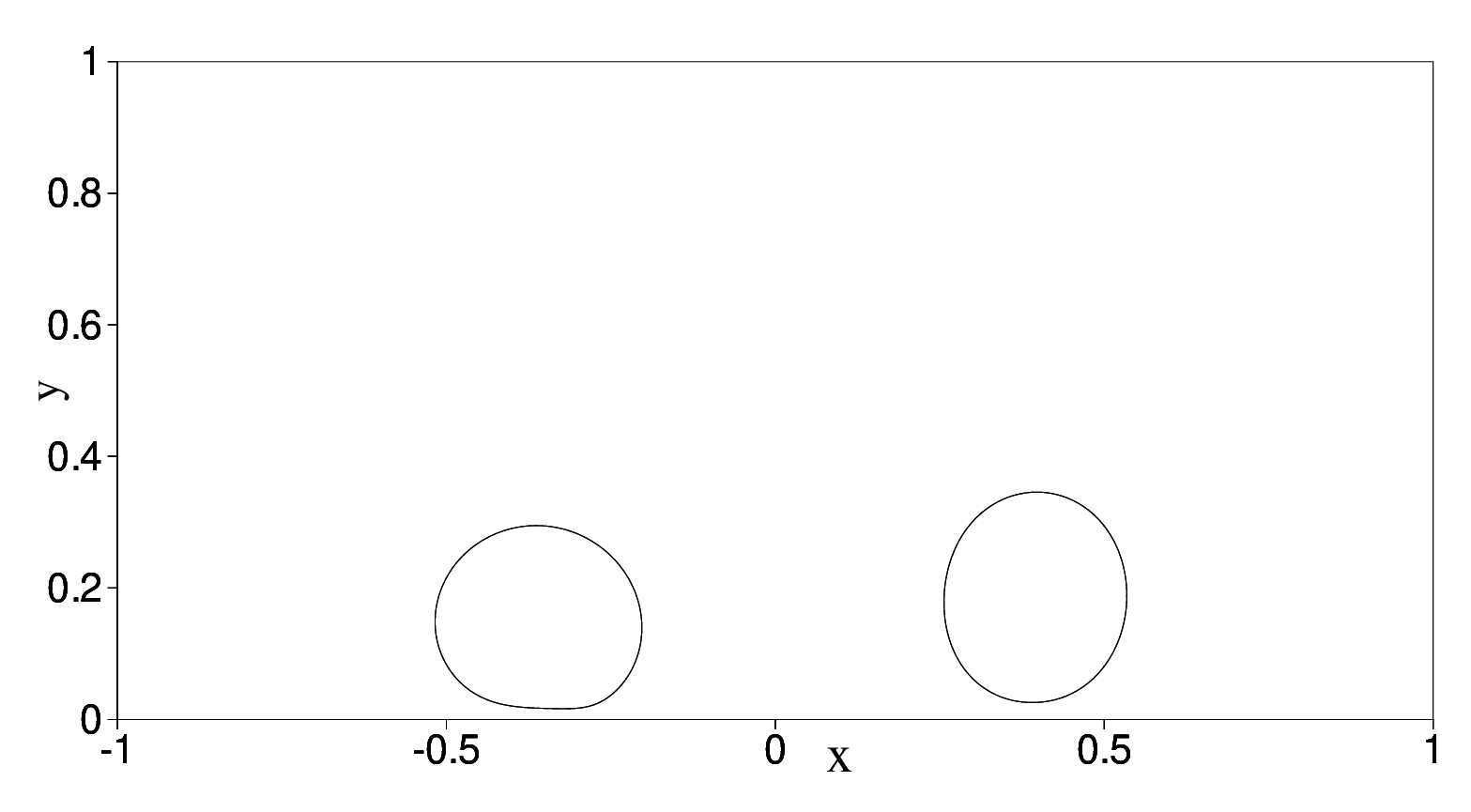}(n)
\includegraphics[width=2in]{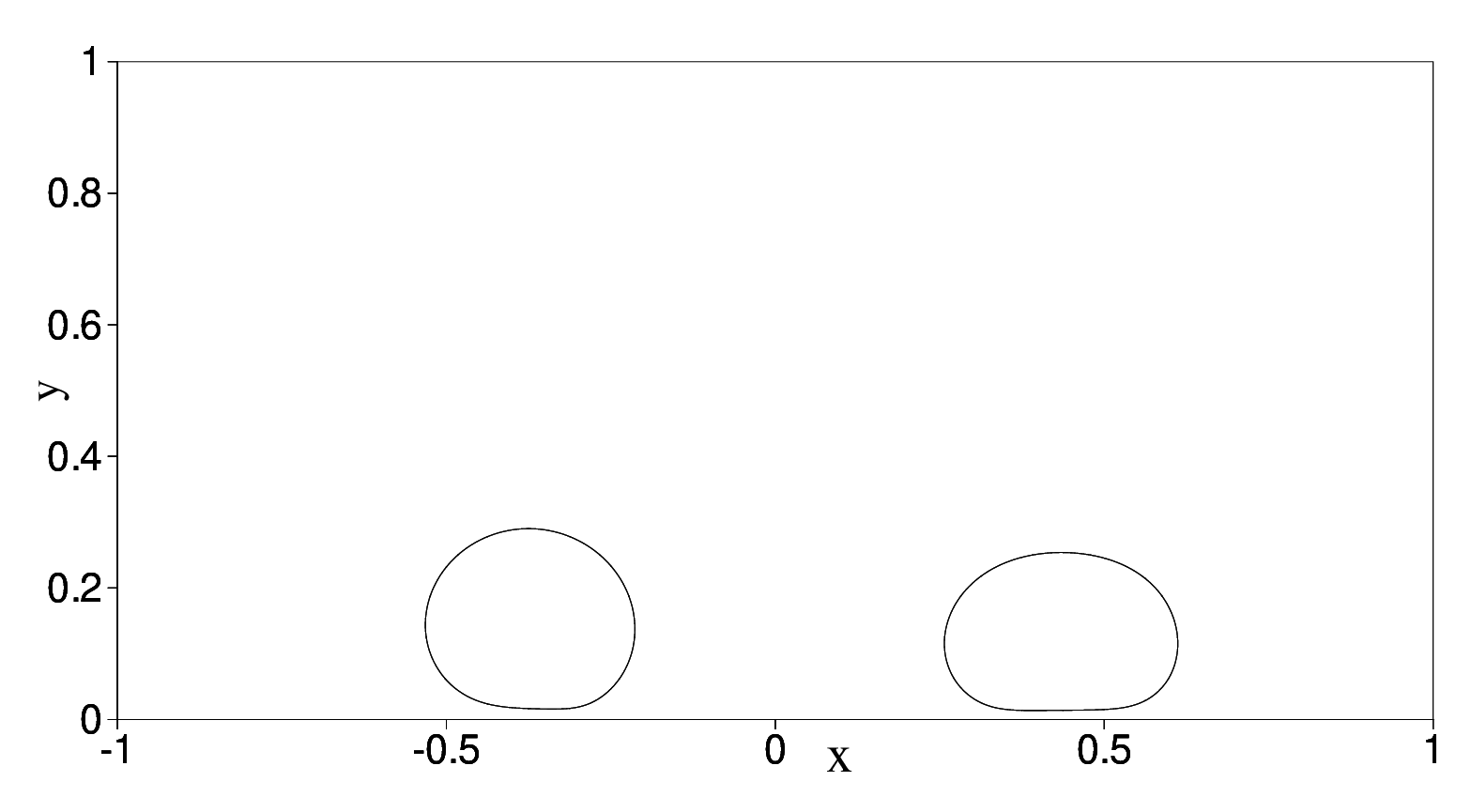}(o)
}
\caption{
Impact of water/oil drops on horizontal wall with contact angles
($\theta_{ao},\theta_{aw})=(120^0,150^0)$: temporal sequence of snapshots
of drop profiles at time instants
(a) $t=0.0125$,
(b) $t=0.1125$,
(c) $t=0.175$,
(d) $t=0.225$,
(e) $t=0.2875$,
(f) $t=0.3375$,
(g) $t=0.375$,
(h) $t=0.4375$,
(i) $t=0.4625$,
(j) $t=0.5$,
(k) $t=0.525$,
(l) $t=0.5625$,
(m) $t=0.625$,
(n) $t=0.6875$,
(o) $t=0.75$.
}
\label{fig:phase_profile_150_120}
\end{figure}

% contact angles: (150,120)

In the second set of tests, the air-water contact angle is $\theta_{aw}=150^0$
(superhydrophobic wall surface), and the air-oil contact angle is the same as
in the first set.
Figure \ref{fig:phase_profile_150_120} shows a temporal sequence of snapshots
of the fluid interfaces  with this set of tests. 
The data depict a scenario similar to that of Figure \ref{fig:phase_profile_120_120},
but with some main difference.
After the impact, the water drop is observed to bounce off the wall and
lift into the air while moving sideways in this case; see
Figures \ref{fig:phase_profile_150_120}(j)-(n).
In contrast, in the first test case with $\theta_{aw}=120^0$ the water drop
is observed to stay in contact with the wall and never bounce off after
the impact; see Figures \ref{fig:phase_profile_120_120}(j)-(n).
Because the surface is more hydrophobic in the this case,
small water drops can bounce off the wall upon impact, which is consistent
with the observations in the two-phase bouncing water drop studies
(see e.g.~\cite{Dong2012,RichardQ2000}).

\section{Concluding Remarks}
\label{sec:summary}

% what have we done?
% what are the main points of results?
% what are the implications of these results?
% what are the outstanding questions?
% 

In this paper we have explored 
the formulation and simulation of wall-bounded multiphase flows
consisting of $N$ ($N\geqslant 2$) immiscible incompressible fluids
with different densities, viscosities and pairwise surface tensions.
We have looked into the implications of 
several reduction consistency properties  between the $N$-phase formulation
and the $M$-phase ($2\leqslant M\leqslant N-1$) formulations.
%on the physical formulation, governing equations, and
%boundary conditions. 
We observe that the reduction consistency has profound implications
on the N-phase governing equations, boundary conditions,
and the form of the potential energy density function.

Our contributions can be specifically summarized 
in terms of the following aspects.
First, we have presented a modified 
thermodynamically consistent phase field model 
for an N-fluid mixture that is more general than that of
previous work~\cite{Dong2014}.
Second, by imposing the consistency property on the N-phase formulation
we have obtained an explicit form for the mobility matrix
in the phase field model.
Third, by requiring the N-phase governing equations
to be reduction consistent we have arrived at a set of
conditions for the potential energy density function
to ensure consistency.
Fourth, we have proposed an N-phase contact-angle
boundary condition that is reduction consistent
with $M$-phase ($2\leqslant M\leqslant N-1$) systems.

% specific N-phase formulation

The potential energy density function adopted in
the current paper (originally suggested by~\cite{BoyerM2014})
ensures only a subset, yet an important subset, of
the consistency conditions obtained herein.
With this potential energy function, the N-phase 
formulation presented herein is fully consistent
with two-phase formulations. In other words,
if only a pair (for any pair) of fluid phases are present in the system,
the N-phase physical formulation presented here will 
exactly reduce to the corresponding two-phase formulation.
Compared with the
previous N-phase formulations~\cite{Dong2014,Dong2015}),
this is a great improvement, noting that
both the current and the previous formulations are
thermodynamically consistent. 

% wall bounded N-phase flows

Additionally, we have presented a numerical algorithm for solving
the N-phase governing equations obtained herein.
Employing this algorithm,
we have performed extensive numerical experiments
with several flow problems involving multiple
flow components and partially wettable walls.
Our simulation results show that the wall wettability 
and the contact angles have a profound influence on
the equilibrium configuration and the dynamics of the multiphase
system.  In particular, we have compared our simulation results
with the de Gennes theory~\cite{deGennesBQ2003}
about the contact angle effects
on the spreading of a water drop and an oil drop  on 
partially wettable wall surfaces.
The comparison demonstrates that our method produces physically
accurate results.

% implications of results
% what is the use of method?

The method developed herein provides an effective technique
for studying the interactions among multiple types of fluid
interfaces and solid-wall surfaces, wettability effects, surface tension effects,
and the dynamics and interactions of multiple types of contact lines.
We anticipate that it will be useful to microfluidics, 
functional surfaces, materials processing, and applications in energy and petroleum
industries.

% consistency conditions for potential energy

The set of conditions on the potential energy density function
(see \eqref{equ:condition_H1}--\eqref{equ:condition_H3})
ensures the consistency of the governing equations
between the $N$-phase system and the $M$-phase systems
($2\leqslant M\leqslant N-1$).
%Does there exist a potential energy density function 
%that fully satisfies these conditions?
%How does one construct such a potential energy density function?
%These are interesting open problems at the moment.
How does one construct a potential energy density function 
that satisfies all these conditions to ensure the full consistency?
Is there some other way to ensure the reduction consistency
between $N$ phases and $M$ phases ($2\leqslant M\leqslant N-1$)?
These are interesting open questions to contemplate for
future research.

% what else to discuss here?

%%
\section*{Acknowledgement}
This work was partially supported by
NSF (DMS-1318820, DMS-1522537), ONR (N000141110028), and NSFC(11571157).
% and ONR (N000141110028).
%Useful discussion with Professor P. Yue
%(Virginia Tech) at the early stage of this project
%is gratefully acknowledged.

\section*{Appendix A. Proof of Theorem \ref{thm:thm_1} 
}

In light of assumption ($\mathcal{A}$1) about $H(\vec{\phi})$, it suffices to show that
the term $\sum_{i,j=1}^{N-1} \frac{\lambda_{ij}}{2}\nabla\phi_i\cdot\nabla\phi_j$
in \eqref{equ:free_energy}
satisfies the consistency property ($\mathcal{C}$2).
Equation \eqref{equ:lambda_mat_gop} implies 
$ %\begin{equation}
\sum_{i,j=1}^{N-1} \frac{\lambda_{ij}}{2}\nabla\phi_i\cdot\nabla\phi_j
= \sum_{i,j=1}^{N-1} \frac{\Lambda_{ij}}{2}\nabla c_i\cdot\nabla c_j
$ (see \cite{Dong2015}), %\end{equation}
where $\phi_i$ ($1\leqslant i\leqslant N-1$) are a
general set of order parameters defined by \eqref{equ:gop_def}.
So we  only need to show that the term
\begin{equation}
K^{(N)}(\vec{c}^{(N)}) = \sum_{i,j=1}^{N-1}
\frac{\Lambda_{ij}^{(N)}}{2}\nabla c_i^{(N)}\cdot\nabla c_j^{(N)},
\ \text{with} \ 
\Lambda_{ij}^{(N)} = \frac{3}{\sqrt{2}}\eta\left(
  \sigma_{iN}^{(N)} + \sigma_{jN}^{(N)} - \sigma_{ij}^{(N)}
\right), 
\ 1\leqslant i,j\leqslant N-1,
\label{equ:kinetic_energy}
\end{equation}
satisfies  ($\mathcal{C}$2),
where 
$
\vec{c}^{(N)} = (c_1^{(N)}, c_2^{(N)}, \dots, c_{N-1}^{(N)}),
$
and the superscript in $(\cdot)^{(N)}$ accentuates the point that 
the quantity is for the $N$-phase system.

It suffices to show that, if one fluid phase is absent 
from the N-phase system then $K^{(N)}$ reduces 
to the corresponding form for the ($N-1$)-phase system.
Let fluid $k$ ($1\leqslant k\leqslant N$) be the phase
that is absent from the N-phase system, i.e.
$
c_k^{(N)} \equiv 0.
$
We distinguish two cases: 
(i) $1\leqslant k\leqslant N-1$, and
(ii) $k=N$.

In the first case ($1\leqslant k\leqslant N-1$),
the first ($k-1$) fluids of the N-phase system have the same IDs
in the ($N-1$)-phase system, and the last ($N-k$) fluids of the N-phase system will have
their IDs reduced by one in the  ($N-1$)-phase system. Therefore,
the following relations hold
\begin{equation}
c_i^{(N-1)} = \left\{
\begin{array}{ll}
c_i^{(N)}, & 1\leqslant i\leqslant k-1, \\
c_{i+1}^{(N)}, & k\leqslant i\leqslant N-1,
\end{array}
\right.
%\quad 1\leqslant i\leqslant N-1,
\label{equ:relation_cN_cN1}
\end{equation}
\begin{equation}
\sigma_{ij}^{(N-1)} = \left\{
\begin{array}{ll}
\sigma_{ij}^{(N)}, & 1\leqslant i\leqslant k-1, \ 1\leqslant j\leqslant k-1, \\
\sigma_{i,j+1}^{(N)}, & 1\leqslant i\leqslant k-1, \ k\leqslant j\leqslant N-1, \\
\sigma_{i+1,j}^{(N)}, & k\leqslant i\leqslant N-1, \ 1\leqslant j\leqslant k-1, \\
\sigma_{i+1,j+1}^{(N)}, & k\leqslant i\leqslant N-1, \ k\leqslant j\leqslant N-1,
\end{array}
\right.
\label{equ:surften_N_N1}
\end{equation}
\begin{multline}
\Lambda_{ij}^{(N-1)} 
= \frac{3}{\sqrt{2}}\eta\left(\sigma_{i,N-1}^{(N-1)}
+ \sigma_{j,N-1}^{(N-1)} - \sigma_{i,j}^{(N-1)} \right) \\
= \left\{
\begin{array}{ll}
\frac{3}{\sqrt{2}}\eta\left(\sigma_{iN}^{(N)}
+ \sigma_{jN}^{(N)} - \sigma_{ij}^{(N)} \right) =
\Lambda_{ij}^{(N)}, & 1\leqslant i\leqslant k-1, \ 1\leqslant j\leqslant k-1, \\
\frac{3}{\sqrt{2}}\eta\left(\sigma_{iN}^{(N)}
+ \sigma_{j+1,N}^{(N)} - \sigma_{i,j+1}^{(N)} \right) =
\Lambda_{i,j+1}^{(N)}, & 1\leqslant i\leqslant k-1, \ k\leqslant j\leqslant N-2, \\
\frac{3}{\sqrt{2}}\eta\left(\sigma_{i+1,N}^{(N)}
+ \sigma_{jN}^{(N)} - \sigma_{i+1,j}^{(N)} \right) =
\Lambda_{i+1,j}^{(N)}, & k\leqslant i\leqslant N-2, \ 1\leqslant j\leqslant k-1, \\
\frac{3}{\sqrt{2}}\eta\left(\sigma_{i+1,N}^{(N)}
+ \sigma_{j+1,N}^{(N)} - \sigma_{i+1,j+1}^{(N)} \right) =
\Lambda_{i+1,j+1}^{(N)}, & k\leqslant i\leqslant N-2, \ k\leqslant j\leqslant N-2.
\end{array}
\right.
\end{multline}
Therefore,
\begin{equation}
\begin{split}
& K^{(N)}(\vec{c}^{(N)}) 
= \sum_{i,j=1}^{N-1}\frac{\Lambda_{ij}^{(N)}}{2}\nabla c_i^{(N)}\cdot\nabla c_j^{(N)} \\
= &
\sum_{i,j=1}^{k-1} \frac{\Lambda_{ij}^{(N)}}{2}\nabla c_i^{(N)}\cdot\nabla c_j^{(N)}
+ \sum_{i=k+1}^{N-1}\sum_{j=1}^{k-1} \frac{\Lambda_{ij}^{(N)}}{2}\nabla c_{i}^{(N)}\cdot\nabla c_j^{(N)} \\
& \quad
+ \sum_{i=1}^{k-1}\sum_{j=k+1}^{N-1} \frac{\Lambda_{ij}^{(N)}}{2}\nabla c_i^{(N)}\cdot\nabla c_{j}^{(N)}
+ \sum_{i=k+1}^{N-1}\sum_{j=k+1}^{N-1} \frac{\Lambda_{ij}^{(N)}}{2}\nabla c_{i}^{(N)}\cdot\nabla c_{j}^{(N)} \\
=&
\sum_{i,j=1}^{k-1} \frac{\Lambda_{ij}^{(N)}}{2}\nabla c_i^{(N)}\cdot\nabla c_j^{(N)}
+ \sum_{i=k}^{N-2}\sum_{j=1}^{k-1} \frac{\Lambda_{i+1,j}^{(N)}}{2}\nabla c_{i+1}^{(N)}\cdot\nabla c_j^{(N)} \\
& \quad 
+ \sum_{i=1}^{k-1}\sum_{j=k}^{N-2} \frac{\Lambda_{i,j+1}^{(N)}}{2}\nabla c_i^{(N)}\cdot\nabla c_{j+1}^{(N)}
+ \sum_{i=k}^{N-2}\sum_{j=k}^{N-2} \frac{\Lambda_{i+1,j+1}^{(N)}}{2}\nabla c_{i+1}^{(N)}\cdot\nabla c_{j+1}^{(N)} \\
=&
\sum_{i,j=1}^{k-1} \frac{\Lambda_{ij}^{(N-1)}}{2}\nabla c_i^{(N-1)}\cdot\nabla c_j^{(N-1)} 
+ \sum_{i=k}^{N-2}\sum_{j=1}^{k-1} \frac{\Lambda_{ij}^{(N-1)}}{2}\nabla c_i^{(N-1)}\cdot\nabla c_j^{(N-1)} \\
& \quad
+ \sum_{i=1}^{k-1}\sum_{j=k}^{N-2} \frac{\Lambda_{ij}^{(N-1)}}{2}\nabla c_i^{(N-1)}\cdot\nabla c_j^{(N-1)}
+ \sum_{i=k}^{N-2}\sum_{j=k}^{N-2} \frac{\Lambda_{ij}^{(N-1)}}{2}\nabla c_i^{(N-1)}\cdot\nabla c_j^{(N-1)} \\
= & \sum_{i,j=1}^{N-2}\frac{\Lambda_{ij}^{(N-1)}}{2}\nabla c_i^{(N-1)}\cdot\nabla c_j^{(N-1)} \\
= & K^{(N-1)}(\vec{c}^{(N-1)}).
\end{split}
\end{equation}

In the second case ($k=N$), the first ($N-1$) fluids of the N-phase system have
the same IDs in the ($N-1$)-phase system. Therefore, 
the following relations hold
\begin{equation}
c_i^{(N-1)} = c_i^{(N)}, \quad 1\leqslant i\leqslant N-1,
\label{equ:relation_cN_1}
\end{equation}
\begin{equation}
\sum_{i=1}^{N-2} c_i^{(N)} + c_{N-1}^{(N)} = 1, \quad
\nabla c_{N-1}^{(N)} = -\sum_{i=1}^{N-2} \nabla c_i^{(N)},
\label{equ:relation_cN_2}
\end{equation}
\begin{equation}
\sigma_{ij}^{(N-1)} = \sigma_{ij}^{(N)}, \quad
1\leqslant i,j\leqslant N-1,
\label{equ:relation_sigma_N}
\end{equation}
\begin{equation}
\begin{split}
 \Lambda_{ij}^{(N)} & - \Lambda_{N-1,j}^{(N)} - \Lambda_{i,N-1}^{(N)}
+ \Lambda_{N-1,N-1}^{(N)}
= \frac{3}{\sqrt{2}}\eta\left(
  \sigma_{i,N-1}^{(N)} + \sigma_{j,N-1}^{(N)} - \sigma_{ij}^{(N)}
\right) \\
& = \frac{3}{\sqrt{2}}\eta\left(
  \sigma_{i,N-1}^{(N-1)} + \sigma_{j,N-1}^{(N-1)} - \sigma_{ij}^{(N-1)}
\right)
= \Lambda_{ij}^{(N-1)},
\quad 
1\leqslant i,j\leqslant N-2.
\end{split}
\end{equation}
Therefore,
\begin{equation}
\begin{split}
& K^{(N)}(\vec{c}^{(N)}) = 
\sum_{i,j=1}^{N-1}\frac{\Lambda_{ij}^{(N)}}{2}\nabla c_i^{(N)}\cdot\nabla c_j^{(N)} \\
= &
\sum_{i,j=1}^{N-2}\frac{\Lambda_{ij}^{(N)}}{2}\nabla c_i^{(N)}\cdot\nabla c_j^{(N)}
+ \sum_{j=1}^{N-2}\frac{\Lambda_{N-1,j}^{(N)}}{2}\nabla c_{N-1}^{(N)}\cdot\nabla c_j^{(N)} \\
& \quad
+\sum_{i=1}^{N-2}\frac{\Lambda_{i,N-1}^{(N)}}{2}\nabla c_i^{(N)}\cdot\nabla c_{N-1}^{(N)}
+ \frac{\Lambda_{N-1,N-1}^{(N)}}{2}\nabla c_{N-1}^{(N)}\cdot\nabla c_{N-1}^{(N)} \\
=&
\sum_{i,j=1}^{N-2}\frac{1}{2}\left[
  \Lambda_{ij}^{(N)} - \Lambda_{N-1,j}^{(N)} 
  - \Lambda_{i,N-1}^{(N)} + \Lambda_{N-1,N-1}^{(N)}
\right] \nabla c_i^{(N)}\cdot\nabla c_j^{(N)} \\
=&
\sum_{i,j=1}^{N-2} \frac{1}{2}\Lambda_{ij}^{(N-1)} \nabla c_i^{(N-1)}\cdot\nabla c_j^{(N-1)} \\
=& 
K^{(N-1)}(\vec{c}^{(N-1)}).
\end{split}
\end{equation}

\section*{Appendix B. Proof of Theorem \ref{thm:thm_2}}

Suppose fluid $k$ (for any $1\leqslant k\leqslant N$) is
absent from the N-phase system.
Then we have the following relations
\begin{equation}
\tilde{\rho}_i^{(N-1)}=\left\{
\begin{array}{ll}
\tilde{\rho}_i^{(N)}, & 1\leqslant i\leqslant k-1 \\
\tilde{\rho}_{i+1}^{(N)}, & k\leqslant i\leqslant N-1,
\end{array}
\right.
\tilde{\mu}_i^{(N-1)}=\left\{
\begin{array}{ll}
\tilde{\mu}_i^{(N)}, & 1\leqslant i\leqslant k-1 \\
\tilde{\mu}_{i+1}^{(N)}, & k\leqslant i\leqslant N-1.
\end{array}
\right.
\label{equ:relation_B_1}
\end{equation}

According to  \eqref{equ:density_expr},
%equation \eqref{equ:rho_consist} is satisfied because
\begin{equation}
\begin{split}
\rho^{(N)} & = \sum_{i=1}^N \tilde{\rho}_i^{(N)} c_i^{(N)}
= \sum_{i=1}^{k-1}\tilde{\rho}_i^{(N)} c_i^{(N)}
+ \sum_{i=k+1}^N \tilde{\rho}_i^{(N)} c_i^{(N)}
= \sum_{i=1}^{k-1}\tilde{\rho}_i^{(N-1)} c_i^{(N-1)}
+ \sum_{i=k+1}^N \tilde{\rho}_{i-1}^{(N-1)} c_{i-1}^{(N-1)} \\
& = \sum_{i=1}^{k-1}\tilde{\rho}_i^{(N-1)} c_i^{(N-1)}
+ \sum_{i=k}^{N-1} \tilde{\rho}_{i}^{(N-1)} c_{i}^{(N-1)}
= \sum_{i=1}^{N-1} \tilde{\rho}_i^{(N-1)} c_i^{(N-1)}
= \rho^{(N-1)}.
\end{split}
\end{equation}
So equation \eqref{equ:rho_consist} holds.

It can be similarly shown that
equation \eqref{equ:mu_consist} holds in light of
equation \eqref{equ:mu_expr},
\begin{equation}
\begin{split}
\mu^{(N)} & = \sum_{i=1}^N \tilde{\mu}_i^{(N)} c_i^{(N)}
= \sum_{i=1}^{k-1}\tilde{\mu}_i^{(N)} c_i^{(N)}
+ \sum_{i=k+1}^N \tilde{\mu}_i^{(N)} c_i^{(N)}
= \sum_{i=1}^{k-1}\tilde{\mu}_i^{(N-1)} c_i^{(N-1)}
+ \sum_{i=k+1}^N \tilde{\mu}_{i-1}^{(N-1)} c_{i-1}^{(N-1)} \\
& = \sum_{i=1}^{k-1}\tilde{\mu}_i^{(N-1)} c_i^{(N-1)}
+ \sum_{i=k}^{N-1} \tilde{\mu}_{i}^{(N-1)} c_{i}^{(N-1)}
= \sum_{i=1}^{N-1} \tilde{\mu}_i^{(N-1)} c_i^{(N-1)}
= \mu^{(N-1)}.
\end{split}
\end{equation}

In Appendix A  we have shown that the term 
$
K^{(N)} = \sum_{i,j=1}^{N-1}\frac{1}{2}\Lambda_{ij}^{(N)}
\nabla c_i^{(N)}\cdot \nabla c_j^{(N)}
$ 
(see \eqref{equ:kinetic_energy})
reduces to the corresponding form $K^{(N-1)}$
of the ($N-1$)-phase system
if one fluid phase is absent.
Using a procedure parallel to that for
$K^{(N)}$, one can show that 
$
\sum_{i,j=1}^{N-1}\Lambda_{ij}^{(N)}\nabla c_i^{(N)}\otimes\nabla c_j^{(N)}
=\sum_{i,j=1}^{N-2}\Lambda_{ij}^{(N-1)}\nabla c_i^{(N-1)}\otimes\nabla c_j^{(N-1)}.
$
Therefore relation
 \eqref{equ:surften_consist} holds.

We next show that the  relation
\eqref{equ:J_consist}  holds
provided that the conditions \eqref{equ:condition_H1}--\eqref{equ:condition_H3}
 are satisfied by $H(\vec{\phi})$.
We use the volume fractions as
the order parameters as defined by \eqref{equ:gop_volfrac}.
Based on equations \eqref{equ:J_expr_1} and \eqref{equ:Li_def}
and noting 
$
\sum_{i=1}^{N-1}\left(1-\frac{N}{\Gamma}\tilde{\gamma}_i  \right)a_{ij}
=\tilde{\rho}_j-\tilde{\rho}_N,
$
where $a_{ij}$ are given by \eqref{equ:volfrac_aij}, we have
\begin{equation}
\begin{split}
\tilde{\mathbf{J}}^{(N)} & = -m_0\sum_{j=1}^{N-1}\left[
  \sum_{i=1}^{N-1}\left(1-\frac{N}{\Gamma^{(N)}}\tilde{\gamma}_i^{(N)} \right)
  a_{ij}^{(N)}
\right]
\nabla\left(
-\nabla^2 c_j^{(N)} + L_j^{(N)}
\right) \\
& =-m_0\sum_{j=1}^{N-1}\left(\tilde{\rho}_j^{(N)}-\tilde{\rho}_N^{(N)}  \right)
\nabla\left(
-\nabla^2 c_j^{(N)} + L_j^{(N)}
\right).
\end{split}
\label{equ:J_expr_2}
\end{equation}
Let $c_k^{(N)}\equiv 0$ for some $k$ ($1\leqslant k\leqslant N$),
and assume that the conditions \eqref{equ:condition_H1}--\eqref{equ:condition_H3}
are satisfied by $H(\vec{\phi})$.
We distinguish two cases: 
(i) $1\leqslant k\leqslant N-1$, and
(ii) $k=N$.

In the first case,
we have the correspondence relations given by \eqref{equ:relation_B_1},
 \eqref{equ:relation_cN_cN1}
and \eqref{equ:condition_H2}--\eqref{equ:condition_H3}.
Therefore,
\begin{equation}
\begin{split}
\tilde{\mathbf{J}}^{(N)} & = 
-m_0\left( \sum_{j=1}^{k-1} +\sum_{j=k+1}^{N-1} \right)\left(\tilde{\rho}_j^{(N)}-\tilde{\rho}_N^{(N)}  \right)
\nabla\left(
-\nabla^2 c_j^{(N)} + L_j^{(N)}
\right) \\
&=
-m_0\sum_{j=1}^{k-1}\left(\tilde{\rho}_j^{(N)}-\tilde{\rho}_N^{(N)}  \right)
\nabla\left(
-\nabla^2 c_j^{(N)} + L_j^{(N)}
\right)
-m_0\sum_{j=k}^{N-2}\left(\tilde{\rho}_{j+1}^{(N)}-\tilde{\rho}_N^{(N)}  \right)
\nabla\left(
-\nabla^2 c_{j+1}^{(N)} + L_{j+1}^{(N)}
\right) \\
&=
-m_0\left( \sum_{j=1}^{k-1} +\sum_{j=k}^{N-2} \right)\left(\tilde{\rho}_j^{(N-1)}-\tilde{\rho}_{N-1}^{(N-1)}  \right)
\nabla\left(
-\nabla^2 c_j^{(N-1)} + L_j^{(N-1)}
\right) \\
&= 
\tilde{\mathbf{J}}^{(N-1)},
\end{split}
\end{equation}
where we have used \eqref{equ:condition_H1}.

In the second case ($k=N$) we have 
the relations given by \eqref{equ:relation_cN_1}--\eqref{equ:relation_cN_2}
and
\begin{equation}
\tilde{\rho}_i^{(N-1)} = \tilde{\rho}_i^{(N)}, \quad
1\leqslant i\leqslant N-1.
\end{equation}
Combining \eqref{equ:condition_H1} and \eqref{equ:Li_def} leads to
\begin{equation}
0 = \nabla L_N^{(N)} 
= -\sum_{i=1}^{N-2} \nabla L_i^{(N)} - \nabla L_{N-1}^{(N)},
\quad
\nabla L_{N-1}^{(N)} = -\sum_{i=1}^{N-2}\nabla L_i^{(N)}.
\label{equ:relation_Li_1}
\end{equation}
Therefore,
\begin{equation}
\begin{split}
\tilde{\mathbf{J}}^{(N)} & =
-m_0\sum_{j=1}^{N-2}\left(\tilde{\rho}_j^{(N)}-\tilde{\rho}_N^{(N)}  \right)
\nabla\left(
-\nabla^2 c_j^{(N)} + L_j^{(N)}
\right)
-m_0\left(\tilde{\rho}_{N-1}^{(N)}-\tilde{\rho}_N^{(N)}  \right)
\nabla\left(
-\nabla^2 c_{N-1}^{(N)} + L_{N-1}^{(N)}
\right) \\
& =
-m_0\sum_{j=1}^{N-2}\left(\tilde{\rho}_j^{(N)}-\tilde{\rho}_N^{(N)}  \right)
\nabla\left(
-\nabla^2 c_j^{(N)} + L_j^{(N)}
\right)
+m_0\sum_{j=1}^{N-2}\left(\tilde{\rho}_{N-1}^{(N)}-\tilde{\rho}_N^{(N)}  \right)
\nabla\left(
-\nabla^2 c_j^{(N)} + L_j^{(N)}
\right) \\
& =
-m_0\sum_{j=1}^{N-2}\left(\tilde{\rho}_j^{(N)}-\tilde{\rho}_{N-1}^{(N)}  \right)
\nabla\left(
-\nabla^2 c_j^{(N)} + L_j^{(N)}
\right) \\
& =
-m_0\sum_{j=1}^{N-2}\left(\tilde{\rho}_j^{(N-1)}-\tilde{\rho}_{N-1}^{(N-1)}  \right)
\nabla\left(
-\nabla^2 c_j^{(N-1)} + L_j^{(N-1)}
\right) \\
& = 
\tilde{\mathbf{J}}^{(N-1)},
\end{split}
\end{equation}
where we have used \eqref{equ:condition_H1}--\eqref{equ:condition_H2},
\eqref{equ:relation_cN_1}--\eqref{equ:relation_cN_2},
and \eqref{equ:relation_Li_1}.

\section*{Appendix C. Proof of Theorem \ref{thm:thm_3}}

Suppose fluid $k$ and fluid $l$ ($1\leqslant k<l\leqslant N$) are
the only fluid phases present in the N-phase system. Then
the system is characterized by the conditions listed in \eqref{equ:nphase_spec_config}.

%Let us next show that $H(\vec{\phi})$ given by \eqref{equ:potential_energy}
%indeed satisfies \eqref{equ:condition_2p_1}--\eqref{equ:condition_2p_4}
%if only two fluids $k$ and $l$ ($1\leqslant k<l \leqslant N$)
%are present in the system, i.e. the condition given
%by \eqref{equ:nphase_spec_config} holds.
For the two-phase system consisting of fluids
$k$ and $l$, we have thses relations,
\begin{equation}
\left\{
\begin{split}
&
c_1^{(2)} = c_k, \quad
c_2^{(2)} = c_l, \quad
\tilde{\rho}_1^{(2)} = \tilde{\rho}_k, \quad
\tilde{\rho}_2^{(2)} = \tilde{\rho}_l, \quad
\sigma_{12}^{(2)} = \sigma_{kl}, \quad \\
&
\Lambda_{11}^{(2)} = \frac{6}{\sqrt{2}}\eta\sigma_{12}^{(2)}, \quad
\Theta_{11}^{(2)} = \frac{1}{\Lambda_{11}^{(2)}}
  = \frac{\sqrt{2}}{6\eta} \frac{1}{\sigma_{12}^{(2)}}, \\
&
H^{(2)} = \beta\sigma_{12}^{(2)} f(c_1^{(2)}),
\end{split}
\right.
\end{equation}
where $\beta$ is given by \eqref{equ:beta_expr},
the superscript in $(\cdot)^{(2)}$ refers to the variable for the corresponding two-phase system, 
and we have used \eqref{equ:potential_energy},
\eqref{equ:lambda_ij_volfrac} and \eqref{equ:theta_ij_volfrac}.
As a result,
\begin{equation}
\left\{
\begin{split}
&
L_1^{(2)} = \Theta_{11}^{(2)}\frac{\partial H^{(2)}}{\partial c_1^{(2)}}
= \frac{\sqrt{2}}{6\eta} \beta f^\prime(c_1^{(2)}) 
= \frac{1}{\eta^2} f^\prime(c_k) \\
&
L_2^{(2)} = -L_1^{(2)} = -\frac{1}{\eta^2} f^\prime(c_k),
\end{split}
\right.
\label{equ:Li_2phase}
\end{equation}
where we have used \eqref{equ:Li_def}.

For the N-phase system characterized by \eqref{equ:nphase_spec_config},
we first note several simple facts that would be
useful for the following discussions
(where $f(c)$ is given in \eqref{equ:potential_energy}):
\begin{subequations}
\begin{align}
&
f(c) = f(1-c), \quad
f^\prime (c) = -f^\prime (1-c), \quad
f(0) = f(1) = 0, \quad
f^\prime (0) = f^\prime (1) =0; 
\label{equ:fact_1}
\\
&
c_k+c_l = 1, \quad
f(c_k) = f(c_l), \quad
f^\prime(c_k) = -f^\prime(c_l), \quad
f(c_k+c_l) = f^\prime (c_k+c_l) = 0; 
\label{equ:fact_2}
\\
&
f(c_i) = f^\prime (c_i) = 0, \ \text{if} \ i\neq k \ 
\text{and} \ i\neq l, \ \text{for} \ 1\leqslant i\leqslant N.
\label{equ:fact_3}
\end{align}
\end{subequations}
In light of \eqref{equ:potential_energy} we get
\begin{equation}
\frac{2}{\beta}\frac{\partial H}{\partial c_i}
= \sigma_i f^\prime (c_i) - \sigma_N f^\prime(c_N)
- \sum_{j=1}^N \sigma_{ij} f^\prime(c_i+c_j)
+ \sum_{j=1}^N \sigma_{Nj}f^\prime(c_N+c_j),
\quad 1\leqslant i\leqslant N-1,
\end{equation}
where 
\begin{equation}
\sigma_i = \sum_{j=1}^N \sigma_{ij}, \quad 
1\leqslant i\leqslant N.
\end{equation}
Note that when computing $\frac{\partial H}{\partial c_i}$
($1\leqslant i\leqslant N-1$) one must treat $c_N$
as a variable dependent on $c_i$ ($1\leqslant i\leqslant N-1$), 
i.e.~$
c_N = 1-\sum_{i=1}^{N-1} c_i.
$
Note also that we have not included the superscript $(N)$ for brevity in
the above expressions, and in subsequent discussions we will also drop
this superscript if no confusion arises.
Therefore,
\begin{equation}
\begin{split}
\frac{2}{\beta}L_i^{(N)} & = 
\frac{2}{\beta}\sum_{j=1}^{N-1}\Theta_{ij}\frac{\partial H}{\partial c_j} \\
& = \sum_{j=1}^{N-1} \Theta_{ij}\sigma_j f^\prime(c_j)
- \Theta_i \sigma_N f^\prime(c_N) 
- \sum_{j=1}^{N-1}\sum_{s=1}^N \Theta_{ij}\sigma_{js}f^\prime(c_j+c_s)
+ \Theta_i\sum_{j=1}^{N}\sigma_{Nj} f^\prime(c_N+c_j), \\
&
\quad 1\leqslant i\leqslant N-1,
\end{split}
\label{equ:Li_trans_1}
\end{equation}
where 
\begin{equation}
\Theta_i = \sum_{j=1}^{N-1}\Theta_{ij}, 
\quad 1\leqslant i\leqslant N-1.
\end{equation}

We distinguish two cases:
(i) $1\leqslant k<l\leqslant N-1 $, and
(ii) $1\leqslant k<l=N$.
In the first case, 
\begin{equation}
\sum_{j=1}^{N-1}\sum_{s=1}^N \Theta_{ij}\sigma_{js} f^\prime(c_j+c_s)
= \Theta_{ik}\sigma_k f^\prime(c_k)
+ \Theta_{il}\sigma_l f^\prime(c_l)
+ \sum_{j=1}^{N-1}\Theta_{ij}\sigma_{jk} f^\prime(c_k)
+ \sum_{j=1}^{N-1}\Theta_{ij}\sigma_{jl} f^\prime(c_l),
\end{equation}
where we have used 
\eqref{equ:nphase_spec_config}, 
and \eqref{equ:fact_2}--\eqref{equ:fact_3}.
Equation \eqref{equ:Li_trans_1} is then transformed into
\begin{equation}
\frac{2}{\beta}L_i^{(N)}
= \Theta_i\sigma_{Nk} f^\prime(c_k)
+ \Theta_i\sigma_{Nl} f^\prime(c_l)
- \sum_{j=1}^{N-1}\Theta_{ij}\sigma_{jk}f^\prime(c_k)
- \sum_{j=1}^{N-1}\Theta_{ij}\sigma_{jl} f^\prime(c_l),
\quad 1\leqslant i\leqslant N-1,
\label{equ:Li_trans_2}
\end{equation}
where we have used \eqref{equ:fact_2} and \eqref{equ:fact_3}.
In light of \eqref{equ:theta_ij_volfrac}, we have the relations
\begin{equation}
\left\{
\begin{split}
&
\delta_{ij} = \sum_{s=1}^{N-1}\Theta_{is}\Lambda_{sj}
= \frac{3}{\sqrt{2}}\eta\sum_{s=1}^{N-1} \Theta_{is}
\left(
  \sigma_{sN} + \sigma_{jN} - \sigma_{sj}
\right), \\
&
\sum_{s=1}^{N-1}\Theta_{is}\sigma_{sj} =
\Theta_i\sigma_{jN} + \sum_{s=1}^{N-1}\Theta_{is}\sigma_{sN}
-\frac{\sqrt{2}}{3\eta}\delta_{ij},
\quad 1\leqslant i,j\leqslant N-1,
\end{split}
\right.
\label{equ:theta_lambda_relation}
\end{equation}
where we have used \eqref{equ:lambda_ij_volfrac}
and $\delta_{ij}$ is the Kronecker delta.
Therefore, equation \eqref{equ:Li_trans_2}
is transformed into
\begin{equation}
\frac{2}{\beta}L_i^{(N)} = 
\frac{\sqrt{2}}{3\eta}\delta_{ik} f^\prime(c_k)
+ \frac{\sqrt{2}}{3\eta}\delta_{il} f^\prime(c_l)
= \frac{\sqrt{2}}{3\eta}(\delta_{ik} - \delta_{il}) f^\prime(c_k),
\quad 1\leqslant i\leqslant N-1.
\end{equation}
As a result,
\begin{equation}
\left\{
\begin{split}
&
L_i^{(N)} = \frac{1}{\eta^2}(\delta_{ik} - \delta_{il}) f^\prime(c_k),
\quad \quad 1\leqslant i\leqslant N-1, \\
&
L_N^{(N)} = -\sum_{i=1}^{N-1}L_i^{(N)} = -L_k^{(N)}-L_l^{(N)} 
= -\frac{1}{\eta^2}f^\prime(c_k) +\frac{1}{\eta^2}f^\prime(c_k)  = 0,
\end{split}
\right.
\label{equ:Li_nphase_1}
\end{equation}
where we have used \eqref{equ:beta_expr} and \eqref{equ:Li_def}.
Comparing equations \eqref{equ:Li_nphase_1} and \eqref{equ:Li_2phase},
we conclude that the relations 
\eqref{equ:condition_2p_1}--\eqref{equ:condition_2p_3} hold
for this case.

In the second case ($1\leqslant k<l=N$),
\begin{equation}
\sum_{j=1}^{N-1}\sum_{s=1}^N \Theta_{ij}\sigma_{js}
f^\prime(c_j+c_s) = 
\Theta_{ik}\sigma_k f^\prime(c_k)
+\sum_{j=1}^{N-1}\Theta_{ij}\sigma_{jk} f^\prime(c_k)
+ \sum_{j=1}^{N-1}\Theta_{ij}\sigma_{jN} f^\prime(c_N),
\end{equation}
where we have used \eqref{equ:fact_2}--\eqref{equ:fact_3},
\eqref{equ:nphase_spec_config} and
\eqref{equ:surften_property}.
Therefore, equation \eqref{equ:Li_trans_1} is transformed into
\begin{equation}
\frac{2}{\beta}L_i^{(N)} = 
-\sum_{j=1}^{N-1}\Theta_{ij}\sigma_{jk}f^\prime(c_k)
-\sum_{j=1}^{N-1}\Theta_{ij}\sigma_{jN}f^\prime(c_N)
-\Theta_i\sigma_{Nk} f^\prime(c_N),
\quad 1\leqslant i\leqslant N-1,
\label{equ:Li_trans_A1}
\end{equation}
where we have used \eqref{equ:fact_2} and \eqref{equ:fact_3}.
In light of \eqref{equ:theta_lambda_relation}, we have
\begin{equation}
\sum_{j=1}^{N-1}\Theta_{ij}\sigma_{jk}
=\Theta_i\sigma_{kN}
+ \sum_{j=1}^{N-1}\Theta_{ij}\sigma_{jN}
- \frac{\sqrt{2}}{3\eta}\delta_{ik}.
\end{equation}
Substitute this expression into \eqref{equ:Li_trans_A1},
and we get
\begin{equation}
\left\{
\begin{split}
&
L_i^{(N)} = 
\frac{\beta}{2}\frac{\sqrt{2}}{3\eta}\delta_{ik}f^\prime(c_k)
= \frac{1}{\eta^2}\delta_{ik}f^\prime(c_k), 
\quad 1\leqslant i\leqslant N-1,
\\
& 
L_N^{(N)} = -\sum_{i=1}^{N-1}L_i^{(N)} = -L_k^{(N)}
= -\frac{1}{\eta^2}f^\prime(c_k),
\end{split}
\right.
\label{equ:Li_nphase_2}
\end{equation}
where we have used \eqref{equ:fact_2}, \eqref{equ:beta_expr}
and \eqref{equ:Li_def}.
Comparing the equations \eqref{equ:Li_nphase_2}
and \eqref{equ:Li_2phase}, we conclude that
the relations \eqref{equ:condition_2p_1}--\eqref{equ:condition_2p_3}
hold for the second case.

From the above discussions we note the following relation
\begin{equation}
L_k^{(N)} + L_l^{(N)} = 0
\label{equ:relation_L_1}
\end{equation}
for the N-phase system characterized by \eqref{equ:nphase_spec_config}.

Finaly let us consider the relation \eqref{equ:condition_2p_4}. We again distinguish two cases:
(i) $1\leqslant k<l\leqslant N-1$, (ii) $1\leqslant k<l=N$.
In the first case,
according to \eqref{equ:J_expr_2},
\begin{equation}
\begin{split}
\tilde{\mathbf{J}}^{(N)} & = -m_0\sum_{j=1}^{N-1}\left(\tilde{\rho}_j-\tilde{\rho}_N  \right)
\nabla\left(
-\nabla^2 c_j + L_j^{(N)} 
\right) \\
&= -m_0\left(\tilde{\rho}_k-\tilde{\rho}_N  \right)
\nabla\left(
-\nabla^2 c_k + L_k^{(N)} 
\right)
-m_0\left(\tilde{\rho}_l-\tilde{\rho}_N  \right)
\nabla\left(
-\nabla^2 c_l + L_l^{(N)} 
\right) \\
&= -m_0\tilde{\rho}_k \nabla\left(
-\nabla^2 c_k + L_k^{(N)} 
\right)
-m_0\tilde{\rho}_l \nabla\left(
-\nabla^2 c_l + L_l^{(N)} 
\right) \\
& = -m_0\left(\tilde{\rho}_k-\tilde{\rho}_l  \right)
\nabla\left(
-\nabla^2 c_k + L_k^{(N)} 
\right) \\
&= -m_0\left(\tilde{\rho}_1^{(2)}-\tilde{\rho}_2^{(2)}  \right)
\nabla\left(
-\nabla^2 c_1^{(2)} + L_1^{(2)} 
\right) \\
&= \tilde{\mathbf{J}}^{(2)},
\end{split}
\end{equation}
where we have used \eqref{equ:condition_2p_1}--\eqref{equ:condition_2p_3} and \eqref{equ:relation_L_1}.
In the second case,
\begin{equation}
\begin{split}
\tilde{\mathbf{J}}^{(N)} & = -m_0\sum_{j=1}^{N-1}\left(\tilde{\rho}_j-\tilde{\rho}_N  \right)
\nabla\left(
-\nabla^2 c_j + L_j^{(N)} 
\right) \\
&= -m_0\left(\tilde{\rho}_k-\tilde{\rho}_N  \right)
\nabla\left(
-\nabla^2 c_k + L_k^{(N)} 
\right)
 \\
& = -m_0\left(\tilde{\rho}_k-\tilde{\rho}_l  \right)
\nabla\left(
-\nabla^2 c_k + L_k^{(N)} 
\right) \\
&= -m_0\left(\tilde{\rho}_1^{(2)}-\tilde{\rho}_2^{(2)}  \right)
\nabla\left(
-\nabla^2 c_1^{(2)} + L_1^{(2)} 
\right) \\
&= \tilde{\mathbf{J}}^{(2)},
\end{split}
\end{equation}
where we have used \eqref{equ:condition_2p_1}--\eqref{equ:condition_2p_3}.

\section*{Appendix D. Algorithm for N-Phase Momentum Equations}

In this Appendix we provide a summary of the algorithm
we developed in \cite{Dong2015}
for the N-phase momentum equations, 
\eqref{equ:nse_final} and \eqref{equ:continuity},
together with the velocity boundary condition,
\eqref{equ:dbc}.
We assume that the phase field variables $\phi_i^{n+1}$ 
($1\leqslant i\leqslant N-1$) are already computed using
the algorithm from Section \ref{sec:algorithm}, and our goal here
is to compute the velocity and pressure from
\eqref{equ:nse_final} and \eqref{equ:continuity}.

Given ($\mathbf{u}^n$, $P^n$, $\phi_i^{n+1}$),
we solve \eqref{equ:nse_final} and \eqref{equ:continuity}
by successively computing $P^{n+1}$ and $\mathbf{u}^{n+1}$
in a de-coupled fashion as follows: \\[0.1in]
\noindent\underline{For $P^{n+1}$:}
\begin{subequations}
\begin{equation}
\begin{split}
\frac{\gamma_0\tilde{\mathbf{u}}^{n+1}-\hat{\mathbf{u}}}{\Delta t}
+ \mathbf{u}^{*,n+1}\cdot\nabla\mathbf{u}^{*,n+1}
& + \frac{1}{\rho^{n+1}}\tilde{\mathbf{J}}^{n+1}\cdot\nabla\mathbf{u}^{*,n+1}
+ \frac{1}{\rho_0}\nabla P^{n+1}
= 
\left(\frac{1}{\rho_0}-\frac{1}{\rho^{n+1}}  \right)\nabla P^{*,n+1} \\
& - \frac{\mu^{n+1}}{\rho^{n+1}}\nabla\times\nabla\times\mathbf{u}^{*,n+1}
+ \frac{1}{\rho^{n+1}}\nabla\mu^{n+1}\cdot\mathbf{D}(\mathbf{u}^{*,n+1}) \\
& - \frac{1}{\rho^{n+1}}\sum_{i,j=1}^{N-1}\lambda_{ij}\nabla^2\phi_j^{n+1}\nabla\phi_i^{n+1}
+ \frac{1}{\rho^{n+1}}\mathbf{f}^{n+1},
\end{split}
\label{equ:pressure_1}
\end{equation}
\begin{equation}
\nabla\cdot\tilde{\mathbf{u}}^{n+1} = 0,
\label{equ:pressure_2}
\end{equation}
\begin{equation}
\left.\mathbf{n}\cdot\tilde{\mathbf{u}}^{n+1}\right|_{\partial\Omega}
= \mathbf{n}\cdot\mathbf{w}^{n+1}.
\label{equ:pressure_3}
\end{equation}
\end{subequations}
\\
\noindent\underline{For $\mathbf{u}^{n+1}$:}
\begin{subequations}
\begin{equation}
\frac{\gamma_0\mathbf{u}^{n+1}-\gamma_0\tilde{\mathbf{u}}^{n+1}}{\Delta t}
 - \nu_0 \nabla^2\mathbf{u}^{n+1}
= \nu_0 \nabla\times\nabla\times\mathbf{u}^{*,n+1},
\label{equ:velocity_1}
\end{equation}
\begin{equation}
\left.\mathbf{u}^{n+1}  \right|_{\partial\Omega} = \mathbf{w}^{n+1}.
\label{equ:velocity_2}
\end{equation}
\end{subequations}

% what do all the symbols mean?

In the above equations all the symbols follow
the notation outlined in Section \ref{sec:algorithm}.
$\mathbf{u}^{*,n+1}$ and $P^{*,n+1}$
are defined by \eqref{equ:var_star_def}.
$\hat{\mathbf{u}}$ and $\gamma_0$ are defined
by \eqref{equ:var_hat_def}.
$\rho^{n+1}$ and $\mu^{n+1}$ are
given by \eqref{equ:rho_mu},
and also \eqref{equ:rho_mu_clamp} in case of large density
ratios among the N fluids.
$\tilde{\mathbf{J}}^{n+1}$ is 
given by \eqref{equ:J_expr_1}.
$\mathbf{f}^{n+1}$ is the external body force
evaluated at time step ($n+1$).
$\mathbf{n}$ is the outward-pointing unit vector
normal to $\partial\Omega$.
$\tilde{\mathbf{u}}^{n+1}$ is an auxiliary
velocity that approximates $\mathbf{u}^{n+1}$.
$\rho_0$ is a chosen constant that must satisfy 
the condition % \cite{Dong2014}
\begin{equation}
0 < \rho_0 \leqslant \min(\tilde{\rho}_1,\tilde{\rho}_2,\dots,\tilde{\rho}_N).
\label{equ:rho_0_condition}
\end{equation}
$\nu_0$ in \eqref{equ:velocity_1} is a chosen positive
constant that is sufficiently large.
%A conservative
%condition for $\nu_0$ is given in \cite{Dong2014}.
In the current paper we  employ an $\nu_0$ value
with the following,
\begin{equation}
\nu_0 \geqslant \max\left(
\frac{\tilde{\mu}_1}{\tilde{\rho}_1},
\frac{\tilde{\mu}_2}{\tilde{\rho}_2},
\cdots,
\frac{\tilde{\mu}_N}{\tilde{\rho}_N},
\right).
\label{equ:nu_0_condition}
\end{equation}

% some comments on algorithm, compare with Dong(2014)

The above algorithm employs a velocity
correction-type idea \cite{DongS2010,DongKC2014,Dong2015clesobc}  to de-couple
the computations for the pressure and the velocity.
It can be noted that the variable density $\rho$
and the variable dynamic viscosity $\mu$ have been
treated with a reformulation of
the pressure term $\frac{1}{\rho}\nabla P$
and a reformulation of the 
viscous term $\frac{\mu}{\rho}\nabla^2\mathbf{u}$,
%similar to \cite{Dong2014}, 
so that the linear
algebraic systems resulting from the discretization
involve only {\em constant} and {\em time-independent} 
coefficient matrices. 
The ideas for the reformulations stem from the original developments for
two-phase flows \cite{DongS2012,Dong2012,Dong2014obc}.

% weak forms, implementation issues

Le us next derive the weak forms for the pressure and
the velocity in order to facilitate the implementation
using $C^0$ spectral elements.
% (or finite elements).
Let $q\in H^1(\Omega)$ denote the test function, and let
\begin{multline}
\mathbf{G}^{n+1} = 
\frac{1}{\rho^{n+1}}\mathbf{f}^{n+1}
- \left(
     \mathbf{u}^{*,n+1} 
     + \frac{1}{\rho^{n+1}} \tilde{\mathbf{J}}^{n+1}
  \right)\cdot\nabla\mathbf{u}^{*,n+1}
+ \frac{\hat{\mathbf{u}}}{\Delta t}
+ \left(\frac{1}{\rho_0} - \frac{1}{\rho^{n+1}}  \right)\nabla P^{*,n+1} \\
+ \frac{1}{\rho^{n+1}}\nabla\mu^{n+1}\cdot\mathbf{D}(\mathbf{u}^{*,n+1})
- \frac{1}{\rho^{n+1}}\sum_{i,j=1}^{N-1}\lambda_{ij}\nabla^2\phi_j^{n+1}\nabla\phi_i^{n+1}
+ \nabla\left( \frac{\mu^{n+1}}{\rho^{n+1}} \right) \times \bm{\omega}^{*,n+1},
\label{equ:G_expr}
\end{multline}
where $\bm{\omega} = \nabla\times \mathbf{u}$ is
the vorticity.
Take the $L^2$ inner product between equation \eqref{equ:pressure_1}
and $\nabla q$, and we get the weak form about $P^{n+1}$,
\begin{equation}
\int_{\Omega} \nabla P^{n+1} \cdot\nabla q
= \rho_0 \int_{\Omega} \mathbf{G}^{n+1}\cdot\nabla q
- \rho_0 \int_{\partial\Omega} \frac{\mu^{n+1}}{\rho^{n+1}} \mathbf{n}\times\bm{\omega}^{*,n+1}\cdot\nabla q
- \frac{\gamma_0\rho_0}{\Delta t}\int_{\partial\Omega}\mathbf{n}\cdot\mathbf{w}^{n+1} q,
\ \
\forall q\in H^1(\Omega)
\label{equ:p_weakform}
\end{equation}
where we have used integration by part,
equations \eqref{equ:pressure_2} and \eqref{equ:pressure_3},
the divergence theorem,
and the identity
$ %\begin{equation*}
\frac{\mu}{\rho}\nabla\times\bm{\omega}\cdot\nabla q
= \nabla\cdot\left(
  \frac{\mu}{\rho} \bm{\omega}\times\nabla q
  \right)
- \nabla\left( \frac{\mu}{\rho} \right)\times\bm{\omega}\cdot\nabla q.
$ %\end{equation*} 

Adding together the equations \eqref{equ:pressure_1} and
\eqref{equ:velocity_1}, we get
\begin{equation}
\frac{\gamma_0}{\Delta t}\mathbf{u}^{n+1} - \nu_0\nabla^2\mathbf{u}^{n+1}
= \mathbf{G}^{n+1} 
- \nabla\left( \frac{\mu^{n+1}}{\rho^{n+1}} \right) \times \bm{\omega}^{*,n+1}
- \frac{1}{\rho_0}\nabla P^{n+1}
- \left( \frac{\mu^{n+1}}{\rho^{n+1}} - \nu_0 \right) \nabla\times\bm{\omega}^{*,n+1}
\label{equ:velocity_1_reform}
\end{equation}
Let 
$
H^1_0(\Omega) = \left\{ \
v \in H^{\Omega} \ : \
v|_{\partial\Omega} = 0
\ \right\},
$
and $\varphi \in H_0^1(\Omega)$ denote
the test function.
Taking the $L^2$ inner product between
equation \eqref{equ:velocity_1_reform} and $\varphi$,
one can get the weak form about $\mathbf{u}^{n+1}$,
\begin{multline}
\int_{\Omega}\nabla\varphi\cdot\nabla\mathbf{u}^{n+1}
+ \frac{\gamma_0}{\nu_0\Delta t}\int_{\Omega}\varphi\mathbf{u}^{n+1}
= \frac{1}{\nu_0}\int_{\Omega}\left(
    \mathbf{G}^{n+1} - \frac{1}{\rho_0}\nabla P^{n+1}
  \right) \varphi \\
- \frac{1}{\nu_0}\int_{\Omega} \left(
    \frac{\mu^{n+1}}{\rho^{n+1}} - \nu_0
  \right) 
  \bm{\omega}^{*,n+1} \times \nabla\varphi,
\qquad
\forall \varphi \in H_0^1(\Omega),
\label{equ:vel_weakform}
\end{multline}
where we have used integration by part,
the divergence theorem, the identity ($\chi$ denoting
a scalar function)
\begin{equation*}
\int_{\Omega} \chi\nabla\times\bm{\omega}\varphi
= \int_{\partial\Omega} \chi\mathbf{n}\times\bm{\omega}\varphi
- \int_{\Omega} \nabla\chi\times\bm{\omega}\varphi
+ \int_{\Omega} \chi \bm{\omega}\times\nabla\varphi,
\end{equation*}
and the fact that the surface integrals of type 
$
\int_{\partial\Omega} \chi \varphi
$
vanish because $\varphi\in H^1_0(\Omega)$.

The weak forms for the pressure and the
velocity, \eqref{equ:p_weakform} and
\eqref{equ:vel_weakform}, can be
discretized in space using $C^0$ spectral 
elements 
%or $C^0$ finite elements 
in a straightforward
fashion. Note that the terms $\nabla^2\phi_i^{n+1}$ 
($1\leqslant i\leqslant N-1$) involved in
the $\mathbf{G}^{n+1}$ expression \eqref{equ:G_expr}
and in the $\tilde{\mathbf{J}}^{n+1}$ expression
(see \eqref{equ:J_expr_2} and \eqref{equ:J_expr_1})
must be computed based on equation \eqref{equ:lap_phi}.

Solving the N-phase momentum equations
\eqref{equ:nse_final} and \eqref{equ:continuity}
amounts to the following two successive operations. First,
solve equation \eqref{equ:p_weakform} for
pressure $P^{n+1}$. Then, solve
equation \eqref{equ:vel_weakform}, together with
the Dirichlet condition \eqref{equ:velocity_2} on $\partial\Omega$,
for $\mathbf{u}^{n+1}$.

% what else to discuss here?
% algorithm characteristics?

\bibliographystyle{plain}
\bibliography{nphase,obc,mypub,nse,sem,contact_line,interface,multiphase}

\end{document}